\documentclass[traditabstract]{aa} 

\usepackage{graphicx}
\usepackage{txfonts}
\usepackage{natbib}
\usepackage[colorlinks=true,urlcolor=blue,citecolor=dblue,linkcolor=dblue]
           {hyperref}

  \def\Td{T_{\hspace*{-0.2ex}\rm dust}}
  \def\Tg{T_{\hspace*{-0.2ex}\rm gas}}
  \def\TPAH{T_{\hspace*{-0.2ex}\rm PAH}}

  \def\amin{{a_{\rm min}}}
  \def\amax{{a_{\rm max}}}
  \def\apow{{a_{\rm pow}}}
  \def\kabs{{\kappa_\nu^{\rm abs}}}
  \def\ksca{{\kappa_\nu^{\rm sca}}}
  \def\HH{{\rm\langle H\rangle}}
  \def\nH{n_{\HH}}
  \def\etal{${\rm \hspace*{0.8ex}et\hspace*{0.7ex}al.\hspace*{0.7ex}}$}
  
  \def\ie{i.\,e.\ }
  \def\eg{e.\,g.\ }

\newenvironment{my_itemize}%
  {\vspace*{-1mm}\begin{itemize}%
    \setlength{\itemsep}{2pt}%
    \setlength{\parskip}{0pt}%
    \setlength{\topsep}{0pt}%
    \setlength{\parsep}{0pt}%
    \setlength{\partopsep}{-30pt}}%
  {\end{itemize}\vspace*{-1mm}}

\begin{document}
\definecolor{GreenYellow}  {cmyk}{0.15,0,0.69,0}
\definecolor{Yellow}{cmyk}{0,0,1,0}
\definecolor{Goldenrod}{cmyk}{0,0.10,0.84,0}
\definecolor{Dandelion}{cmyk}{0,0.29,0.84,0}
\definecolor{Apricot}  {cmyk}{0,0.32,0.52,0}
\definecolor{Peach}    {cmyk}{0,0.50,0.70,0}
\definecolor{Melon}    {cmyk}{0,0.46,0.50,0}
\definecolor{YellowOrange}  {cmyk}{0,0.42,1,0}
\definecolor{Orange}   {cmyk}{0,0.61,0.87,0}
\definecolor{BurntOrange}   {cmyk}{0,0.51,1,0}
\definecolor{Bittersweet}   {cmyk}{0,0.75,1,0.24}
\definecolor{RedOrange}{cmyk}{0,0.77,0.87,0}
\definecolor{Mahogany} {cmyk}{0,0.85,0.87,0.35}
\definecolor{Maroon}   {cmyk}{0,0.87,0.68,0.32}
\definecolor{BrickRed} {cmyk}{0,0.89,0.94,0.28}
\definecolor{Red} {cmyk}{0,1,1,0}
\definecolor{OrangeRed}{cmyk}{0,1,0.50,0}
\definecolor{RubineRed}{cmyk}{0,1,0.13,0}
\definecolor{WildStrawberry}{cmyk}{0,0.96,0.39,0}
\definecolor{Salmon}   {cmyk}{0,0.53,0.38,0}
\definecolor{CarnationPink} {cmyk}{0,0.63,0,0}
\definecolor{Magenta}  {cmyk}{0,1,0,0}
\definecolor{VioletRed}{cmyk}{0,0.81,0,0}
\definecolor{Rhodamine}{cmyk}{0,0.82,0,0}
\definecolor{Mulberry} {cmyk}{0.34,0.90,0,0.02}
\definecolor{RedViolet}{cmyk}{0.07,0.90,0,0.34}
\definecolor{Fuchsia}{cmyk}{0.47,0.91,0,0.08}
\definecolor{Lavender} {cmyk}{0,0.48,0,0}
\definecolor{Thistle}{cmyk}{0.12,0.59,0,0}
\definecolor{Orchid}{cmyk}{0.32,0.64,0,0}
\definecolor{DarkOrchid}{cmyk}{0.40,0.80,0.20,0}
\definecolor{Purple}{cmyk}{0.45,0.86,0,0}
\definecolor{Plum}{cmyk}{0.50,1,0,0}
\definecolor{Violet} {cmyk}{0.79,0.88,0,0}
\definecolor{RoyalPurple} {cmyk}{0.75,0.90,0,0}
\definecolor{BlueViolet}{cmyk}{0.86,0.91,0,0.04}
\definecolor{Periwinkle}{cmyk}{0.57,0.55,0,0}
\definecolor{CadetBlue}{cmyk}{0.62,0.57,0.23,0}
\definecolor{CornflowerBlue}{cmyk}{0.65,0.13,0,0}
\definecolor{MidnightBlue}{cmyk}{0.98,0.13,0,0.43}
\definecolor{NavyBlue} {cmyk}{0.94,0.54,0,0}
\definecolor{RoyalBlue}{cmyk}{1,0.50,0,0}
\definecolor{Blue}{cmyk}{1,1,0,0}
\definecolor{Cerulean} {cmyk}{0.94,0.11,0,0}
\definecolor{Cyan}{cmyk}{1,0,0,0}
\definecolor{ProcessBlue} {cmyk}{0.96,0,0,0}
\definecolor{SkyBlue}{cmyk}{0.62,0,0.12,0}
\definecolor{Turquoise}{cmyk}{0.85,0,0.20,0}
\definecolor{TealBlue} {cmyk}{0.86,0,0.34,0.02}
\definecolor{Aquamarine}{cmyk}{0.82,0,0.30,0}
\definecolor{BlueGreen}{cmyk}{0.85,0,0.33,0}
\definecolor{Emerald}{cmyk}{1,0,0.50,0}
\definecolor{JungleGreen} {cmyk}{0.99,0,0.52,0}
\definecolor{SeaGreen} {cmyk}{0.69,0,0.50,0}
\definecolor{Green}{cmyk}{1,0,1,0}
\definecolor{ForestGreen} {cmyk}{0.91,0,0.88,0.12}
\definecolor{PineGreen}{cmyk}{0.92,0,0.59,0.25}
\definecolor{LimeGreen}{cmyk}{0.50,0,1,0}
\definecolor{YellowGreen} {cmyk}{0.44,0,0.74,0}
\definecolor{SpringGreen} {cmyk}{0.26,0,0.76,0}
\definecolor{OliveGreen}{cmyk}{0.64,0,0.95,0.40}
\definecolor{RawSienna}{cmyk}{0,0.72,1,0.45}
\definecolor{Sepia}{cmyk}{0,0.83,1,0.70}
\definecolor{Brown}{cmyk}{0,0.81,1,0.60}
\definecolor{Tan} {cmyk}{0.14,0.42,0.56,0}
\definecolor{Gray}{cmyk}{0,0,0,0.50}
\definecolor{Black}{cmyk}{0,0,0,1}
\definecolor{White}{cmyk}{0,0,0,0}

\definecolor{StaubBraun}{cmyk}{0,0.61,0.87,0.20}
\definecolor{lyellow}{cmyk}{0,0,0.2,0}
\definecolor{lred}{cmyk}{0,0.4,0.4,0}
\definecolor{lgray}{cmyk}{0,0,0,0.15}
\definecolor{dgreen}{rgb}{0.05,0.5,0.1}
\definecolor{dred}{rgb}{0.85,0.0,0.0}
\definecolor{dblue}{rgb}{0.0,0.0,0.6}
\definecolor{brown}{rgb}{0.6,0.1,0.1}

   \title{Consistent dust and gas models for protoplanetary disks}
   \subtitle{I. Disk shape, dust settling, opacities, and PAHs}
   
   \author{P.~Woitke\inst{1}, 
           M.~Min\inst{4},
           C.~Pinte\inst{2,6},
           W.-F.~Thi\inst{8},
           I.~Kamp\inst{3},
           C.~Rab\inst{5},
           F.~Anthonioz\inst{2},
           S.~Antonellini\inst{3},
           C.~Baldovin-Saavedra\inst{5},
           A.~Carmona\inst{7,11,12},
           C.~Dominik\inst{4},
           O.~Dionatos\inst{5},
           J.~Greaves\inst{1},
           M.~G{\"u}del\inst{5},
           J.\,D.~Ilee\inst{13,1},
           A.~Liebhart\inst{5}, 
           F.~M{\'e}nard\inst{2,6},
           L.~Rigon\inst{1},
           L.\,B.\,F.\,M.~Waters\inst{4},
           G.~Aresu\inst{10},
           R.~Meijerink\inst{9},
           M.~Spaans\inst{3}
   }
   
   \titlerunning{Consistent dust and gas models for protoplanetary disks}
   \authorrunning{P.~Woitke et al.}

   \institute{ 
             SUPA, School of Physics \& Astronomy, University of St.~Andrews,
             North Haugh, St.~Andrews KY16 9SS, UK
          \and 
              UJF-Grenoble 1 / CNRS-INSU, Institut de Plan{\'e}tologie et
             d'Astrophysique (IPAG) UMR 5274, Grenoble, F-38041, France
          \and 
             Kapteyn Astronomical Institute, Postbus 800,
             9700 AV Groningen, The Netherlands
          \and 
             Astronomical Institute ``Anton Pannekoek'', University of 
             Amsterdam, PO Box 94249, 1090 GE Amsterdam, The Netherlands
          \and 
             University of Vienna, Department of Astrophysics, 
             T{\"u}rkenschanzstrasse 17, 1180 Vienna, Austria
          \and 
             UMI-FCA, CNRS/INSU France (UMI 3386), and Departamento de 
             Astronom{\'i}ca, Universidad de Chile, Santiago, Chile
          \and 
             Departamento de F\'isica Te\'orica, Universidad Autonoma
             de Madrid, Campus Cantoblanco, 28049, Madrid, Spain
          \and 
             Max Planck Institute for Extraterrestrial Physics,
             Giessenbachstrasse, 85741 Garching, Germany
          \and 
             Leiden Observatory, Leiden University, P.O. Box 9513, 
             2300 RA Leiden, The Netherlands
          \and 
             INAF, Osservatorio Astronomico di Cagliari, Via della 
             Scienza 5, Selargius (CA), 09047, Italy
          \and 
             Konkoly Observatory, Hungarian Academy of Sciences, 
             H-1121 Budapest, Konkoly Thege Mikl\'os \'ut 15-17, Hungary
          \and 
             Universit\'e de Toulouse, UPS-OMP, IRAP, 14 avenue
             E.~Belin, Toulouse F-31400, France
          \and 
             Institute of Astronomy, University of Cambridge, 
             Madingley Road, Cambridge CB3 0HA, UK 
             }

   \date{Received May\,15, 2015; accepted November\,4, 2015}

   \abstract{We propose a set of standard assumptions for the
     modelling of Class II and III protoplanetary disks, which
     includes detailed continuum radiative transfer, thermo-chemical
     modelling of gas and ice, and line radiative transfer from
     optical to cm wavelengths. The first paper of this series focuses
     on the assumptions about the shape of the disk, the dust
     opacities, dust settling, and PAHs. In particular, we propose new
     standard dust opacities for disk models, we present a simplified
     treatment of PAHs in radiative equilibrium which is sufficient to
     reproduce the PAH emission features, and we suggest using a
     simple yet physically justified treatment of dust settling.  We
     roughly adjust parameters to obtain a model that predicts
     continuum and line observations that resemble typical
     multi-wavelength continuum and line observations of Class\,II
     T\,Tauri stars. We systematically study the impact of each model
     parameter (disk mass, disk extension and shape, dust settling,
     dust size and opacity, gas/dust ratio, etc.) on all mainstream
     continuum and line observables, in particular on the SED,
     mm-slope, continuum visibilities, and emission lines including
     [OI]\,63\,$\mu$m, high-$J$ CO lines, (sub-)mm CO isotopologue
     lines, and CO fundamental ro-vibrational lines.  We find that
     evolved dust properties, i.e.\ large grains, often needed to
     fit the SED, have important consequences for disk chemistry and
     heating/cooling balance, leading to stronger near- to far-IR
     emission lines in general.  Strong dust settling and missing disk
     flaring have similar effects on continuum observations, but
     opposite effects on far-IR gas emission lines.  PAH molecules can
     efficiently shield the gas from stellar UV radiation because of
     their strong absorption and negligible scattering opacities in
     comparison to evolved dust.  The observable millimetre-slope of
     the SED can become significantly more gentle in the case of cold disk
     midplanes, which we find regularly in our T\,Tauri models. We
     propose to use line observations of robust chemical tracers of
     the gas, such as O, CO, and H$_2$, as additional constraints to
     determine a number of key properties of the disks, such as disk
     shape and mass, opacities, and the dust/gas ratio, by
     simultaneously fitting continuum and line observations.}

   \keywords{ Stars: formation --
              stars: circumstellar matter --  
              radiative transfer --
              astrochemistry --
              line: formation --
              methods: numerical}
   \maketitle

\section{Introduction}

Disk models are widely used by the community to analyse and interpret
line and continuum observations from protoplanetary disks, such as
photometric fluxes, low- and high-resolution spectroscopy, images and
visibility data, from X-ray to centimetre wavelengths. Historically,
disk models could be divided into {\sl continuum radiative transfer
  models}, such as {\sl HOCHUNK3D} \citep{Whitney2003}, {\sl MC3D}
\citep{Wolf2003}, {\sl RADMC} \citep{Dullemond2004}, {\sl TORUS}
\citep{Harries2004}, {\sl MCFOST} \citep{Pinte2006} and {\sl MCMax}
\citep{Min2009}, to explore the disk shape, dust temperature and grain
properties, and {\sl thermo-chemical models}, see
e.g. \citep{Henning2013} for a review, and Table~\ref{tab:other}.  The
thermo-chemical models usually include chemistry and UV and X-ray
physics to explore the temperature and chemical properties of the gas,
with particular emphasis on the outer disk as traced by (sub-)mm line
observations.

However, this distinction is becoming more and more obsolete, because 
new disk models try to combine all modelling components and techniques,
either by developing single, stand-alone modelling tools like {\sl
  ProDiMo} \citep{Woitke2009a} and {\sl DALI} \citep{Bruderer2014}, or
by coupling separate continuum and gas codes to achieve a similar
level of consistency.

\begin{table*}
\caption{Assumptions about disk shape, grain size, opacities, 
  dust settling and PAHs in different thermo-chemical disk models.}
\label{tab:other}
\vspace*{-2mm}
\resizebox{\textwidth}{!}{\hspace*{-1mm}
\begin{tabular}{p{3.8cm}|p{4.2cm}|c|p{2.9cm}|p{1.8cm}|p{2.1cm}|p{2.9cm}}
\hline
&&&&&&\\[-2.2ex]
reference &      
model setup \& disk shape & $\!\!$radial range$\!\!$ & 
  grain size & $\!$dust opacities$\!\!$ & dust settling & PAHs\\
&&&&&&\\[-2.2ex]
\hline
\hline
&&&&&&\\[-1.8ex]
\citet{Semenov2011},\newline see \citet{Semenov2006} &
adopted from \citet{Dalessio1998}, $\Tg\!=\!\Td$ &
$\!\!(10-700)$\,AU$\!\!$ &
uniform 0.1$\,\mu$m&
n.a. &
well-mixed &
n.a. \\
&&&&&&\\[-1.2ex]
\citet{Gorti2008}$\!\!$ & 
powerlaw $\Sigma(r)$,\newline modified CG97 & 
$\!\!(0.5-200)$\,AU$\!\!$ &
powerlaw,\newline $(0.005-50)\,\mu$m &
n.a. &
well-mixed &
reduced ISM abundance, PAHs in heating and chemistry\\
&&&&&&\\[-1.2ex]
\citet{Dutrey2011}, see\newline also \citet{Semenov2010} &
series of 1D vertical slabs, based on \citet{Hersant2009}, $\Tg\!=\!\Td$ &
$(40-300)$\,AU &
uniform 0.1$\,\mu$m&
n.a. &
well-mixed &
n.a.\\
&&&&&&\\[-1.2ex]
\citet{Walsh2014}, based\newline on \citet{Nomura2005}$\!\!\!$ &
$\Sigma(r)$ from $\alpha$-model, vertical hydrostatic equilibrium &
$(1-300)$\,AU &
MRN, details see\newline \citep{Nomura2005} &
mix of AS,\newline graphite and water ice$\!\!\!$ &
well-mixed &
n.a.\\
&&&&&&\\[-1.2ex]
\citet{Du2014}, based$\!\!\!$\newline on \citet{Bethell2011}$\!\!\!$ &
powerlaw $\Sigma(r)$ with self-similar tapered outer edge, 
parametric &
$(1-140)$\,AU &
2 powerlaws:\newline 
$C_1$:\,$(0.01-1)\,\mu$m,\newline 
$C_2$:\,$(1-100)\,\mu$m$\!\!$ &
7:3 mixture\newline of AS and\newline graphite & 
$C_1$ well-mixed,$\!\!\!$\newline $C_2$ reduced $H$ &
reduced ISM abundance, for heating\\
&&&&&&\\[-1.2ex]
\citet{Mathews2013},\newline based on \citet{Qi2011} &
powerlaw $\Sigma(r)$ with self-similar tapered outer edge, 
modified parametric &
complete disk &
2 powerlaws:\newline 
$C_1$:\,$(0.005\!-\!0.25)\,\mu$m$\hspace*{-4mm}$,\newline 
$C_2$:\,$0.005\,\mu{\rm m}-\!1$\,mm$\!\!$ &
3:2 mixture\newline of AS and\newline graphite &
$C_1$ well-mixed,$\!\!\!$\newline $C_2$ reduced $H$ &
n.a.\\
&&&&&&\\[-1.2ex]
\citet{Akimkin2013} &
viscous disk evolution, vertical hydrostatic equilibrium &
$\!\!(10-550)$\,AU$\!\!$ &
dust evolution from\newline initial MRN dist.,\newline 
$(0.003-200)\,\mu$m &
AS &
included in\newline dust evolution &
reduced ISM abundance, for heating\\
&&&&&&\\[-1.2ex]
\citet{Bruderer2013} &
powerlaw $\Sigma(r)$ with self-similar tapered outer edge, 
parametric &
complete disk &
2 powerlaws:\newline 
$C_1$:\,$(0.005\!-\!1)\,\mu$m$\hspace*{-4mm}$,\newline 
$C_2$:\,$0.005\,\mu{\rm m}-\!1$\,mm$\!\!$ &
mixture\newline of AS and\newline graphite &
$C_1$ well-mixed,$\!\!\!$\newline $C_2$ reduced $H$ &
reduced ISM abundance, in heating, chemistry and RT\\
&&&&&&\\[-1.2ex]
\citet{Woitke2009a} &
powerlaw $\Sigma(r)$, vertical hydrostatic equilibrium &
$\!\!(0.5-500)$\,AU$\!\!$ &
powerlaw $(0.1-10)\,\mu{\rm m}$ &
AS &
well-mixed &
reduced ISM abundance for heating\\
&&&&&&\\[-1.2ex]
this work\newline (more details in Sect.\,\ref{sec:Model}) &
two zones, powerlaw $\Sigma(r)$ with tapered outer edge, parametric &
complete disk &
powerlaw,\newline $0.05\,\mu{\rm m}-3\,{\rm mm}$ &
lab.~silicates\newline mixed with\newline AC,\ \ \ DHS &
\citet{Dubrulle1995}, about\newline 100 size bins$\!\!\!$ &
reduced ISM abundance, in heating, chemistry and RT
\\[-1.2ex]
&&&&&&\\
\hline
\end{tabular}}\\[1mm]
\tiny{
  {\bf CG97: } two-layer model according to \citet{Chiang1997};\ \ \ 
  {\bf parametric: } $\rho(r,z)\!\propto\!\exp(-z^2/[2H_{\rm g}(r)^2])$ with 
              prescribed gas scale height $H_{\rm g}(r)$;\ \ \ 
  {\bf modified parametric: } parametric with more slowly declining
              tail into the upper regions, additional shape parameter 
              for puffed-up inner rim;\ \ \ 
  {\bf $\mathbf{\alpha}$-model: } the surface density distribution
              $\Sigma(r)$ is
              derived from the stellar mass, a constant disk 
              mass accretion rate $\dot{M}$, and the parametrised
              kinematic viscosity $\alpha$ \citep{Shakura1973}; \ \ \
  {\bf complete disk: } from inner rim (dust sublimation
              temperature) to some large distance where the column 
              density becomes vanishingly small;\ \ \ 
  {\bf RT: } 2D continuum radiative transfer;\ \ \ 
  {\bf MRN: } powerlaw size distribution $f(a)\!\propto\!a^{-3.5}$
              between $\amin\!=\!0.005\,\mu$m and
              $\amax\!=\!0.25\,\mu$m \citep{Mathis1977};\ \ \  
  {\bf AS: }  smoothed UV astronomical silicate 
             \citep{Draine1984,Laor1993};\ \ \ 
  {\bf lab.~silicates mixed with AC: } optical properties from
              laboratory measurements of silicates and amorphous
              carbon, see Sect.~\ref{sec:dustopac} for deatils;\ \ \ 
  {\bf DHS: } distribution of hollow spheres \citep{Min2005}; \ \ \
  {\bf reduced ISM abundance: } PAH abundance lower than ISM standard
              (see Eq.\,\ref{eq:PAHabun}); \ \ \
  {\bf dust evolution: } detailed numerical simulations including
              growth, radial drift and settling according 
              to \citet{Birnstiel2010}.
 }
\end{table*}

Observational data from protoplanetary disks obtained with a single
observational technique in a limited wavelength interval can only
reveal certain information about the physical properties at particular
radii and at particular vertical depths in the disk. Therefore, in
order to derive an overall picture of protoplanetary disks, it is
essential to combine all observational data, and to make consistent
predictions for all continuum and line observables in a large range of
wavelengths on the basis of a single disk model.

However, this {\em \,holistic modelling approach\,} does not come
without a price.  The number of free parameters in such models is
large (around 20), and the computational time required to run one
model can exceed days, weeks or even months
\citep[e.g.][]{Semenov2006}.  These limitations have resulted in quite
limited parameter space being explored in such models.

Therefore, previous chemical models have not fully explored the role
of disk shape and dust opacities.  In Table\,\ref{tab:other}, we list
assumptions made in differnt thermo-chemical disk models about
the shape of the disk, the dust size distribution, opacities, dust
settling, and PAHs. The selection of models is not exhaustive in
Table\,\ref{tab:other}, for a more comprehensive overview of modelling
techniques and assumptions see Table~3 in \citep{Henning2013}.
Table\,\ref{tab:other} shows the diversity of modelling assumptions 
currently used by different disk modelling groups. These models often
focus on the outer disk, consider small dust particles,
and use different approaches for dust settling and PAHs.

All these assumptions have crucial impacts on the modelling results,
not only with regard to the predicted continuum observations, as known
from SED fitting, but also on chemical composition and line
predictions. Therefore, to compare modelling results from a
large number of protoplanetary disks, a set of consistent
{\sl\,standard modelling assumptions\,} is required.


In this paper, we explore what could be a minimum set of physical
assumptions about the star, the disk geometry, the dust and PAH
opacities, dust settling, gas \& ice chemistry, gas heating \&
cooling, and line transfer, needed to capture the most commonly
observed multi-wavelength properties of Class II and III
protoplanetary disks. We aim at a consistent, coupled modelling of
dust and gas, and we want to predict the full suite of observations
simultaneously and consistently from one model.

In Sect.\ref{sec:DIANA} we introduce our aims and basic approach, in
Sect.~\ref{sec:Model} we describe the details of our model, and in
Sect~\ref{sec:Implementation} we cross-check and verify the
implementation of our assumptions into our three main modelling tools
{\sl ProDiMo}, {\sl MCFOST}, and {\sl MCMax}. In
Sect~\ref{sec:Results}, we show first results for a simple model of a
Class II T\,Tauri star, and systematically study the impact of
all modelling parameters on the various predicted observables. We
summarise these results and conclude in Sect.~\ref{sec:Conclusions}.

In the Appendices, we collate a number of auxiliary information. We
explain our fitting routine of stellar parameters including UV and
X-ray properties (App.~\ref{app:StellarPara}), and describe our
assumptions concerning interstellar UV and IR background radiation
fields.  We compare some results obtained with the simplified
treatment of PAHs (see Sect.~\ref{sec:PAHs}) against models using the
full stochastic quantum heating method in App.~\ref{app:PAHtemp}.
Appendix~\ref{app:tdep} compares the results obtained with
time-dependent chemistry against those obtained in kinetic chemical
equilibrium.  We detail how certain observable key properties are
computed from the models in App.~\ref{app:integrated}.
Appendix~\ref{app:ThickLines} discusses the behaviour of optically
thick emission lines.  In App.~\ref{app:NumConv}, we discuss the
convergence of our results as function of the model's spatial grid resolution.

Two forthcoming papers will continue this paper series, to study the
impacts of chemical rate networks (Kamp\etal in prep.,
Paper~II) and element abundances (Rab\etal 2015, submitted,
Paper~III) on the resulting chemical abundances and emission
lines. Kamp\etal introduce a simplified, small chemical rate network
and a more exhaustive, large chemical network, henceforth called the
{\em\,small DIANA chemical standard\,} and the {\em\,large DIANA
  chemical standard}, respectively.

\section{The DIANA project}
\label{sec:DIANA}

The European FP7 project {\sl DiscAnalysis} (or {\sl DIANA}) was
initiated to bring together different aspects of dust and gas
modelling in disks, alongside multiwavelength datasets, in order to
arrive at a common set of agreed physical assumptions that can be
implemented in all modelling software, which is a precondition 
to cross-correlate modelling results for different objects.

The {\sl DIANA} goal is to combine dust continuum and gas emission
line diagnostics to infer the physical and chemical structure of Class
II and III protoplanetary disks around M-type to A-type stars from
observations, including dust, gas and ice properties. The project aims
at a uniform modelling of a statistically relevant sample of
individual disks with coherent observational data sets, from X-rays to
centimetre wavelengths.

In protoplanetary disks, various physical and chemical mechanisms are
coupled with each other in complicated, at least two-dimensional ways.
Turbulence, disk flaring, dust settling, the shape of the inner rim,
UV and X-ray irradiation, etc., lead to an
intricate interplay between gas and dust physics.  Therefore,
consistent gas and dust models are required. Each of the processes
listed above can be expected to leave specific fingerprints in form of
observable continuum and line emissions that can possibly be used for 
their identification and diagnostics. 

However, to include all relevant physical and chemical effects in a
single disk model is a challenging task, and we have decided to
include only processes which we think are the most important
ones, clearly with some limitations. We also want to avoid approaches
that are too complicated and pure theoretical concepts that are not yet
verified by observations. The level of complexity in the models should
be limited by the amount and quality of observational data we have to
check the results.  The result of these efforts are our {\sl \,disk
  modelling standards\,} that we are proposing to the community in
this paper series.



New challenges for disk modelling have emerged with the advances in
high-resolution imaging (\eg ALMA, NACO, SPHERE and GPI).  These
observations show evidence for non-axisymmetric structures such as
spiral waves, warps, non-aligned inner and outer disks, and
horseshoe-like shapes in the sub-mm.  In the future, 3D models are
clearly required to model such structures, but these challenges go beyond the
scope of this paper. This paper aims at setting new 2D disk modelling
standards as foundation for the {\sl DIANA} project, sufficient to
reproduce the majority of the observations, simple to implement, yet
physically established, and sufficiently motivated by observations.  We
will offer our modelling tools and collected data sets to the
community\footnote{see {\sl http://www.diana-project.com}.}.

\section{Standard disk modelling approach}
\label{sec:Model}

\begin{figure*}[t]
\begin{tabular}{ccc}
\hspace*{-7mm}\includegraphics[width=66mm]{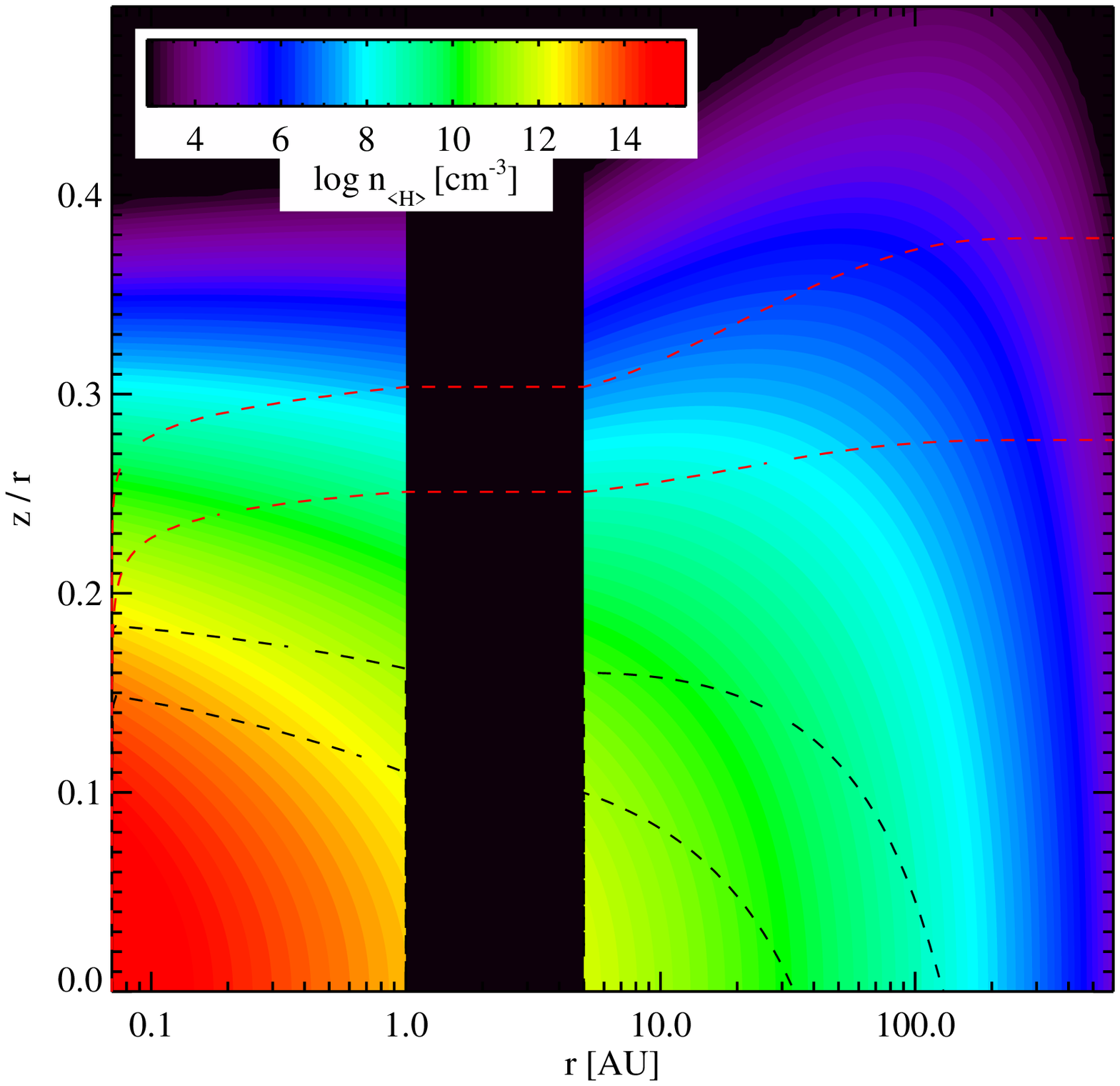} &
\hspace*{-7mm}\includegraphics[width=66mm]{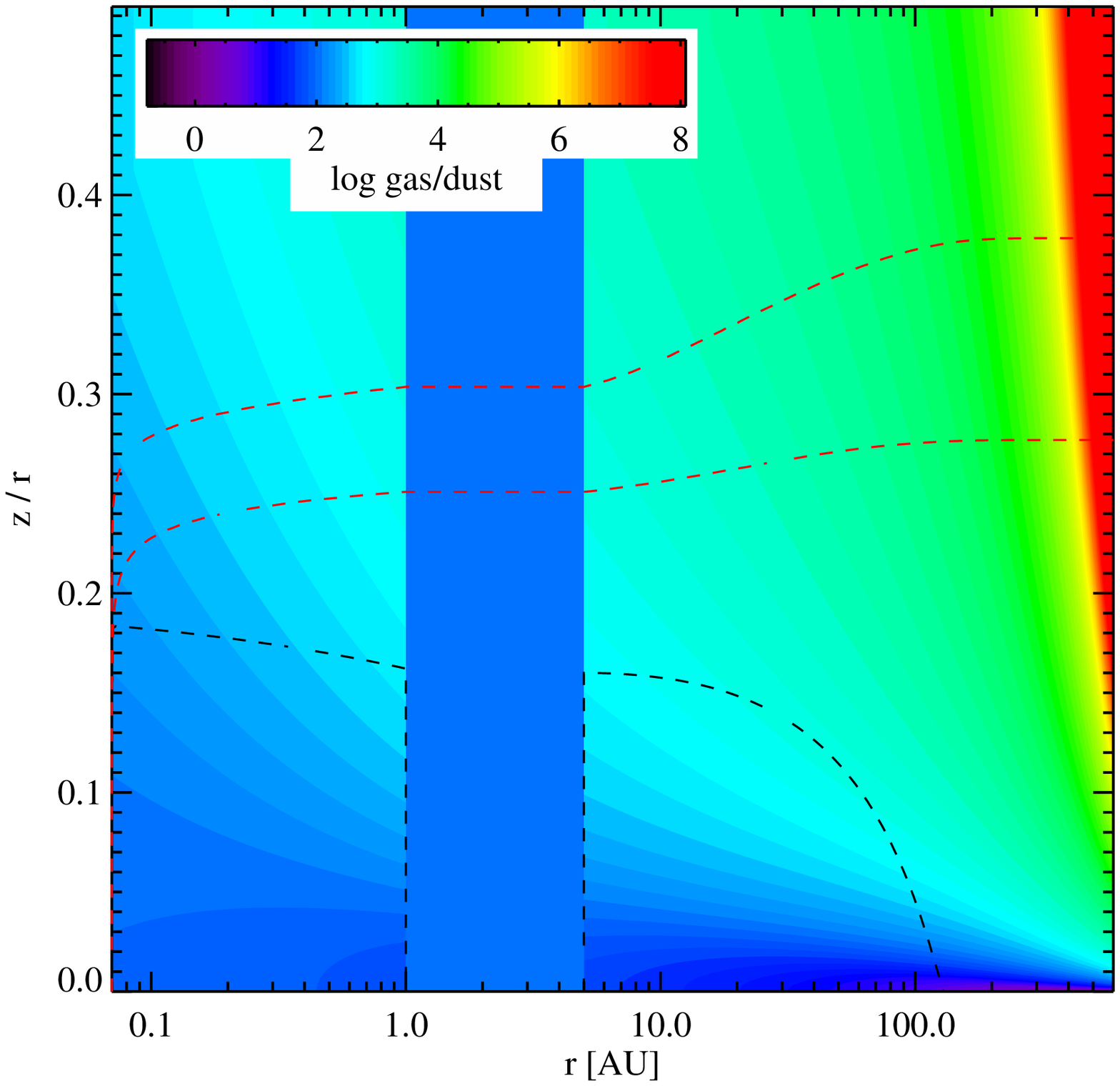} &  
\hspace*{-7mm}\includegraphics[width=66mm]{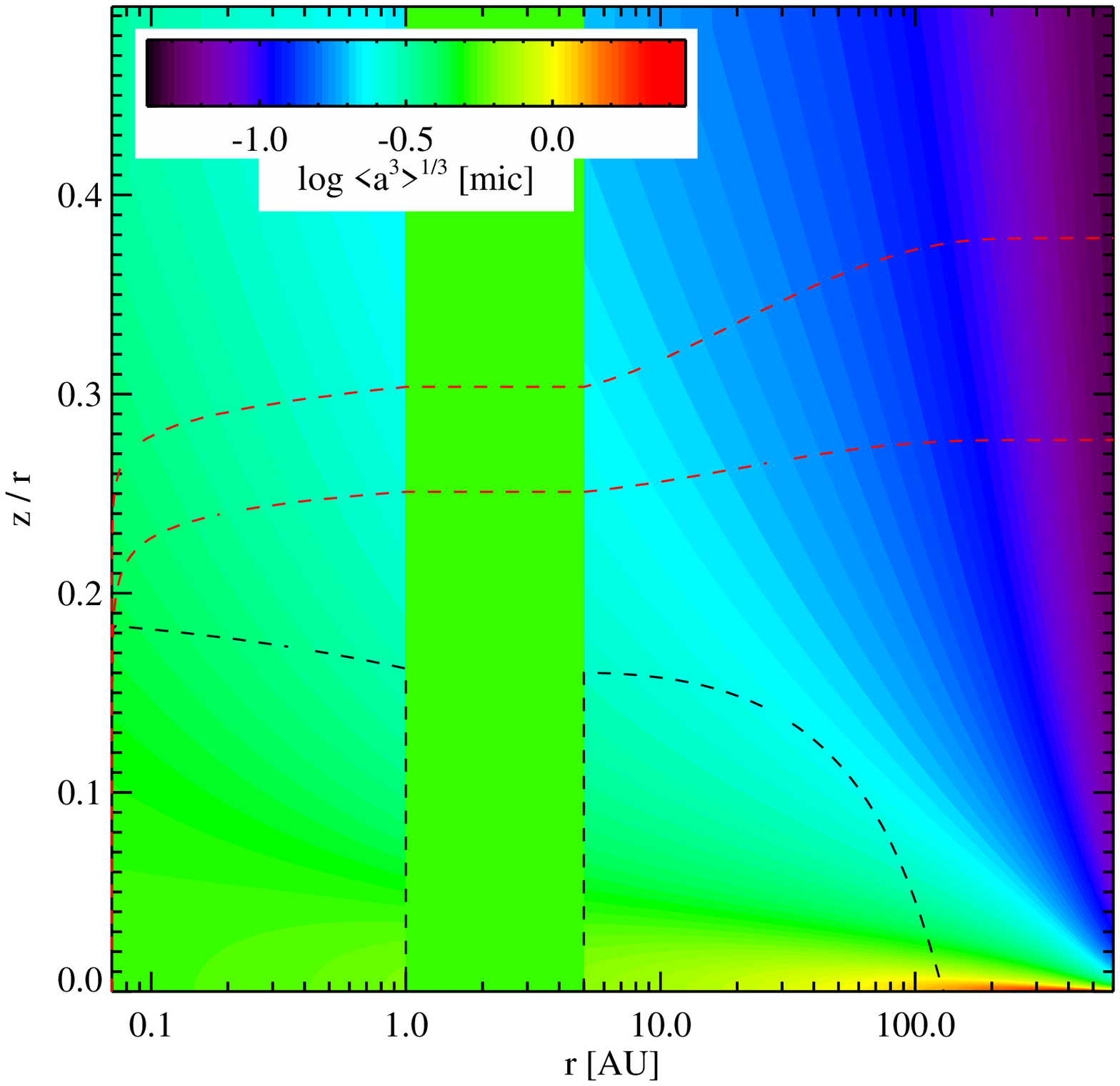} \\
\end{tabular}
\vspace*{-2mm}
\caption{Three figures visualising our disk modelling approach
  concerning disk shape and dust settling. This particular model has
  two radial zones, with a (dust and gas-free) gap between
  $r\!=\!1\,$AU to $r\!=\!5\,$AU. The outer zone is featured by a
  tapered outer edge. The left plot shows the hydrogen nuclei
  particle density $\nH(r,z)$. The middle and right plots show the
  local gas/dust ratio, and the mean dust particle size,
  respectively. These two properties are not constant throughout the
  disk, but depend on $r$ and $z$ through dust settling, see
  Sect.~\ref{sec:settling}. From top to bottom, the two red
  dashed contour lines show the radial optical depth, in terms of the
  visual extinction, $A_{V,\rm rad}\!=\!0.01$ and $A_{V,\rm
    rad}\!=\!1$, and the two dashed black contours show the vertical
  optical depths $A_V\!=\!1$ and $A_V\!=\!10$. In the middle and
  r.h.s.\ plots, the vertical $A_V\!=\!10$ contour line has been
  omitted.}
\label{fig:diskshape}
\end{figure*}

\subsection{Stellar parameters and irradiation}

To model Class\,II protoplanetary disks, we need to specify the
stellar and interstellar irradiation at all wavelengths. 
This requires determining the photospheric parameters of the central
star, \ie the stellar luminosity $L_\star$, the effective stellar
temperature $T_\star$ and the stellar mass $M_\star$. From these
properties, the stellar radius $R_\star$ and the stellar surface
gravity $\log(g)$ can be derived. Fitting these stellar properties to
photometric observations is essential for modelling individual disks,
which requires knowing the distance $d$ and determining the
interstellar extinction $A_V$. More details about our procedure for
fitting the photospheric stellar properties are explained in
Appendix~\ref{app:StellarPara}.

Young stars are known to be strong UV and X-ray emitters.  This
additional, non-photospheric, high-energy disk irradiation can be
neglected for pure dust continuum models, but is essential for the
modelling of the chemistry and energy balance of the gas in
protoplanetary disks. In Appendices~\ref{app:StellarUV} and
\ref{app:StellarXrays}, we explain how observed UV spectra and
measured X-ray data can be used to prescribe this additional
high-energy irradiation in detail. If such detailed data is not
available, we propose a 4-parameter prescription, the relative UV
luminosity $f_{\rm UV}\!=\!L_{\rm UV}/L_\star$, a UV powerlaw index
$p_{\rm UV}$, the X-ray luminosity $L_X$ and the X-ray emission
temperature $T_X$, see Appendices~\ref{app:StellarUV} and
\ref{app:StellarXrays} for details.  All stellar irradiation
components (photosphere, UV, X-rays) are treated by one point source
in the centre of the disk.

In addition to the stellar irradiation, the disk is also exposed to
interstellar irradiation: the interstellar UV field, infrared
background radiation, and the $2.7\,$K cosmic background. All these
types of irradiation are treated by an additional isotropic
irradiation, approaching the disk from all sides (see
Appendix~\ref{app:ISM}).

\subsection{Disk mass and column density structure}

The gas column density structure $\Sigma(r)\rm\,[g/cm^2]$ is assumed
to be given by a radial powerlaw with index $\epsilon$, modified by an
exponential tapering off factor
\begin{equation}
  \Sigma(r) \propto r^{\,-\epsilon} 
          \exp\left(-\,\Big(\frac{r}{R_{\rm tap}}\Big)^{2-\gamma}\right) \ ,
  \label{eq:Sigma}
\end{equation}
where $r$ is the radius and $R_{\rm tap}$ the tapering-off
radius. This approach can naturally explain the often somewhat larger
spectral appearance of protoplanetary disks in (sub-)mm molecular
lines as compared to millimetre continuum images. If the disk has a
tenous continuation, the lines remain optically thick to quite large
radii, where the optically thin continuum signal already vanishes in
the background noise.  For example, the CO\,$J\!=\!2\!\to\!1$ line at
1.3\,mm only requires a hydrogen nuclei column density of about
$N_{\rm H}\!=\!\Sigma/(1.4\,{\rm
  amu})\!\approx\!10^{\,21}\rm\,cm^{-2}$ in our models to
become optically thick, whereas the 1.3\,mm continuum requires $N_{\rm
  H}\!\approx\!2\times10^{\,24}\rm\,cm^{-2}$.  In our reference model
(see Sect.~\ref{sec:refmodel}), these column densities correspond to
radii of about 450\,AU and 20\,AU, respectively.

While a tapering-off column density structure according to
Eq.\,(\ref{eq:Sigma}), with constant dust/gas ratio, seems sufficient
to explain the different apparent sizes of gas and dust in many cases,
for example \citep{Isella2007,Tilling2012} for HD\,163296 and
\citep{Panic2009} for IM\,Lupi, recent high S/N ALMA observations of
TW\,Hya \citep{Andrews2012} and HD\,163296 \citep{deGregorio2013}
suggest that the radial extension of millimeter-sized grains can be
significantly smaller, with a sharp outer edge, \ie a varying dust/gas
ratio. However, since the predicted timescales for radial migration
are in conflict with current observations
\citep[e.g.][]{Birnstiel2014}, and not much is known quantitatively
about migration yet, we assume a constant dust/gas ratio in this
paper, to avoid the introduction of additional free parameters.

The default choice for the tapering-off exponent is
$\gamma\!=\!\epsilon$\, \citep[self-similar
  solution,][]{Hartmann1998}. However, here the two exponents are kept
independent, in order to avoid $\epsilon$ to be determined by $\gamma$
in cases where high-quality sub-mm image data allow for a precise
determination of $\gamma$.  We think that the inner disk structure
should rather be constrained by observations originating from the
inner regions, for example near-IR excess, IR interferometry,
ro-vibrational CO lines, etc..  Radial integration of
Eq.\,(\ref{eq:Sigma}), from $R_{\rm in}$ to $R_{\rm out}$, results in
the total disk mass $M_{\rm disk}$, which is used to fix the
proportionality constant in Eq.\,(\ref{eq:Sigma}). The inner rim is
assumed to be sharp and positioned at $R_{\rm in}$. The outer radius
$R_{\rm out}\!\gg\!R_{\rm tap}$ is chosen large enough to ensure that
$\Sigma(R_{\rm out})$ is small enough to be neglected, \eg $N_{\rm
  H}(R_{\rm out})\!\approx\!10^{20}\rm\,cm^{-2}$.

\subsection{Vertical gas stratification}

We assume a Gaussian vertical gas distribution with parametric
gas scale height $H_{\rm g}$ as function of $r$ as
\begin{equation}
  \rho(r,z)\;\propto\;\exp\left(-\frac{z^2}{2H_{\rm g}(r)^{\,2}}\right)  
  \quad\mbox{with}\quad 
  H_{\rm g}(r) = H_0\,\left(\frac{r}{r_0}\right)^{\,\beta} \ ,
  \label{eq:H}
\end{equation}
where $\rho(r,z)$ is the gas mass density in cylindrical coordinates,
$H_0$ is the reference gas scale height at radius $r_0$, and
$\beta$ is the flaring exponent. Vertical integration of the gas
density $\rho(r,z)$ results in $\Sigma(r)$ which is used to fix the
proportionality constant in Eq.\,(\ref{eq:H}). We have chosen this
simple approach to be most flexible with our fits of near-IR excess,
far-IR excess, visibilities and gas lines. An alternative approach
would be to assume vertical hydrostatic equilibrium, either using the
calculated dust or gas temperature as input, but these models have
some issues reproducing T\,Tauri disk observations, see
Sect.~\ref{sec:hydrostat}. Also, such models take about 5$\times$ to
100$\times$ more computational time to complete, because an iteration
between radiative transfer, gas physics, and vertical structure is
required. This approach is not appropriate when considering the
calculation of a large number of models, as is required when fitting
observations.

\subsection{Dust size distribution}
\label{sec:dustsize}

We assume a powerlaw dust size distribution $f_0(a)$ $\rm[cm^{-4}]$ as
function of particle radius $a\,\rm[cm]$ as
\begin{equation}
  f_0(a)\;\propto\;a^{-\apow} 
  \quad\mbox{with}\quad
  a\in[\amin,\amax] \ .
  \label{eq:dustsize}
\end{equation}
Equation~(\ref{eq:dustsize}) prescribes the dust size distribution
function in the disk ``before settling''.  Dust settling concentrates
the larger grains toward the midplane, and therefore, the local dust
size distribution $f(a,r,z)$ $\rm[cm^{-4}]$ will, in general, deviate
from $f_0(a)$.  Since dust settling only re-distributes the dust
particles vertically in a given column, vertical integration over that
column must again result in $f_0(a)\!=\!\int
f(a,r,z)\,dz\,/\int\!dz$. The local dust mass density $\rm[g/cm^3]$,
before settling, is given by
$\rho\times\delta\!=\!\frac{4\pi}{3}\rho_{\rm d}\int_{\amin}^{\amax}
f_0(a)\,a^3\,da$, where $\rho_{\rm d}$ is the dust material density,
$\rho$ the gas density and $\delta$ the assumed unsettled dust/gas
mass ratio.  This condition is used to fix the proportionality
constant in Eq.\,(\ref{eq:dustsize}).

The minimum and maximum dust size, $\amin$ and $\amax$, are set by the
following simple considerations: sub-micron sized particles are
directly seen in scattered light images, high above the midplane.
They are abundant in the interstellar medium \citep[larger grains up
  to a few $\mu$m seem to already exist in dense cores,
  see][]{Lefevre2014}, and disks are primordially made of such dust.
Millimetre sized grains do also exist in protoplanetary disks, as
indicated by the observed SED slope at mm-wavelengths. Therefore, a
powerlaw covering the entire size range seems to be the most simple,
straightforward option. We will explore the effects of $\amin$,
$\amax$ and $\apow$ on the various continuum and line observations in
Sect.~\ref{sec:dusteffect}.

\subsection{Dust settling}
\label{sec:settling}

Dust settling is included according to \citet{Dubrulle1995}, assuming
an equilibrium between upward turbulent mixing and downward
gravitational settling. The result is a size and density-dependent
reduction of the dust scale heights $H_{\rm d}(r,a)$ with
respect to the gas scale height $H_{\rm g}(r)$,
\begin{eqnarray}
  \left(\frac{H_{\rm d}(r,a)}{H_{\rm g}(r)}\right)^{\!2} &=& 
            \frac{(1+\gamma_0)^{-1/2}\,\alpha_{\rm settle}}
                 {\tau_{\!f}(r,a)\;\Omega(r)} 
  \label{eq:settle}\\[3mm]
  \tau_{\!f}(r,a) &=& \frac{\rho_{\rm d}\;a}{\rho_{\rm mid}(r)\;c_T(r)}
  \ ,
  \label{eq:tauf}
\end{eqnarray}
where $\Omega(r)$ is the Keplerian orbital frequency,
$\gamma_0\!\approx\!2$ for compressible turbulence, and
$\tau_{\!f}(r,a)$ is the frictional timescale in the Stokes
regime. $\rho_{\rm mid}(r)$ is the midplane gas density, and $c_T$ the
midplane sound speed. To avoid iterations involving the midplane
temperature as computed by dust radiative transfer, we use
$c_T(r)\!=\!H_{\rm g}(r)\,\Omega(r)$ here, where $H_{\rm g}(r)$ is the gas scale
height from Eq.\,(\ref{eq:H}). $\alpha_{\rm settle}$ is the
dimensionless viscosity parameter describing the strength of the
turbulent mixing.  The l.h.s.\ of Eq.\,(\ref{eq:settle}) is smoothly
limited to a maximum value of one by $y^2\!\to\!y^2/(1+y^2)$ with
$y\!=\!H_{\rm d}(r,a)/H_{\rm g}(r)$.  Technically, in every disk column,
Eq.\,(\ref{eq:settle}) is computed for a number of (about 100) dust
size bins. Starting from the unsettled dust size distribution, the
dust particles in each size bin are re-distributed in $z$-direction
according to $f(a,r,z)\propto\exp(-z^2/[2H_{\rm d}(r,a)^{\,2}])$, building up
a numerical representation of the local dust size distribution
function at every point in the disk $f(a,r,z)$.

We consider dust settling as a robust physical effect that should
occur rapidly in any disk \citep{Dullemond2004a}, with the dust grains
relaxing quickly toward a vertical equilibrium distribution as
described by Eq.\,(\ref{eq:settle}). Therefore, we think this
important effect should be included in radiative transfer as well as
in thermo-chemical disk models. Equations\,(\ref{eq:settle}) and
(\ref{eq:tauf}) offer an easy-to-implement, yet physically
well-justified method to do so, with just a single parameter
$\alpha_{\rm settle}$.

\subsection{Radial zones, holes, and gaps}

There is increasing evidence that protoplanetary disks are frequently
sculptured by the planets forming in them, which results in the
formation of various shape defects, in particular disk gaps
\citep[e.g.][]{Forrest2004, Andrews2011, Kraus2013}. The gaps are apparently
mostly devoid of dust, but may still contain gas as traced by CO
rotational and ro-vibrational emission lines
\citep[e.g.][]{Bruderer2013,Carmona2014}.  Such objects are classified as
transitional disks, with a strong deficiency of near-IR to mid-IR
flux. Understanding the SEDs of transitional disks requires to
position the inner wall of the (outer) disk at much larger radii than
expected from the dust sublimation temperature or the co-rotation
radius \citep[e.g.][]{Espaillat2014}. The physical mechanisms
responsible for gap formation and disk truncation are still debated,
for example planet formation and migration
\citep{Lin1986,Trilling1998,Nelson2000,Zhang2014}, and/or
photo-evaporation winds \citep{Font2004,Alexander2006,Gorti2009}, but
one observational fact seems to have emerged: disk shape defects are
common.  In fact, single-zone, continuous protoplanetary disks
\citep[\eg FT\,Tau, see][]{Garufi2014} could
be a rather rare class \citep[e.g.][]{Maaskant2014}, and for the
modelling of individual protoplanetary disks we need an additional
option. For an archetypal shape defect, we consider two distinct
radial disk zones, with a gap in-between. In such a case, all disk
shape, dust and settling parameters come in two sets, one for the 
inner zone, and one for the outer zone.

\subsection{Standard dust opacities}
\label{sec:dustopac}

As we will show in this paper, the assumptions about the dust
opacities have a crucial impact not only on the predicted continuum
observations, but also for chemistry and emission lines. Therefore,
the authors of this paper have agreed on a new common approach, which
includes a number of robust facts and requirements that are essential
to model disks. We will explain these new {\em \,standard dust
  opacities\,} carefully in the following, because we think that the
dust in disks is different from the dust in the interstellar medium
and standard dust opacities so far only exist for the interstellar
medium \citep[e.g.][]{Draine1984,Laor1993}.

\begin{figure}
  \hspace*{-3mm} \includegraphics[width=93mm,height=78mm,
                        trim= 30 30 37 38,clip]{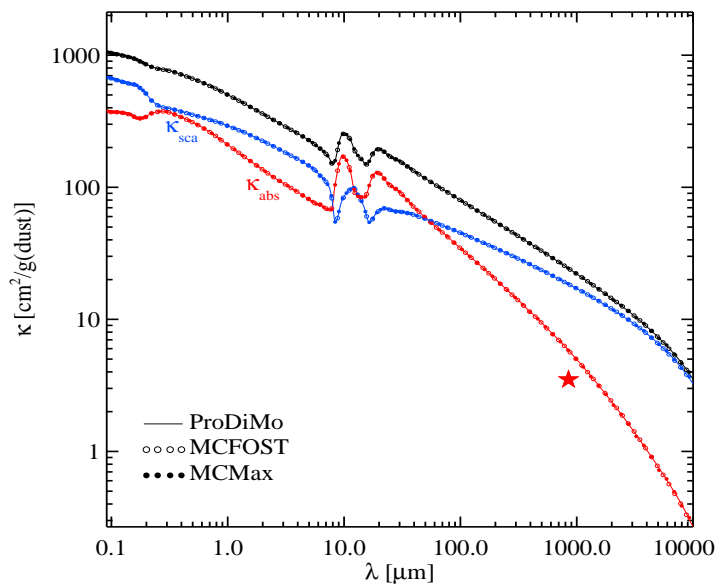}\\[-5mm]
  \caption{Dust opacities used for the standard disk radiative
    transfer modelling, with parameters $\amin\!=\!0.05\,\mu$m,
    $\amax\!=\!3\,$mm, $\apow\!=\!3.5$, and 15\% amorphous carbon by
    volume. The extinction per dust mass is shown in black, absorption
    in red, and scattering in blue.  Results shown with lines for {\sl
      ProDiMo}, open symbols for {\sl MCMax}, and full symbols for
    {\sl MCFOST} -- all of which agree. The red star represents the
    value of $3.5\rm\,cm^2/g(dust)$ at 850\,$\mu$m used by
    \citet{Andrews2005} to determine disk masses from sub-mm fluxes.}
  \label{fig:dustopac}
  \vspace*{-2mm}
\end{figure}

\begin{figure*}
\centering
\vspace*{-1mm}
\resizebox{16cm}{!}{\begin{tabular}{ll}
\hspace*{-7mm}\includegraphics[height=53mm,trim= 0 20 0 0,clip]{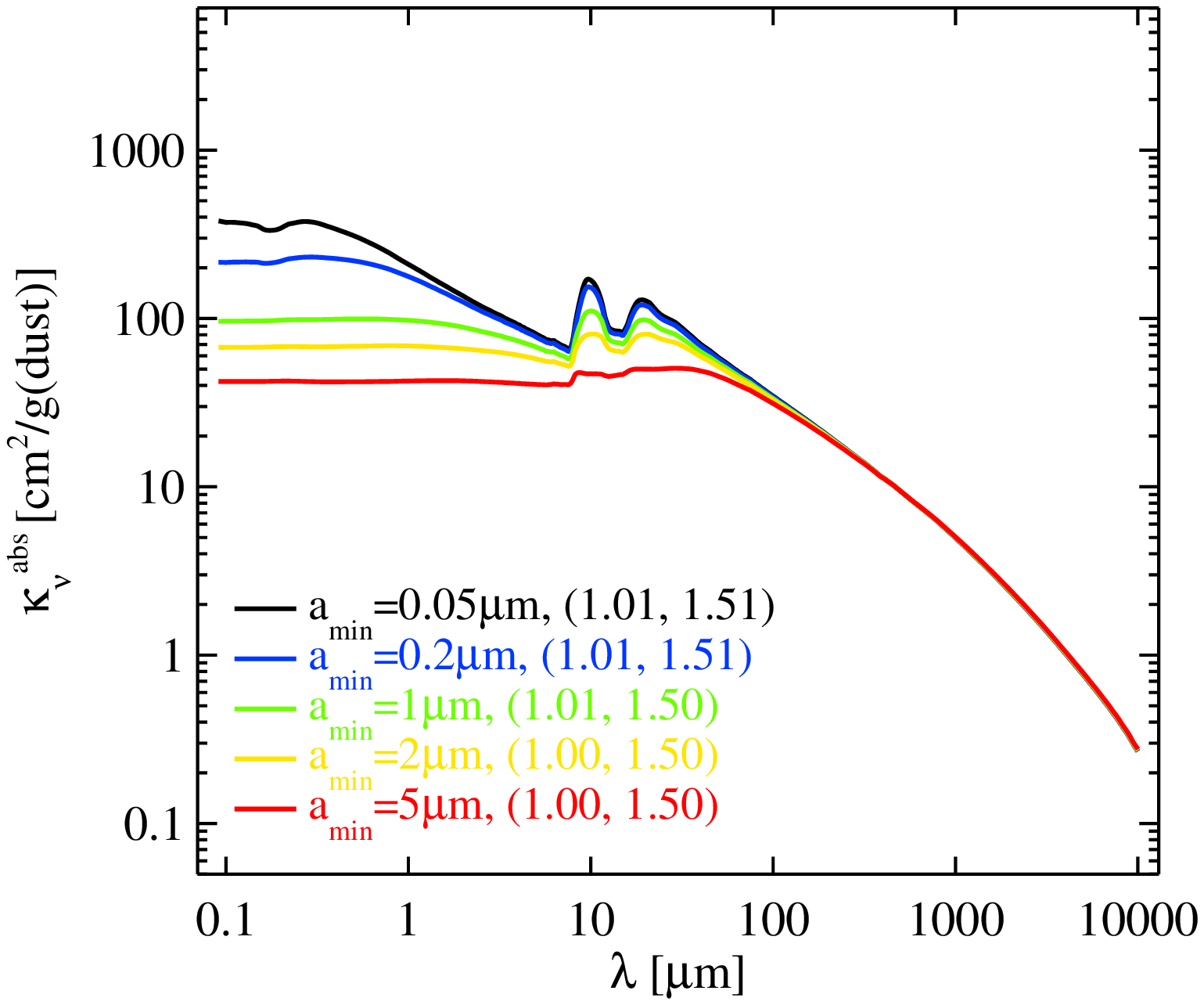} &
\hspace*{-7mm}\includegraphics[height=53mm,trim=39 20 0 0,clip]{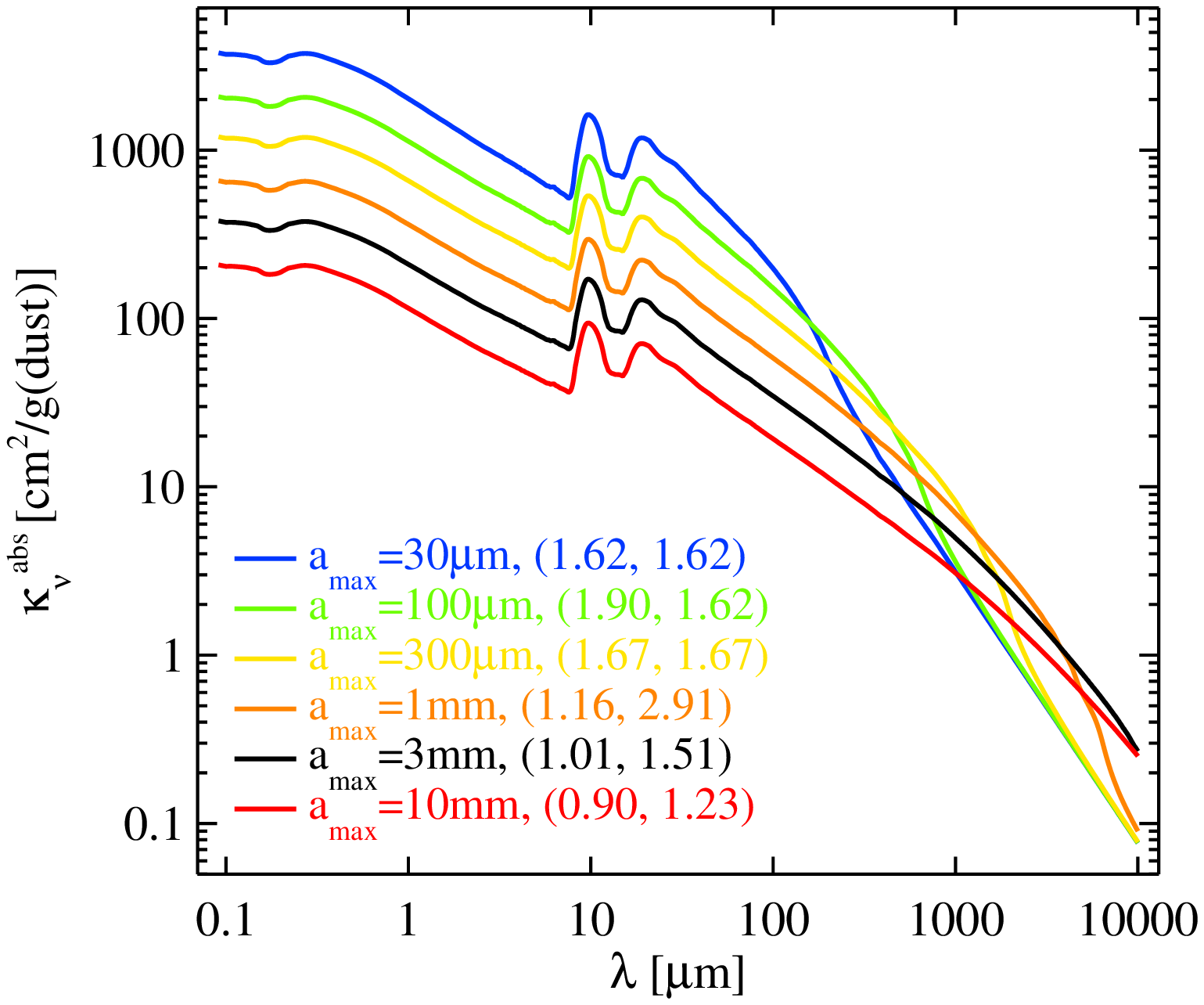} \\[-1mm]
\hspace*{-7mm}\includegraphics[height=56mm,trim= 0 0 0 0,clip]{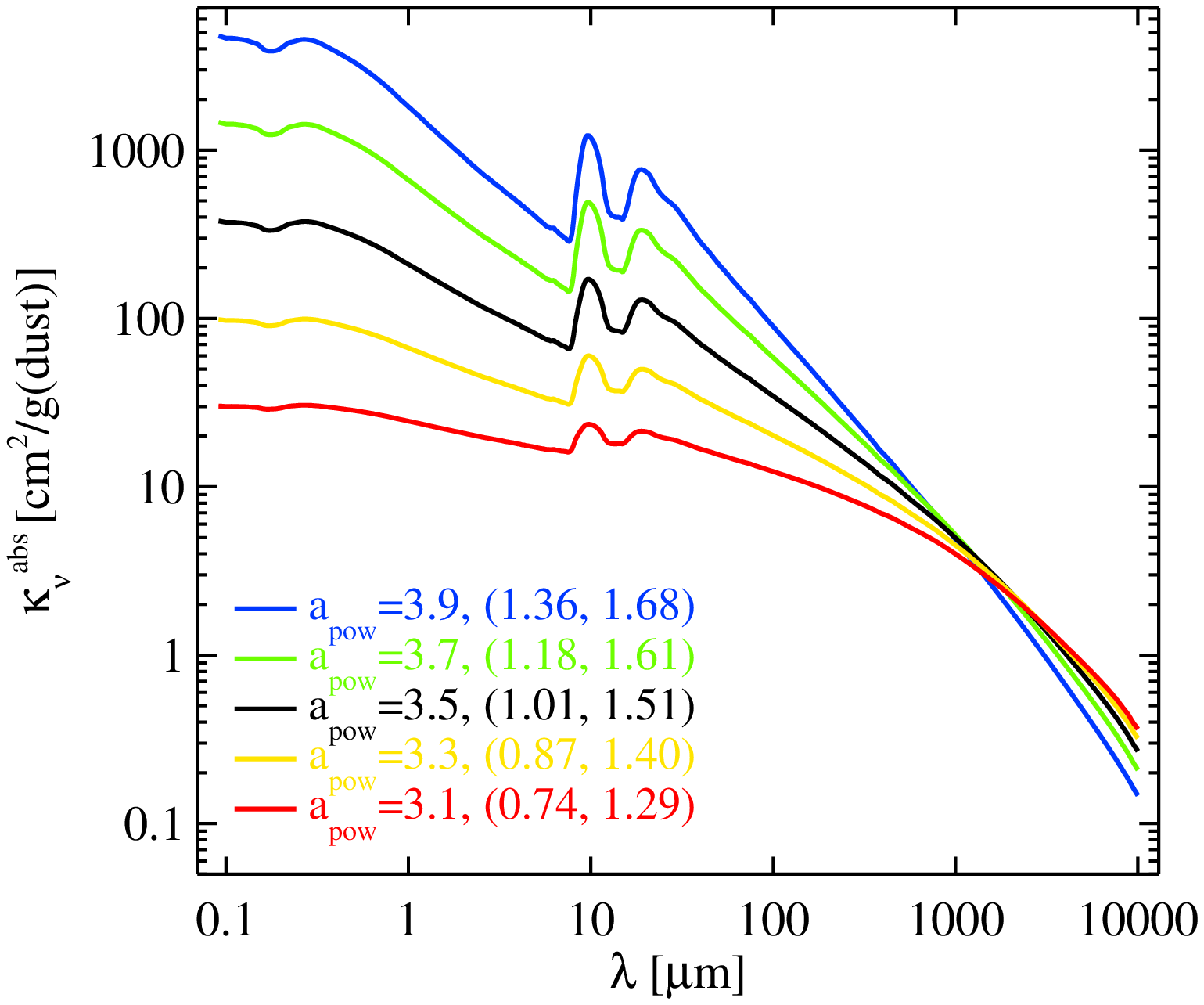} &
\hspace*{-7mm}\includegraphics[height=56mm,trim=39 0 0 0,clip]{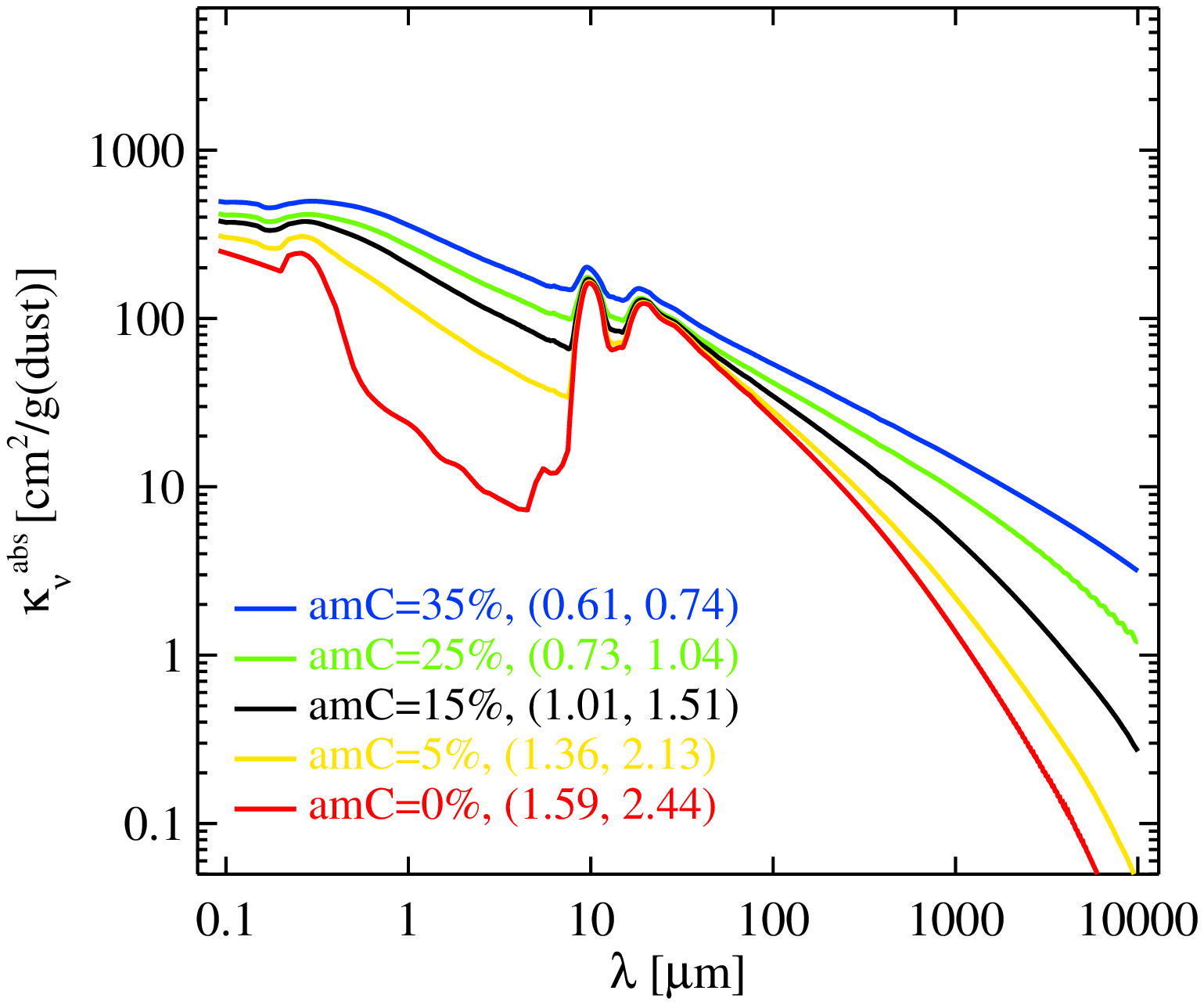} \\
\end{tabular}}
\vspace*{-2mm}
\caption{Dust absorption coefficient per dust mass as function of dust
  size and material parameters. The black line is identical in every part
  plot, with parameter values as used in the reference model, our {\em
    \,dust standard opacities\,}, see Table~\ref{tab:refmodel}. The upper two
  figures show the dependencies on minimum and maximum particle size,
  $\amin$ and $\amax$. The lower two plots show the dependencies on
  dust size powerlaw index $\apow$ and on the volume fraction of
  amorphous carbon. 25\% porosity and maximum hollow volume ratio
  $V_{\rm hollow}^{\rm max}\!=\!0.8$ are assumed throughout. The two
  numbers in brackets represent the log-log dust absorption opacity
  slopes between 0.85\,mm and 1.3mm, and between 5\,mm and 1\,cm,
  see Appendix~\ref{app:integrated}.}
\label{fig:dustabs}
\vspace*{-1mm}
\end{figure*}

Our assumptions for the dust opacity treatment are guided by a study
of multi-wavelength optical properties of dust aggregates
\citep{Min2015} where the Discrete Dipole Approximation (DDA) is used
to compute the interaction of light with complexly shaped,
inhomogeneous aggregate particles. These opacity calculations are
computationally too expensive to be applied in complex disk models,
but Min\etal have developed a simplified, fast numerical treatment
that allows us to reproduce these opacities reasonably well.

We consider a mixture of amorphous laboratory silicates \citep[][$\rm
  Mg_{0.7}Fe_{0.3}SiO_3$]{Dorschner1995} with amorphous carbon
\citep[][BE-sample]{Zubko1996}. Pure laboratory silicates are
``glassy'' particles with almost negligible absorption cross sections
in the near-IR, just where T\,Tauri stars are most luminous. Such
particles would rather scatter the incident light away from the disk,
which would lead to substantial problems in explaining the near-IR
excess of T\,Tauri stars.  Therefore, the inclusion of a conductive,
hence highly opaque, albeit featureless material in the near-IR is
necessary. However, it is unclear whether amorphous carbon, metallic iron,
or \eg troilite (FeS) should be used. Our simplified material
composition is inspired by the past disk modelling expertise of the
team, and by the solar system composition proposed by \citet{Min2011}.

The dust grains are assumed to be composed of 60\% silicate, 15\%
amorphous carbon, and 25\% porosity, by volume, well-mixed on small
scales. The effective refractory index of this porous material is
calculated by applying the \citet{Bruggeman1935} mixing rule. In
contrast, just adding opacities (assuming separate grains of pure
materials with the same size distribution and equal temperatures) does
not seem physically justified at all, and does not reproduce the
optical properties of aggregate particles equally well \citep{Min2015}.

We use a distribution of hollow spheres (DHS) with a maximum hollow
volume ratio $V_{\rm hollow}^{\rm max}\!=\!0.8$. This approach avoids
several artefacts of Mie theory (spherical resonances) and can account
for the most important shape effects, see details in \citep{Min2015}.
In combination with our choice of the amorphous carbon
optical constants from \citet[][BE-sample]{Zubko1996}, this approach
captures the ``antenna-effect'' observed from the aggregate particles,
where irregularly shaped inclusions of conducting materials result in
a considerable increase of mm-cm absorption opacities.

Table~\ref{tab:refmodel} summarises our standard choices of dust
parameter values, and Fig.~\ref{fig:dustopac} shows the resulting
opacities including scattering and extinction.
Figure~\ref{fig:dustabs} shows the dependencies of the dust absorption
opacities on the remaining free dust size and material parameters. We
will continue to discuss these results and their impact on predicted
continuum and line observations in Sect.~\ref{sec:dusteffect}. Our
standard dust opacities feature
\begin{my_itemize}
\item a FUV-dust extinction opacity of about 1000\,cm$^2$/g(dust),
  which is about 100 times less than in standard ISM models with MRN
  size distribution \citep{Mathis1977},
\item a dust albedo of 64\% at FUV wavelengths, and 58\% at $1\,\mu$m,
\item a dust absorption cross section of about 5.8\,cm$^2$/g(dust) at
  850\,$\mu$m, which is about a factor of 1.6 larger than the value of
  3.5\,cm$^2$/g(dust) used by \citet{Andrews2005}
  to determine disk masses from sub-mm fluxes\footnote{This critical
    opacity value traces back to \citet{Beckwith1990} who proposed
    $\approx$\,10\,cm$^2$/g(dust) at 300\,$\mu$m (1000\,GHz).},
\item a millimetre dust absorption slope of about 1, and
\item a centimetre dust absorption slope of about 1.5.
\end{my_itemize}
The dust opacities described above have been determined from the
well-mixed dust size distribution function $f_0(a)$, see
Sect.~\ref{sec:dustsize}.  Since we have dust settling in the disk,
the dust opacities need to be computed from the local settled dust
size distribution function $f(a,r,z)$, and will hence not only depend
on $\lambda$, but also on the location in the disk $(r,z)$.  Simply
put, dust settling leads to a strong concentration of the (sub-)mm
opacity in the midplane, in particular in the outer disk regions, but
only to a mild reduction of the UV opacities in the upper and inner
layers. According to Fig.~\ref{fig:dustabs}, the main effects are:
\begin{my_itemize}
\item[1)] Large $\amin$ values reduce the optical and UV opacities, and
  destroy the 10\,$\mu$m / 20\,$\mu$m silicate emission features.

\item[2)] Large $\amax$ values reduce the UV, optical, near-IR and far-IR
  opacities considerably, because the available dust mass is spread
  over a larger size range, and the big particles do not
  contribute much to the dust opacities at those wavelengths. Here,
  increasing $\amax$ has similar consequences as lowering the dust/gas
  mass ratio.

  At longer wavelengths, $\amax$ determines where the final transition
  to the Rayleigh-limit takes place ($\lambda_{\rm
    trans}\!\approx\!2\pi\amax$), beyond which
  ($\lambda\!>\!\lambda_{\rm trans}$) the opacity changes to a steeper
  slope. For larger $\amax$, this transition occurs at longer
  wavelengths. The choice of our reference value of $\amax\!=\!3\,$mm
  is motivated by cm-observations which usually do not show such
  breaks. However, Greaves\etal (in prep.) report on first evidence of
  such a steepening toward cm wavelengths after removal of the
  free-free emission component, based on new Green Bank Telescope data
  up to 8.6\,mm.

\item[3)] The powerlaw size index $\apow$ determines the mixing ratio
  of small and large grains. Since the smaller particles are
  responsible for the short wavelength opacities, and the large grains
  for the long wavelength opacities, $\apow$ determines the general
  opacity slope, and the mm and cm-slopes in particular.

\item[4)] A large volume fraction of amorphous carbon reduces the
  10\,$\mu$m, 20\,$\mu$m silicate emission features, fills in the
  opacity deficits of the major solid-state
  silicate resonances (up to about 8\,$\mu$m), and flattens the absorption
  opacities at millimetre and centimetre wavelengths.
\end{my_itemize}
In order to facilitate the adoption of these opacities in other work,
a Fortran-90 package to compute the DIANA standard dust opacities is
available at {\sl
  http://www.diana-project.com/data-results-downloads}.

\subsection{PAHs}
\label{sec:PAHs}

Polycyclic Aromatic Hydrocarbon molecules (PAHs) play an important
role in our disk models via (i) continuum radiative transfer effects,
(ii) photoelectric heating of the gas, and (iii) chemical effects.
The chemical effects include the charging of the PAHs, the release and
consumption of free electrons via photo-ionisation and recombination,
and further effects due to charge exchange reactions. We study (i) and
(ii) for neutral PAHs in this paper, but do not include
the chemical effects as we are using here the small DIANA chemical
standard.  In contrast, the large DIANA chemical standard (Paper\,II)
has the PAHs included in the selection of chemical specimen, and
therefore accounts for (iii) as well.

PAHs are observed via their strong mid-IR emission bands in many
Herbig Ae/Be stars \citep[e.g.][]{Maaskant2014}, whereas detection
rates in T\,Tauri stars are much lower \citep{Geers2006}, possibly
because T\,Tauri stars generate much less blue and soft UV stellar
radiation to heat the PAHs. PAHs in Herbig Ae/Be disks seem to have
sizes of at least 100 carbon atoms \citep{Visser2007}.
The PAH abundance in the disk is assumed to be given by the standard
abundance in the interstellar medium \citep{Tielens2008},
modified by factor $f_{\rm PAH}$
\begin{equation}
 \frac{n_{\rm PAH}}{\nH} = 3\times 10^{-7}\,f_{\rm
   PAH}\,\frac{50}{N_{\rm C}} \ .
 \label{eq:PAHabun}
\end{equation}
$f_{\rm PAH}\!=\!1$ corresponds to the interstellar medium (ISM)
standard\footnote{For a gas/dust mass ratio of 100, $f_{\rm
    PAH}\!=\!1$ corresponds to a PAH/dust mass ratio of about
  0.013.}. Here, $n_{\rm PAH}\rm\,[cm^{-3}]$ is the PAH particle
density, $\nH\rm\,[cm^{-3}]$ is the hydrogen nuclei density and
$N_{\rm C}$ is the number of carbon atoms in the PAH.  The actual PAH
abundance in disks is disputed
\citep[e.g.][]{Geers2006,Visser2007}. Values $f_{\rm
  PAH}\!\approx\!0.1$ or lower seem typical in Herbig\,Ae disks
\citep{Geers2006}.

We assume $N_{\rm C}\!=\!54$ carbon atoms and $N_{\rm H}\!=\!18$
hydrogen atoms (``circumcoronene'') in the reference model, resulting
in a PAH mass of 667\,amu and a PAH radius of
4.87\AA\ \citep{Weingartner2001}. $N_{\rm C}$ and $f_{\rm PAH}$ are
free model parameters, as well as a decision whether to select the
neutral or charged PAH opacities. Circumcoronene IR and UV spectra
have been directly measured by \citet{Bauschlicher2000}.
However, in the DIANA framework, we use "synthetic" PAH
opacities of neutral and charged PAHs are calculated according to
\citet{Li2001} with updates from \citet{Draine2007}, including the
``graphitic'' contribution in the near-IR and the additional
``continuum'' opacities of charged PAHs. 

In comparison to the low UV opacities of evolved dust in disks
(Sect.~\ref{sec:dustopac}), PAHs can easily dominate the blue and UV
opacities, see Fig.~\ref{fig:PAHmodels}.  This happens in the
well-mixed case for $f_{\rm PAH}\!\ga\!0.1$ in our disk models.  The
dominance of the PAH opacities in the UV is even stronger in the upper
disk regions because of dust settling (we assume that PAH molecules do
not settle).  For Herbig\,Ae disks, where the maximum of the stellar
radiation is released around 400\,nm, $f_{\rm PAH}\!\ga\!1$ would
imply that the stellar photons are predominantly absorbed by the PAHs
rather than by the dust.  The absorbed energy would then be re-emitted
via the strong PAH mid-IR resonances, and it is this mid-IR PAH
emission that would predominantly heat the disk.  Furthermore, the
stellar UV usually reaches the line forming regions in a disk
indirectly, via scattering on dust particles from above.  For $f_{\rm
  PAH}\!\ga\!0.1$, the PAHs would effectively shield the disk from UV
radiation, because UV scattering by PAHs is extremely inefficient.
These two effects have large implications on our models for both the
internal dust and gas temperature structure in a disk.

The treatment of PAHs in the Monte Carlo programs {\sl MCFOST} and
{\sl MCMax} is standard, using a quantum heating formalism with
stochastic PAH temperature distribution. This mechanism was first
proposed for small grains by \citet{Desert1986} and later applied to PAHs
in the interstellar medium by
\citet{Manske1998,Guhathakurta1989,Siebenmorgen1992}. The Monte Carlo
programs offer additional options to take into account \eg a PAH
size distribution and an internal determination of the PAH charge, by
balancing the basic photo-ionisation and recombination rates
\citep{Maaskant2014}.  However, these options involve some quite
specific simplifications, for example no negative PAHs, no charge
exchange reactions, and an assumed electron concentration, which we
need to avoid for reasons of consistency for the DIANA modelling
efforts.

While looking for a fast, simplified, and robust way to treat the most
important effects of PAHs equally well in all our disk models, we
discovered that a simplified treatment of the PAHs in radiative
equilibrium, according to
\begin{eqnarray}
 S_{\!\nu} &=& \frac{\kappa_\nu^{\rm dust,abs}B_\nu(\Td)
                +\kappa_\nu^{\rm PAH,abs}B_\nu(\TPAH)
                +\ksca J_\nu}
                {\kappa_\nu^{\rm dust,abs}+\kappa_\nu^{\rm PAH,abs}+\ksca} 
 \label{eq:source}
\end{eqnarray}
\vspace*{-4mm}
\begin{eqnarray}
 \int \kappa_\nu^{\rm dust,abs}\big(B_\nu(\Td)-J_\nu\big)\,d\nu
 &=& 0
 \label{eq:dustRE}\\
 \int \kappa_\nu^{\rm PAH,abs}\big(B_\nu(\TPAH)-J_\nu\big)\,d\nu
 &=& 0 \ ,
 \label{eq:PAHRE}
\end{eqnarray}
leads to quite accurate results in comparison to the stochastic PAH
treatment, see App.~\ref{app:PAHtemp} and
Fig.~\ref{fig:PAHtemp}. Here, $S_\nu$ is the source function, $\kabs$
and $\ksca$ are the absorption and scattering opacities (of dust and
PAHs as annotated), $B_\nu(T)$ is the Planck function and $J_\nu$ is
the mean intensity.  The scattering term $\ksca J_\nu$ is here
simplifyingly written for isotropic scattering.
Equations~(\ref{eq:dustRE}) and (\ref{eq:PAHRE}) express the
conditions of radiative equilibrium with separate dust and PAH
temperatures, $\Td$ and $\TPAH$, respectively.

The quantum heating formalism is appropriate for ISM conditions, where
PAHs are only sometimes heated by rare FUV photons. In contrast, we
show in Appendix~\ref{app:PAHtemp} that the PAHs in the inner regions
of protoplanetary disks, which are responsible for the observable
mid-IR PAH emission features, are situated in an intense optical and
infrared radiation field created by the star and by the dust and the
PAHs in the disk, which keeps the stochastic PAH temperature
distribution high and quite narrow, \ie close to the analytical
treatment in radiative equilibrium. \citet{Li2012} found similar
results for nano grains acquiring an equilibrium temperature when
exposed to intense starlight.

\subsection{Chemistry and heating/cooling balance}

Based on the results of the continuum radiative transfer as described
in the previous sections, the gas phase and ice chemistry is
calculated in kinetic chemical equilibrium, coupled to the gas
heating/cooling balance. These parts of the model have been described
elsewhere, see \citep{Woitke2009a} for the basic model,
\citep{Thi2011} for continuum radiative transfer, \citep{Woitke2011}
for updates concerning non-LTE treatment and heating \& cooling, and
\citep{Aresu2011} for X-ray heating and chemistry, and are not
discussed in this paper.  In this paper we use the {\em \,small
  chemical network\,}, as proposed in Paper\,II. We carefully select
100 gas phase and ice species (see Paper\,II), and take into account
altogether 1288 reactions. All 1065 gas phase and UV reactions among
the selected species are taken into account from the UMIST\,2012
database \citep{McElroy2013}, including the old collider reactions. We
replace the treatment of photo-reactions by individual photo
cross-sections from the Leiden {\sc Lamda} database \citep{Lamda2005}
where possible, add 145 X-ray reactions \citep{Aresu2011}, 40 ice
adsorption and thermal, UV photo and cosmic ray desorption reactions,
and 38 auxiliary reactions including those of vibrationally excited
molecular hydrogen H$_2^\star$, see details in Paper\,II.  We take the
(gas\,$+$\,ice) element abundances from Table~2 in Paper\,III.
Appendix~\ref{app:tdep} discusses the validity of our approach to use
the time-independent solution of our chemical rate network to compute
the chemical composition of the disk and the gas emission lines.

\subsection{Line radiative transfer}
\label{sec:lineRT}

After the continuum radiative transfer, gas and dust temperature
structure, chemistry and non-LTE level populations have been
determined, a formal solution of line and continuum radiative transfer
is carried out in 3D, using a bundle of 356x144 parallel rays towards
the observer at distance $d$ under inclination angle $i$, see
\citep[][appendix A.7 therein]{Woitke2011} for details. These
computations result in observable quantities like line fluxes, line
velocity-profiles, molecular line maps and channel maps.

\begin{figure*}
  \vspace*{-1mm}
  \hspace*{-3mm}\includegraphics[width=188mm]{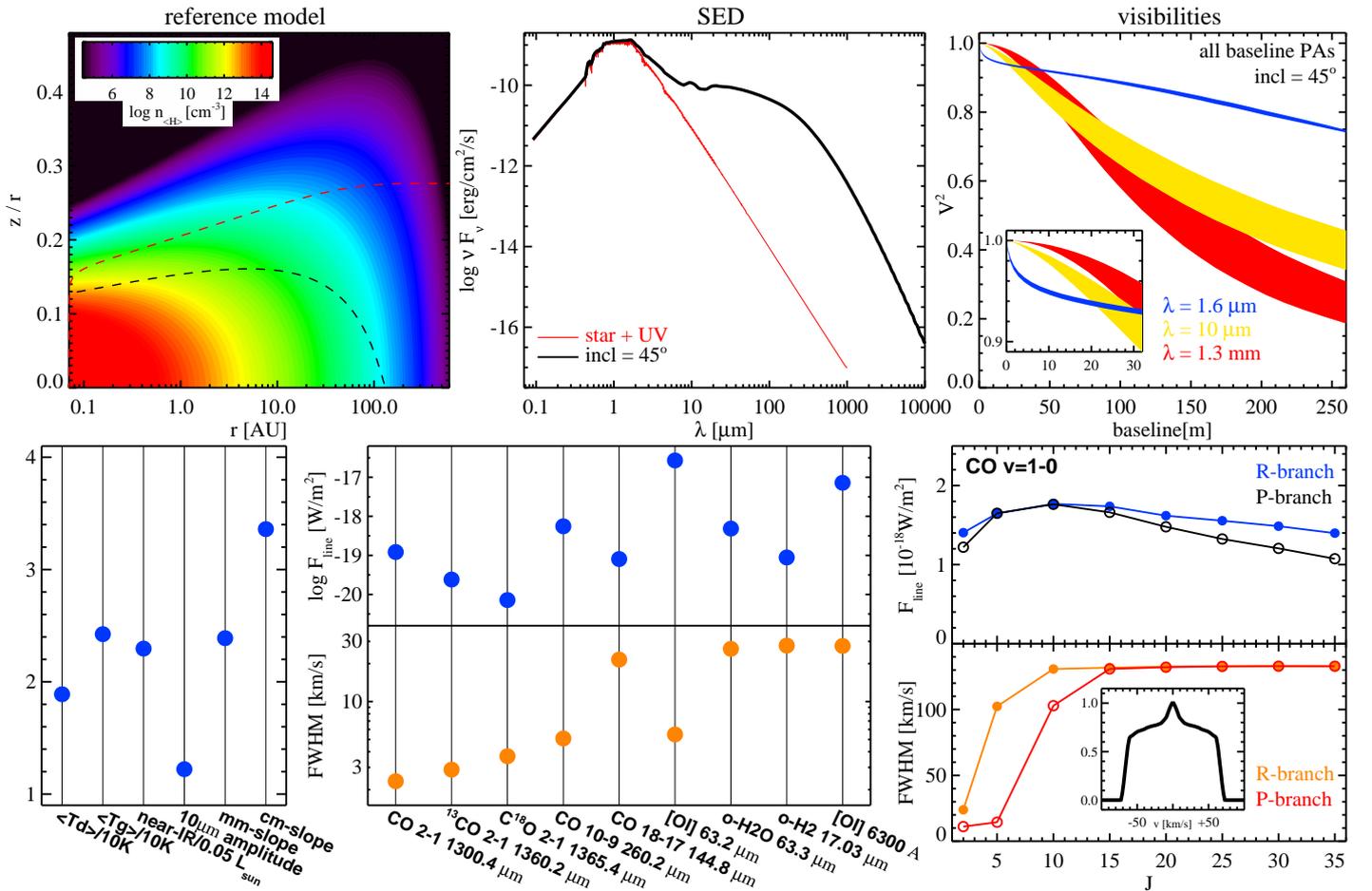}
  \caption{Summary of results from the reference model.  {\bf\ \ Top
      row:} assumed gas density structure $\nH(r,z)$ with overplotted
    radial (red) and vertical (black) optical depths $A_V\!=\!1$
    dashed contours, computed SED, and visibilities. In the visibility
    plot, the coloured areas show $V^2$ for all baseline
    orientations at 3 different wavelengths, with a zoom-in on the
    first 30\,m.  {\bf\ \ Lower row:} other resulting quantities. The
    left plot shows the mean dust and gas temperatures (in units of
    10\,K), the near-IR excess (in units of 0.05\,$L_\odot$) and the
    logarithmic SED slopes at mm and cm wavelengths. The
    centre plot shows calculated line fluxes and full widths at half
    maximum (FWHM). The right plot shows some results for the CO
    fundamental ro-vibrational line emissions, line fluxes as function
    of rotational quantum number $J$ for the $R$-branch and the
    $P$-branch, as well as computed FWHM for those lines. The inserted
    figure shows the line profile averaged over all emission lines,
    scaled from 0 (continuum) to 1 (maximum).}
  \label{fig:refmodel}
  \vspace*{-1mm}
\end{figure*}

\section{Model implementation and verification}
\label{sec:Implementation}

The DIANA standard modelling assumptions summarised in
Sect.~\ref{sec:Model} (stellar and interstellar irradiation, disk
shape, dust opacities, dust settling, treatment of PAHs) have been
implemented into {\sl MCFOST} \citep{Pinte2006,Pinte2009} {\sl MCMax}
\citep{Min2009} and {\sl ProDiMo} \citep{Woitke2009a}.  The
independent implementation of our modelling assumptions has allowed us
to perform stringent checks on our computational methods and numerical
results. Figure~\ref{fig:dustopac} shows a validation of our dust
opacity implementation.  Figure~\ref{fig:PAHmodels} compares the
assumed gas calculated settled dust densities, the resulting dust and
PAH temperatures, as well as the SEDs. Apart from some minor
temperature deviations in the optically thick midplane regions, which
are irrelevant for the predicted observations, we achieve an excellent
agreement concerning the physical state of the disk and all predicted
observations. In particular, the upper right part of
Figure~\ref{fig:PAHmodels} shows that we obtain very similar SED and
spectral shape of the PAH features no matter whether we use {\sl
  ProDiMo}, {\sl MCFOST}, or {\sl MCMax}.

Further verification tests (not shown here) have been undertaken for
disk models with gaps, where the numerical resolution of the inner
wall of the outer disk is particularly important, and for
MC\,$\to$\,{\sl ProDiMo} ``chain models''. In these chain models, we
use the Monte Carlo codes to compute the disk structure, the dust and
PAH temperatures, and the internal radiation field $J_\nu(r,z)$, and
then pass these results on to {\sl ProDiMo} to compute the gas
temperature structure, the chemical composition of ice and gas, and
the emission lines.

The advantages of using the MC\,$\to$\,{\sl ProDiMo} chain models are
(i) the Monte-Carlo technique is computationally faster, (ii) the
temperature iteration scheme is more robust, in particular at high
optical depths, and (iii) the Monte-Carlo technique allows for a more
detailed implementation of radiation physics, in particular
anisotropic scattering, PAHs with stochastic quantum heating, and
polarisation.  For the effects of anisotropic scattering, see
Fig.~\ref{fig:aniso} in Appendix~\ref{app:figures}. These more
sophisticated options are not used in this paper, in order to
facilitate comparisons to the results obtained with pure {\sl
  ProDiMo}.  The pre-existing interface between {\sl MCFOST} and {\sl
  ProDiMo} \citep{Woitke2010} has been generalised and implemented in
{\sl MCMax}, such that now all {\sl MCFOST} and {\sl MCMax} users are
able to use {\sl ProDiMo} to predict chemical and line results on top
of their continuum models.

Appendix~\ref{app:NumConv} discusses the numerical convergence of our
models as function of numerical grid resolution, for both the pure
{\sl ProDiMo} and the MC\,$\to$\,{\sl ProDiMo} chain models.  Our
conclusion here is that we need about $100\times100$ grid points in
both {\sl ProDiMo} and MC models, to achieve an accuracy $<10\%$ for
all continuum observables and line flux predictions.

\section{Results}
\label{sec:Results}

The results of our disk models are presented in the following way.  We
first introduce a simple single-zone {\em \,reference model\,} in
Sect.~\ref{sec:refmodel} which roughly fits a number of typical
continuum and line observations of Class\,II T\,Tauri stars.  In the
following two sub-sections, we then study the impact of our model
parameters on the various continuum and line observables by looking at
how our model predictions change with respect to the reference
model. In Sect.~\ref{sec:ParaObs}, we study the impact of selected
model parameters on all observables at a time, and in
Sect.~\ref{sec:ObsPara}, we discuss particular observables separately.

\subsection{The reference model}
\label{sec:refmodel}

\begin{figure*}
  \vspace*{-2mm}
  \includegraphics[width=180mm]{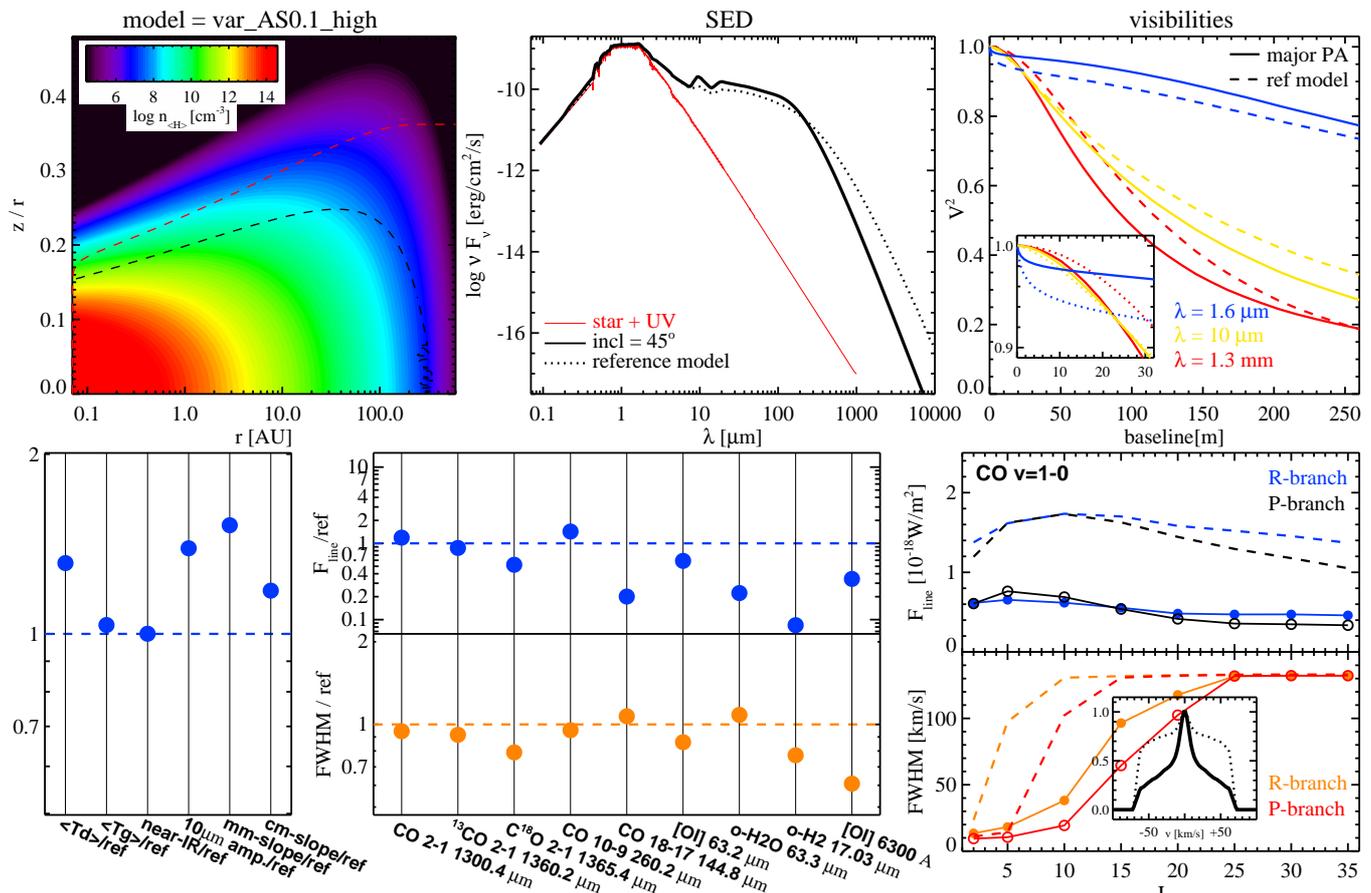}\\
  \vspace*{-6mm}
  \caption{Results from the T\,Tauri reference model, but assuming uniform
    0.1\,$\mu$m sized dust particles and astronomical silicate Mie
    opacities.  Depicted quantities are shown with respect to the
    reference model, and explained in the caption of
    Fig~\ref{fig:refmodel}.}
  \label{fig:0.1mic}
  \vspace*{-4mm}
\end{figure*}

Table~\ref{tab:refmodel} summarises our model parameters, and lists
the values used for the reference model. The resulting spectral energy
distribution (SED), visibilities, line observations, and some
integrated properties are shown in Fig.~\ref{fig:refmodel}. Concerning
the integrated properties, we calculate the mean gas temperature in
the disk $\langle \Tg\rangle$, the mean dust temperature $\langle
\Td\rangle$, the near-IR excess, the $10\,\mu$m SED amplitude, the
mm-slope and the cm-slope as explained in
Appendix~\ref{app:integrated}, see Eqs.~(\ref{eq:Tgmean}) to
(\ref{eq:cm-slope}). The reference model is characterised by
\begin{my_itemize}
\item a near-IR excess of about $0.12\rm\,L_\odot$,
\item clearly visible silicate dust emission features around
  10\,$\mu$m and 20\,$\mu$m,
\item a descending SED-slope $d\log(\nu F_\nu)/d\log\lambda\!<\!0$
  beyond 20\,$\mu$m, as is typical for continuous (\ie
  non-transitional) T\,Tauri disks,
\item a 1.3\,mm continuum flux of 60\,mJy with an apparent radius
  (semi-major axis) of about 100\,AU (0.75$\arcsec$ at a distance of
  140\,pc) -- typical observed values are about $20-200\,$mJy, and
  $0.25\arcsec-1.4\arcsec$, see \citep{Guilloteau2011},
\item a mm-slope of about 2.4 -- typical observed values are about
  $1.9-2.7$, see \citep{Ricci2010,Ricci2012},
\item a [OI]\,63\,$\mu$m line flux of $25\times10^{-18}\rm\,W/m^2$
  -- typical observed values for non-outflow sources are
    about $(3-50)\times10^{-18}\rm\,W/m^2$, see \citep{Howard2013},
\item a $^{12}$CO $J\!=\!2\!\to\!1$ line flux of about 15 Jy\,km/s,
  and a $^{12}$CO/$^{13}$CO line ratio of about 5 -- for typical
    values, see \citep{Williams2014}, 
\item an apparent radius (semi-major axis) in the $^{12}$CO
  $J\!=\!2\!\to\!1$ line of about 450\,AU 
 (3.5$\arcsec$ at 140\,pc), typical observed
    values are about $1\arcsec-5\arcsec$, see \citep{Williams2014} .
\item weak CO ro-vibrational lines with a broad, box-shaped emission
  profile mostly emitted from the far side of the inner rim, which are
  not very typical with respect to observations. The ``central nose''
  on top of the averaged line profile is a contribution from low-$J$
  lines which are also emitted from more distant disk regions, see
  Fig.~\ref{fig:COrovib_profiles}. For typical CO ro-vibrational
  observations, see Sect.~\ref{sec:COrovib}.
\end{my_itemize}
All other emission lines in the IR to far-IR spectral region are
rather weak, which would likely result in non-detections with current
instruments (for example o-H$_2$O 63.3\,$\mu$m, CO
$J\!=\!18\!\to\!17$, o-H$_2$\,17.03\,$\mu$m), maybe except for the optical
[OI]\,6300\,\AA\ line (model flux $7\times10^{-18}\rm\,W/m^2$).  The
CO fundamental ro-vibrational lines are also rather weak in the reference model
(of order $2\times10^{-18}\rm\,W/m^2$ at $\rm FWHM\!=\!130\,km/s)$,
which is below the detection limit of \eg the CRIRES spectrograph
(about $10^{-18}\rm\,W/m^2$ at $\rm FWHM\!=\!20\,km/s)$.

\subsection{Impact of selected model parameters}
\label{sec:ParaObs}

\subsubsection{Dust size and opacity parameters}
\label{sec:dusteffect}

\begin{table}
  \vspace*{4mm}
  \begin{center}
  \caption{Unsettled dust properties in the reference model in
    comparison to a MRN size distribution and 
    uniform $a\!=\!0.1\,\mu$m dust particles.}
  \label{tab:dustprop}
  \vspace*{-5mm}
  \resizebox{90mm}{!}{\begin{tabular}{rl|c|c|c}
  \hline
            & & $\!\!$ref.~model$\!\!$  & MRN & $0.1\,\mu$m\\
  \hline
  \hline
  & & & \\[-1.8ex]
  $\!\!\!\!\!$dust material density $\rho_{\rm d}$ 
  & $\!\!\!\!\!\rm[g/cm^3]\!\!$   
              & 2.09             & 3.0            & 3.0\\
  mean dust size $\langle a\rangle$ 
  & $\!\!\!\!\![\mu$m] 
              & 0.083            & 0.0083         & 0.1\\
  mean dust size $\langle a^2\rangle^{1/2}$ 
  & $\!\!\!\!\![\mu$m]
              & 0.11             & 0.010          & 0.1\\
  mean dust size $\langle a^3\rangle^{1/3}$ 
  & $\!\!\!\!\![\mu$m]
              & 0.53             & 0.016          & 0.1\\
  particle density $n_{\rm d}/\nH$ 
  & 
              & $1.7(-14)$ 
              & $4.9(-10)$ 
              & $1.9(-12)$ \\
  surface $n_{\rm d}\,4\pi\langle a^2\rangle/\nH$ 
  & $\!\!\!\!\!\rm[cm^2]$
              & $\!\rm 2.7(-23)\!\!$
              & $\!\rm 6.6(-21)\!\!$
              & $\!\rm 2.3(-21)\!\!$ \\
  $\!\!\!\!\!$FUV extinct.\ $\kappa^{\,\rm ext}_{\,912\AA}/\nH$ 
  & $\!\!\!\!\!\rm[cm^2]$
              & $\!\rm 2.5(-23)\!\!$
              & $\!\rm 2.8(-21)\!\!$
              & $\!\rm 1.2(-21)\!\!$ \\
  FUV dust albedo 
  & 
              & 64\%             & 33\%           & 47\%\\
  $\!\!$mm opacity $\kappa^{\,\rm abs}_{1.3{\rm mm}}/\rho$ 
  & $\!\!\!\!\!\rm[cm^2/g]\!\!$
              & $\!\!0.038\!\!\!$
              & $\!\!0.0018\!\!\!$
              & $\!\!0.0018\!\!\!$\\[0.4ex]
  \hline
  \end{tabular}}
  \end{center}
  \vspace*{-3mm}
  \tiny{Notation $\rm a\,(-b)$ means $\rm a\times 10^{\,-b}$.
    The MRN model assumes
    $f(a)\!\propto\!a^{-3.5}$ from $\amin\!=\!0.005\,\mu$m to
    $\amax\!=\!0.25\,\mu$m. For the MRN and
    $a\!=\!0.1\,\mu$m models, we use Mie opacities for astronomical
    silicates \citep{Draine1984,Laor1993}.  All models assume
    $\rho\!=\!(1.4\,{\rm amu})\times\nH$ and a dust/gas ratio of
    $\delta\,=\!0.01$.}
  \vspace*{-4mm}
\end{table}

\begin{table}
\begin{center}
\caption{Model parameters, and values for the reference model.}
\vspace*{-0.5mm}
\label{tab:refmodel}
\resizebox{88mm}{!}{\begin{tabular}{l|c|c}
\\[-3.8ex]
\hline
 quantity & symbol & value\\
\hline 
\hline 
&&\\[-2.2ex]
stellar mass                      & $M_{\star}$      & $0.7\,M_\odot$\\
effective temperature             & $T_{\star}$      & $4000\,$K\\
stellar luminosity                & $L_{\star}$      & $1\,L_\odot$\\
UV excess                         & $f_{\rm UV}$     & $0.01$\\
UV powerlaw index                 & $p_{\rm UV}$     & $1.3$\\
X-ray luminosity                  & $L_X$           & $10^{30}\rm erg/s$\\
X-ray emission temperature        & $T_{X,\rm fit}$  & $2\times10^7$\,K\\
\hline
&&\\[-2.2ex]
strength of interstellar UV       & $\chi^{\rm ISM}$ & 1\\
strength of interstellar IR       & $\chi^{\rm ISM}_{\rm IR}$ & 0\\
cosmic ray H$_2$ ionisation rate  & $\zeta_{\rm CR}$   
                                  & $\!\!1.7\times 10^{-17}$~s$^{-1}\!\!\!$\\
\hline
&&\\[-2.2ex]
disk mass$^{(1)}$                 & $M_{\rm disk}$   & $0.01\,M_\odot$\\
dust/gas mass ratio$^{(1)}$       & $\delta$        & 0.01\\
inner disk radius                 & $R_{\rm in}$     & 0.07\,AU\\
tapering-off radius               & $R_{\rm tap}$    & 100\,AU\\
column density power index        & $\epsilon$      & 1\\
reference scale height         & $H_{\rm g}(100\,{\rm AU})$ & 10\,AU\\
flaring power index               & $\beta$         & 1.15\\ 
\hline
&&\\[-2.2ex]
minimum dust particle radius      & $a_{\rm min}$         & $0.05\,\mu$m\\
maximum dust particle radius      & $a_{\rm max}$         & $3\,$mm\\
dust size dist.\ power index      & $a_{\rm pow}$         & 3.5\\
turbulent mixing parameter        & $\alpha_{\rm settle}$ & 0.01\\
max.\ hollow volume ratio         & $V_{\rm hollow}^{\rm max}$   & 80\%\\
dust composition                  & $\rm Mg_{0.7}Fe_{0.3}SiO_3$ & 60\%\\
(volume fractions)                & amorph.\,carbon            & 15\%\\
                                  & porosity                   & 25\%\\
\hline
&&\\[-2.2ex]
PAH abundance rel. to ISM         & $f_{\rm PAH}$        & 0.01\\
chemical heating efficiency       & $\gamma^{\rm chem}$  & 0.2\\
\hline   
&&\\[-2.2ex]
distance                          & $d$ & 140\,pc\\
disk inclination                  & $i$ & 45\degr\\
\hline
\end{tabular}}
\end{center}
\vspace*{-3mm}
\resizebox{90mm}{!}{
\begin{minipage}{100mm}{$^{(1)}$: The total disk dust mass
      $M_{\rm dust}$ and the dust/gas ratio $\delta$ are the primary
      parameters in this paper, whereas the total disk mass $M_{\rm
        disk}\!=\!M_{\rm gas}$ is derived from $M_{\rm dust}$ and
      $\delta$.  Changing $M_{\rm disk}$ means to change $M_{\rm
        dust}$ as well, by the same factor, whereas changing $\delta$
      (or the inverse called ``gas/dust'' later) means to change
      $M_{\rm disk}$, but not to change $M_{\rm dust}$. The chemical
      heating efficiency $\gamma^{\rm chem}$ is an efficiency by which
      exothermal chemical reactions are assumed to heat the gas, see
      \citep[][see Appendix A.8 therein]{Woitke2011} for details.}
\end{minipage}}
\vspace*{-1mm}
\end{table}

It is important to realise that dust grains in protoplanetary disks
are likely to be very different from the tiny dust particles in the
diffuse interstellar medium (ISM) for which the astronomical silicate
opacities have been constructed by \citet{Draine1984}, only
considering $\lambda\!<\!1\,\mu$m, and using MRN \citep{Mathis1977}
size parameters.  In contrast, we expect the dust grains in disks to
be much larger, up to mm-sizes, which reduces the UV dust opacities by
a large factor (about 100) depending on parameters $\amin$, $\amax$
and $\apow$, see Table~\ref{tab:dustprop} and Fig.~\ref{fig:dustabs}.

This simple and straightforward fact distinguishes our disk models
from other chemical models (Table.~\ref{tab:other}).  In our models,
the dust is much more transparent in the UV, allowing the UV to
penetrate deeper into the disk, which increases the importance of
molecular self-shielding, and reduces the importance of X-rays
relative to the UV. 

Table~\ref{tab:dustprop} shows that a disk-typical dust size
distribution can be expected to have additional substantial impacts on
disk chemistry, see also \citet{Vasyunin2011}. The total grain surface
area per H nucleus, important for surface chemistry, H$_2$ formation
and photoelectric heating, is reduced by a factor of about 250 with
respect to the MNR dust model, and the dust particle concentration
$n_{\rm d}/\nH$ is only of order $10^{-14}$, where $n_{\rm d}\!=\!\int
f_0(a)\,da$ is the (unsettled) total dust particle density.  This
implies, for example, that even if every dust grain was negatively
charged once, there would be hardly any effect on the midplane
electron concentration.

Figure~\ref{fig:0.1mic} shows the results of the model when switching
to uniform $0.1\,\mu$m sized dust particles and astronomical silicate
opacities. The SED is now featured by stronger 10\,$\mu$m / 20\,$\mu$m
silicate emission features, higher far-IR continuum fluxes, and a quite
sudden kink around 200\,$\mu$m, followed by a steeper decline toward
millimetre wavelengths. The apparent size of the disk is smaller at
$1.6\,\mu$m, but larger at $10\,\mu$m and 1.3\,mm.  The disk is now
warmer in dust, but cooler in gas, in fact mean dust and gas
temperatures are more equal. Most emission lines in the model with
uniform $0.1\,\mu$m dust particles show weaker fluxes, by up to a
factor of ten. However, the CO\,$J\!=\!10\!\to\!9$ line doesn't follow
this general trend.

Understanding the impact of the dust size and material parameters on
gas temperature and emission lines can be isolated to the effect of a
single quantity, namely the dust optical depths at UV wavelengths
$\tau_{\rm UV}$, see Sects.~\ref{sec:OI63}, \ref{sec:COhighJ} and
\ref{sec:COrovib}. All dust parameter alterations that result in lower
$\tau_{\rm UV}$ will generally lead to a deeper penetration of UV into
the disk, causing an increase of the thickness of the warm molecular
gas layer, and this leads to stronger emission lines at optical to
far-IR wavelengths. The impact is less pronounced on (sub-)mm lines,
see Sect.~\ref{sec:mmCO}, although secondary temperature and chemical
effects are important to understand the (sub-)mm line ratios.

\begin{figure*}
  \vspace*{-2mm}
  \hspace*{-3mm}\begin{tabular}{c}
  \hline
  \\[-3mm]
  \resizebox{!}{2.5mm}{\bf simplified hydrostatic model}\\
  \hline
  \includegraphics[width=170mm]{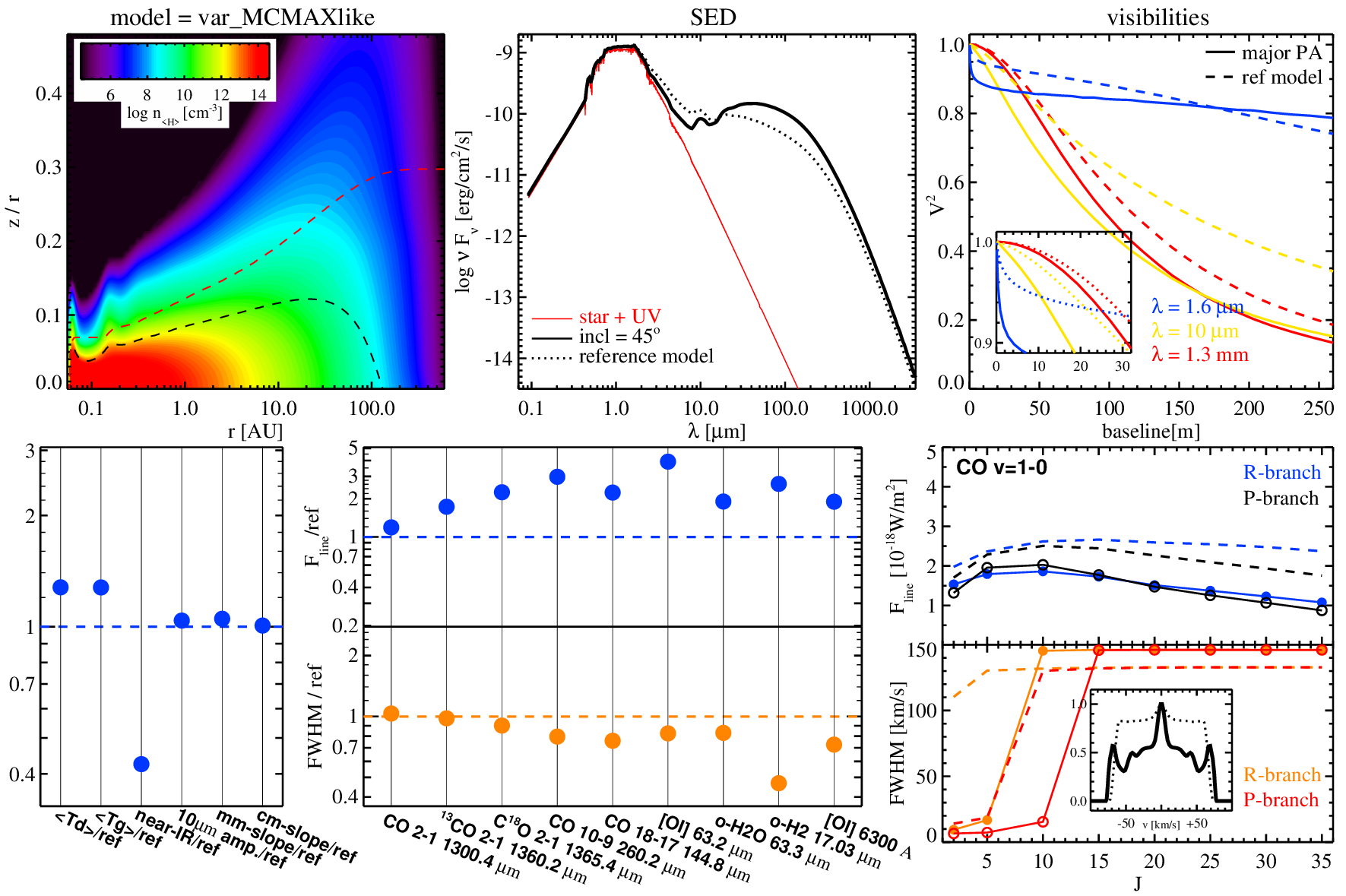}\\
  \hline
  \\[-3mm]
  \resizebox{!}{2.5mm}{\bf full hydrostatic model}\\
  \hline
  \includegraphics[width=170mm]{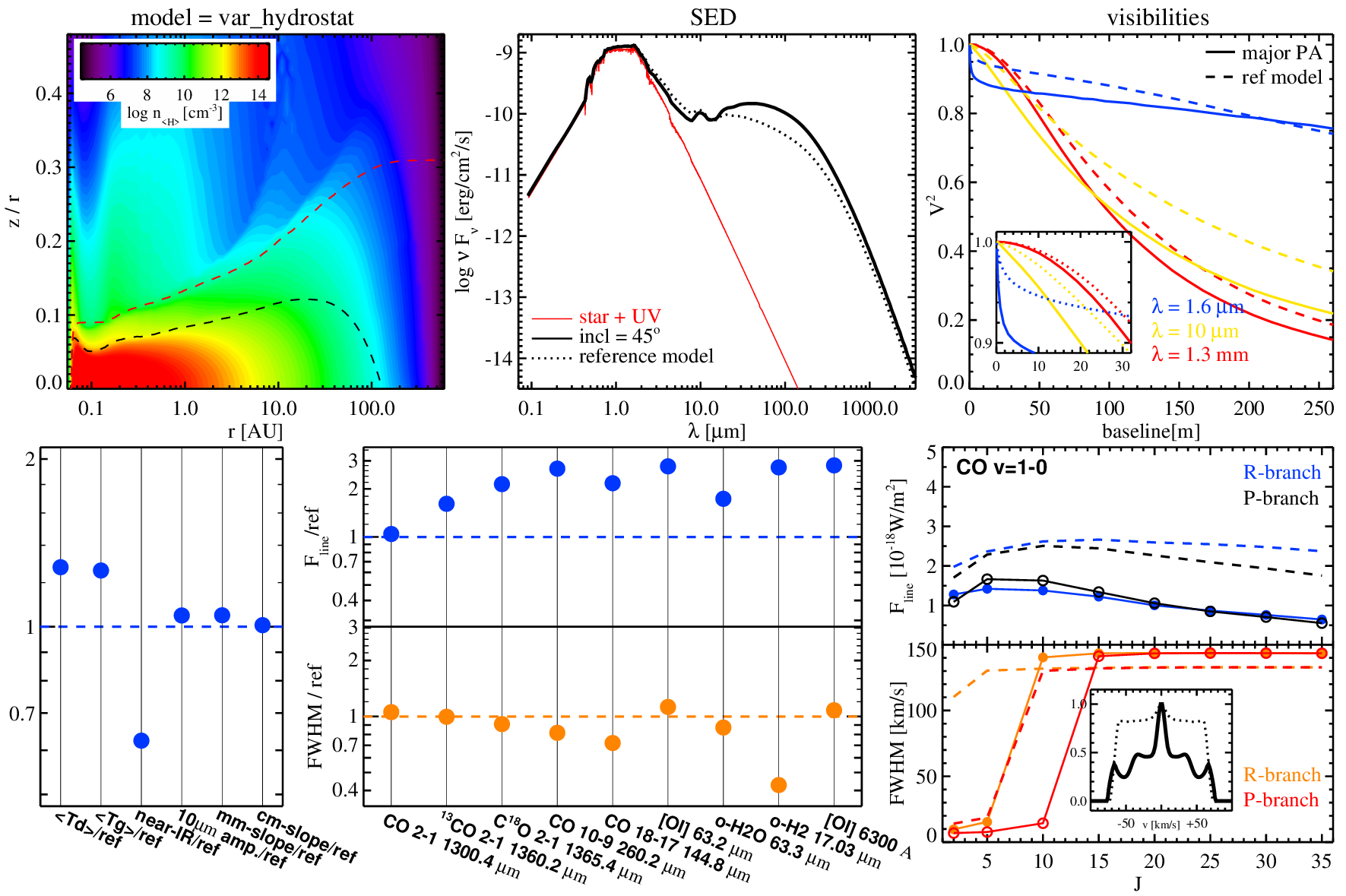}
  \end{tabular}
  \vspace*{-2mm}
  \caption{Two variants of $(1\!+\!1)$D hydrostatic disk
    models. Depicted quantities are explained in
    Fig~\ref{fig:refmodel}, but shown here in comparison to the
    reference model. The upper half shows the results for the
    simplified hydrostatic model based on the dust temperature $\Td$
    and a constant molecular weight, \ie sound speed
    $c_T^2\!=\!k\Td/(2.3\,{\rm amu})$. The lower half shows the full
    hydrostatic model based on the gas temperature structure $\Tg$ and
    the proper mean molecular weight $\mu$ as resultant from the
    chemistry and heating \& cooling balance, \ie
    $c_T^2\!=\!k\Tg/\mu$.}
  \label{fig:hydrostat}
\end{figure*}

\subsubsection{Hydrostatic disk models}
\label{sec:hydrostat}

Figure~\ref{fig:hydrostat} summarises the results of two variants of
(1+1)D hydrostatic models, where the vertical disk extension at any
radius is computed from the condition of hydrostatic equilibrium,
which requires an iterative approach of radiative transfer, chemistry
and gas heating/cooling balance, see \citep{Woitke2009a} for
details. The lower model is the proper hydrostatic solution, where the
pressure $p\!=\!\rho\,k\Tg/\mu$ is calculated according to the local
gas temperatures $\Tg$ and mean molecular weight $\mu$ resulting from
chemistry. The upper model is a simplified version thereof, where the
dust temperature is used instead ($\Tg\!\approx\!\Td$), and the mean
molecular weight is assumed to be constant ($\mu\!\approx\!2.3\,{\rm
  amu}$). The observable properties of these hydrostatic models are as
follows.
\begin{my_itemize}
\item The SEDs of both types of hydrostatic models cannot explain the
  observed levels of near-IR excess for T\,Tauri stars, because the
  inner rim is quite low. 
\item Between 20$\,\mu$m and 50$\,\mu$m, the SED displays an {\em
  \,increasing\,} slope, which is caused by the strong flaring of the
  outer disk \citep[see also Fig.~2 in][]{Meijerink2012}.
\item The models are substantially warmer, both in gas and dust, as
  compared to the reference model, again because of the flaring of the
  outer disk. 
\item The far-IR to mm emission lines are all stronger as compared to
  the reference model. The [OI]\,63.2\,$\mu$m flux is $(7-10)\times
  10^{-17}\rm\,W/m^2$, which is out 4 times stronger as in the reference
  model, and quite high with respect to observations. The
  $^{12}$CO/$^{13}$CO line ratio is as large as 8.
\item The ro-vibrational CO ro-vibrational lines are weaker, because of
  the low inner rim, and have complicated multi-component line
  profiles.
\end{my_itemize}
The two variants of hydrostatic models have quite similar observable
properties, including the spectral lines, although the disk gas
density structure is remarkably different. The proper hydrostatic
model displays larger amounts of extended hot gas high above the disk 
in the inner regions.  However, this highly extended hot gas is
purely atomic, hence it does not emit in molecular lines, one has to use
atomic tracers to detect it, such as [OI]\,6300\,\AA\ or maybe
[NeII]\,12.82\,$\mu$m. In summary, hydrostatic passive disk models
have some issues explaining the observed SED and line
properties of T\,Tauri stars.

\begin{figure*}
  \vspace*{-2mm}
  \hspace*{-3mm}\begin{tabular}{c}
  \hline
  \\[-3mm]
  \resizebox{!}{2.5mm}{\bf model with little disk flaring 
                       \ \ $\beta\!=\!1.05$}\\
  \hline
  \includegraphics[width=173mm]{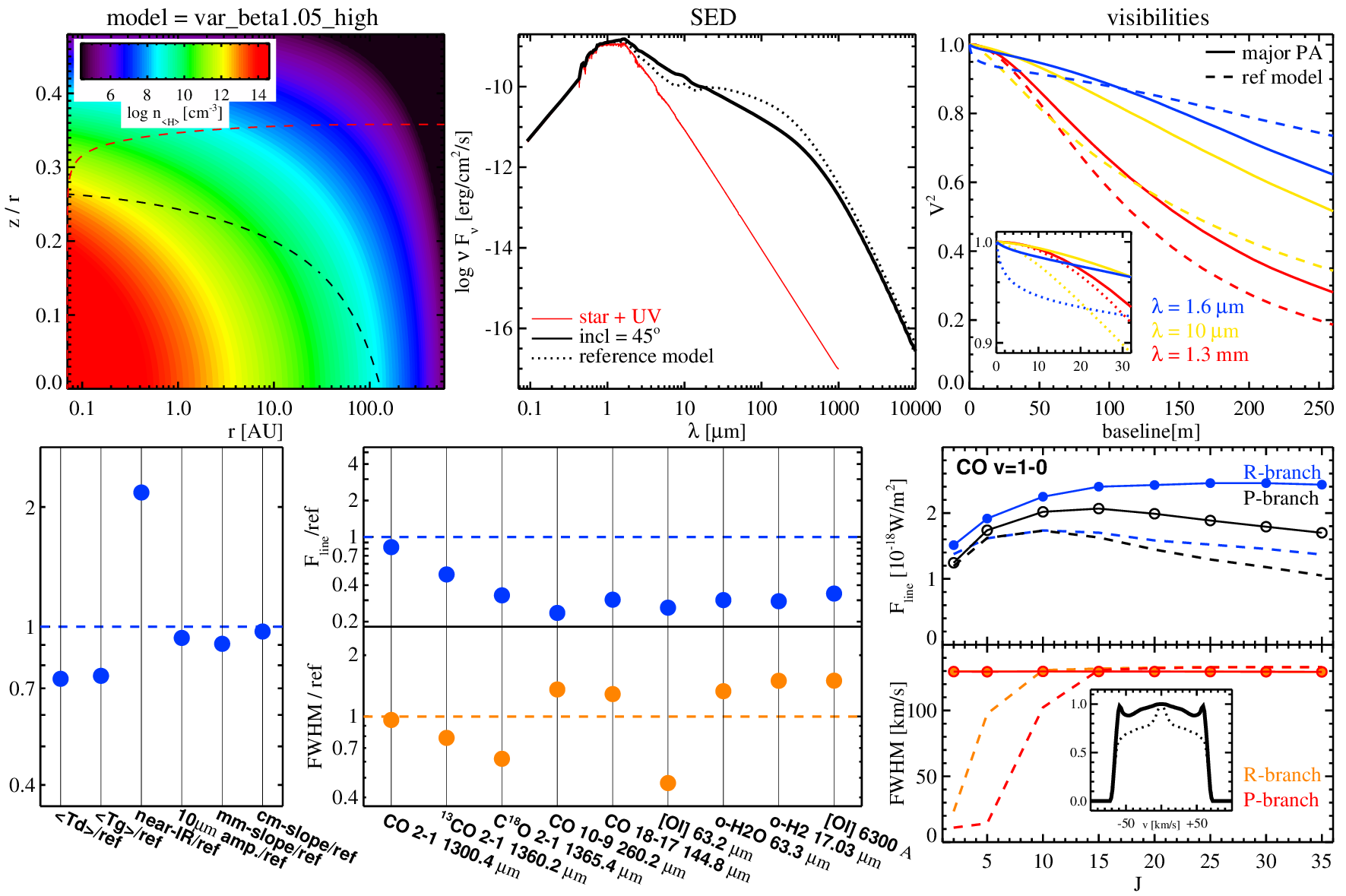}\\
  \hline
  \\[-3mm]
  \resizebox{!}{3.0mm}{\bf model with strong dust settling
                       \ \ $\alpha_{\rm settle}\!=\!10^{-4}$}\\
  \hline
  \includegraphics[width=173mm]{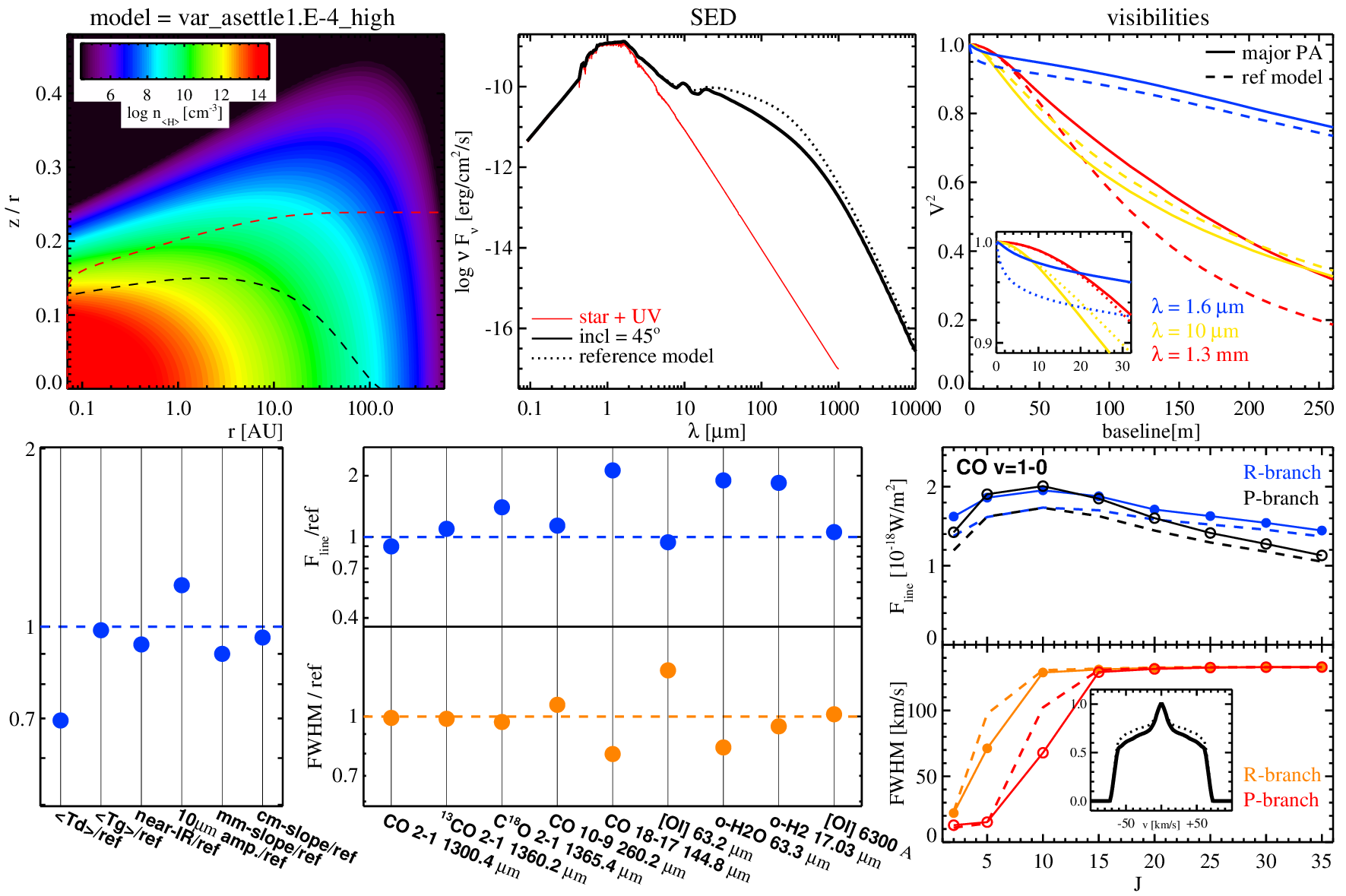}
  \end{tabular}
  \vspace*{-2mm}
  \caption{Effects of disk flaring and dust settling on all
    observables, with respect to the reference model. The upper set of
    figures shows a model with less flaring as compared to the
    reference model $\beta\!=\!1.05$, and the lower half shows a model
    with stronger dust settling $\alpha_{\rm settle}\!=\!10^{-4}$.
    See Fig.~\ref{fig:hydrostat} for further explanations.}
  \label{fig:settle}
\end{figure*}

\subsubsection{Disk flaring and/or dust settling?}
\label{sec:flaring}

Little disk flaring and strong dust settling have similar effects on
the SED, see Fig.~\ref{fig:settle}. The $\beta\!=\!1.05$ and
$\alpha_{\rm settle}\!=\!10^{-4}$ models have practically
indistinguishable SEDs beyond 20\,$\mu$m, with a more steeply
decreasing slope as compared to the reference model around
$\sim\!50\,\mu$m. The reason for that steeper slope is that a more
self-shadowed dust configuration leads to less interception of star
light per disk radius interval, hence to a steeper decline of the dust
temperatures as function of radius
\citep[e.g.][]{Beckwith1990,Chiang1997}.  We can see from
Fig.~\ref{fig:settle} that indeed the mass averaged dust temperature
$\langle\Td\rangle$, which is dominated by the outer disk regions, has
fallen from about 19\,K in the reference model to about 13\,K in both
cases.  Lacking disk flaring and strong dust settling both cause very
low dust temperatures in the midplane, of order 4\,K already at
$r\!=\!150\,$AU in the $\alpha_{\rm settle}\!=\!10^{-4}$ model, only
limited by CMB and other background radiation. These very cold disk
models have particular mm and cm-properties, because the cold dust is
not entirely emitting in the Rayleigh-Jeans limit even at millimetre
wavelengths, see Sect.~\ref{sec:mm-slope}.

\begin{figure*}
  \vspace*{-1mm}
  \hspace*{4mm}\resizebox{185mm}{!}{
  \begin{tabular}{rrr}
  \hspace*{-7.0mm}\includegraphics[height=41mm,trim= 0 40 0 0,clip]{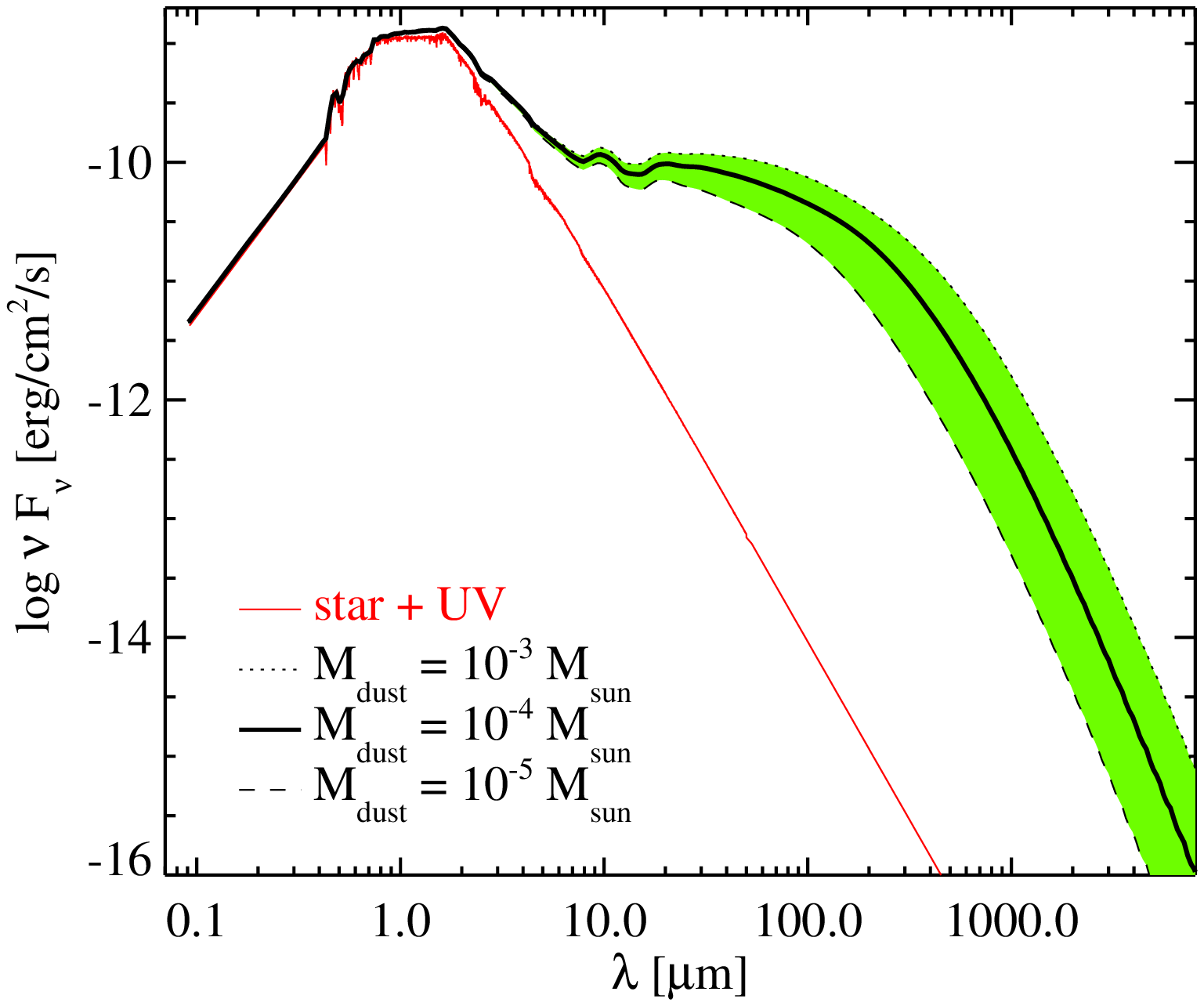}&
  \hspace*{-6.3mm}\includegraphics[height=41mm,trim=58 40 0 0,clip]{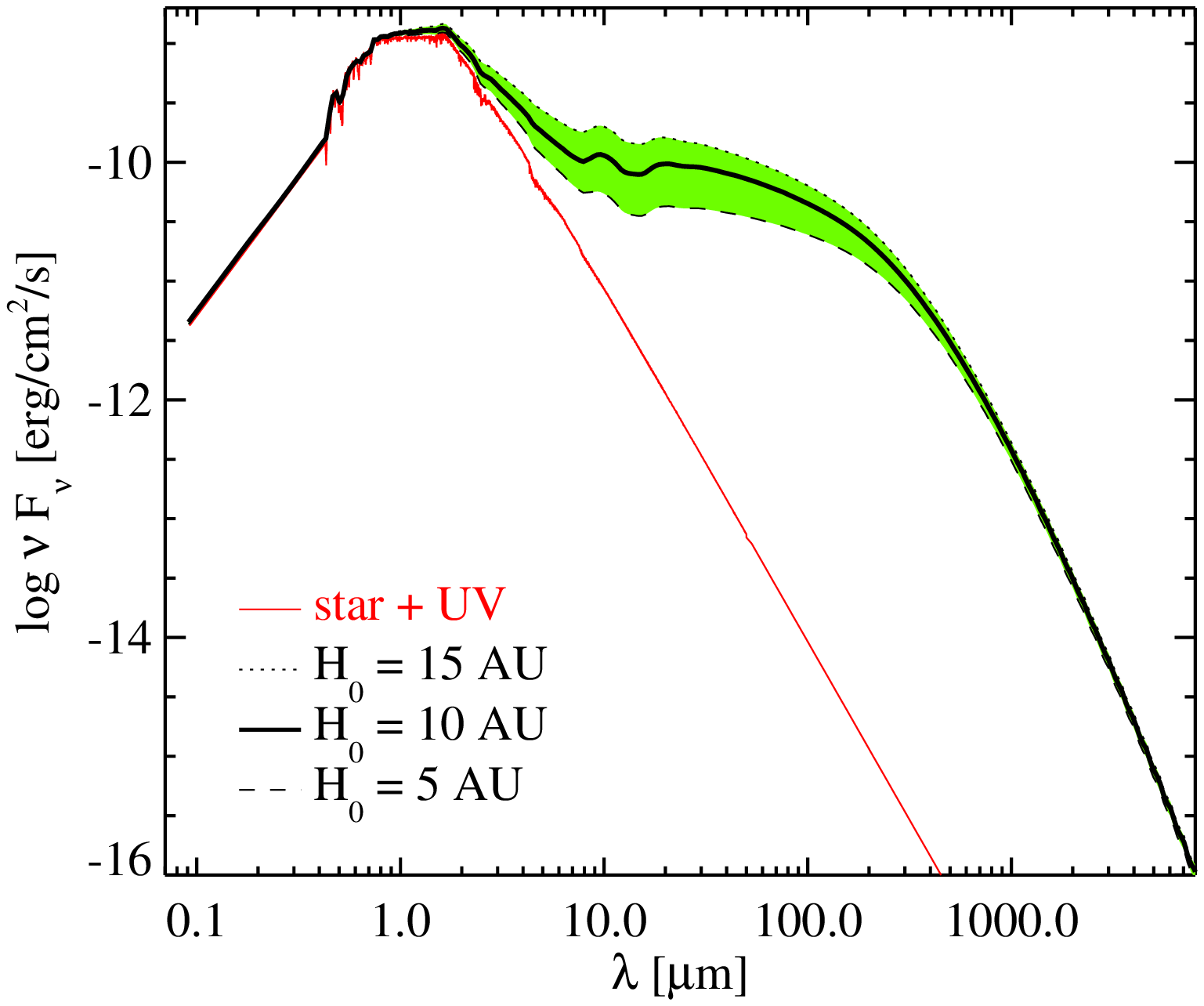} &
  \hspace*{-6.3mm}\includegraphics[height=41mm,trim=58 40 0 0,clip]{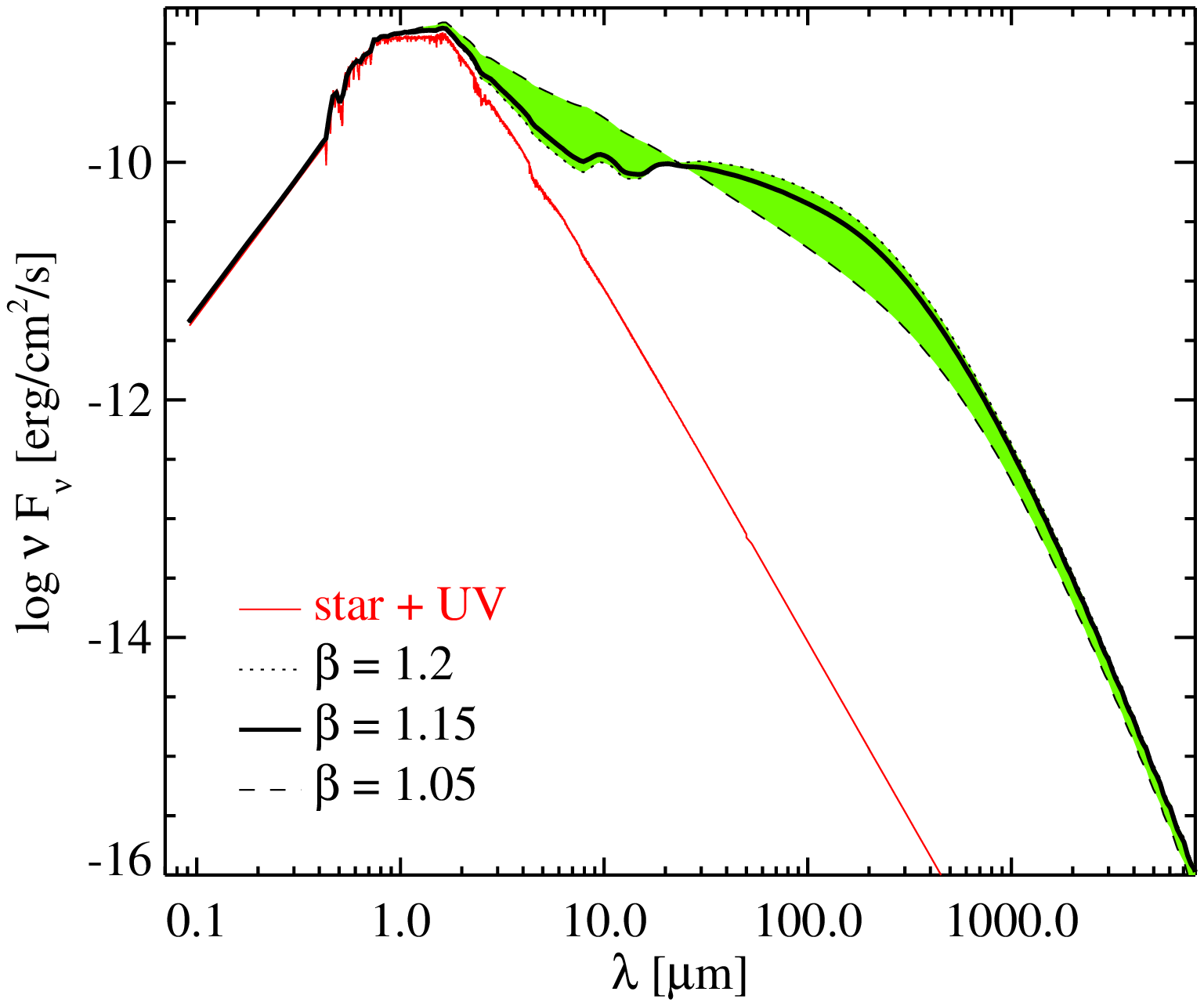} 
  \\[-40mm]
  \resizebox{!}{2.3mm}{\bf disk mass}\hspace*{4mm} & 
  \resizebox{!}{2.3mm}{\bf scale height}\hspace*{4mm} & 
  \resizebox{!}{2.2mm}{\bf flaring exponent}\hspace*{4mm}\\[33.9mm]
  \hspace*{-7.0mm}\includegraphics[height=41mm,trim= 0 40 0 0,clip]{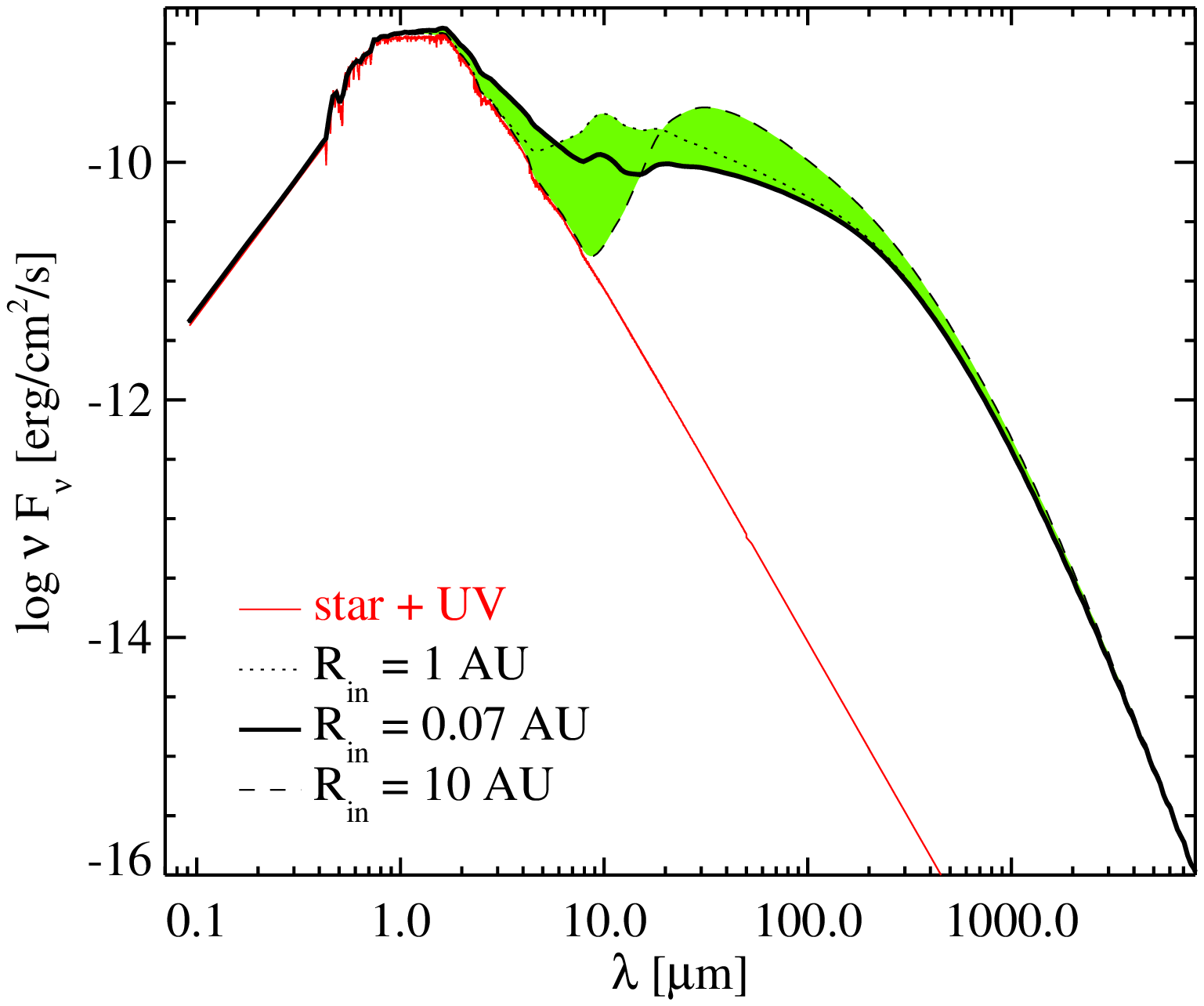} &
  \hspace*{-6.3mm}\includegraphics[height=41mm,trim=58 40 0 0,clip]{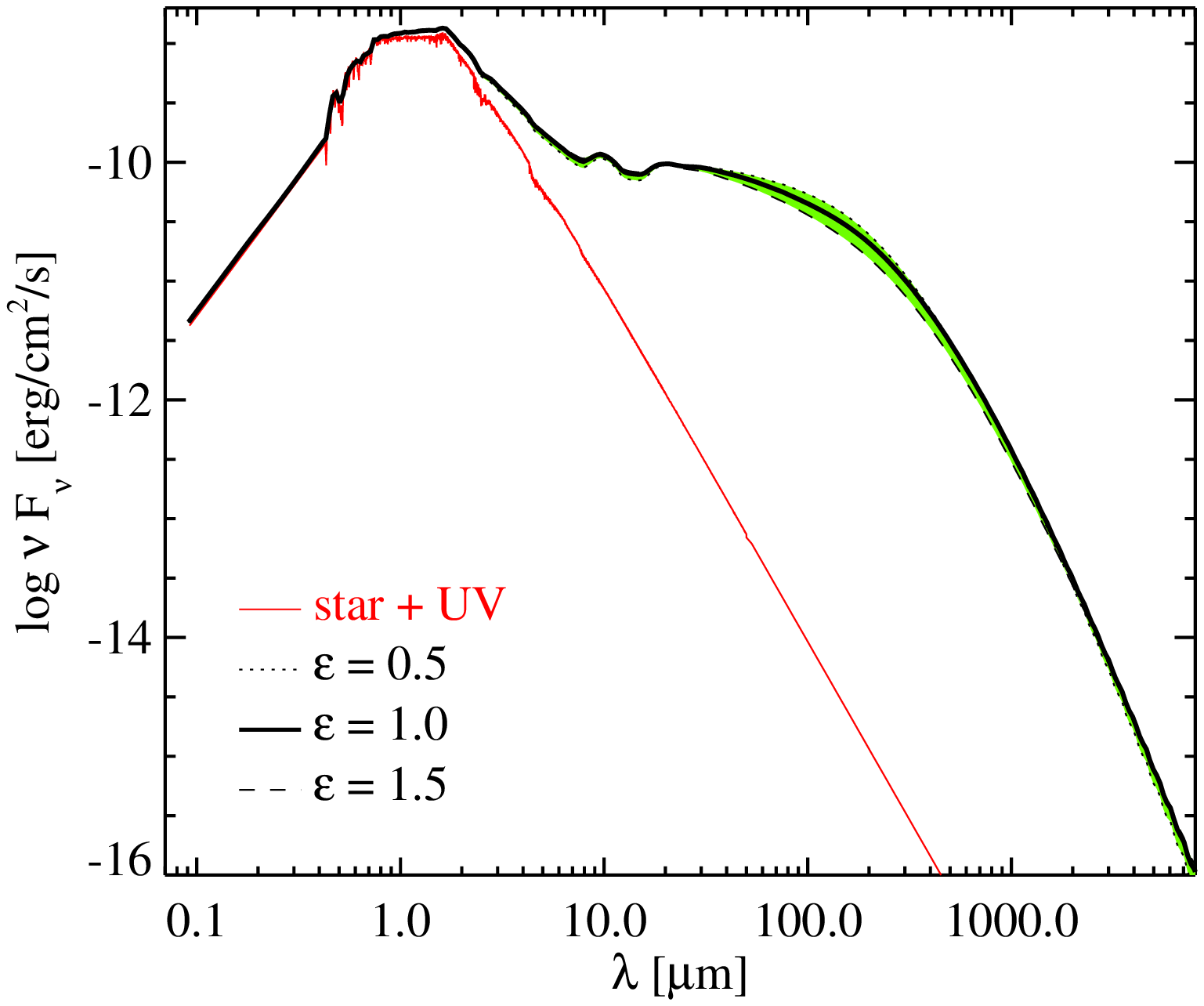} &
  \hspace*{-6.3mm}\includegraphics[height=41mm,trim=58 40 0 0,clip]{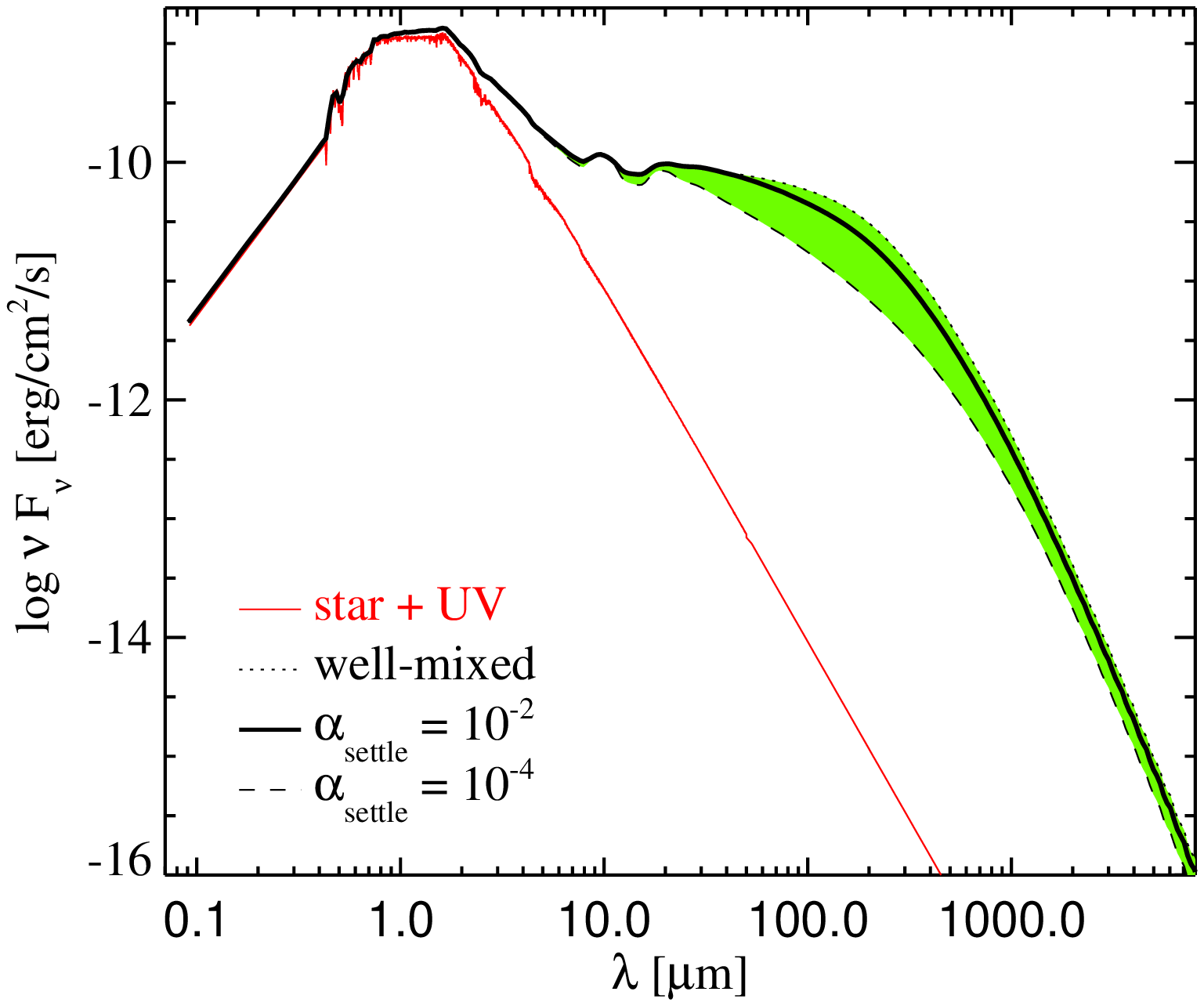} 
  \\[-40mm]
  \resizebox{!}{2.3mm}{\bf inner radius}\hspace*{4mm} & 
  \resizebox{!}{1.9mm}{\bf col.\ dens.\ power index}\hspace*{3mm} & 
  \resizebox{!}{2.3mm}{\bf dust settling}\hspace*{4mm}\\[33.9mm]
  \hspace*{-7.0mm}\includegraphics[height=46mm,trim= 0 0 0 0,clip]{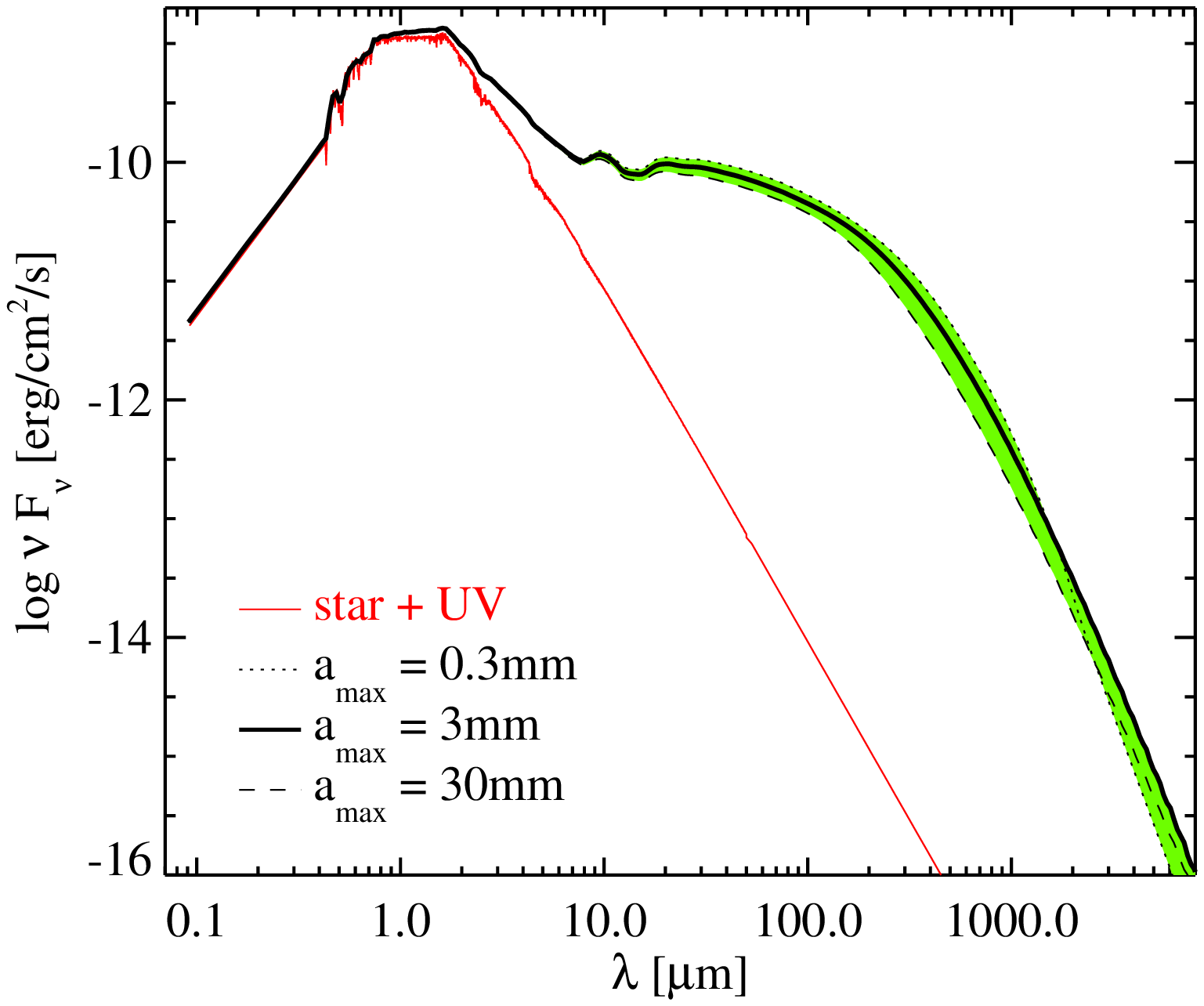} &
  \hspace*{-6.3mm}\includegraphics[height=46mm,trim=58 0 0 0,clip]{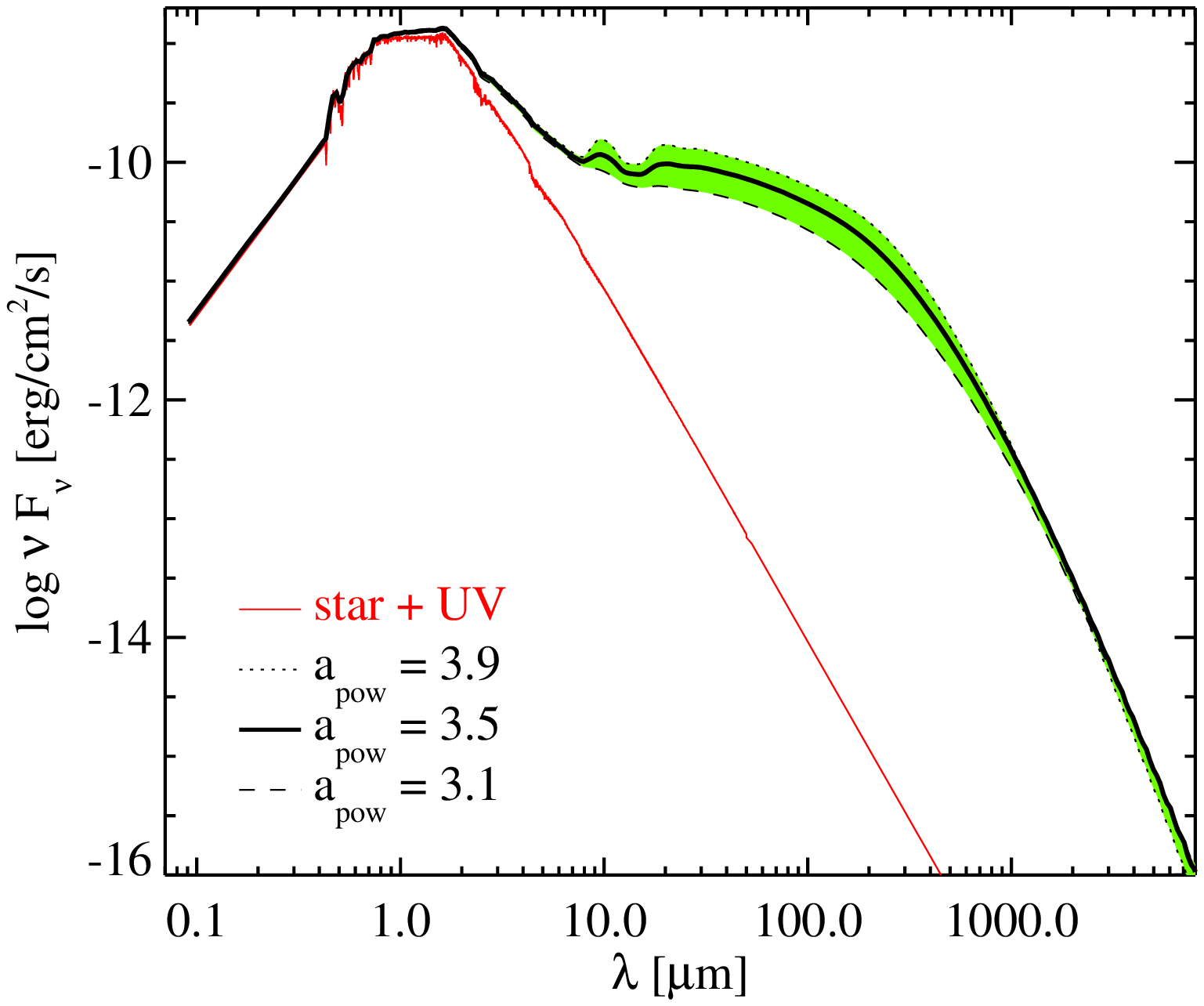} &
  \hspace*{-6.3mm}\includegraphics[height=46mm,trim=58 0 0 0,clip]{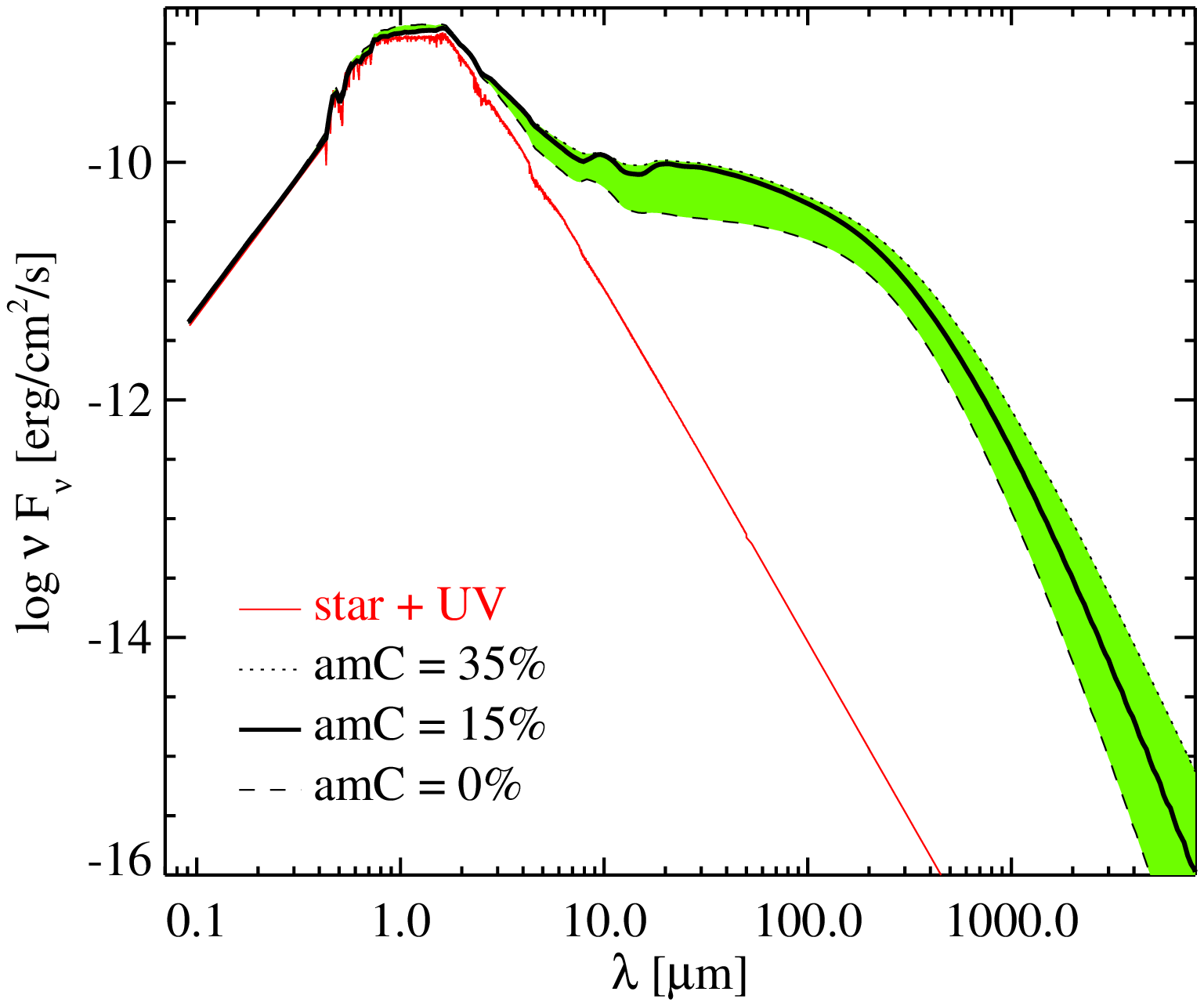}
  \\[-45mm]
  \resizebox{!}{2.3mm}{\bf max.\ dust size}\hspace*{4mm} & 
  \resizebox{!}{1.9mm}{\bf dust size power index}\hspace*{3mm} & 
  \resizebox{!}{2.3mm}{\bf amorph.\ carbon}\hspace*{4mm}\\[40mm]
  \end{tabular}}
  \caption{Effect of dust and disk parameters on model SED at distance
    140\,pc and inclination $45^\circ$. The thick full black line is
    the reference model (identical in every part figure), whereas the
    green shaded area indicates the effect of a single parameter on
    the SED, where the dashed and dotted lines correspond to the
    changed parameter values as annotated.  {\bf\ \ Top row:} dust
    mass $M_{\rm dust}$, scale height $H_0$, and flaring exponent
    $\beta$.  {\bf\ \ 2$^{\bf nd}$ row:} inner radius $R_{\rm in}$,
    column density powerlaw index $\epsilon$, and dust settling
    parameter $\alpha_{\rm settle}$.  {\bf\ \ 3$^{\bf rd}$ row:}
    maximum grain size $a_{\rm max}$, dust size powerlaw index $a_{\rm
      pow}$, and volume ratio of amorphous carbon.  The
    dependencies of the SED on the tapering-off radius $R_{\rm tap}$
    (not shown), on the outer radius $R_{\rm out}$ (not shown), and on
    the minimum dust particle size $a_{\rm min}$ (not shown) are less
    than the one shown for $\epsilon$.}
  \label{fig:SEDeffect}
  \vspace*{-2mm}
\end{figure*}

\begin{figure*}
  \hspace*{2mm}\resizebox{186mm}{!}{
  \begin{tabular}{lll}
  \hspace*{-7mm  }\includegraphics[height=46mm  ,trim= 0 42 0  0,clip]
                  {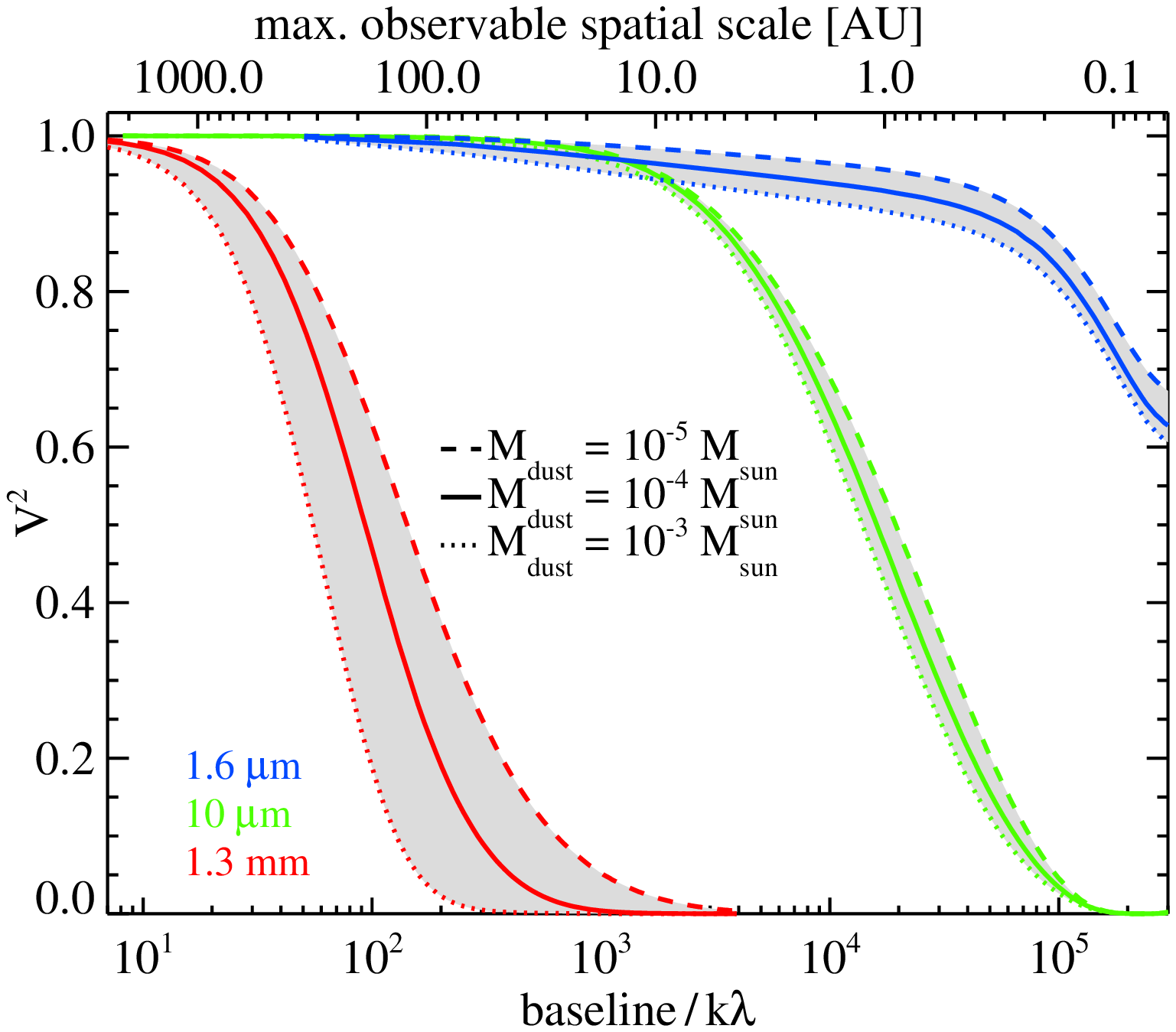}&
  \hspace*{-5.5mm}\includegraphics[height=46mm  ,trim=45 42 0  0,clip]
                  {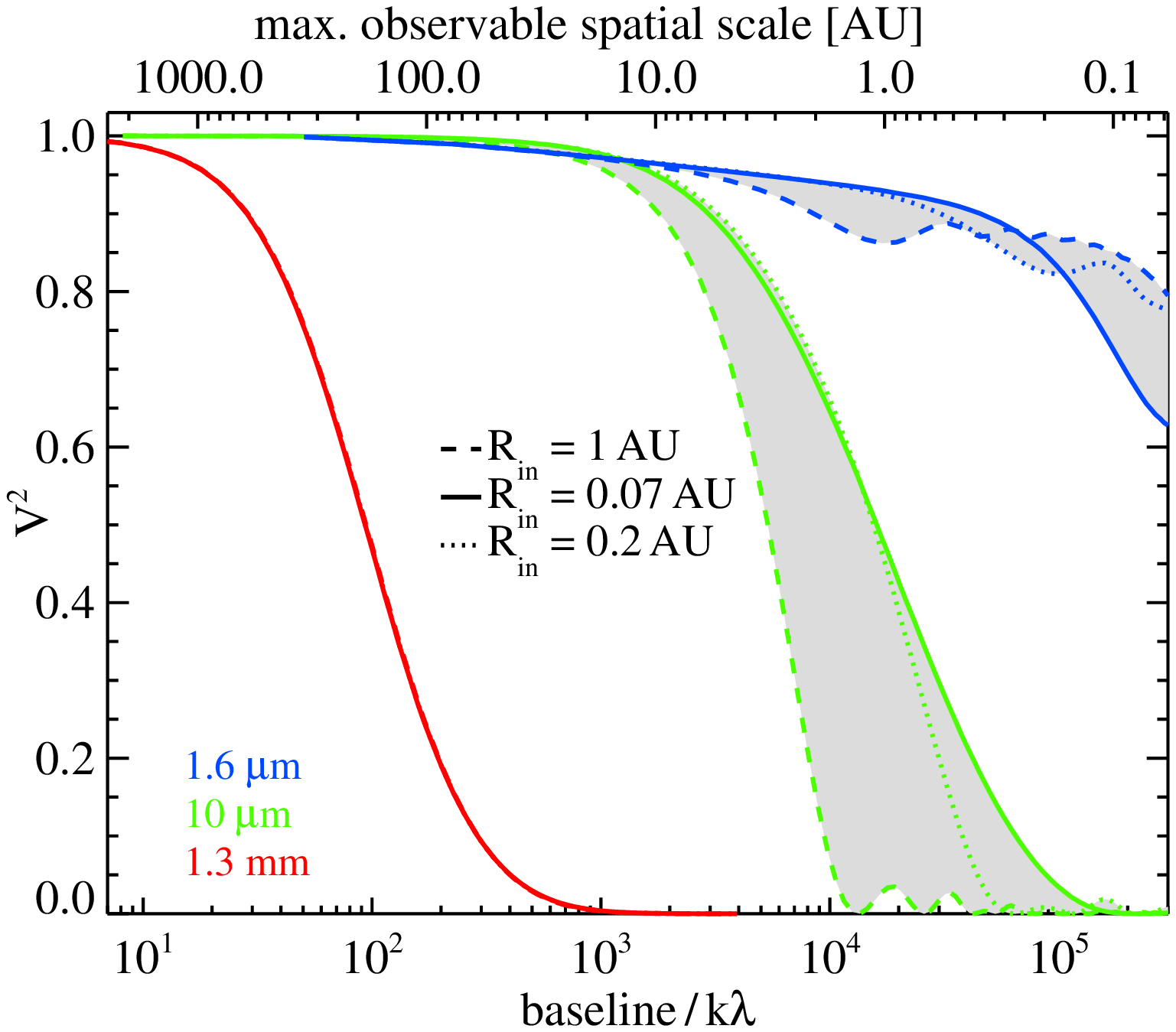}&
  \hspace*{-5.5mm}\includegraphics[height=46mm  ,trim=45 42 0  0,clip]
                  {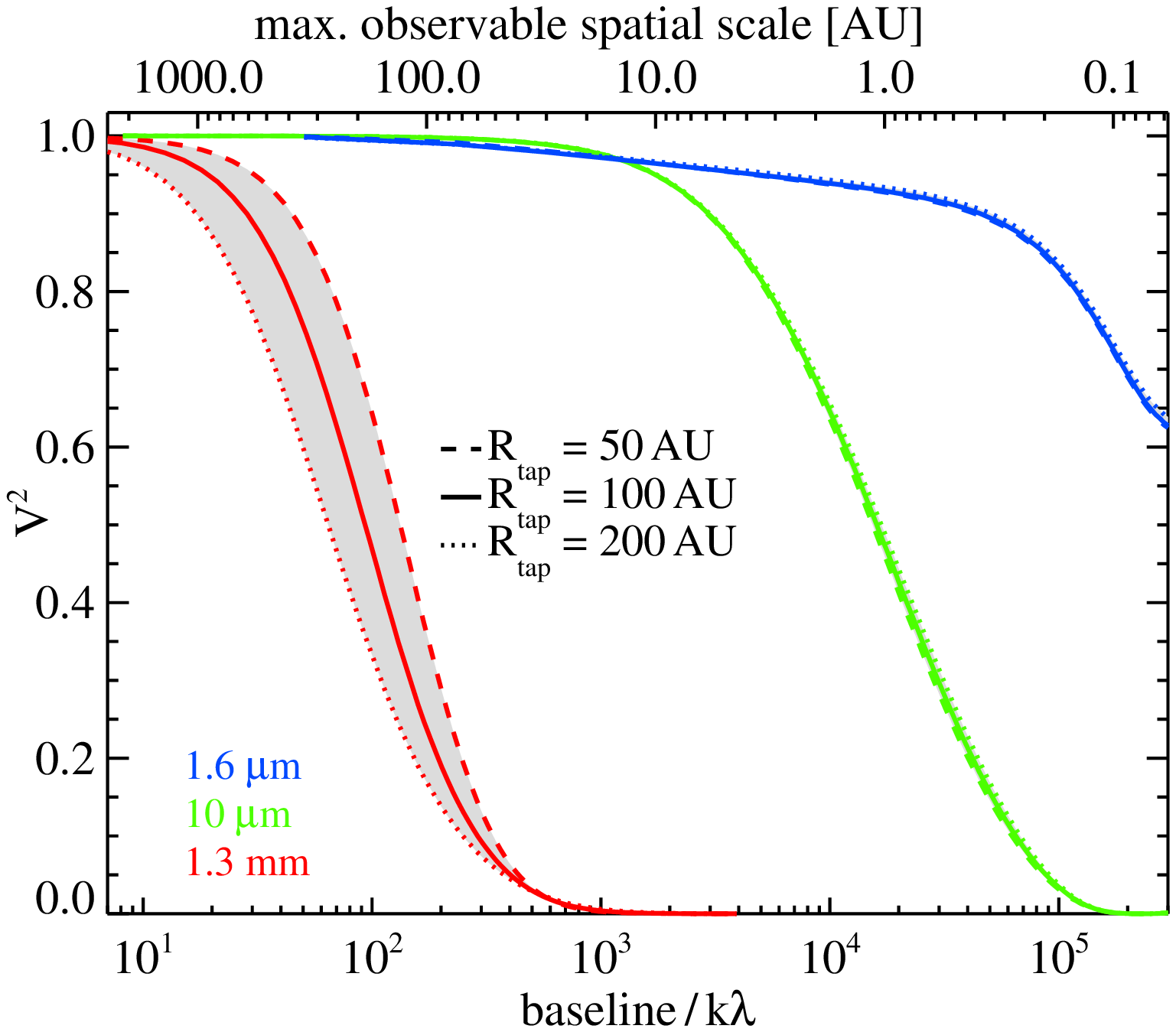}
  \\[-30mm]
  \hspace*{13mm}\resizebox{!}{2.0mm}{\bf dust mass} & 
  \hspace*{7mm}\resizebox{!}{2.0mm}{\bf inner radius} & 
  \hspace*{8mm}\resizebox{!}{2.0mm}{\bf tapering-off radius}\\[24.6mm]
  \hspace*{-7mm  }\includegraphics[height=40.1mm,trim= 0 42 0 46,clip]
                  {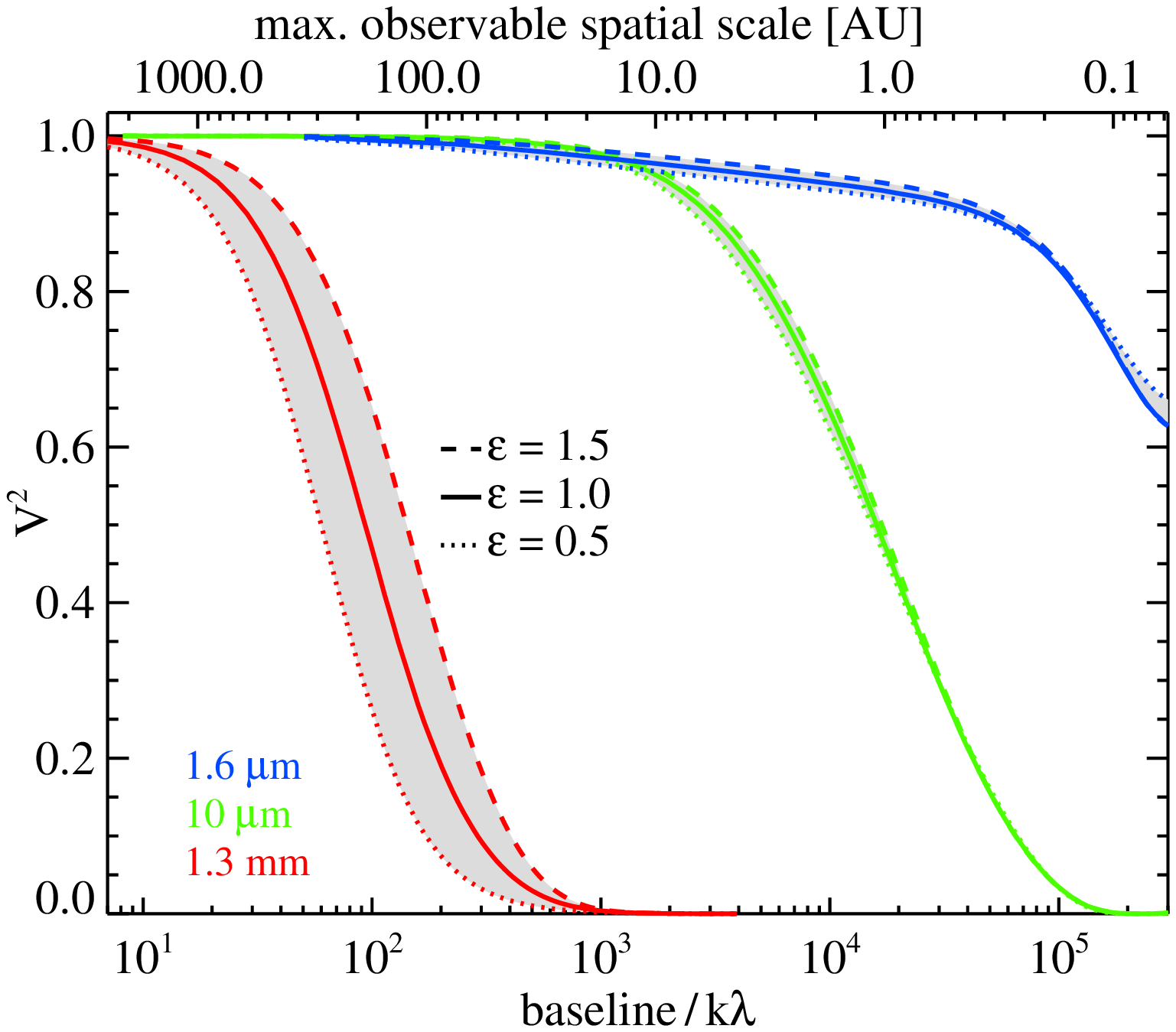}&
  \hspace*{-5.5mm}\includegraphics[height=40.1mm,trim=45 42 0 46,clip]
                  {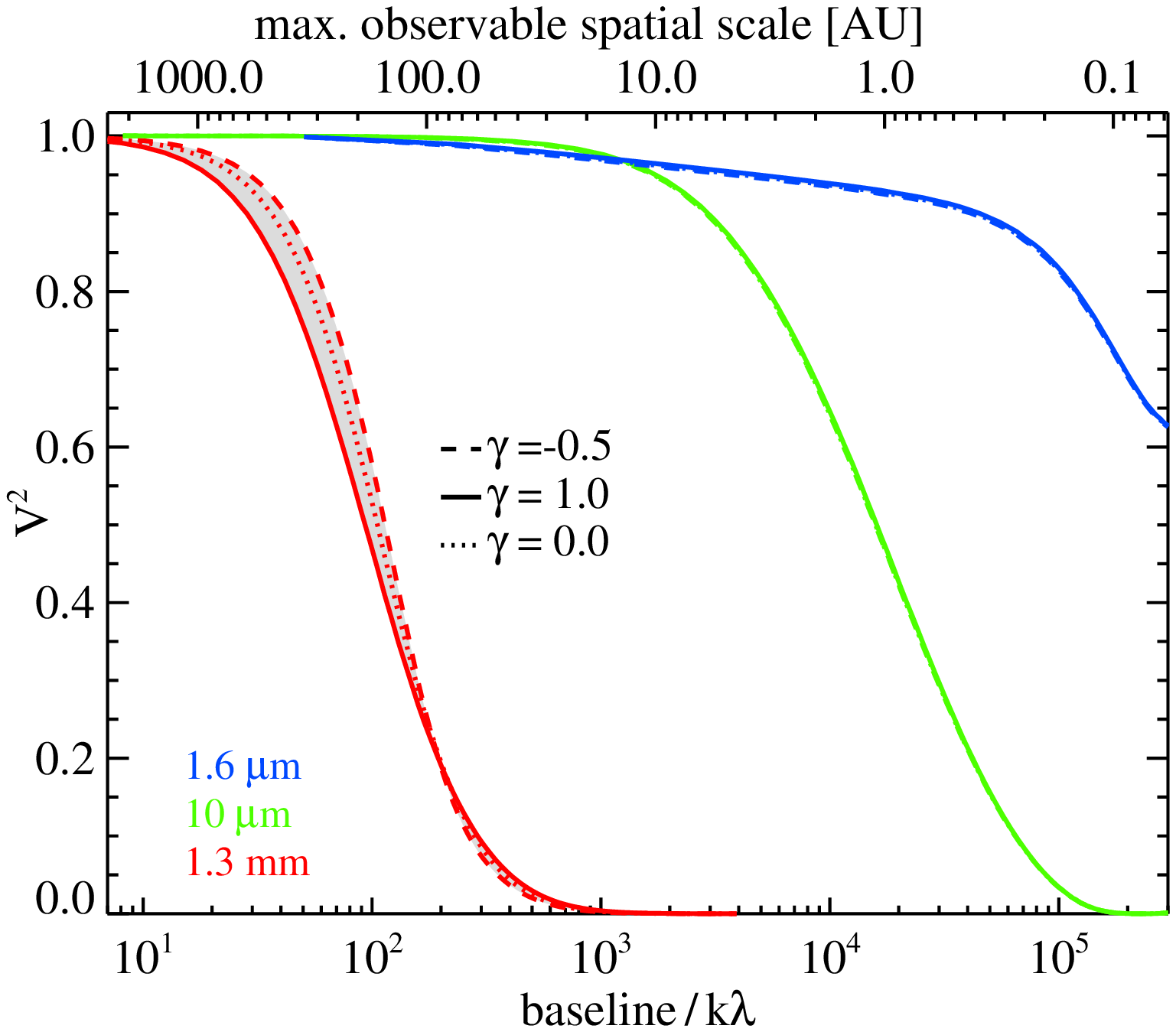}&
  \hspace*{-5.5mm}\includegraphics[height=40.1mm,trim=45 42 0 46,clip]
                  {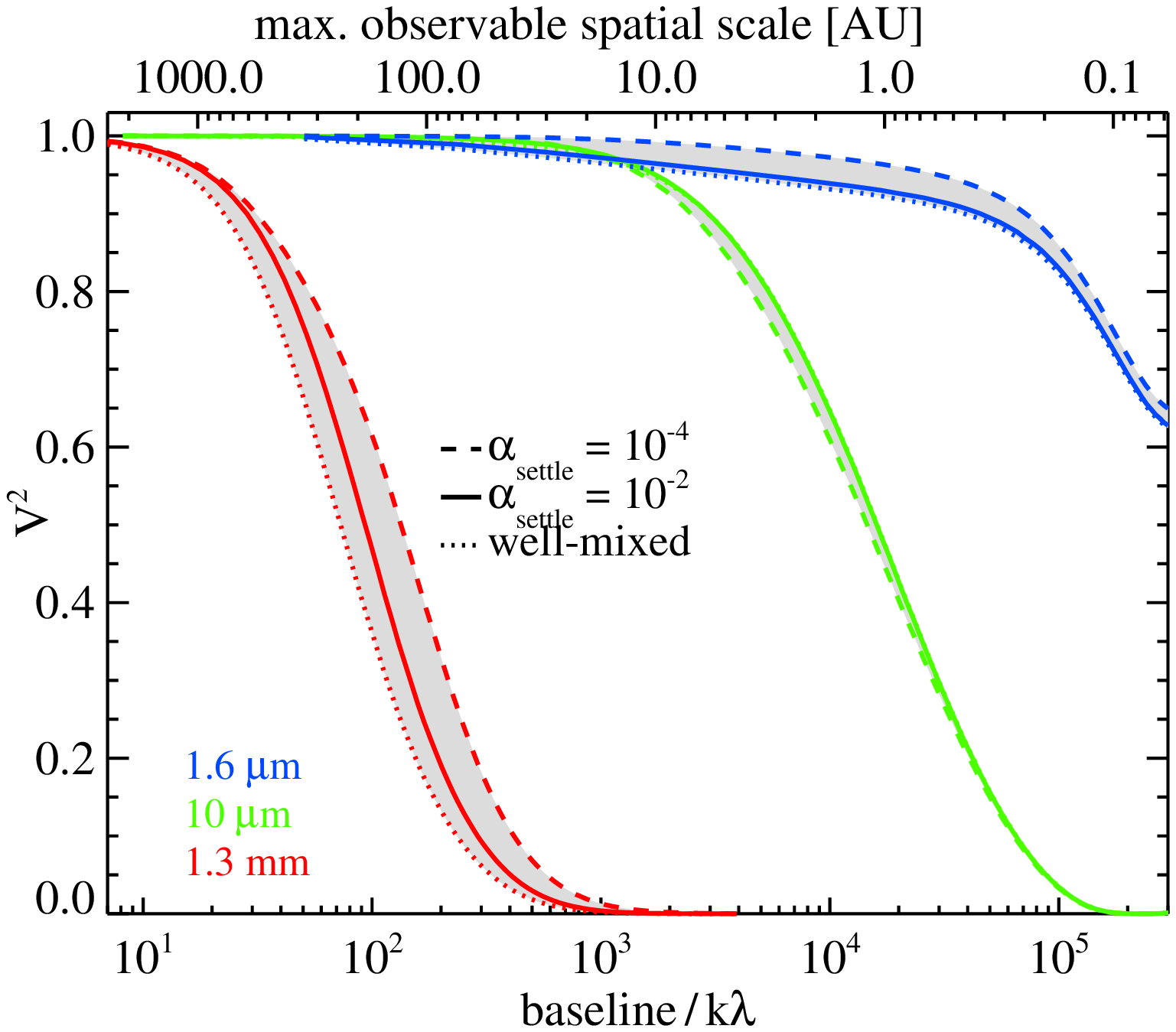}
  \\[-30mm]
  \hspace*{13mm}\resizebox{!}{2.0mm}{\bf col.\,dens.\,power-index} & 
  \hspace*{7mm}\resizebox{!}{2.0mm}{\bf tapering-off exponent} & 
  \hspace*{8mm}\resizebox{!}{2.0mm}{\bf dust settling}\\[24.6mm]
  \hspace*{-7mm  }\includegraphics[height=40.1mm,trim= 0 42 0 46,clip]
                  {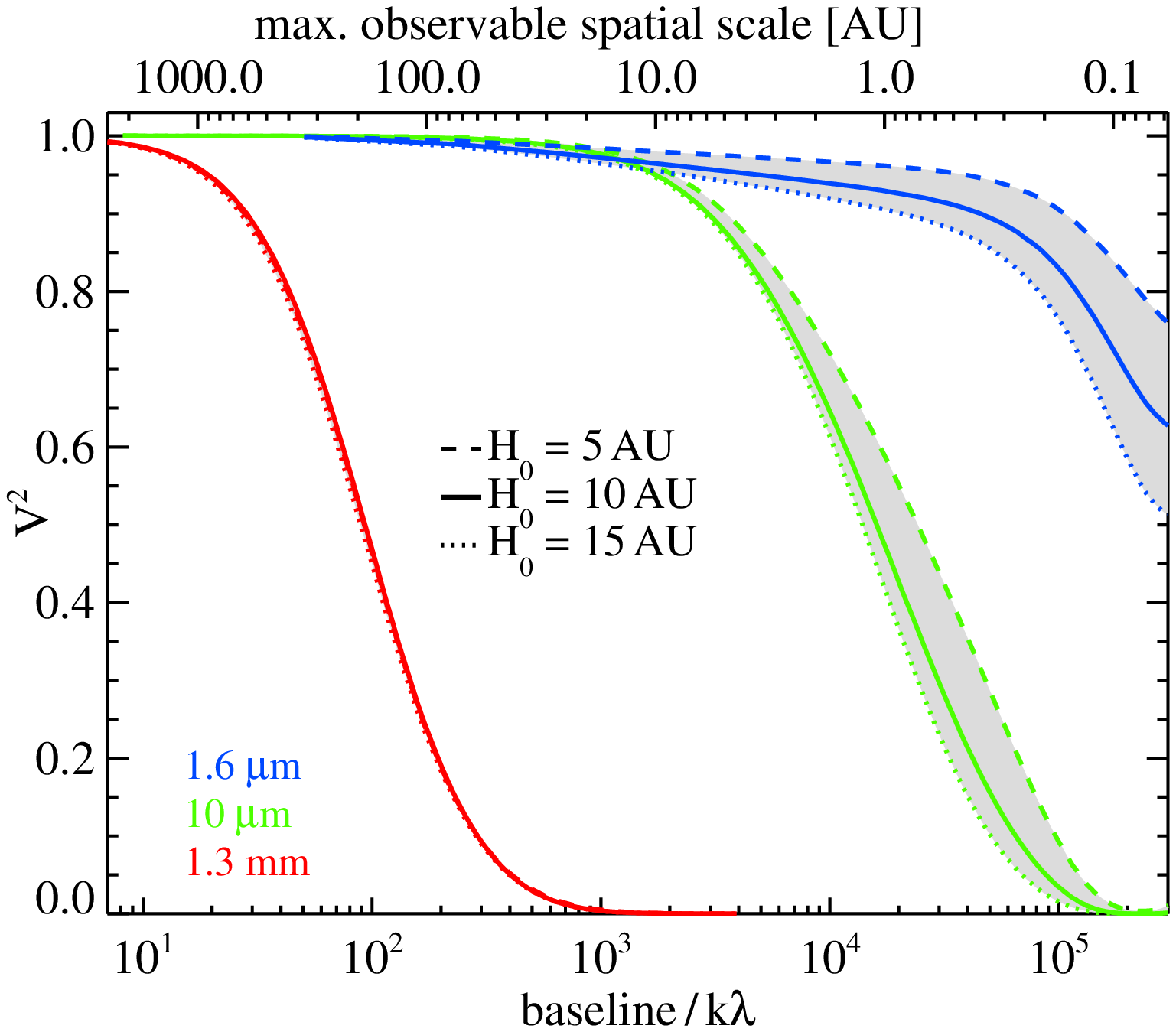}&
  \hspace*{-5.5mm}\includegraphics[height=40.1mm,trim=45 42 0 46,clip]
                  {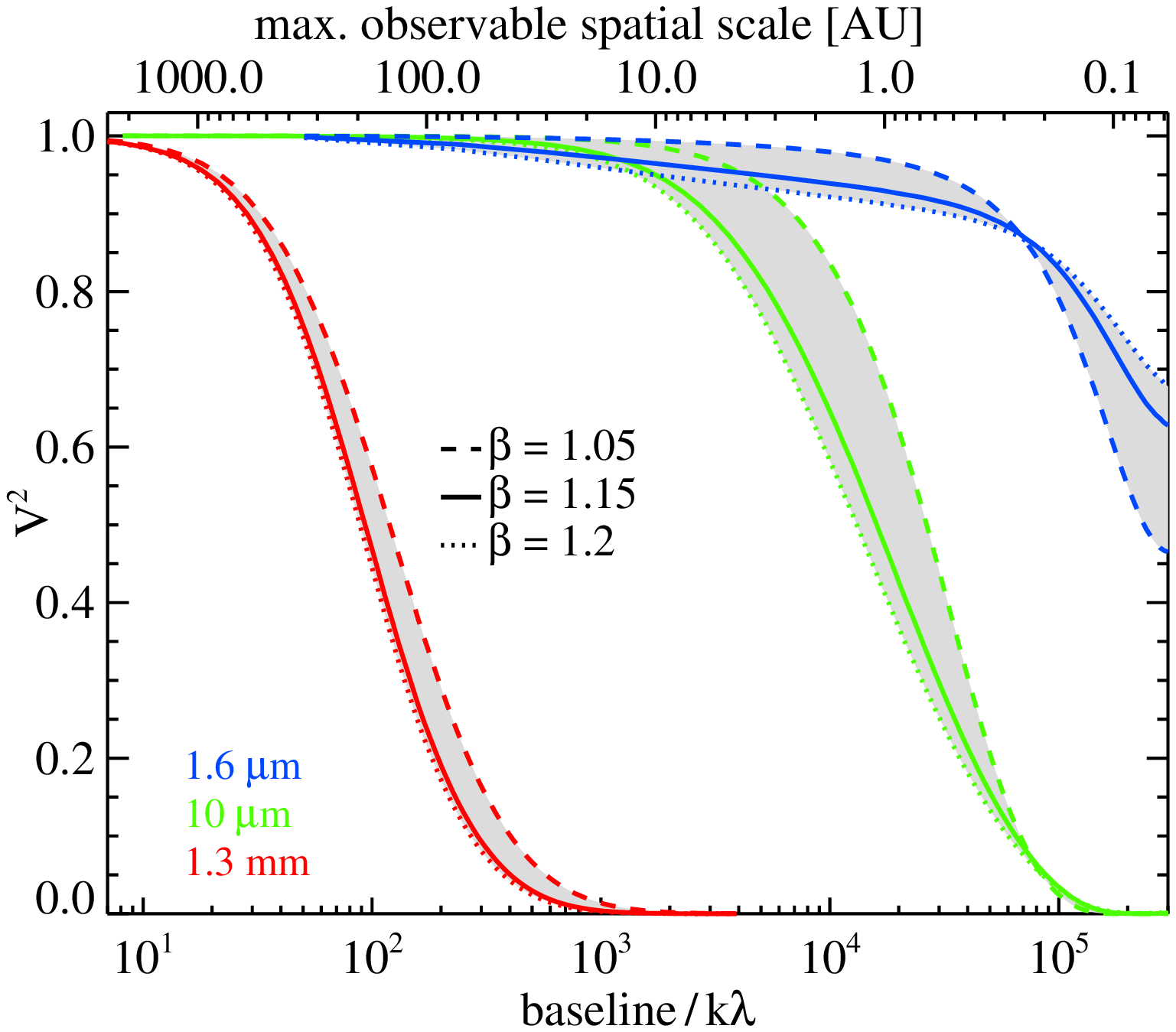}&
  \hspace*{-5.5mm}\includegraphics[height=40.1mm,trim=45 42 0 46,clip]
                  {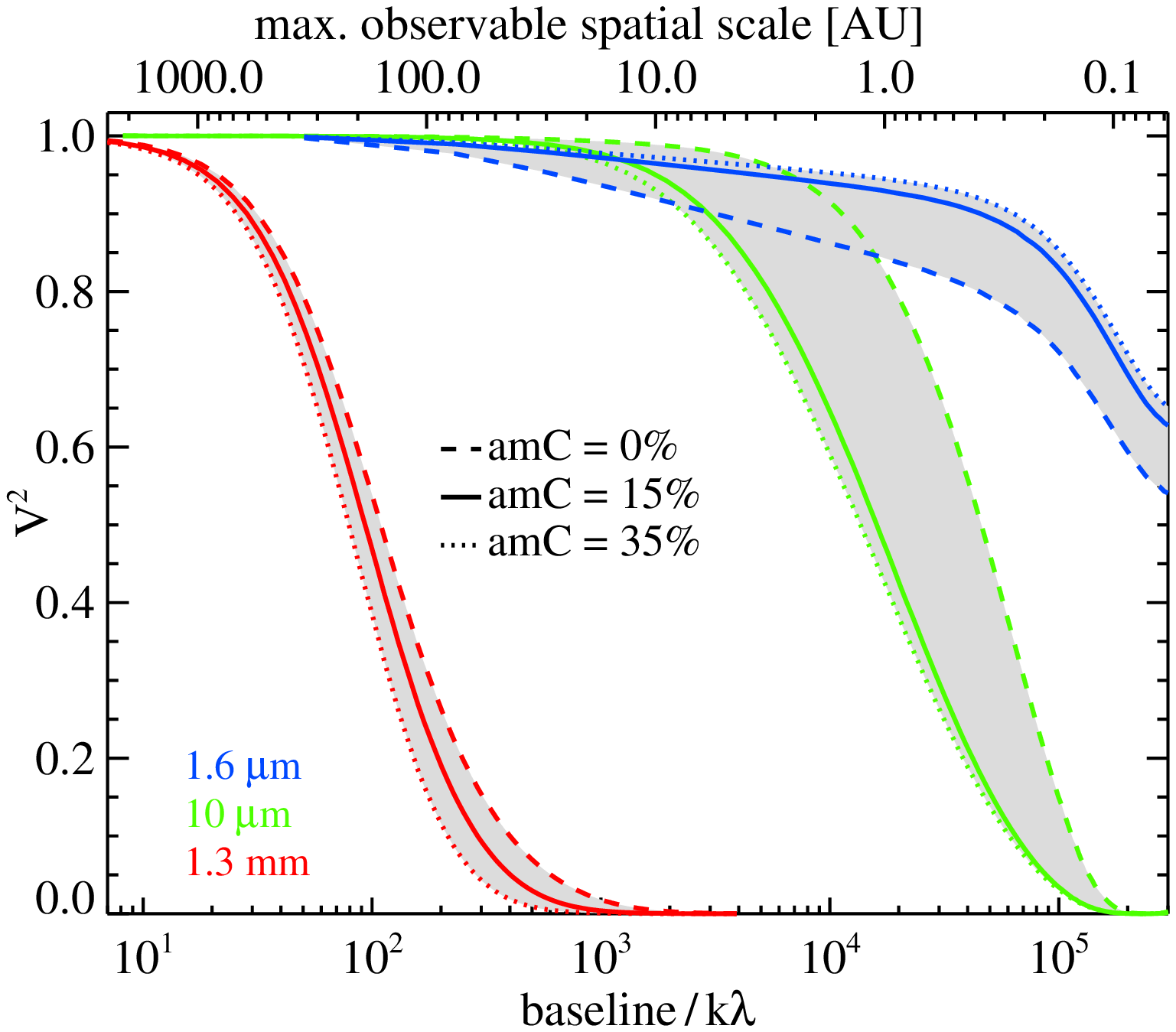}
  \\[-30mm]
  \hspace*{13mm}\resizebox{!}{2.0mm}{\bf scale height} & 
  \hspace*{9mm}\resizebox{!}{2.0mm}{\bf flaring} & 
  \hspace*{7mm}\resizebox{!}{2.0mm}{\bf amorph.\,carbon}\\[24.6mm]
  \hspace*{-7mm  }\includegraphics[height=45.4mm,trim= 0  0 0 46,clip]
                  {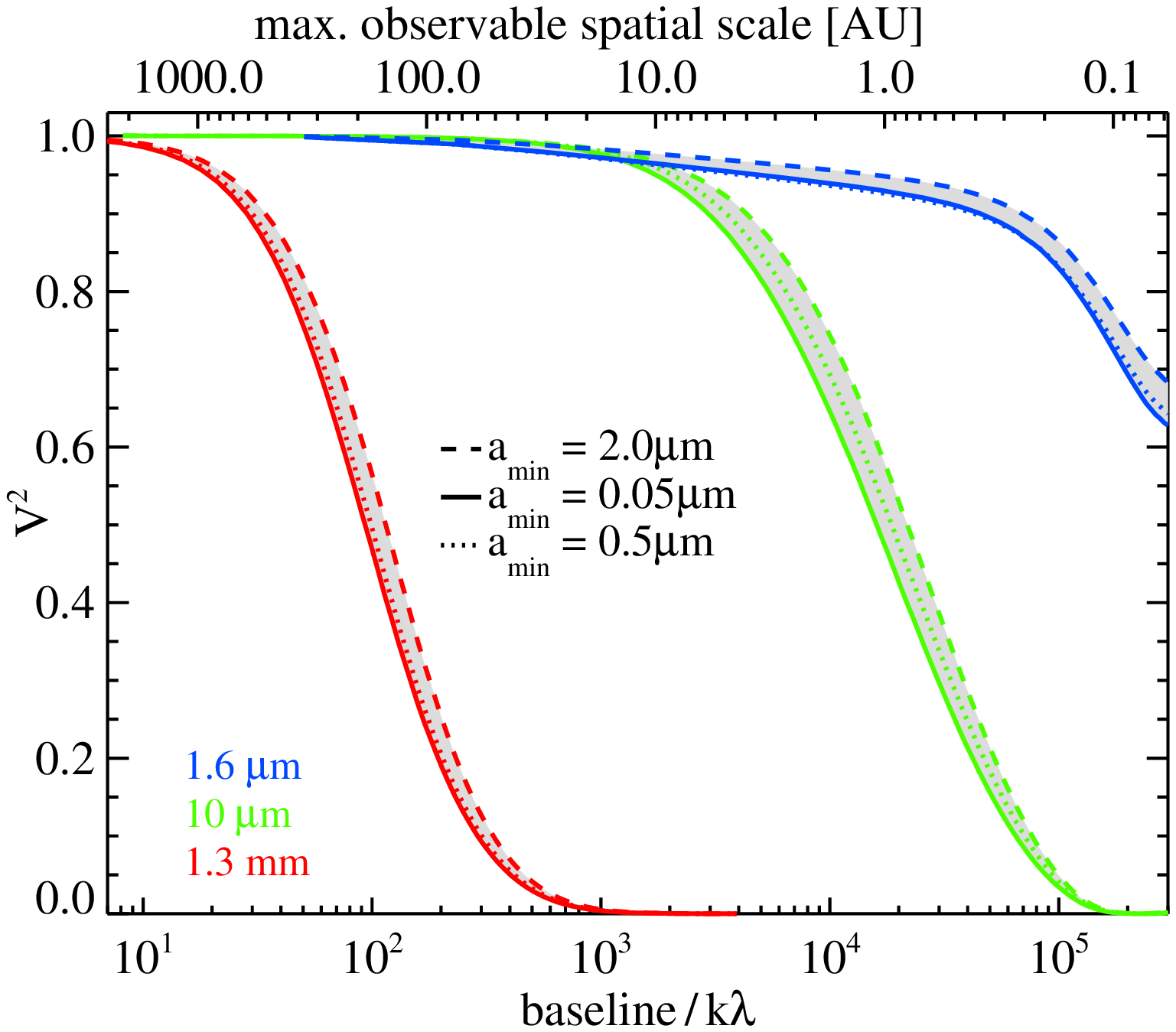}&
  \hspace*{-5.5mm}\includegraphics[height=45.4mm,trim=45  0 0 46,clip]
                  {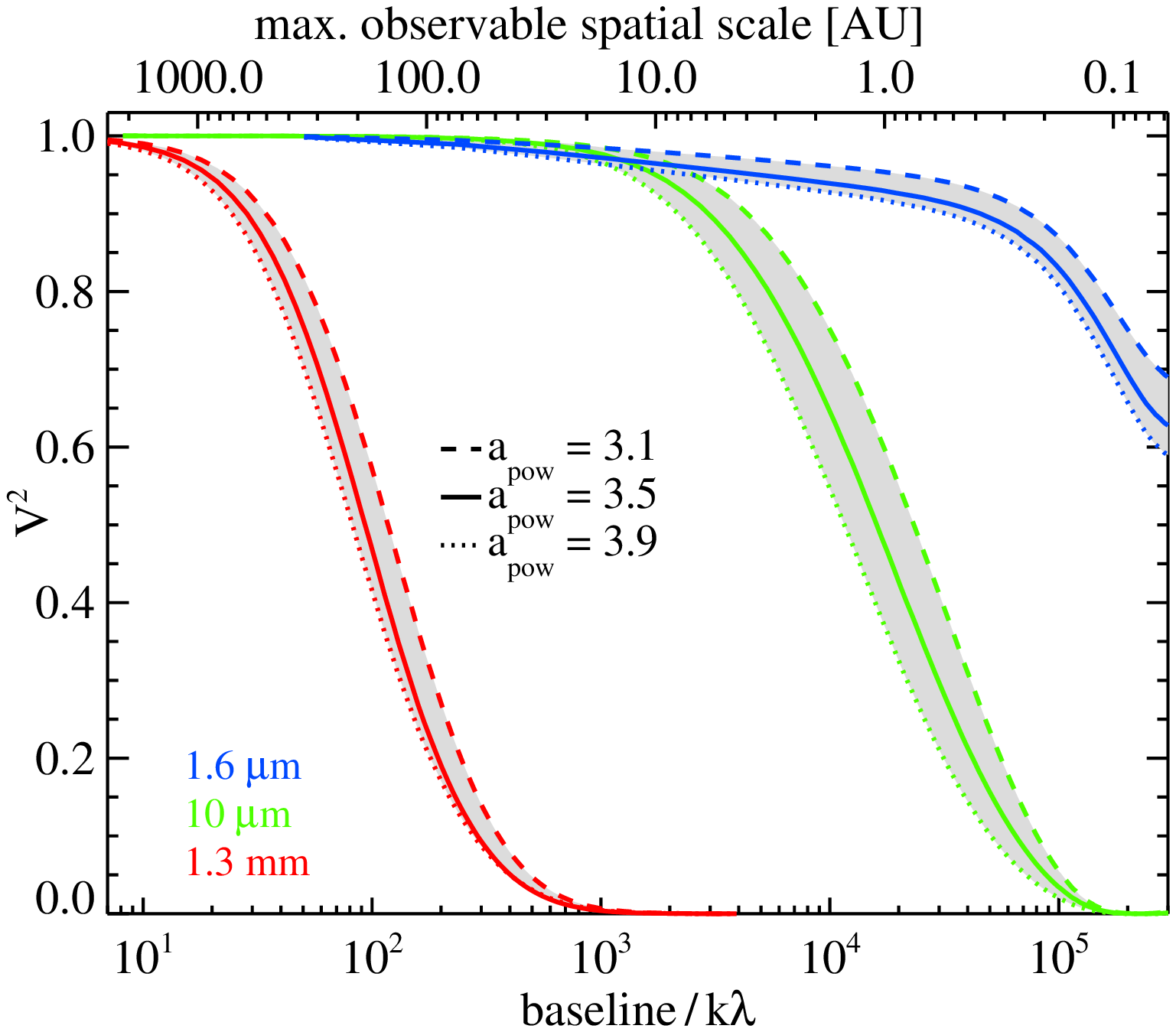}&
  \hspace*{-5.5mm}\includegraphics[height=45.4mm,trim=45  0 0 46,clip]
                  {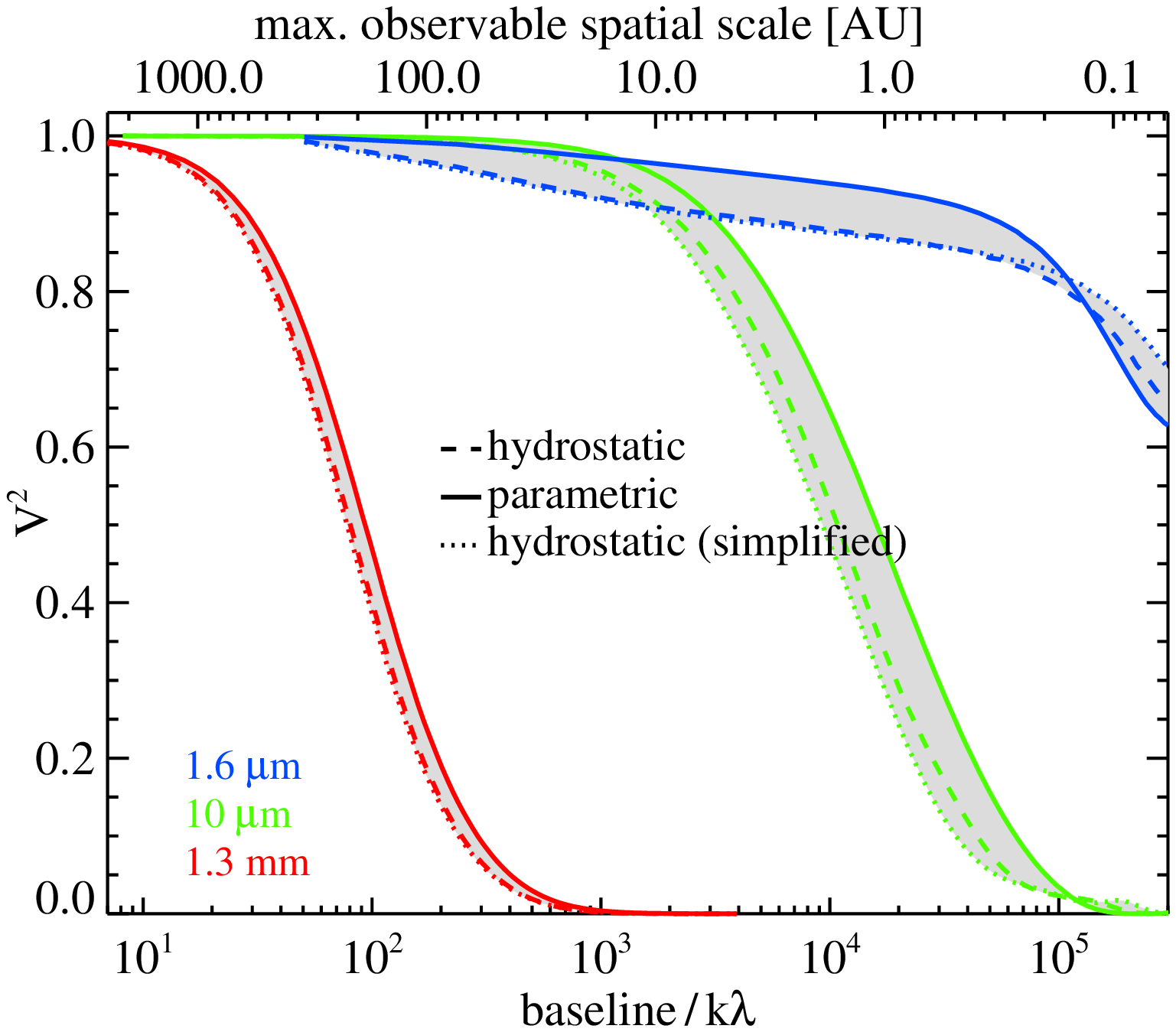}
  \\[-35mm]
  \hspace*{13mm}\resizebox{!}{2.0mm}{\bf min.\,dust size} & 
  \hspace*{7mm}\resizebox{!}{2.0mm}{\bf dust size powerlaw}\\[30mm]
  \end{tabular}}
  \caption{Effects of selected dust and disk shape parameters on
    continuum visibilities at 1.6\,$\mu$m (blue), 10\,$\mu$m (green)
    and 1.3\,mm (red).  Distance is 140\,pc.  The squared visibility
    $V^2$ (fraction of correlated flux) is shown as function of $\rm
    baseline[cm]/1000/\lambda[cm]$ for baseline orientation along the
    major axis of the disk on the sky. The abscissa on top is the
    corresponding maximum observable spatial scale\,[cm] $=0.6\cdot
    d{\rm [cm]}/{\rm baseline\,[k\lambda]}/1000$. The full lines show
    the reference model, identical in every part figure.  The dashed
    and dotted lines correspond to the changed parameter values as
    annotated.  Non-depicted parameters have less influence on the
    visibilities, for example $a_{\rm max}$. See
    Table~\ref{tab:refmodel} for explanations of parameter symbols.}
  \label{fig:VIS_effect}
\end{figure*}

Since the scale height is anchored at $r\!=\!100\,$AU in the model,
little flaring (the $\beta\!=\!1.05$ model) implies a tall inner
disk which causes a strong near-IR excess. Assuming strong dust
settling instead avoids these artefacts, because the impact of dust
settling is strongest where the gas densities are lowest, \ie in the
outer regions, whereas the inner regions are only affected a little.

When looking at the impact on gas temperature and emission lines,
however, the two models show just opposite effects. Lacking disk flaring
moves the gas into the disk shadow, causing lower gas temperatures and
weaker emission lines. Dust settling, in contrast, leaves the gas bare
and exposed to the stellar UV radiation, leading to higher gas
temperatures and stronger gas emission lines in general.

Therefore, in order to diagnose lacking disk flaring and/or strong
dust settling, the far-IR SED slope around 50\,$\mu$m is crucial, but
to distinguish between disk flaring and dust settling, the simultaneous
observation of far-IR gas lines is the key.

\subsection{Parameter impact on selected observables}
\label{sec:ObsPara}

\subsubsection{SED}
\label{sec:SED}

Figure~\ref{fig:SEDeffect} shows the impact of our model parameters on
the calculated spectral energy distribution (SED). Some parameter
dependencies have already been discussed and explained in
Sects.~\ref{sec:dusteffect} and \ref{sec:flaring}, but we repeat the
essence here to give a comprehensive overview of all important effects.
 
The {\bf dust mass} $M_{\rm dust}$ shifts the SED up and down at long
wavelength, where the disk is predominantly optically thin. Its
influence diminishes at $\lambda\la 100\,\mu$m, but even at
$20\,\mu$m, where the disk is massively optically thick, a change of
$M_{\rm dust}$ still produces noticeable changes. This is because
more mass increases the height at which the disk becomes radially
optically thick, which has similar consequences as increasing the
scale height. 

The reference {\bf scale height} $H_0$ affects the SED at all shorter
wavelength $\lambda\!\la\!200\,\mu$m, but not the
Rayleigh-Jeans tail of the SED.  Larger scale heights mean to
intersect more star light, and to produce a warmer disk interior which
re-emits more thermal radiation from all optically thick disk regions.

The {\bf flaring index} $\beta$ rotates the SED around a point at 
$\lambda\!\approx\!20\,\mu$m here, depending on the model and on the
  choice of the reference radius $H_0$ in Eq.\,(\ref{eq:H}). Large
$\beta$ values mean that we have a flared disk with a low inner rim
but with tall outer regions, which produce less near-IR but more
far-IR excess. Small $\beta$ lead to a ``self-shadowed'' disk
structure with very cold dust in the outer parts.

The {\bf inner radius} $R_{\rm in}$ regulates the maximum temperature
of the dust grains at the inner rim.  Larger $R_{\rm in}$ therefore
result in less near-IR emission (``transitional disks''). However, the
total amount of excess luminosity is not changing. For large $R_{\rm
  in}$, the luminosity excess merely shifts from the near-IR to the
mid-IR region, and beyond.

{\bf Dust settling}, with parameter $\alpha_{\rm settle}$ describing
the strength of the turbulent mixing, almost exclusively affects the
long wavelength parts of the SED ($\lambda\!\ga\!20\,\mu$m). According
to the Dubrulle prescription (see Eq.\,\ref{eq:tauf}), dust settling
is much more effective at large radii where the densities are low, in
which case the dust grains cannot be easily dragged along turbulent
gas motions.  Consequently, the outer disk parts become flat as seen
in dust, although the gas still extends high up. Therefore, strong
settling has similar consequences as lacking disk flaring at long
wavelengths. For a well-mixed dust/gas mixture, the mm-grains tend to
cover all spectral features produced by the small grains with their
flat, greyish opacity, washing out the $10\,\mu$m and $20\,\mu$m
silicate emission features.  Dust settling removes the
large grains from the disk surface, and therefore amplifies the
silicate emission features, which seems necessary in many cases to
re-produce the observed shape of the silicate emission features.

The {\bf maximum dust particle size} $a_{\rm max}$ has a similar
influence as $M_{\rm dust}$ at long wavelengths. Increasing $a_{\rm
  max}$ effectively means to put more dust mass into very
large particles which have almost no opacity at shorter
wavelengths. However, beyond about 1\,mm in this model, where the largest
particles do provide the dominating opacities, the SED starts to
change slope depending on the value of $a_{\rm max}$. 

The {\bf dust size distribution powerlaw index} $a_{\rm pow}$
regulates the mixture of small and large dust particles in the disk.
It thereby changes, in particular, the mm and cm-slopes. Larger
$a_{\rm pow}$ values also amplify the $10\,\mu$m and $20\,\mu$m
silicate emission features, because the grayish opacity of the large
grains is mostly removed from the model.

The {\bf volume fraction of amorphous carbon} has a surprisingly large
impact on the SED at all wavelengths. As discussed in
Sect.~\ref{sec:dustopac}, pure laboratory silicates are very effective
scatterers, keeping the stellar radiation out of the disk, but they
will hardly absorb it. Therefore, disks made of pure silicates 
are much cooler and emit less near-IR and far-IR excess. 
The fraction of amorphous carbon also changes substantially the mm and cm
slopes through opacity effects.

The {\bf surface density powerlaw index} $\epsilon$ has practically no
influence on the SED, same with the {\bf tapering-off radius} $R_{\rm
  tap}$ (not depicted), the {\bf outer radius} $R_{\rm out}$
\citep[not depicted, see also][]{Bouy2008} and the {\bf minimum dust
  size} $a_{\rm min}$ (as long as $a_{\rm min}\la 0.5\,\mu$m, not
depicted). The dependencies of the SED on those non-depicted
parameters are less than those shown for $\epsilon$.

From the shape of the SED changes caused by the nine parameters
depicted in Fig.~\ref{fig:SEDeffect}, one can easily imagine how
degenerate pure SED fitting can be \citep[e.g.][]{Robitaille2007},
just consider, for example, a combination of lower dust mass 
with more amorphous carbon.

\begin{figure*}[!t]
  \vspace*{-2mm}
  \hspace*{3mm}\begin{tabular}{ll}
  \hspace*{-2mm}\includegraphics[height=89mm]{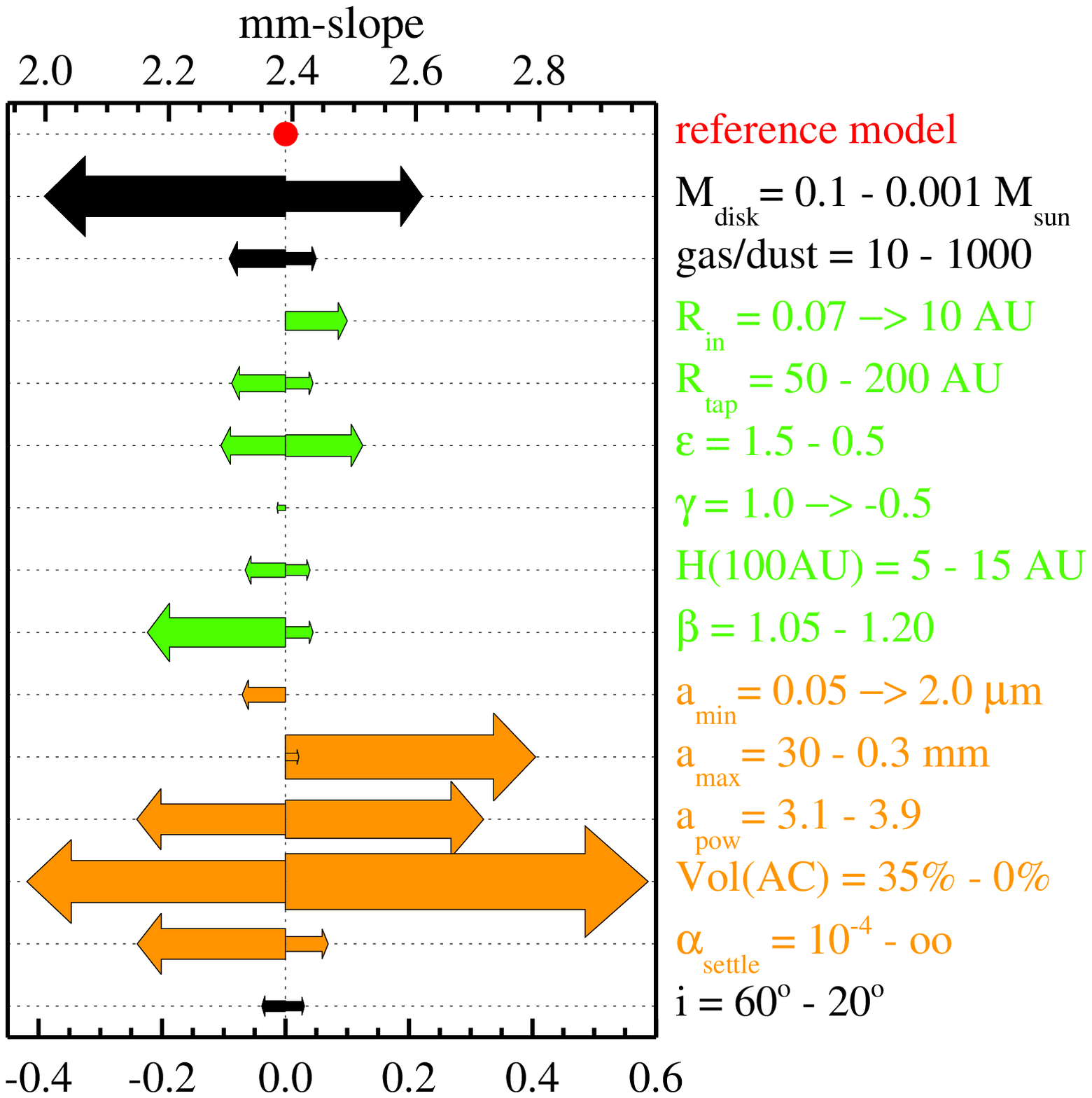} &
  \hspace*{-2mm}\includegraphics[height=89mm]{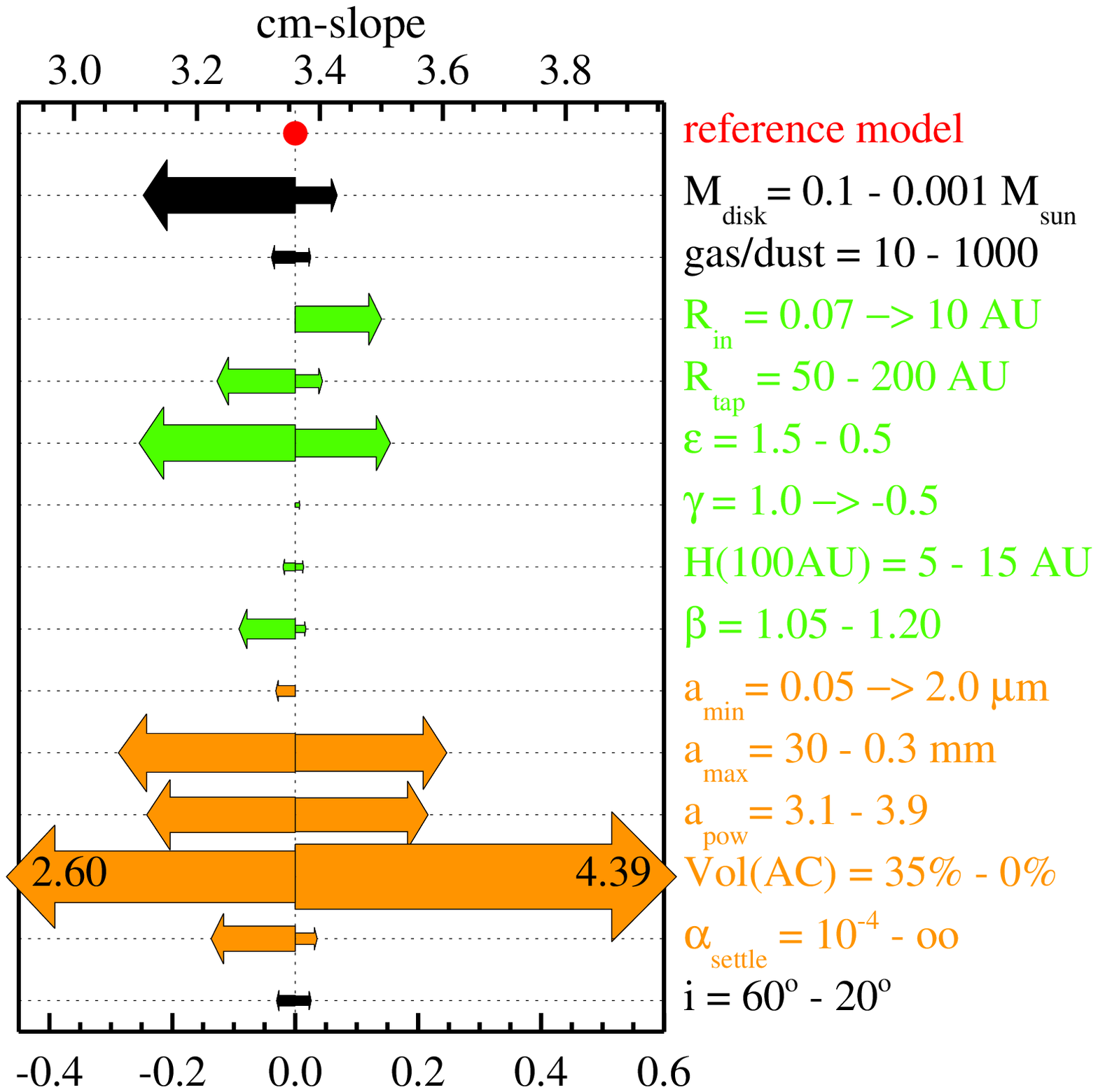} 
  \end{tabular}
  \caption{Impact of model parameters on the millimetre SED slope
    (left) and centimetre SED slope (right), as defined by
    Eqs.\,(\ref{eq:mm-slope}) and (\ref{eq:cm-slope}), respectively.
    The model parameters which have been varied are listed to the
    right of each plot (see Table~\ref{tab:refmodel} for explanations
    of the symbols), along with the ranges explored. The colours
    indicate different groups of model parameters. Gas and dust masses
    are shown in black, disk shape parameters in green, and dust size,
    material and settling parameters in orange.  The changes of the
    observable quantity, here e.g. the mm-slope, caused by varying a
    particular model parameter, are shown with arrows.  The original
    value of the observable quantity is shown by the red point marked
    with ``reference model'' (for example, 2.4 for the mm-slope),
    where the top $x$-axis provides an absolute scale, and the bottom
    $x$-axis provides a relative scale with respect to the value
    obtained by the reference model.  The arrows indicate the
    direction and magnitude of changes caused.  Leftward arrows
    indicate a flattening of the SED, rightward arrows a
    steepening. The corresponding parameter values are shown to the
    left and right of the '--' to the right of each plot.  For
    example, increasing the disk mass $M_{\rm disk}$ from the
    reference value to $0.1\,M_{\odot}$ results in a flatter SED
    $\alpha^{\rm mm}_{\rm SED}\!=\!2.0$, whereas decreasing $M_{\rm
      disk}$ to $0.001\,M_{\odot}$ leads to a slightly steeper SED
    $\alpha^{\rm mm}_{\rm SED}\!=\!2.6$.  If there is a '$-\!\!>$' on
    the r.h.s., it means that only one parameter direction has been
    explored and that there is only one corresponding arrow.  In those
    cases, the parameter value to the left of '$-\!\!>$' is the value
    in the reference model.  In one case (the mm-slope as function of
    $a_{\rm max}$) both directions of parameter changes resulted in
    a steepening, here the small arrow belongs to $a_{\rm
      max}\!=\!30\,$mm.}
  \label{fig:mm-slope}
\end{figure*}

\subsubsection{Visibilities}
\label{sec:visi}

Figure~\ref{fig:VIS_effect} shows the impact of model parameters on
the calculated visibilities at 1.6\,$\mu$m (\eg PIONIER), at
10\,$\mu$m (\eg MIDI), and at 1.3\,mm (\eg CARMA, ALMA). The
mm-visibility \citep[see e.g.][]{Guilloteau2011} probes the apparent
spatial extension of the disk, limited by minimum optical depth
requirements to produce a detectable signal at those wavelengths. This
apparent size is most directly influenced by the tapering-off radius
$R_{\rm tap}$. The ``sharpness'' of the outer edge $\gamma$ is
reflected by the steepness of the $V^2$-decline. However, more dust in
the outer regions (larger $M_{\rm dust}$, smaller $\epsilon$) also
increases the optical depths in these regions, which leads to larger
apparent sizes as well. Strong dust settling $\alpha_{\rm
  settle}\!\to\!10^{-4}$ moves the grains toward the midplane into the
disk shadow where they are substantially cooler (see
Sect.~\ref{sec:mm-slope} and Fig.~\ref{fig:settle}), so cold that some
fraction of the dust grains does not emit in the Rayleigh limit at
1.3mm, thus producing less extended flux, which leads to a smaller
apparent size\footnote{We note that radial dust migration is not
  included in these models, which can potentially lead to a reduction 
of the apparent disk sizes at millimetre wavelengths as well.}.

The 10\,$\mu$m visibilities reflect the radial extension of
warm dust in the disk surface layer producing the 10\,$\mu$m silicate
emission feature (about 1\,AU in the reference model). This extension
is larger for warm, \eg flared disks. In contrast, parameter choices
which lead to cooler conditions at 1\,AU cause a smaller appearance of
the disk at 10\,$\mu$m. Such parameter choices include smaller scale
heights $H_0$, dust size parameter variations that increase the
mean dust size (larger $a_{\rm min}$, smaller $a_{\rm pow}$), and
lacking amorphous carbon. Dust settling plays no significant role
here.

The 1.6\,$\mu$m visibilities are more difficult to understand, see
\citep{Anthonioz2015}. They have three components: the star,
scattering and emission from the inner rim, and extended scattering.
The extended scattering leads to a slight tilt of the $V^2$-curves
beyond $\rm baselines\,[k\lambda]\!\ga 100$, before $V^2$ drops to
much lower values at baselines which corresponds to the inner rim of
the disk (0.07\,AU in the model). At even longer baselines, the
interferometer would start to resolve the star
($R_\star\!=\!0.0097\,$AU), and the visibility would drop sharply, but
such long baselines are currently not accessible, and not included in
Figure~\ref{fig:VIS_effect}).  The 1.6\,$\mu$m visibilities hence
probe the relative contributions of these three components and their
spatial extensions. Most remarkably is the influence of flaring and
settling, which powers/suppresses the extended scattering component,
and the fraction of amorphous carbon which changes the albedo of the
dust particles. Pure silicate dust particles ($\rm amC\!=\!0$), for
example, are almost perfect scatterers at 1.6\,$\mu$m, leading to a
much more pronounced extended scattering component.

Noteworthy, the hydrostatic disk models show a stronger extended
scattering component as well (because of the strong flaring of the
outer disk), and less contributions from the inner rim, which has a
lower wall height. 

The inner rim radius $R_{\rm in}$ directly affects the second
component of the 1.6\,$\mu$m visibilities directly, namely the
emission and scattering from the inner rim. Large $R_{\rm in}$ can
also limit the radial extension of the 10\,$\mu$m emission region from
the inside, introducing new small scales in form of the ring thickness
and the apparent height of the inner rim wall, with sharp edges, which
leads to more complex visibility shapes.

\subsubsection{The mm-slope and cm-slope}
\label{sec:mm-slope}

Figure~\ref{fig:mm-slope} shows the impact of the model parameters on
the observable SED slopes at millimetre and centimetre wavelengths, as
defined by Eqs.\,(\ref{eq:mm-slope}) and (\ref{eq:cm-slope}), which
are important diagnostics of grain growth in protoplanetary disks, see
\eg \citep{Natta2007} and \citep{Testi2014}.  The SED slopes are
expected to reflect the dust absorption opacity slopes, with some
flattening due to optical depth effects
\begin{equation}
  \alpha_{\rm SED} \approx 2 + \frac{\beta_{\rm abs}}{1+\Delta}\ ,
  \label{eq:Beckwith}
\end{equation}
Equation~(\ref{eq:Beckwith}) was derived by
\citet{Beckwith1990}\footnote{Note that Beckwith\etal consider the
  slope of $L_\nu=\nu F_\nu$.}  for a powerlaw surface density
structure $\Sigma\!\propto\!r^{-p}$ and a vertically isothermal
powerlaw temperature distribution $T\!\propto\!r^{-q}$. This formalism
was later relaxed by \citet{Ricci2010,Ricci2012}, who determined $T(r)$
according to \citet{Chiang1997} and considered a self-similar disk
with tapered outer edge. Ricci\etal found values
$\alpha_{\rm SED}\!\approx\!1.9-2.7$ for the Taurus-Auriga
region, which are significantly lower than what is expected from small
interstellar grains $\beta_{\rm abs}\!\approx\!1.7$
\citep{Draine2006}, suggesting that the dust in protoplanetary disks
must have much smaller $\beta_{\rm abs}\approx 0.3-1.0$ at
$(1\!-\!3)$mm, indicating dust growth.

The standard DIANA opacities have $\beta_{\rm
  abs}^{\rm\,mm}\!\approx\!1.0$ and $\beta_{\rm
  abs}^{\rm\,cm}\!\approx\!1.5$, thus the SED slopes of the reference
model are expected to be $\alpha_{\rm SED}^{\rm mm}\!=\!3.0$ and
$\alpha_{\rm SED}^{\rm cm}\!=\!3.5$ in the optically thin limit
$\Delta\!=\!0$. However, the reference model exhibits $\alpha_{\rm
  SED}^{\rm mm}\!\approx\!2.4$ and $\alpha_{\rm SED}^{\rm
  cm}\!=\!3.35$, in agreement with observations, suggesting optical
depths corrections of $\Delta_{\rm mm}\!\approx\!1.5$ and $\Delta_{\rm
  cm}\!\approx\!0.1$, respectively. 

Closer inspection shows, however, that these derived $\Delta$ do not
agree at all with the expected flattening due to optical depth
effects.  The radius $r_1$ where the vertical dust optical depth
equals unity
\begin{equation}
 \tau_\nu^{\rm abs}(r_1) = \int_{-\infty}^{\infty}
                          \kappa_\nu^{\rm dust,abs}(r_1,z)\,dz = 1 \ , 
\end{equation}
is only $r_1\!\approx\!6.9\,$AU at $\lambda\!=\!1.3\,$mm and
$r_1\!=\!0.9\,$AU at $\lambda\!=\!7\,$mm in the reference model, which
results in tiny corrections, $\Delta_{\rm mm}\!=\!0.04-0.12$ and
  $\Delta_{\rm cm}\!=\!0.01-0.03$, as derived from the equations in
  \citet{Beckwith1990}, depending on what is assumed for the outer
  radius in our tapered-edged models. At both wavelengths, the
  expected $\Delta$-corrections are too small, inconsistent with the
  results obtained from our radiative transfer models.



This conclusion holds for all models computed in this paper. The
strongest optical depth effects occur in massive disks ($M_{\rm
  disk}\!=\!0.1\,M_\odot$), if the mass is more concentrated toward
the centre ($\epsilon\!=\!1.5$), if the disk is small ($R_{\rm
  tap}\!=\!50\,$AU), and/or if the dust size and opacity parameters
lead to larger mm-opacities.  But even in all these cases, the
expected Beckwith\etal $\Delta$-corrections for optical depths effects
at 1.3\,mm stay well below unity, which is insufficient to explain the
gentle mm-slopes obtained from our radiative transfer models. \ \ {\sl
  The mm-slopes from the computed SEDs are more gentle than expected},
and optical depth effects are {\sl not}\ \,the key to explain these
discrepancies\footnote{\citet{Woitke2013} and \citet{Pietu2014}
  have reported on the detection of very small protoplanetary disks
  which could represent a non-negligible fraction of protoplanetary
  disks in general. These disks are likely optically thick,
  where this statement is probably not valid.}.


\citet{Beckwith1990} derived Eq.\,(\ref{eq:Beckwith}) by assuming that
all dust grains emit in the Rayleigh limit. If we ignore optical depth
effects for a moment, the observable flux $F_\nu$ is exactly given by
\begin{equation}
  F_\nu = \frac{\hat{\kappa_\nu}^{\rm abs}}{d^2}
          \int \rho_{\rm dust} B_\nu(\Td)\,dV
        \,=\, \frac{M_{\rm dust}}{d^2}\, 
          \big\langle B_\nu(\Td)\big\rangle\,\hat{\kappa_\nu}^{\rm abs} 
\end{equation}
where $\hat{\kappa_\nu}^{\rm abs}\!=\kappa_\nu^{\rm abs}/\rho_{\rm
  dust}$ is the dust absorption coefficient per dust mass (assumed to
be constant throughout the disk), $\big\langle B_\nu(\Td)\big\rangle
=\int\rho_{\rm dust} B_\nu(\Td)\,dV \Big{/} \int \rho_{\rm dust}\,dV$
is the dust mass averaged Planck function, and $M_{\rm dust}$ is the
total dust mass.  The log-log slope $\alpha_{\rm
  SED}\!=\!-\partial\log F_\nu/\partial\log\lambda$ is then given by
\begin{equation}
   \alpha_{\rm SED} = - \frac{\partial\log\,\langle B_\nu(\Td)\rangle}
                            {\partial\log\lambda} + \beta_{\rm abs}
   \label{eq:mmslope}
\end{equation}
Table~\ref{tab:dlogBnu} shows that the deviations of the Planck
derivative from its limiting value of $2$ can be substantial. At
1.3\,mm, for example, the Planck slope is about 1.35 and not 2, if the
grains emit at 10\,K. At 7\,mm, deviations $\ga\!0.2$\,dex are still
conceivable if the majority of grains would emit at 5\,K. Indeed,
using these deviations from Rayleigh-Jeans regime, \citet{Dutrey2014}
report on vertical mean dust temperatures of 8.5\,K at 300\,AU in the
disk of GGTau.

\begin{table}
\caption{Negative logarithmic derivative of the Planck function,
  $-\partial\log B_\nu(T)/\partial\log\lambda$, as function of
  temperature and wavelength.}
\vspace*{-2mm}
\centering
\resizebox{60mm}{!}{\begin{tabular}{c|cccc}
\hline                                            
              &        &        &       &        \\[-3mm]
              &  40\,K &  20\,K & 10\,K &   5\,K \\[-0.2mm]
\hline                                            
\hline                                            
              &        &        &       &        \\[-3mm]
  850\,$\mu$m &  1.774 &  1.518 & 0.926 & -0.504 \\
  1.3\,mm     &  1.855 &  1.698 & 1.347 &  0.515 \\
  3.5\,mm     &  1.948 &  1.894 & 1.780 &  1.533 \\
   7\,mm      &  1.974 &  1.948 & 1.894 &  1.780 \\
 $\infty$     &  2     &  2     & 2     &  2     \\
\hline                                            
\end{tabular}}
\label{tab:dlogBnu}
\end{table}

The dependencies of $\alpha_{\rm SED}$ on the disk temperature
structure, according to Eq.\,(\ref{eq:mmslope}), can explain 
the results obtained from our radiative transfer models. The
mean dust temperature according to Eq.\,(\ref{eq:Tdmean}) is
$\langle\Td\rangle\!\approx\!19\,$K in the reference model, but this
is a linear mean, and the Planck function is highly non-linear at 
low temperatures $\langle B_\nu(T)\rangle\ll B_\nu(\langle
T\rangle)$. Simply put, a considerable part of the dust in the disk
is so cold that it does not contribute significantly to the
1.3\,mm flux. The minimum dust temperature in the reference model is
about 4.5\,K. The cold dust over total dust mass fraction is 0.12 (for
$\Td\!<\!7\rm\,K$), 0.31 (for $\Td\!<\!10\rm\,K$), 0.58 (for
$\Td\!<\!15\rm\,K$), and 0.75 (for $\Td\!<\!20\rm\,K$). It is about
this fraction, with efficiencies according to Table~\ref{tab:dlogBnu},
that is missing in the observable flux, causing the deviations from
$\alpha_{\rm SED} = 2 + \beta_{\rm abs}$.

\begin{figure}[!t]
  \centering
  \vspace*{-1mm}
  \hspace*{-2mm}\includegraphics[height=60mm,trim=0 45 0 80,clip]
                                {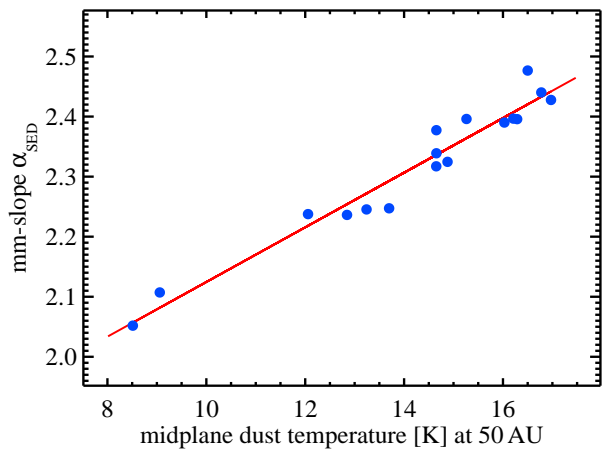}
  \vspace*{0mm}
  \caption{Correlation between computed SED mm-slope (as measured
    between 850\,$\mu$m and 1.3\,mm) and midplane dust temperature at
    $r\!=\!50\,$AU for models with identical dust properties and disk
    mass. Varied parameters include gas/dust (at constant $M_{\rm
      dust}$), $R_{\rm in}$, $R_{\rm tap}$, $\epsilon$, $\gamma$,
    $H$(100\,AU), $\beta$ and $\alpha_{\rm settle}$, \ie the green
    disk shape parameters in Fig.~\ref{fig:mm-slope} and dust
    settling.}
  \label{fig:mmslope_corr}
\end{figure}

These temperature effects explain the qualitative behaviour of the SED
millimetre and centimetre slopes in the models as shown in
Fig.~\ref{fig:mm-slope}.  The dust size and material parameters 
impact the SED slope directly via changing the dust absorption slope
$\beta_{\rm abs}$, compare Fig.~\ref{fig:dustabs}. These parameters
have the strongest impact on the SED slope. The green disk shape
parameters have an influence on the dust temperature structure in the
disk $\Td(r,z)$ and, therefore, have an indirect influence on the SED
slope. Figure~\ref{fig:mmslope_corr} shows that all disk shape
parameters that lead to very cold midplane conditions in the disk
around $r\!\approx\!50\,$AU are well correlated with the more gentle
mm-slopes in the models, in particular little flaring
($\beta\!=\!1.05$) and a steep column-density structure
($\epsilon\!=\!1.5$). However, dust settling has in fact an even stronger
impact ($\alpha_{\rm settle}\!=\!10^{-4}$, leftmost point in
Fig.~\ref{fig:mm-slope}). By settling, the bigger grains are moved to the cooler
midplane, and concentrating the grains toward the midplane also increases
the shadow formation there, making the midplane even cooler.

Our conclusion is that, indeed, the dust absorption opacity
slope $\beta_{\rm abs}$ mostly determines the SED slope.
However, $\alpha_{\rm SED}$ flattens significantly for cold
disks, an effect that has not been reported so far and that  
seems more important that the classical $\Delta$-correction for optical
depth effects \citep{Beckwith1990}.

\subsubsection{The [OI]\,63.2\,$\mu$m emission line}
\label{sec:OI63}

\begin{figure}
  \includegraphics[width=90mm,trim=8 3 8 5,clip]{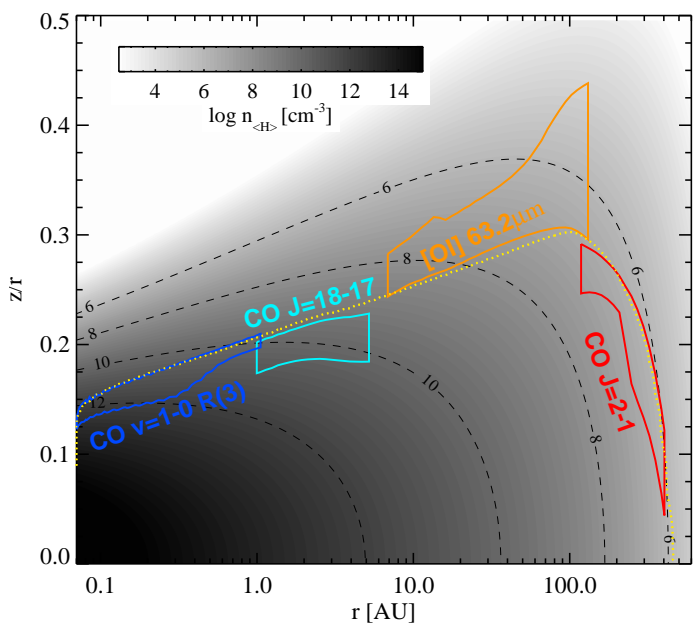}
  \caption{Line emitting regions in the reference model. The
    surrounded areas are responsible for 50\% of the vertically
    emitted line fluxes as annotated. The black dashed contours show
    the hydrogen nuclei particle density $\log\nH(r,z)$. The yellow dotted
    contour line shows the upper boundary of the molecular layer where
    $n_{\rm CO}/\nH\!=\!10^{-5}$.}
  \label{fig:LineOrigin}
  \vspace*{-1mm}
\end{figure}

We now turn our attention from continuum observations to gas emission
lines.  We have selected four representative emission lines for the
discussion in this paper, which are frequently observed, and which
emerge from different disk regions, see Fig.~\ref{fig:LineOrigin}. All
selected lines have chemically robust carriers, namely the O-atom
or the CO-molecule, which are not critically dependent on chemical
details.  These lines are rather influenced by the shape of the disk and
gas temperature distribution in the disk surface layer.

The [OI]\,63.2\,$\mu$m line is usually the strongest disk emission
line throughout the electromagnetic spectrum, compare
Fig.~\ref{fig:refmodel}, and has been detected in 84\% of T\,Tauri
stars with disk dust masses $>\!10^{-5}\rm\,M_\odot$ by the Herschel
open time key program GASPS \citep{Dent2013}, including outflow
sources \citep[see][]{Howard2013}.  With an excitation energy of
227\,K and a critical density of $\sim\!6\times10^5\rm\,cm^{-3}$, this
line originates in disk layers even above the CO containing molecular
layers, see Fig.~\ref{fig:LineOrigin}. In order to excite this line,
the gas needs to be as warm as $\ga\!50\,$K \citep{Kamp2010}. These
conditions are only present in outflows and in the disk up to radial
distances $\la\!100$\,AU, {\rm above} the molecular layers which are
too cold because of molecular line cooling.

\begin{figure}[!t]
  \vspace*{-2mm}
  \includegraphics[width=93mm]{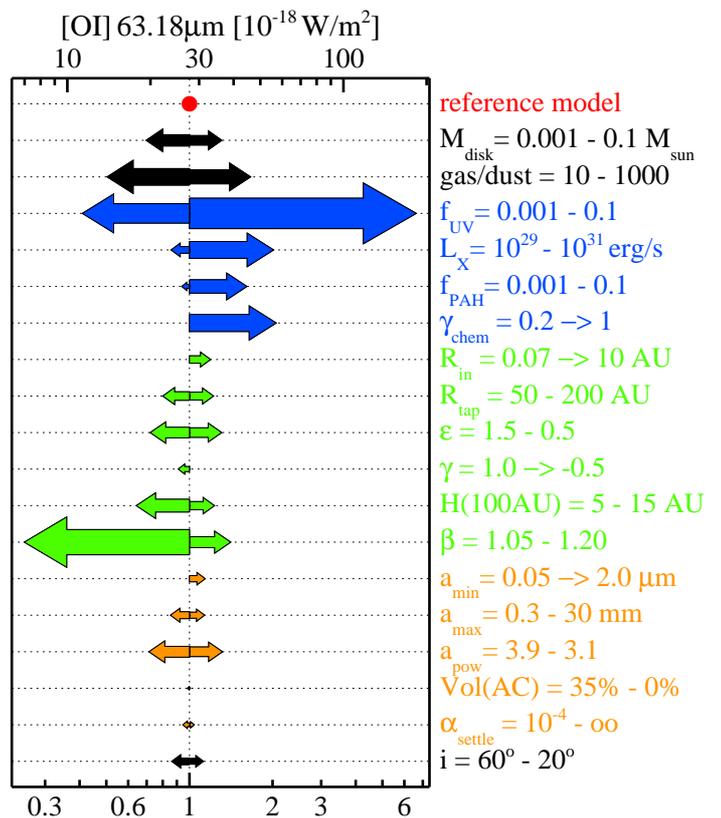}
  \vspace*{-5mm}
  \caption{``Impactogram'' of the $\rm [OI]\,63.2\,\mu$m
    emission line. The absolute line flux is depicted on the top
    abscissa, as well as relative to the reference model at the
    bottom. Leftward arrows indicate a weakening of the line, and
    rightward arrows a strengthening of the line, due to changes of
    single model parameter as indicated on the r.h.s.. The colours indicate
    groups of parameters. Gas and dust masses are shown in black,
    heating parameters in blue, disk shape parameters in green, and
    dust parameters in orange. See Fig.~\ref{fig:mm-slope} for further
    explanations.}
  \label{fig:OI63}
\end{figure}

The impact of the model parameters on the predicted [OI]\,63.2\,$\mu$m
line flux is shown in Fig.~\ref{fig:OI63}. According to the physical
excitation mechanism explained in the previous paragraph, the line
flux increases with all parameters that directly trigger the heating
in the uppermost disk layers, marked in blue in Fig.~\ref{fig:OI63},
namely the stellar UV excess $f_{\rm UV}$, the X-ray luminosity $L_X$,
the PAH concentration $f_{\rm PAH}$ and the efficiency of exothermic
reactions $\gamma_{\rm chem}$. The disk shape parameters (marked in
green in Fig.~\ref{fig:OI63}) also play an important role. For a
self-shadowed disk (\eg for flaring index $\beta\!=\!1.05$), the
distant oxygen gas is situated in the disk shadow, and the
[OI]\,63.2\,$\mu$m line is substantially suppressed.  The line is
massively optically thick ($\tau_{\rm line}\!\approx\!100$ in the
reference model at 100\,AU), so what counts is the {\em \,size of the
  disk surface area with gas temperatures $\Tg\!\ga\!50\,$}K,\, which
depends on the disk mass and shape parameters $H_{\rm g}(100\rm\,AU)$, $\beta$
and $\epsilon$. In comparison, dust size parameters and inclination
play no significant role for this line.

Figure~\ref{fig:OI63} shows that 5 different model parameters
are able to change the [OI]\,63.2\,$\mu$m line flux by at least a
factor of 2 within their reasonable ranges of values, therefore,
finding clear correlations of the [OI]\,63.2\,$\mu$m line flux with
one of these parameters seems quite unlikely, which could explain why
\citet{Meeus2012} and \citet{Aresu2014} have reported on negative 
results concerning such correlations.

Since the [OI]\,63.2\,$\mu$m line is optically thick, one would not
expect the gas mass nor the gas/dust ratio to be important.  However,
for this particular line, there is an interesting energy conservation
mechanism at work. The emission of [OI]\,63.2\,$\mu$m line photons is
the dominant cooling process in the line emitting regions, and
therefore, the line luminosity must roughly equal {\em \,the
  integrated heating rate in the [OI]\,63.2\,$\mu$m line emitting
  volume}.  Consequently, the gas temperature in this volume can be
expected to relax towards an equilibrium value where the
[OI]\,63.2\,$\mu$m line cooling balances the total heating, henceforth
denoted as ``self-regulation mechanism''.  The heating in the line
emitting region is provided by a number of physical processes that
absorb and thermalise fractions of the incoming UV and X-ray photon
energies, namely X-ray Coulomb heating, heating by neutral carbon
photo-ionisation, PAH heating via photo-effect, and heating by
exothermic chemical reactions driven by UV and X-ray reactions. For
larger disk masses, the solid angle (as seen from the star) of the
distant disk regions that absorb the UV and X-ray photons increases,
therefore, the [OI]\,63.2\,$\mu$m line flux increases. The impact of
the gas/dust ratio is actually more significant, as there is a
competition between dust versus gas absorption of UV photons.  For
larger gas/dust ratios, fewer UV photons are absorbed by the dust (and
re-emitted as continuous far-IR radiation), hence more of the
available UV flux is converted into gas heating.

As a consequence of these processes, more gas (less dust) generally
leads to an increase of the size of the disk surface
area where $\Tg\!\ga\!50\,$K. More specifically, an increase of the
gas mass causes the [OI]\,63.2\,$\mu$m line emission region to shift 
upwards \citep{Kamp2010}, which captures more of the impinging UV and
X-ray photons. However, because of the self-regulating energy
conservation mechanisms explained above, effects are rather modest. By
varying the gas mass by a factor of 10, the [OI]\,63.2\,$\mu$m line
only changes by factors of a few.

\subsubsection{CO high-$J$ emission lines}
\label{sec:COhighJ}

As an example for high-$J$ CO emission lines, we have selected the CO
$J\!=\!18\!\to\!17$ line at 144.8\,$\mu$m with an excitation energy of
945\,K and a critical density of $\sim\!2\times10^6\rm\,cm^{-3}$. Due
to its high excitation energy, gas temperatures $\Tg\!\ga\!200\,$K are
required to excite this line, and these conditions are only present in
the CO surface layer inside of $r\!\la\!5\,$AU in the reference model.
The line is optically thick, but approaches $\tau_{\rm line}\!=\!1$ at
the outer boundary of the line emitting region. The line is rather
weak in the reference model, a factor of about 30 lower than the
Herschel/PACS detection limit. In fact, detection rates of this line
for T\,Tauri stars are rather low, about 41\% of the 34 objects
selected by GASPS \citep{Dent2013}\footnote{There was a clear
  selection bias in \citep{Dent2013}, because only objects with bright
  [OI]\,63.2\,$\mu$m line were selected for the CO observations,
  included T\,Tauri stars with outflows.}, most of them being identified
as outflow sources \citep[see][]{Howard2013}.

\begin{figure}
  \vspace*{-1mm}
  \hspace*{-1mm}\includegraphics[width=93mm]{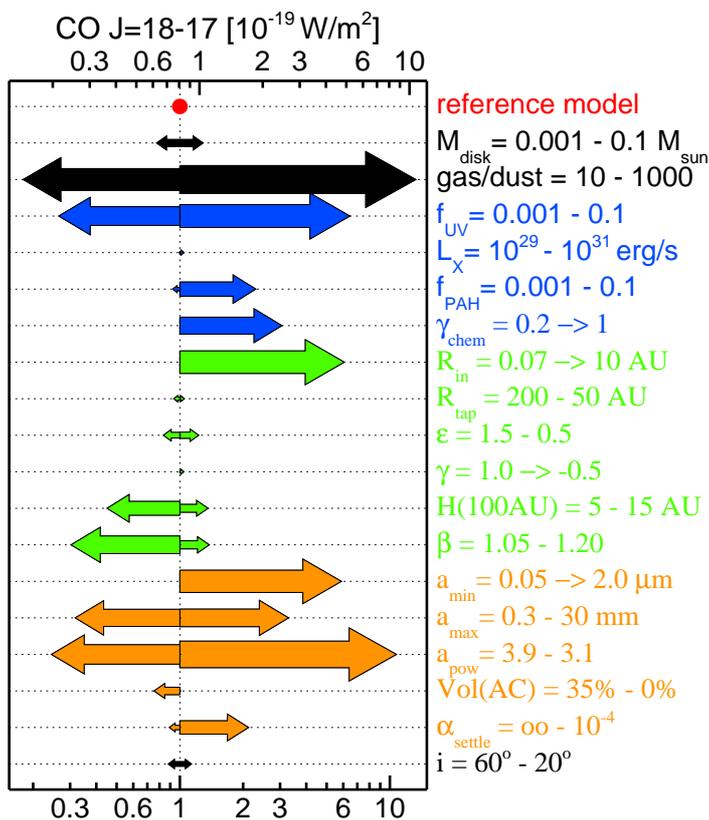}
  \vspace*{-5mm}
  \caption{Impact of model parameters on the CO $J\!=\!18\!\to\!17$ emission
    line at 144.87\,$\mu$m, see Fig.~\ref{fig:mm-slope} for explanations.}
  \label{fig:CO1817}
  \vspace*{-2mm}
\end{figure}

The ``impactogram'' of the CO $J\!=\!18\!\to\!17$ line
(Fig.~\ref{fig:CO1817}) shows many similarities to the behaviour of
the [OI]\,63.2\,$\mu$m line (Fig.~\ref{fig:OI63}), for example the
direction of effects, but the scaling of the abscissa is different. We
are now reporting on effects that can potentially change a line flux
by one order of magnitude.

Since both line and continuum are optically thick, and level
populations are close to LTE, we can use Eq.\,(\ref{eq:Flapprox}), see
Appendix, to estimate the line flux. This demonstrates that for these
optically thick far-IR lines, it is the difference between gas and
dust temperatures (in the disk surface layers between about 1\,AU and
5\,AU) that determines the line flux.

Besides the parameters directly involved in the gas heating (blue), we
can conclude from Fig.~\ref{fig:CO1817} that also the dust size
parameters play an essential role. All parameter changes that imply an
increase of the mean dust particle size (orange) lead to a reduction
of the dust UV opacity, see Sect.~\ref{sec:dusteffect}, and hence to
an increase of the thickness of the layer where the gas temperature is
substantially larger than the dust temperature.  Precisely speaking,
as will be explained in Sect.~\ref{sec:COrovib} and
Fig.~\ref{fig:COvibUnderstand}, it is the ratio $\tau_{\rm
  line}/\tau_{\rm UV}$ which is important. Since the gas temperature
structure is more or less fixed to $\tau_{\rm UV}$, more gas, or
lower UV continuum optical depths, both result in larger CO
$J\!=\!18\!\to\!17$ line fluxes. This also explains the dependence on
dust settling.

Interestingly, when the inner disk radius is increased in the model to
10\,AU, \ie well beyond the outer radius of the line emitting region
in the reference model, the line flux increases. In this case, instead
of radially continuous line emission, the line is preferentially
emitted from the distant inner wall, and the larger emitting area of
that wall seems more relevant than the radial dilution of the
impinging UV and X-ray photons, as long as this high-energy
irradiation is sufficient to cause gas temperatures $\ga\!200\,$K at
the inner wall.

\subsubsection{(sub-)mm CO isotopologue lines}
\label{sec:mmCO}

\begin{figure*}[!t]
  \vspace*{-2mm}
  \begin{tabular}{lll}
  \hspace*{-3mm}\includegraphics[height=86mm]{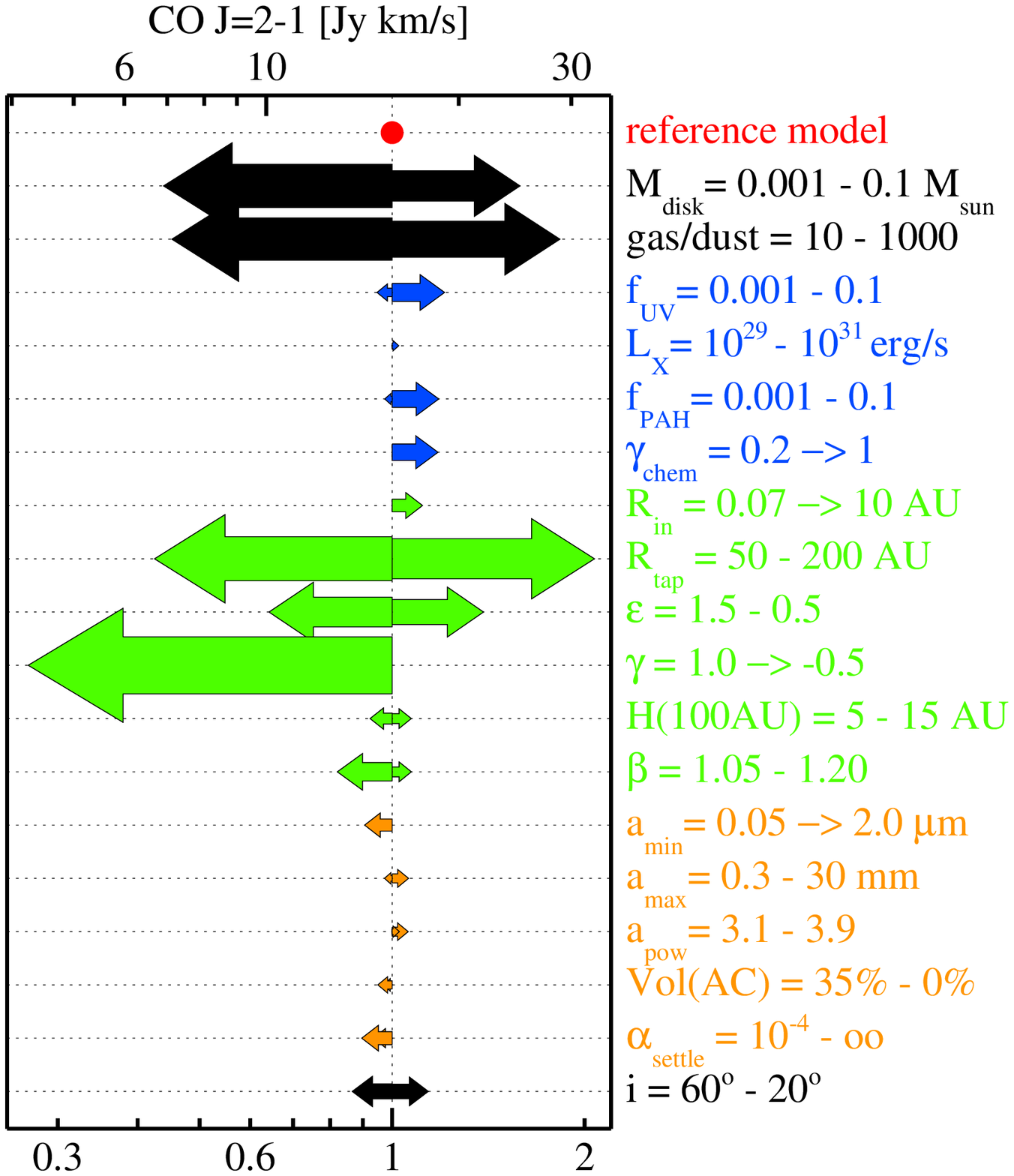} &
  \hspace*{-5mm}\includegraphics[height=86mm,trim=0 0 179 0,clip]{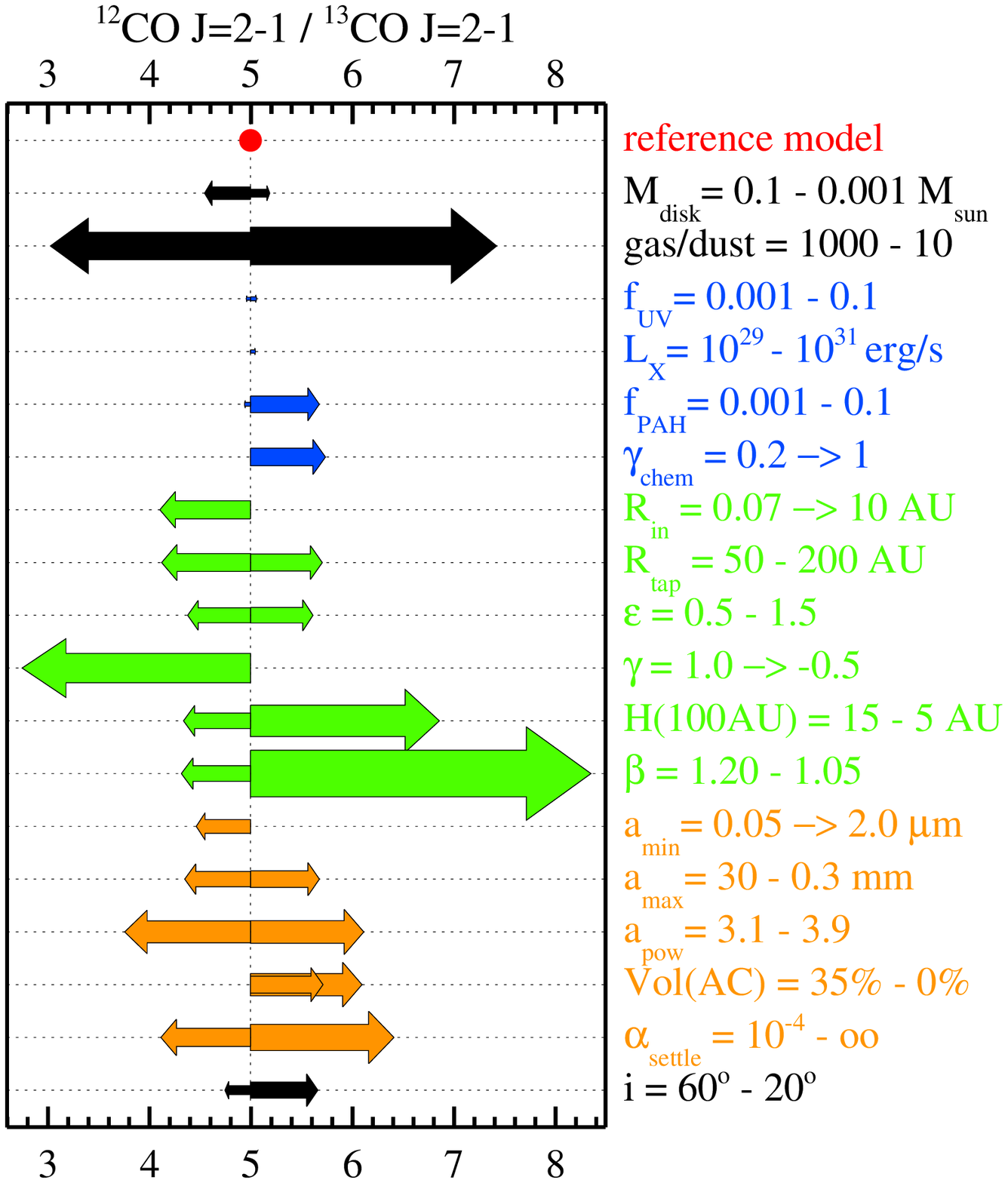} &
  \hspace*{-5.3mm}\includegraphics[height=86mm]{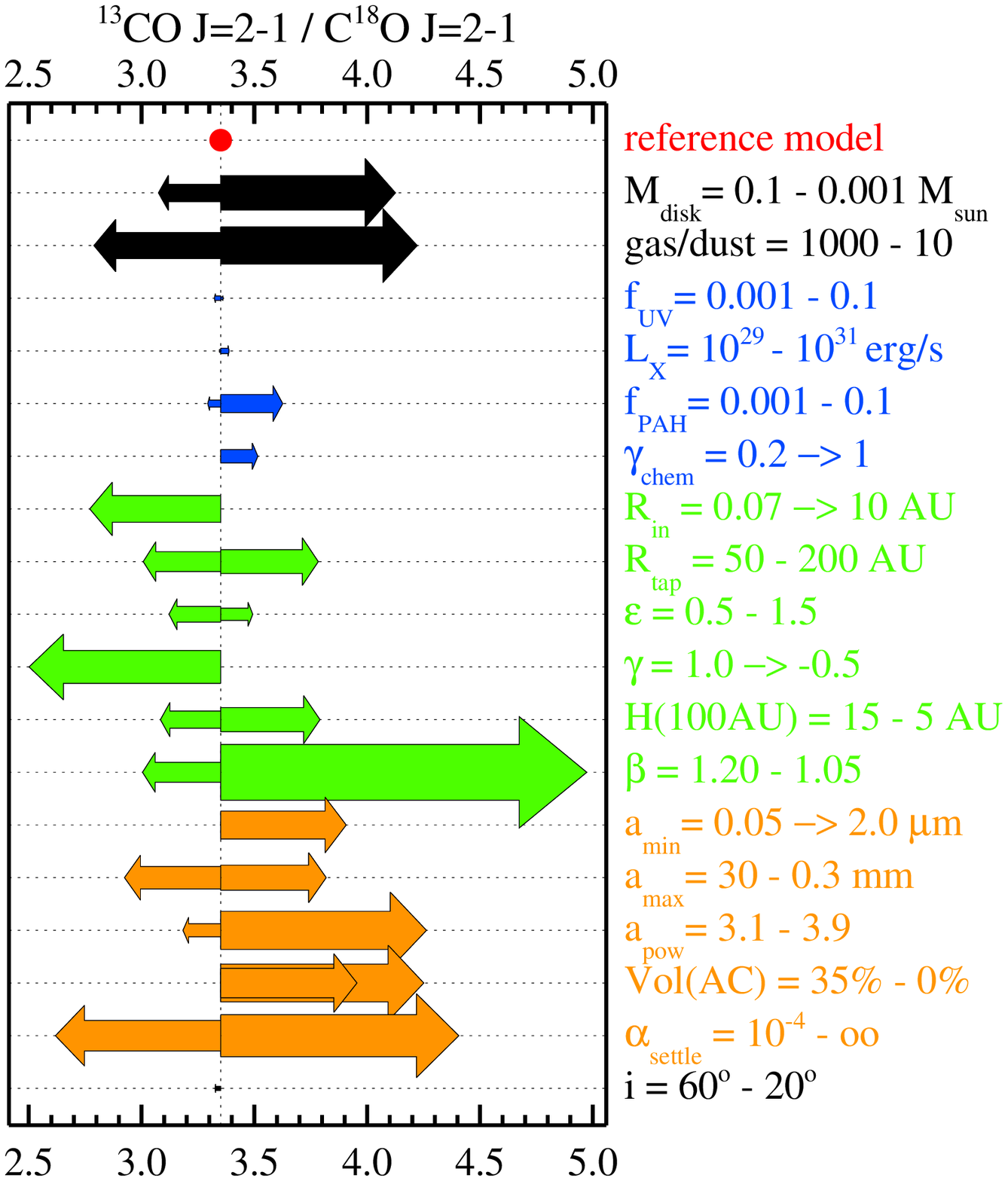}
  \end{tabular}
  \vspace*{-1mm}
  \caption{Predicted behaviour of the $^{12}$CO, $^{13}$CO and
    C$^{18}$O $J\!=\!2\!\to\!1$ isotopologue lines around 1.3\,mm.
    The left plot shows the impact of model parameters on the
    predicted $^{12}$CO line flux, see Fig.~\ref{fig:mm-slope} for
    more explanations. The two figures on the right show the impacts
    of model parameters on the $\rm^{12}CO/^{13}CO$ and
    $\rm^{13}CO/C^{18}O$ line ratios. Note that the direction of
    effects is sometimes inversed on the r.h.s., for example the
    dependency on disk flaring parameter $\beta$. Less flaring leeds
    to weaker CO lines in general, but to larger line ratios. On the
    right, both an increase and a decrease of the amorphous carbon
    dust volume fraction Vol(AC) lead to higher line ratios, here the larger
    arrows correspond to $\rm Vol(AC)\!=\!0$\%.}
  \label{fig:COiso}
\end{figure*}

The CO $J\!=\!2\!\to\!1$ isotopologue lines have excitation energies
of about $17\,$K and critical densities
$\sim\!7\times10^{4}\rm\,cm^{-3}$.  The impact of the model parameters
on our computed CO isotopologue line fluxes is shown in
Fig.~\ref{fig:COiso}. In the following, we discuss these results
obtained by our 3D diagnostic line transfer computations
(Sect.\,\ref{sec:lineRT}), by means of a few simplified equations, to
demonstrate and understand the main dependencies we find.  The
$\rm^{12}CO$ line is optically thick in all models, the $\rm^{13}CO$
is optically thick in most models, and the $\rm C^{18}O$ line is
borderline optically thin, meaning that the radial distances
$R(\tau_{\rm line}\!=\!1)$, up to which the lines are optically thick,
depends on isotopologue. In the reference model, $R(\tau_{\rm
  line}\!=\!1)\!\approx\!450\,$AU for the $\rm^{12}CO$ line,
$\approx\!210\,$AU for the $\rm^{13}CO$ line, and $\approx\!110\,$AU
for the $\rm C^{18}O$ line. The level populations connected to the CO
isotopologue lines in the (sub-)mm regime result to be close to
LTE, and the continuum is optically thin. Under these circumstances,
the line fluxes are approximately given by
\begin{equation}
  F_{\rm line} \approx 2\Delta\Omega\,\Delta\nu\,
                      B_\nu[T_{\rm gas}(\tau_{\rm line}\!=\!1)]\ ,
  \label{eq:lineapprox}
\end{equation}
see Eq.\,(\ref{eq:Fline}) and explanations in
Appendix~\ref{app:ThickLines}.  The mean gas temperature in the CO
(sub-)mm line emission regions (see Fig.\,\ref{fig:LineOrigin})
results to be $\langle T_{\rm gas}\rangle \approx T_{\rm
  gas}(\tau_{\rm line}\!=\!1)\approx 20-35$\,K, and varies only little
throughout the presented models (all isotopologues). This temperature
only affects the line fluxes in a linear way in the Rayleigh-Jeans
limit.  Therefore, the discussion of the CO isotopologue line fluxes
\begin{equation}
  F_{\rm line} \propto \Delta\Omega 
              \approx \frac{\pi R^2(\tau_{\rm line}\!=\!1) \cos(i)}{d^2}\ ,
\end{equation} 
simplifies to a discussion of the radius up to which the CO lines are
optically thick $R(\tau_{\rm line}\!=\!1)$ as follows.
\begin{my_itemize}
\item The $^{12}$CO line fluxes depend only little on all parameters
  which do not change $R(\tau_{\rm line}\!=\!1)$ significantly, in
  particular $f_{\rm UV}$, $L_X$, $f_{\rm PAH}$, dust size and
  settling parameters, scale height $H_0$ and flaring parameter
  $\beta$, as well as parameters only relevant for the inner disk,
  such as $R_{\rm in}$.
\item Most important are those model parameters which directly 
  determine $R(\tau_{\rm line}\!=\!1)$, these are the tapering-off
  radius $R_{\rm tap}$ and the tapering-off exponent $\gamma$.
\item The dependency on $\gamma$ is remarkable. The reference model
  has a tapering-off radius of $R_{\rm tap}\!=\!100$\,AU, a mean CO
  emission temperature of $T_{\rm gas}(\tau_{\rm line}\!=\!1)\approx
  25\,$K, and a $^{12}$CO $J\!=\!2\!\to\!1$ line flux of $\rm
  15.7\,Jy\,km/s$ ($1.2\times10^{-19}\rm\,W/m^2$). According to
  Eq.\,(\ref{eq:lineapprox}), this line flux corresponds to an
  emitting radius of $R(\tau_{\rm line}\!=\!1)\!=\!450\,$AU, which is
  possible only because we assume a smoothly decreasing surface
  density structure beyond $R_{\rm tap}$ with $\gamma\!=\!1$ in the
  reference model.  If the disk has a much sharper outer edge
  ($\gamma\!=\!-\,0.5$) the mean CO emission temperature increases by
  10\,K, but the flux is down to $\rm 4.2\,Jy\,km/s$
  ($3.3\times10^{-20}\rm\,W/m^2$), which then corresponds to a radius
  of only 200\,AU. These numbers are in very good agreement with the
  radii $R(\tau_{\rm line}\!=\!1)$ that we can directly measure in the
  models, see Fig.~\ref{fig:LineOrigin}.
\item There are a number of model parameters which indirectly change
  $R(\tau_{\rm line}\!=\!1)$, among them the disk mass $M_{\rm disk}$,
  the gas/dust ratio, and the column density powerlaw exponent
  $\epsilon$. These parameters change the amount of CO gas in
  the outer regions, which implies that $R(\tau_{\rm line}\!=\!1)$
  changes, too. These indirect influences on the CO line fluxes are
  stronger in models with exponential tapering-off.  Models with a
  sharp outer edge do not show much of those effects.
\end{my_itemize}
To summarise, the $^{12}$CO (sub-)mm lines probe the conditions in the
tapering-off distant gas above and around the disk, well beyond the
radial zones emitting the continuum. These conclusions are robust,
almost purely geometrical, because CO is such an abundant and robust
chemical constituent of the disks. There is no need for any
sophisticated chemical effects to understand these lines.

In contrast, the CO isotopologue line ratios (see r.h.s. of
Fig.~\ref{fig:COiso}) are more difficult to understand. It has been
suggested, \eg by \citet{Williams2014} and \citet{Miotello2014}, that
the CO isotopologue line ratios are an excellent probe of the disk
mass, in particular if rare isotopes like C$^{18}$O and C$^{17}$O can
be observed with high sensitivity\footnote{CO isotopologue chemistry
  is not included in our models, we use fixed abundance ratios as
  $\rm^{12}CO/^{13}CO\!=\!71.4$ and
  $\rm^{12}CO/C^{18}O\!=\!498.7$. See \citet{Miotello2014} for the
  effects of isotope-selective photo-destruction of carbon
  monoxide.}. We see the dependence of isotopologue line ratios on
disk mass clearly in our models, too, but we also see other parameter
dependencies that can be equally important. When looking at line
ratios, the major dependencies (like $F_{\rm line}\!\propto\!R^2$)
cancel, though not completely, and other secondary temperature and
chemical effects come into play. The (sub-)mm CO isotopologue line
ratios show the following effects:
\begin{figure}
  \centering
  \includegraphics[width=75mm]{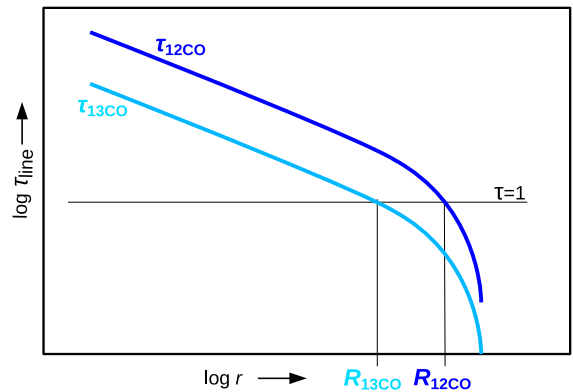}
  \vspace*{-1mm}
  \caption{Understanding (sub-)mm CO isotopologue line
    ratios. The $\rm^{13}CO$ line is weaker, because it approaches
    $\tau_{\rm line}\!=\!1$ at smaller radii.  Similar apparent radii
    (and fluxes) of the $^{12}$CO and $^{13}$CO lines can be obtained
    by making the outer disk edge sharp.}  
  \label{fig:COisoUnderstand}
  \vspace*{-3mm}
\end{figure}
\begin{my_itemize}
\item The dependence of the $\rm^{12}CO/^{13}CO$ line ratio on disk
  mass is almost negligible in our models, consistent with
  \cite{Williams2014}, because both lines are
  optically thick. The dependence of 
  the $\rm^{13}CO/C^{18}O$ line ratio on $M_{\rm disk}$ is more
  pronounced, because the C$^{18}$O line is borderline optically thin.
\item The asymmetric dependence of the $\rm^{13}CO/C^{18}O$ line ratio
  on $M_{\rm disk}$ indicates that we need disk masses as low as
  $10^{-3}\rm\,M_\odot$ to fully be in the optically thin limiting
  case for the C$^{18}$O line (the $\rm^{13}CO$ line is optically
  thick in most models anyway), also evident from the missing
  dependence on inclination.
\item The impact of the sharpness of the outer disk edge $\gamma$
  is remarkable. For a very sharp outer edge ($\gamma\!=\!-0.5$)
  we come close to a configuration where all three CO isotopologue
  lines remain optically thick until the disk ends abruptly, see
  Fig.~\ref{fig:COisoUnderstand}, in which case the line ratios would be
  unity. However, there are secondary temperature-effects (the
  $^{13}$CO and C$^{18}$O lines are emitted from deeper layers which
  are cooler) which always keep the isotopologue line ratios $>\!1$.
\item The dependencies on the gas/dust ratio are much more pronounced
  than on $M_{\rm disk}$, which shows that it is not simply the gas
  mass that counts. In fact, the two models denoted by $M_{\rm
    disk}\!=\!0.001\,M_\odot$ and $\rm gas/dust\!=\!10$ have the same
  gas mass, they only differ in terms of their total dust mass (see
  footnote below Table~\ref{tab:refmodel}). The more dusty model ($\rm
  gas/dust\!=\!10$) has a taller midplane shadow, hence cooler
  conditions even at higher disk layers, and more CO ice. Therefore,
  the $^{13}$CO line becomes borderline optically thin and its flux
  drops quickly, leading to a larger $\rm^{12}CO/^{13}CO$ line ratio.
\item Figure~\ref{fig:COiso} shows a strong impact of disk flaring
  (parameter $\beta$) on the CO isotopologue line ratios, which again
  can be attributed to CO ice formation. The $\beta\!=\!1.05$ model
  produces a self-shadowed disk configuration with very cold midplane
  conditions, favouring CO ice formation. The resulting lower gaseous
  CO column densities weaken in particular the $^{13}$CO line, hence
  the $\rm^{12}CO/^{13}CO$ line ratio becomes larger.
\item There are also noticeable dependencies on dust size parameters and
  dust settling, which are again related to CO ice formation. All dust
  size and opacity parameters which lead to an increase of the dust
  opacities around 1\,$\mu$m lead to a more pronounced midplane shadow,
  cooler conditions around $(100-300)\,$AU, more CO ice, hence larger
  $\rm^{12}CO/^{13}CO$ line ratios.
\end{my_itemize}
To summarise, the analysis of the (sub-)mm CO isotopologue line ratios
requires sophisticated modelling where the sharpness of the outer disk
edge, the strength and thickness of the disk midplane shadow impacting
the CO ice formation, and vertical temperature effects all play a
significant role. Spatially resolved line data with high S/N (\eg
ALMA) are required to disentangle these effects.

\subsubsection{CO $\upsilon=\!1\!\!\to\!0$ emission lines}
\label{sec:COrovib}

The fundamental ro-vibrational CO $\upsilon\!=\!1\!\!\to\!0$ emission
lines\linebreak are regularly detected in T\,Tauri stars
\citep[e.g.][]{Najita2003, Salyk2011, Brown2013}, in Herbig Ae/Be
stars \citep{Brittain2003, Blake2004, vanderPlas2015}, as well as in
transition disks \citep{Goto2006, Salyk2009, Pontoppidan2008}.
Surveys of CO ro-vibrational emission lines in young stars
\cite[e.g.][]{Brown2013} detect $^{12}$CO emission in about
$\sim$80$\%$ of the objects.  The line profiles are generally
double-peaked, but, in many objects, in particular in T\,Tauri stars,
line profiles can also be singly peaked which is usually interpreted
in terms of a slow (few km/s) molecular disk wind
\citep{Pontoppidan2011, Bast2011, Brown2013}.  $^{12}$CO integrated
line fluxes observed are of order 10$^{-17}\rm\,W/m^2$, and
line widths are 12-200\,km/s, mostly reflecting the expected distribution
of inclination.  Disks around early-type Herbig AeBe stars have
narrower line profiles, on average, than T Tauri stars
\cite{Brown2013}.

The lines have excitation energies $(3000-6000)$\,K and critical
densities\footnote{The large optical depths in the CO fundamental
  lines tend to 'quench' non-LTE effects, so the effective critical
  density is actually lower by about another 2 orders of magnitude
  \cite[see, e.g.][]{Woitke1996}.} of order
$(10^{12}-10^{14\,})\rm\,cm^{-3}$, depending on temperature and
depending on whether the gas is H$_2$-rich or atomic, see
\citep{Thi2013} for details.  The low-$J$ fundamental CO lines are
always massively optically thick in our models, with vertical line
centre optical depths of order $10^{\,4}-10^{\,7}$ for $r\!<\!10\,$AU
in the reference model. Figure~\ref{fig:LineOrigin} shows that the CO
$\upsilon\!=\!1\!\!\to\!0$ lines are vertically emitted by a region
that extends radially to about 1\,AU in the reference model, from a
thin horizontal layer at the top of the warm molecular layer.  In addition, the
line can also be emitted from the far side of the directly illuminated
inner rim, if the disk is seen under an inclination angle
$i\!>\!0^{\rm o}$. This contribution is not accounted for in
Fig.~\ref{fig:LineOrigin}, and may in fact domiante under certain
circumstances.

\begin{figure}[!t]
  \centering
  \vspace*{-1mm}
  \begin{tabular}{ll}
  \hspace*{-5.5mm}\includegraphics[height=47.5mm]{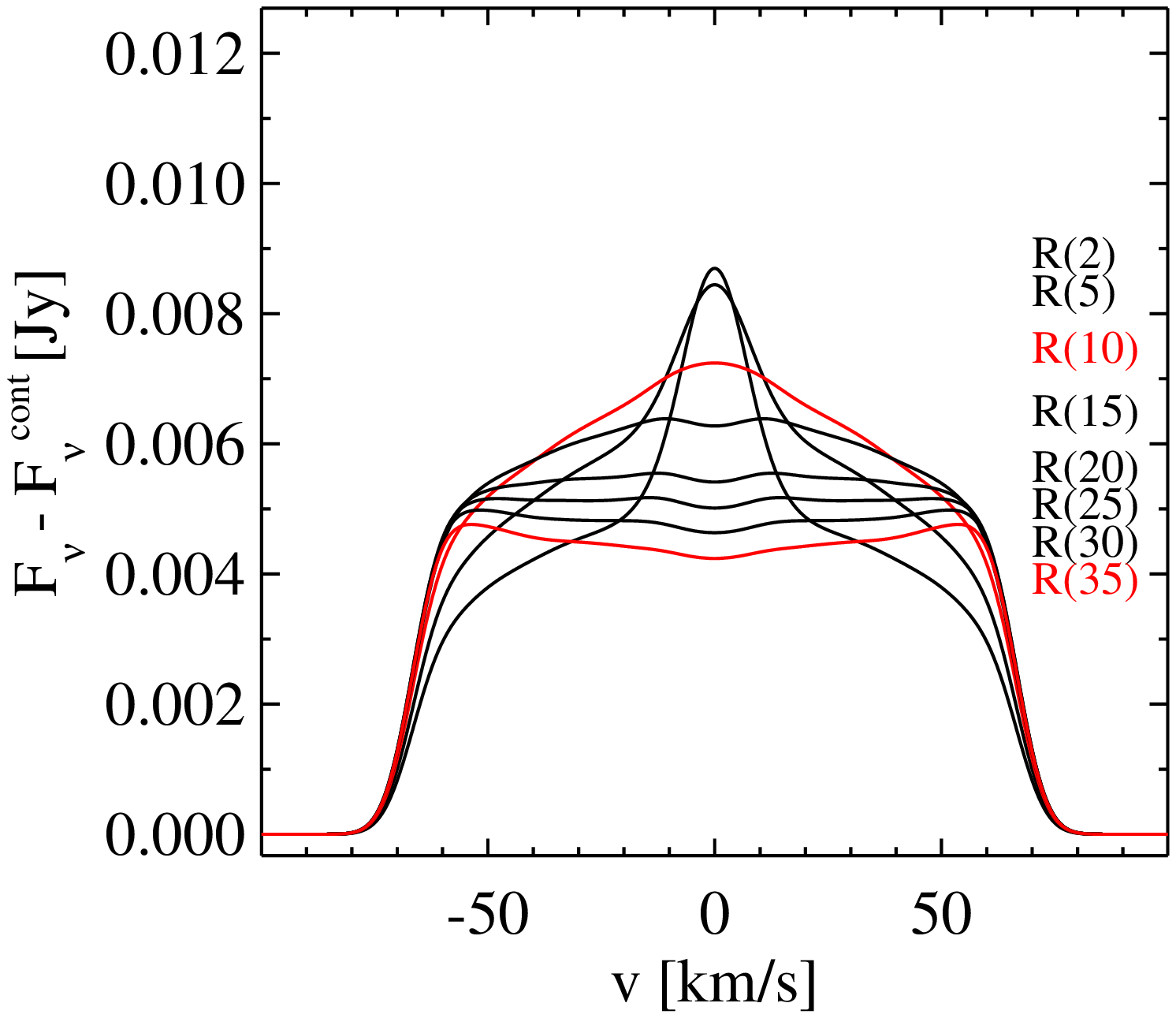} &
  \hspace*{-5.5mm}\includegraphics[height=47.5mm,trim=110 0 0 0,clip]{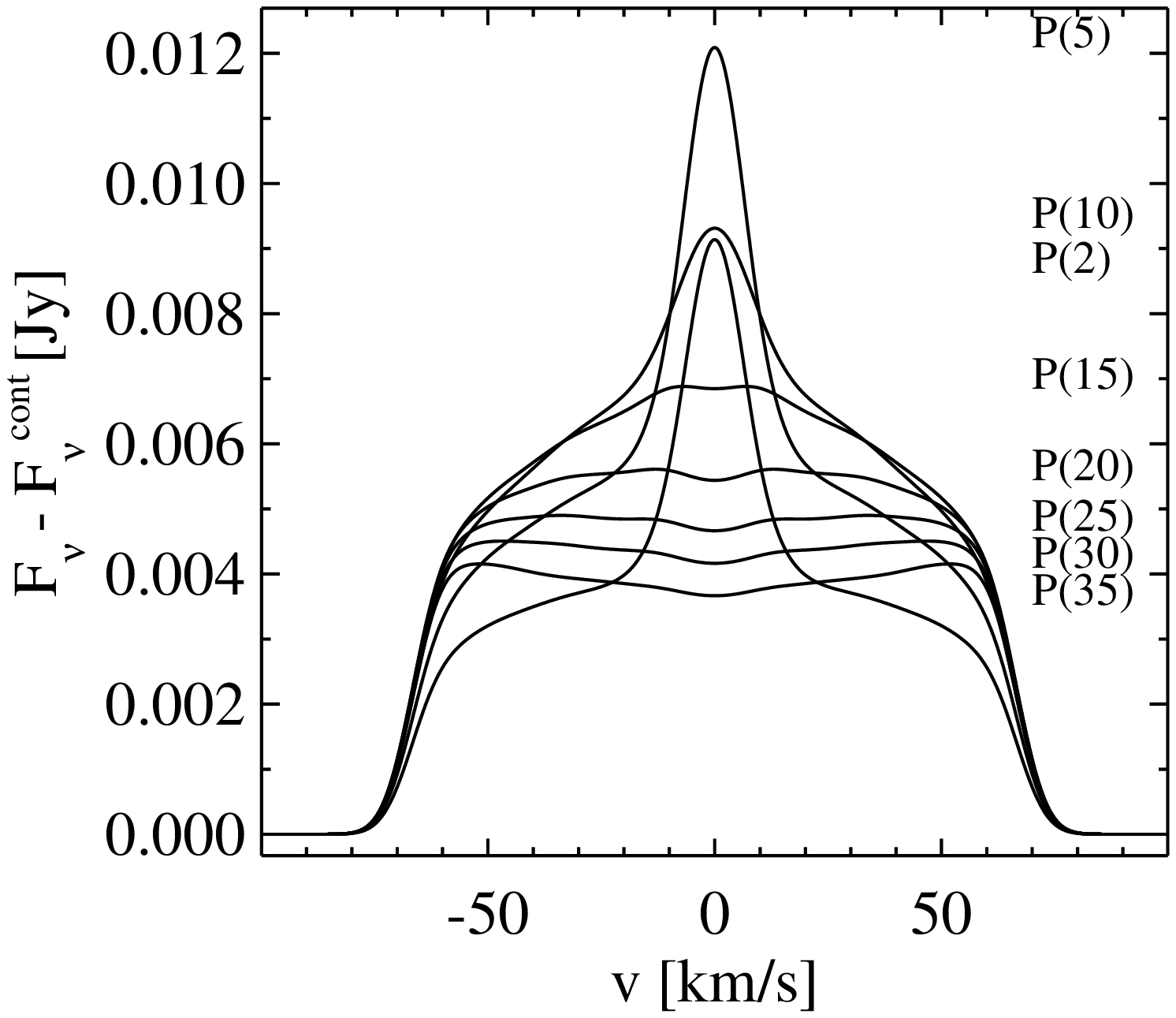} 
  \end{tabular}\\[-4.5mm]
  \caption{CO $\upsilon\!=\!1\!\to\!0$ line velocity profiles predicted
    by the reference model, continuum subtracted and convolved with a
    12\,km/s Gaussian (resolution $R\!\approx\!25000$). The R(10) and
    R(35) lines plotted in red are selected for further study of 
    the model parameters impacts in Fig.~\ref{fig:COrovib_effect1}.}
  \label{fig:COrovib_profiles}
  \vspace*{-2mm}
\end{figure}

\begin{figure*}
  \centering
  \vspace*{-2mm}
  \begin{tabular}{ll}
  \hspace*{-2mm}\includegraphics[height=97mm]{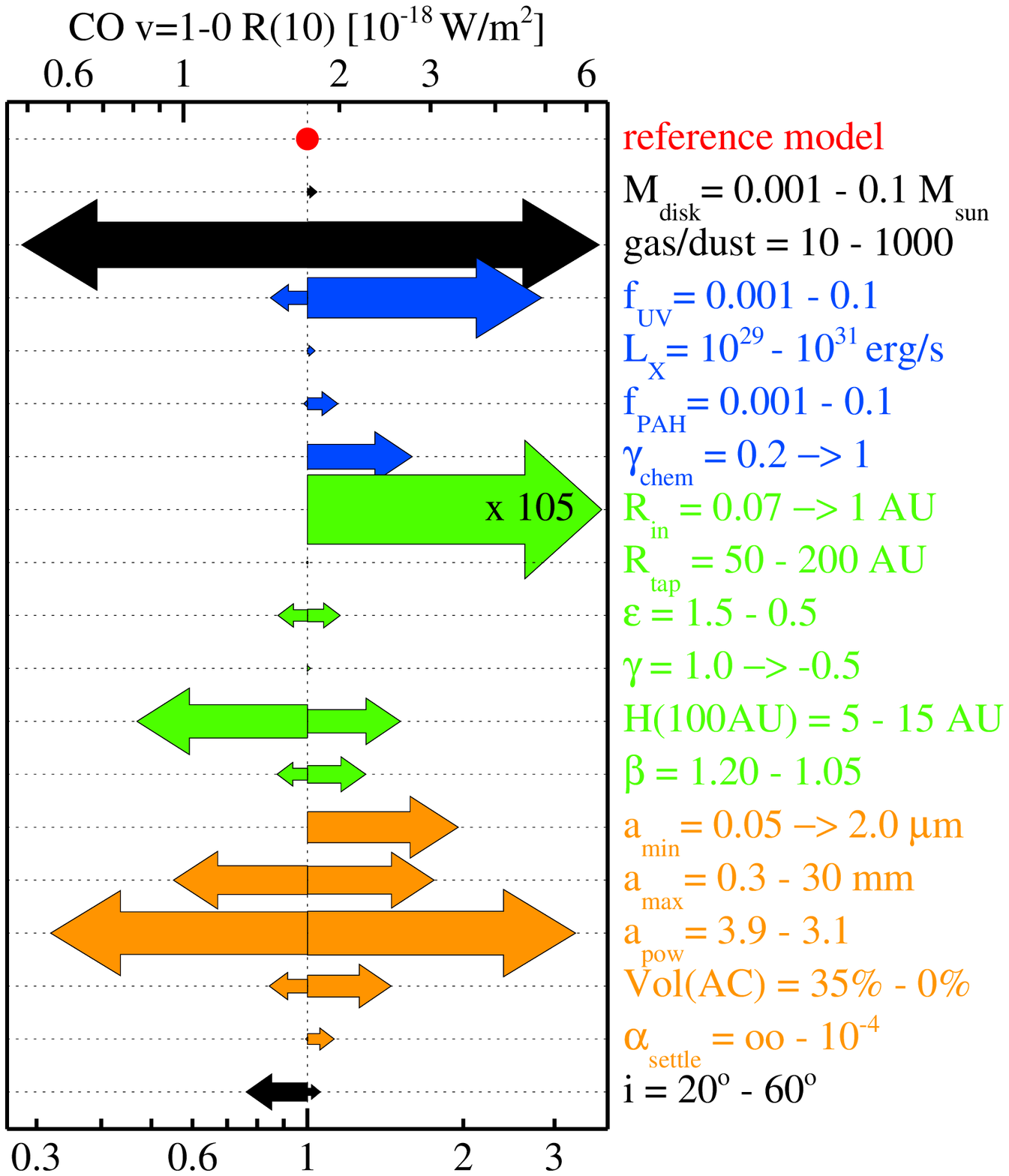} &
  \hspace*{-2mm}\includegraphics[height=97mm]{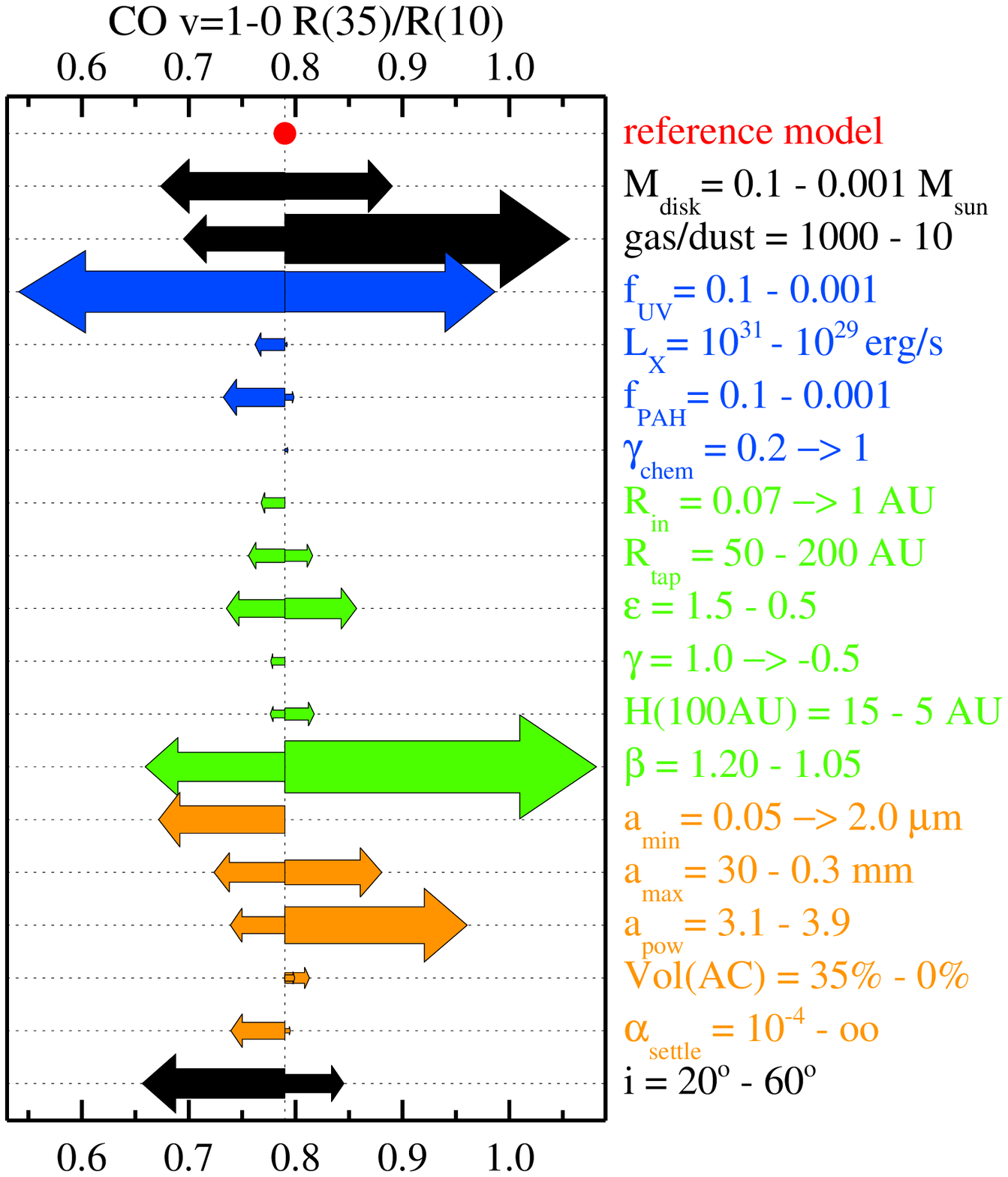} 
  \end{tabular}
  \caption{Impact of model parameters on ro-vibrational CO
    emission. On the l.h.s, the $\upsilon\!=\!1\!\!\to\!0$ $R(10)$
    line has been selected to study the effects of changing model
    parameters on line flux. The r.h.s. shows the $R(35)/R(10)$ line
    ratio, a measure for the 'CO rotational excitation
    temperature'. Large line ratios indicate emission from hot
    CO. Note that the direction of parameter impact is often reversed
    on the r.h.s., \ie weak lines are usually emitted from a tiny area
    of hot gas, whereas strong lines are emitted from an extended area
    of cool gas. See Fig.~\ref{fig:mm-slope} for further
    explanations.}
  \label{fig:COrovib_effect1}
  \vspace*{-1mm}
\end{figure*}

In LTE, the CO gas will emit substantially in the fundamental lines if
(i) the gas temperatures exceed about $500-1000\,$K, and (ii) the gas is
significantly warmer than the local dust temperature, according to
Eq.~\ref{eq:Flapprox} in App.~\ref{app:ThickLines}
\begin{equation}
F_{\rm line} \approx 2 \Delta\Omega\,\Delta\nu \left(B_\nu[T_{\rm
    gas}(\tau_{\rm line}\!=\!1)] -B_\nu[T_{\rm dust}(\tau_{\rm
    cont}\!=\!1)]\right) \ .
\end{equation}
These two conditions are {\em
  \,always\,} fulfilled at the inner rim, or, more precisely speaking,
in the thin hot surface layer that covers the inner rim facing the
star, which forms a very thin radial photodissociation region
(PDR). These emissions from the inner rim creates a broad, box-like
``minimum CO emission profile'' for all $R$-branch and $P$-branch
lines (Fig.~\ref{fig:COrovib_profiles}) with a hot characteristic
emission temperature. However, the CO emission lines created this way
would be quite faint and could not be detected with current
instruments, because of the tiny solid angle $\Delta\Omega$ occupied
by the inner rims of T\,Tauri stars.

In order to create observable line fluxes, the model must fulfil the
above stated two conditions for CO $\upsilon\!=\!1\!\!\to\!0$ emission also
at larger radii, {\em \,which only occurs in some models\,}, depending on
the parameters important for the gas heating, flaring, and dust
shielding (Fig.~\ref{fig:COrovib_effect1} and
Fig.~\ref{fig:COrovib_effect2}). If the extended gas is sufficiently
warm, we have a mixture of narrow (extended) cool CO emission with some
broad, hot emission from the inner rim, 
(Fig.~\ref{fig:COrovib_profiles}). The emissions from the extended
regions are lacking strong high-$J$ lines not only because the gas is
cooler there, but also because the optical depths are smaller for the
high-$J$ lines, \ie the high-$J$ lines probe deeper layers which
are cooler, and where the differences between gas
and dust temperatures start to vanish (Fig.~\ref{fig:COvibUnderstand}). The cool, extended contributions add a
central component to the line profile which is often double-peaked,
but not always (Fig.~\ref{fig:COrovib_effect2}).

The following results have been obtained with the default CO
  model molecule in {\sl ProDiMo} with vibrational quantum numbers
$\rm v\!\leq 2$ and rotational quantum numbers $J\!\leq\!50$, limited
to 110 levels, see \citep{Thi2013} for extended options.
Figure~\ref{fig:COrovib_effect1} shows the impact of the model
parameters on the R(10) line flux and characteristic CO emission
temperature as measured by R(35)/R(10).
Figure~\ref{fig:COrovib_effect2} shows how the mean line profile
changes when selected model parameters are varied. The main effects
are as follows:
\begin{figure}[!b]
  \vspace*{-3mm}
  \centering
  \hspace*{15mm}\begin{tabular}{p{27mm}l}
    \sf increase of $\Delta\Omega$ & \sf breakdown of excitation\\[0mm]
  \includegraphics[width=18mm]{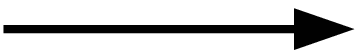} & 
  \includegraphics[width=18mm]{arrow.eps} \\[-1mm]
  \end{tabular}
  \hspace*{0mm}\includegraphics[width=75mm,trim=5 45 10 80,clip]
                               {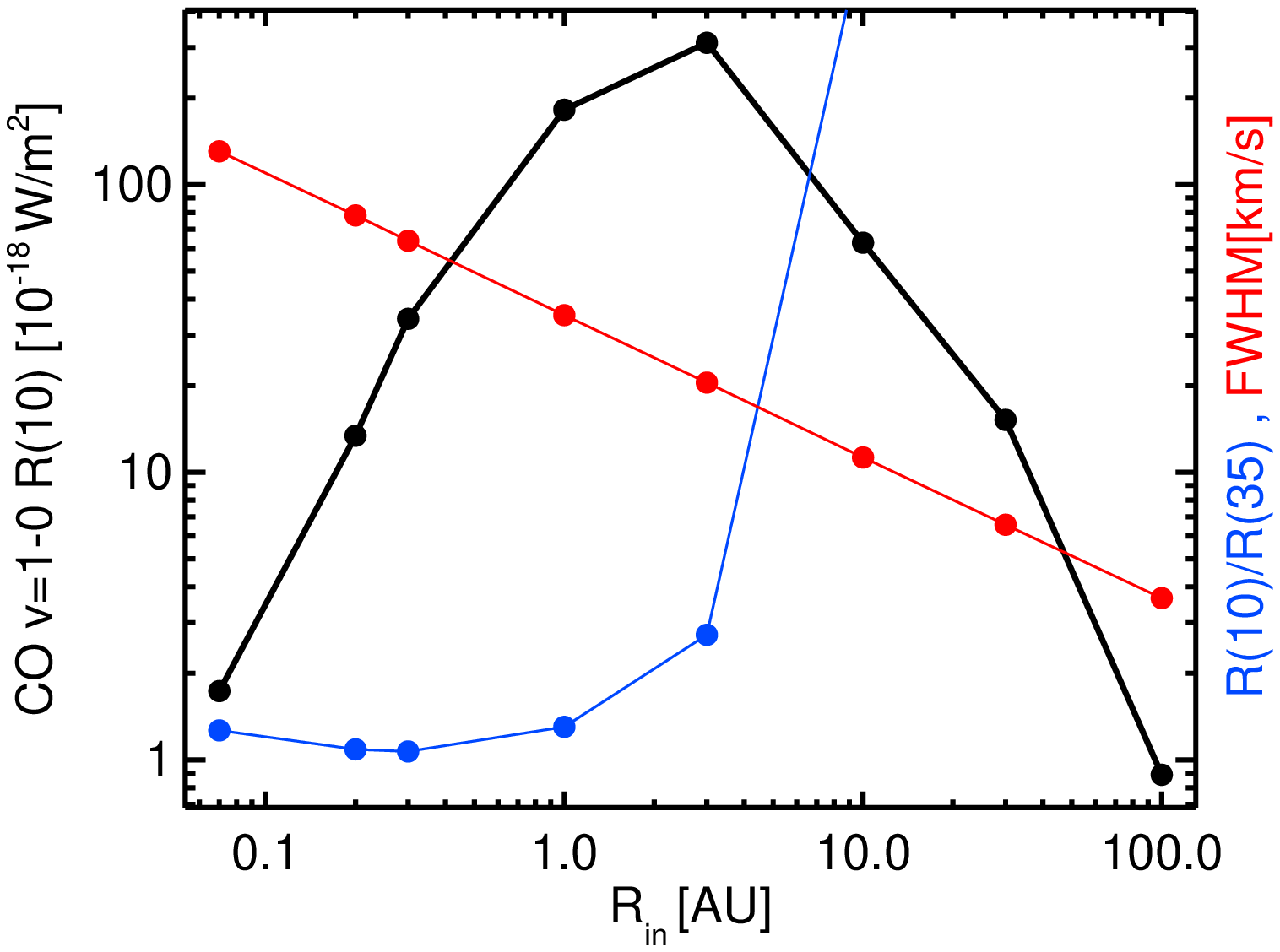}\\[-1mm]
  \caption{Impact of inner disk radius $R_{\rm in}$ on CO 
    $\upsilon\!=\!1\!\to\!0$ line emission fluxes and FWHM. The black line 
    shows the computed $R(10)$ line fluxes for the reference model,
    when varying $R_{\rm in}$ between 0.07\,AU and 100\,AU. The blue
    line shows the $R(10)/R(35)$ line ratio, and the red line shows
    the computed $R(10)$ line widths.}
  \label{fig:COrovib_Rin}
  \vspace*{-1mm}
\end{figure}
\begin{my_itemize}
\item Stronger CO line fluxes are connected with more centrally
  peaked line emissions coming from larger emitting areas.
\item All parameters that increase gas heating will generally
  increase line flux, in particular the UV excess $f_{\rm UV}$ 
  \citep{Garufi2014} and the efficiency of chemical heating $\gamma_{\rm chem}$. 
\item However, the X-ray luminosity has no significant
  influence on the fundamental CO emission lines.
\item Disk flaring makes the CO lines stronger and cooler.
\item A larger inner disk radius $R_{\rm in}$ leads to a larger solid
  angle $\Delta\Omega\!\propto\!R_{\rm in}^2$, hence {\em
    \,substantially\,} stronger CO lines. If $R_{\rm in}$ is
  increased, the heating UV flux
  gets radially diluted $\propto\!1/R_{\rm in}^2$, but can still provide
  sufficient gas heating and CO excitation up to about $R_{\rm
    in}\!\sim\!3\,$AU, depending on stellar UV luminosity. For even
  larger $R_{\rm in}$, however, the excitation conditions for
  fundamental CO line emission break down, see
  Fig.~\ref{fig:COrovib_Rin}.
\item There is a strong impact of the dust size parameters on the CO
  fundamental line emission. This is due to changes in the dust UV
  optical depths $\tau_{\rm UV}$, see Fig.~\ref{fig:COvibUnderstand}.
  Dust parameter choices which favour larger particles (larger $a_{\rm
    min}$, larger $a_{\rm max}$, smaller $a_{\rm pow}$) and less
  amorphous carbon {\em \,reduce\,} the UV dust opacities, \ie the UV
  light can penetrate deeper into the CO line emitting regions, which
  increases the gas temperatures and CO line fluxes.
\item After the inner disk radius, the gas/dust ratio has the largest
  impact on the fundamental CO emission. The reason for this effect is
  the same as for the dust size parameters. Larger gas/dust ratios
  lead to an increase of the $\tau_{\rm line}/\tau_{\rm UV}$ ratio.
\end{my_itemize}
\begin{figure}[!t]
  \centering
  \includegraphics[width=75mm]{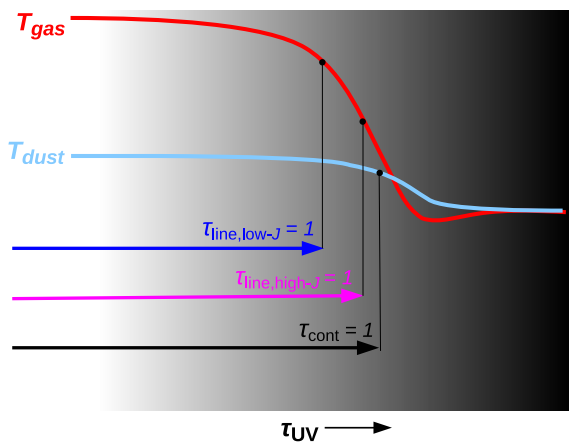}
  \vspace*{-1mm}
  \caption{Understanding optically thick CO ro-vibrational emission
    lines. The sketch can either represent the inner rim horizontally,
    or a distant disk column vertically. The gas and dust temperature
    structure is connected to $\tau_{\rm UV}$, the dust optical depth in the UV.
    High-$J$ and low-$J$ CO emission lines have
    different line optical depths, so their fluxes probe the contrast
    between gas and dust temperatures at different depths.}
  \label{fig:COvibUnderstand}
  \vspace*{-1mm}
\end{figure}
To summarise, the impinging UV flux and the local $\tau_{\rm
  line}/\tau_{\rm UV}$ ratio regulate the strength of the CO
fundamental line emissions, but the disk shape determines how much of
the stellar UV flux reaches a disk region under consideration. The
reference model shows very weak CO fundamental line fluxes with broad,
box-like profiles, which would in fact lead to non-detections, but the
model parameters can be changed to result in observable line flux
levels and more typical CO line profiles.

The strong dependence of the CO $\upsilon\!=\!1\!\!\to\!0$ line fluxes on
$R_{\rm in}$ suggests that any disk shape irregularities in the inner
regions $\la\!30\,$AU, which create highly extended vertical gas
columns that are exposed to fresh UV light from the central star,
could easily dominate the CO fundamental emission lines. Such
irregularities could be, for example, (i) large inner holes, (ii) disk
gaps with highly puffed-up secondary inner walls, (iii) spiral waves
which are warmer than the surrounding gas, causing those regions to
stick out vertically, or (iv) the launching regions of disk winds.

\section{Summary and conclusions}
\label{sec:Conclusions}

The analysis and interpretation of observational data from Class\,II
and III protoplanetary disks is a challenging task. Various
hydrodynamical, chemical, dust, and radiative processes are coupled to
each other in complicated ways, and current disk modelling groups are
using quite different assumptions to setup their 2D radiative transfer
and thermo-chemical models concerning disk shape, dust
size and opacity parameters, treatment of PAHs, and dust settling.

We have systematically investigated the effects of those assumptions in
this paper by have studying the impact of the associated model
parameters on the various continuum and line predictions, using a {\em
  \,holistic disk modelling approach\,} which allows us to calculate all
continuum and line observations on the basis of a single model. The
most important effects are:
\begin{my_itemize}
\item {\bf Disk shape} matters. In particular, scale heights and 
  disk flaring have large impacts on
  the gas and dust temperature distribution in the disk, and hence on
  SED, visibilities and gas emission lines. For very cold,
  self-shadowed disk configurations, ice formation can strongly reduce
  the (sub-mm) emission lines.

\item {\bf Hydrostatic T\,Tauri disk models} produce only little
  near-IR excess, but a strong flaring of the outer disk, which
  leads to a {\em \,re-increase\,} of the spectral flux $\nu F_\nu$
  between $\lambda\!=\!(20-50)\,\mu$m, whereas the opposite is
  typically observed for non-transitional disk.

\item {\bf Dust size} is important, not only for
  modelling the continuum, but also for modelling chemistry and
  emission lines, which is often not discussed in thermo-chemical disk
  models. By extending the dust size distribution to mm sizes, as
  required to fit the SED, the UV dust opacity is reduced by a
  factor of about 100, which means that UV photons can penetrate much
  deeper into the disk with ample effects on chemistry, temperature
  structure and gas emission lines.
 
\item {\bf New dust standard opacities} have been developed. Guided by
  a study of the multi-wavelength optical properties of dust
  aggregates particles \citep{Min2015}, we have developed a
  simplified and fast numerical treatment for dust opacities. We propose
  to use an effective porous mixture of amorphous laboratory silicates
  with amorphous carbon, a powerlaw size distribution with a
  distribution of hollow spheres to capture the most important size,
  material, and shape effects. A Fortran-90 package to compute the
  DIANA standard dust opacities is available at {\sl
    http://www.diana-project.com/data-results-downloads}.

\item {\bf Disk flaring and/or dust settling?\ } Dust settling affects
  primarily the outer disk regions, leading to cooler disks, lower
  continuum fluxes at mid-IR to cm wavelengths, and smaller apparent
  disk sizes in the millimetre continuum. These effects on the
  continuum observables are very similar to those in models with
  little or lacking disk flaring. However, concerning the gas emission
  lines, dust settling has just the opposite effect. Dust settling
  leaves the vertically extended gas bare and exposed to the stellar
  UV radiation, leading to higher gas temperatures and stronger gas
  emission lines in general. Thus, we can expect to distinguish
  between disk flaring and dust settling by observing in particular
  far-IR emission lines.

\item {\bf PAHs} can have important effects on the disk radiative
  transfer, even if the mid-IR PAH features are not visible in the
  SED.  If the PAH abundance reaches about 10\% of the interstellar
  standard, the PAH opacities start to catch up with the dust
  opacities in the UV and blue parts of the spectrum. Since the PAHs
  are not settled, and have negligible scattering opacities, the PAHs
  change the ways in which UV photons reach the disk.  In fact, the
  PAHs can effectively ``shield'' the disk from UV photons.  We are
  proposing a simplified method by treating the PAHs consistently in
  the continuum radiative transfer assuming radiative equilibrium,
  with a PAH temperature independent from the dust temperature.
\end{my_itemize}
With regard to particular observations, we find the
following robust dependencies that can be used for diagnostic
purposes. 
\begin{my_itemize}
\item The {\bf SED mm-slope} is more gentle than expected from the
  dust opacity-slope if the disk is cold in the midplane, which
  happens consistently in all of our T\,Tauri models.

\item The {\bf [OI]\,63.2\,$\mu$m line} is optically thick under all
  explored circumstances, and probes the tenuous layers {\em \,above the
    molecular disk\,} at radii $\sim(10-100)\,$AU. Since this line
  provides the most important cooling process in these layers, it is
  subject to a self-regulation mechanism where the line luminosity must
  equal the spatially integrated heating rate in that
  region, in form of various UV and X-ray processes.

\item The {\bf high-$J$ CO lines} are optically thick and triggered by
  an excess of the gas temperature over the local dust temperature at the
  upper edge of the molecular layer around $1-10\,$AU. The lines
  are stronger for larger $\tau_{\rm line}/\tau_{\rm UV}$ ratios, and
  hence an excellent tracer of the gas/dust ratio. 

\item The {\bf (sub-)mm $^{12}$CO lines} probe the radial extension
  and conditions in the tapering-off distant gas around the disk, well
  beyond the radial zones responsible for the (sub-)mm
  continuum. These conclusions are robust, almost purely geometrical,
  because these lines are always optically thick in our models,
  and because CO is such an abundant and chemically robust constituent
  of the disks.

\item The {\bf (sub-)mm CO isotopologue lines} of $^{13}$CO and
  C$^{18}$O require more sophisticated modelling, where the sharpness
  of the outer disk edge, the strength and vertical extension of the
  disk midplane shadow, and vertical temperature gradients all play a
  significant role.  The $^{12}$CO/$^{13}$CO isotopologue line ratio
  is quite independent from disk mass, because both lines are usually
  optically thick in our models, but increases significantly in
  cold disks where the increased efficiency of CO ice formation
  affetcs the $^{13}$CO lines more than the $^{12}$CO lines.

\item The {\bf CO $\upsilon\!=\!$1-0 fundamental lines} are massively optically
  thick in all our models, and hence provide a clear diagnostic for
  the existence of warm ($\ga\!500$\,K) gas in the inner disk regions,
  with a gas temperature in clear excess to the underlying dust
  temperature. In our standard T\,Tauri disk setup, these lines are
  quite weak, because these conditions are mostly met just at the
  inner rim. Stronger and more extended CO emission lines are
  obtained, however, if the impinging UV flux is larger, and/or\,\ if
  the local optical depth ratio $\tau_{\rm line}/\tau_{\rm UV}$ is
  larger.  In practise, such extended emission occurs in models where
  sub-micron grains are missing, and/or\,\ where the gas/dust ratio is
  larger, leading to more realistic line profiles. Disk shape irregularities,
  like inner holes with diameters of order several AU, can lead to
  much stronger CO fundamental line emissions as well.
   
\end{my_itemize}

\noindent Our results demonstrate that the various continuum and line
observables probe the physical conditions at very different radii and
different heights in the disk. Thus, only a combination of suitable
multi-wavelength dust and gas observations can break the various
degeneracies, for example those in SED modelling, and can lead to
more reliable disk diagnosis.

This paper series aims at setting {\em \,new disk modelling
  standards\,} for the analysis of multi-wavelength continuum and line
observations for protoplanetary disks, with easy to implement, yet
physically grounded, and practical assumptions, which are sufficiently 
motivated by observations. We will continue this series by exploring
the effects of chemical networks and rates in Paper\,II (Kamp\etal in prep.)
and element abundances in Paper\,III (Rab\etal 2015, submitted).

We intend to offer our modelling tools and collected data sets to the
community at the end of the FP7 {\sl DIANA} project, see {\sl
  http://www.diana-project.com}.

\bigskip\noindent {\bf Acknowledgements:\ } We would like to thank
Dr.~Pieter Degroote for compiling a database of updated UV to far-IR
photometric measurements for all our 80 targets, including accurate
filter functions and zero-points, and Dr.~Hein Bertelsen for improving
the discussion about fundamental CO emission. The research leading to
these results has received funding from the European Union Seventh
Framework Programme FP7-2011 under grant agreement no 284405. Ch.~Rab
and C.\,Baldovin-Saavedra acknowledge funding by the Austrian Science
Fund (FWF), project number P24790, and the Austrian Research Promotion
Agency (FFG) under grant agreement FA 538022, respectively.

\bibliography{reference}

\clearpage
\appendix

\section{Stellar parameters and irradiation}
\label{app:StellarPara}

\begin{figure*}
\vspace*{-2mm}
\centering
\begin{tabular}{cc}
 \hspace*{-5mm}\includegraphics[width=82mm]{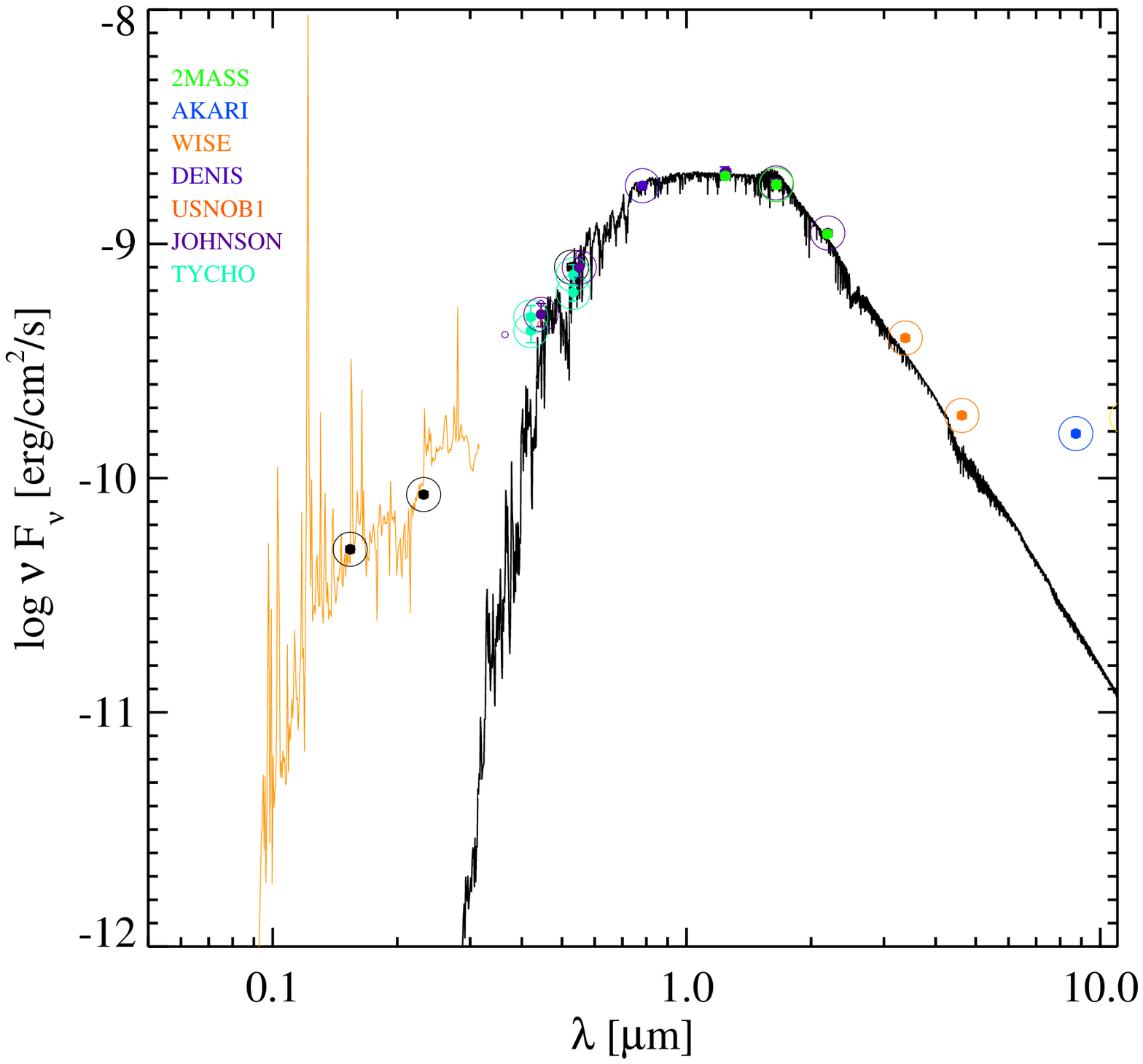}
&\hspace*{-5mm}\includegraphics[width=82mm]{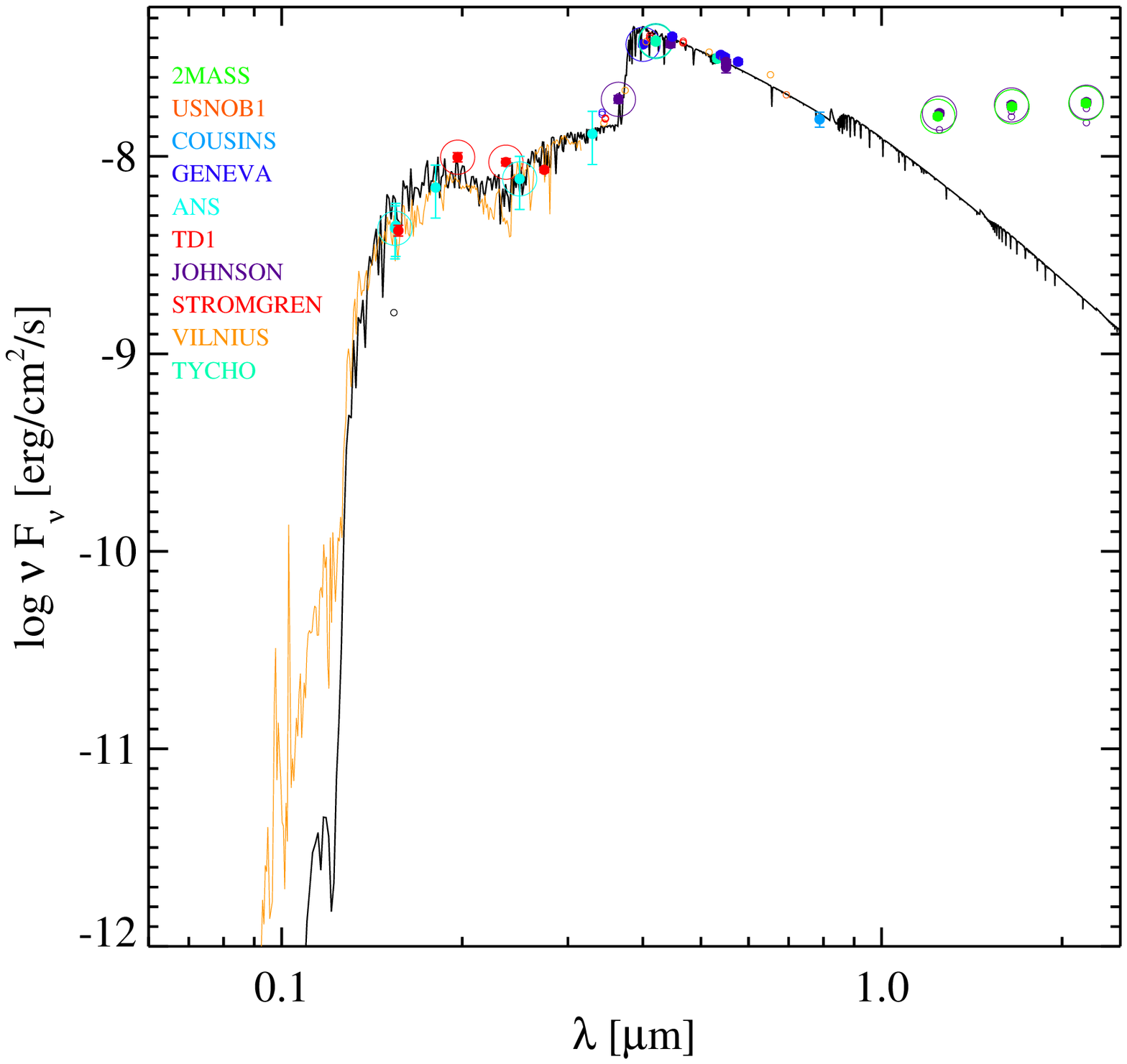}\\[-1ex]
 \hspace*{-6mm}\includegraphics[width=85mm]{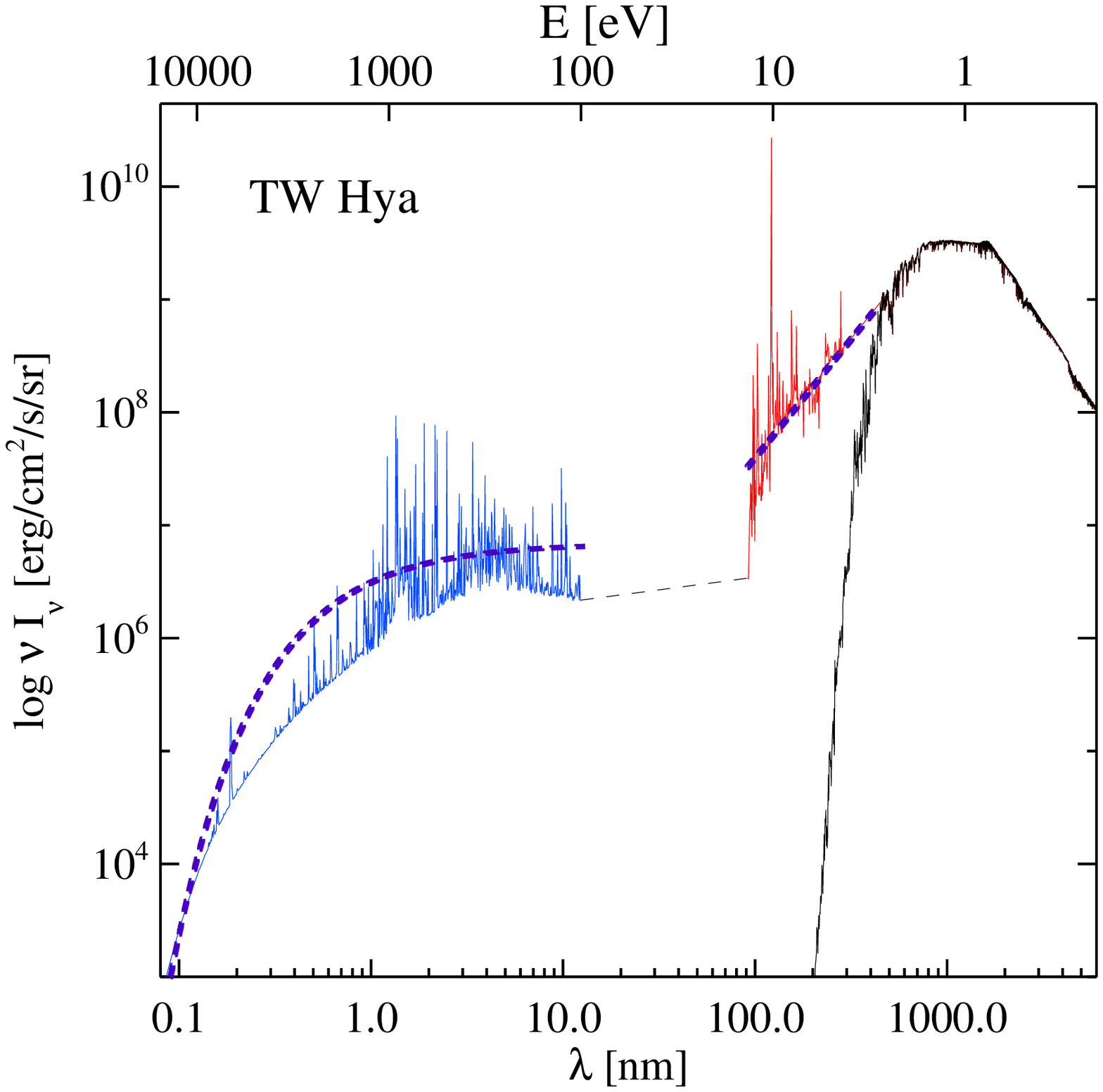}
&\hspace*{-6mm}\includegraphics[width=85mm]{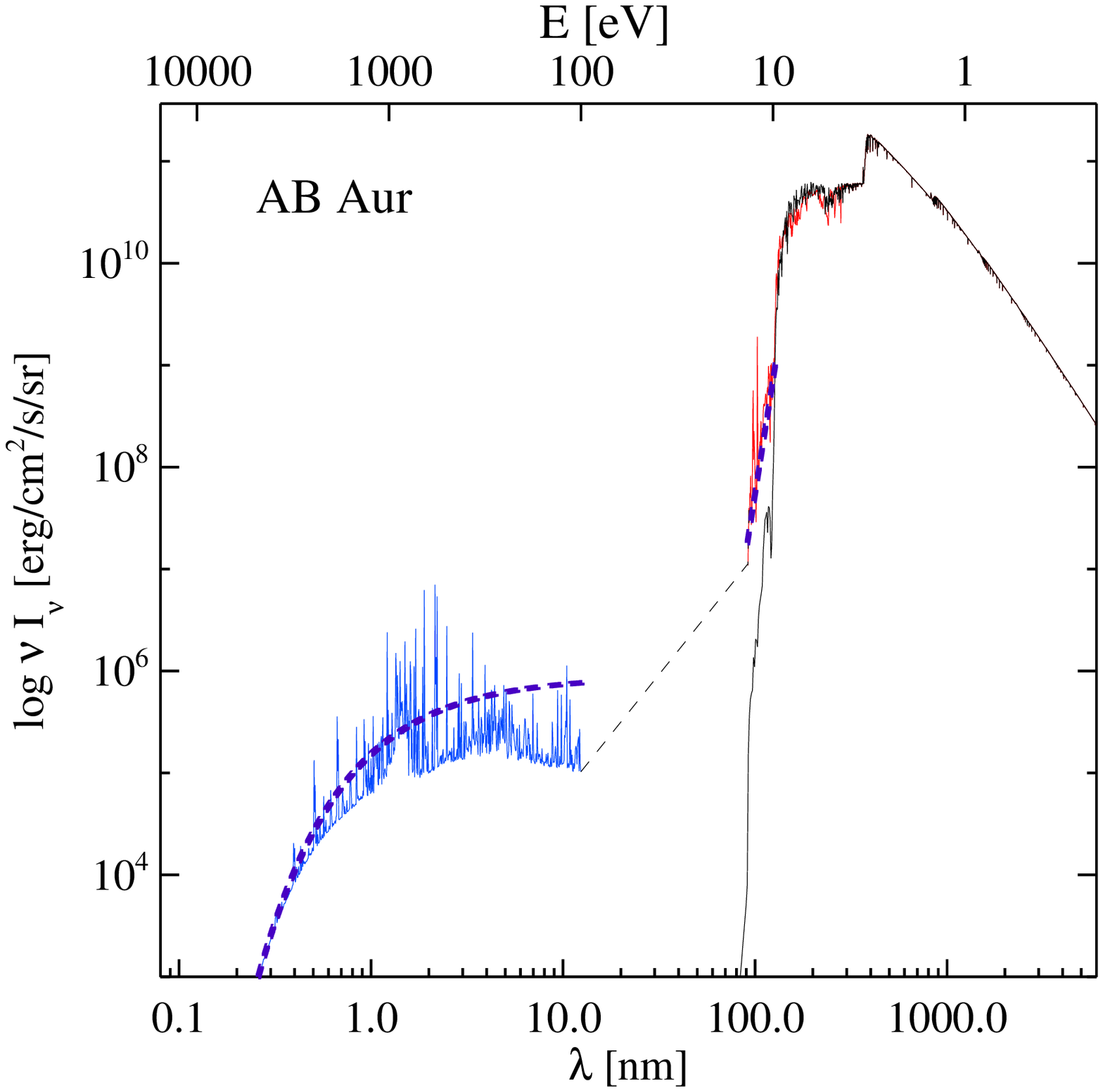}\\[-1ex]
\end{tabular}
  \caption{Fitting stellar parameters, and compilation of the stellar
    irradiation for two examples, TW\,Hya (left) and AB\,Aur
    (right). The upper plots show measured photometric fluxes
    (coloured symbols) as function of wavelength $\lambda$, in
    comparison to our best fitting, reddened {\sl Phoenix} stellar
    atmosphere model spectrum (black), and the averaged observed FUV
    data (orange). The lower plots show the de-reddened surface
    intensities $I_\nu(r\!=\!R_\star)$ (black), and the averaged,
    de-reddened FUV data ($\lambda\!>\!91.2$\,nm, red). The blue lines
    show fits to de-absorbed X-ray data, using a two-component
    X-ray gas emission model (energy $E\!>\!0.1$\,keV). The dashed grey line
    marks the EUV regime in between, which is assumed to be absorbed
    by neutral hydrogen between the star and the disk and hence
    disregarded in the disk model.  The dashed magenta lines show
    quick parametric fits to the UV and X-ray data, see Eqs.~(\ref{UV-fit}) 
    and (\ref{Xray-fit}) and text, not used for the TW\,Hya and AB\,Aur 
    models.}
  \label{fig:irradiation}
\end{figure*}

\subsection{Stellar parameters}

The photospheric component of the stellar emission is characterised by the
effective temperature $T_\star$, the surface gravity $g$, and the
stellar (photospheric) luminosity $L_\star$. Since these properties
are fairly well-known for most of our target objects in the
literature, we have decided not to put too much efforts into this
issue.  Assuming solar abundances for the star, we use standard {\sl
  PHOENIX} stellar atmosphere models \citep{Brott2005} to fit $T_{\rm
  eff}$, $L_\star$ and the interstellar extinction $A_V$, to our
photometric and spectroscopic data (see Fig.\,\ref{fig:irradiation}).
A thorough determination of $T_\star$ requires to fit
high-resolution optical spectra. For most target objects, this has
been done already in the literature, so we can pick $T_\star$
from the literature and only fit $A_V$ and $L_\star$ for the assumed
distance $d$, using at first an estimate of the surface
gravity $g$.

Once $T_\star$ and $L_\star$ are determined, we involve pre-main
sequence stellar evolutionary models \citep{Siess2000} to find the
stellar mass $M_\star$ and the age.  The stellar radius and surface
gravity are then given by $L_\star\!=\!4\pi\,R_\star^2\sigma T_{\rm
  eff}^4$ and $g\!=\!GM_\star/R_\star^2$. We use the resulting value
for $g$ to redo the fitting above. This procedure is found to converge
very quickly, and -- thanks to having fixed $T_\star$ from the
literature -- gives quite unambiguous results, see examples in
Table~\ref{tab:stellar}.

\begin{table}[!b]
\def\z{\hspace*{-2mm}}
\caption{Assumed/derived stellar parameters for two example objects}
\vspace*{-2mm}
\hspace*{-0.5mm}\resizebox{90mm}{!}{
\begin{tabular}{ccc|cc}
\hline
&&&&\\[-2.2ex] 
symbol        & unit        & meaning                 & TW\,Hya & AB\,Aur\\
\hline
\hline
&&&&\\[-2.2ex] 
$d$           & [pc]        & distance$^{(0)}$        & 51      & 144 \\
$T_\star$      & [K]        & effective temperature$^{(0)}$   & 4000    & 9550 \\
$L_\star$     & [$L_\odot$] & \z photospheric luminosity\z & 0.24  & 42 \\
$A_V$         & [mag]       & \z interstellar extinction\z & 0.20  & 0.42\\
\hline
&&&&\\[-2.2ex] 
$M_\star$     & [$M_\odot$] & stellar mass            & 0.75    & 2.5 \\
$R_\star$     & [$R_\odot$] & photospheric radius     & 1.026   & 2.37 \\
\z $\log\,g$\z & [$\rm cm/s^2$]  & surface gravity    & 4.29    & 4.08 \\ 
\z SpTyp\z   & [-]          & spectral type           & K7      & A0 \\
age          & [Myrs]       & stellar age             & 13      & 5  \\ 
\hline
&&&&\\[-2.2ex] 
\z$L_{\rm UV1}/L_\star$\z &[-]& \z band\,1 UV luminosity$^{(1)}$  
                                          & \z 0.0019\z & \z 0.00024\z\\
\z$L_{\rm UV2}/L_\star$\z &[-]& \z band\,2 UV luminosity$^{(2)}$  
                                                 & 0.053 & \z 0.095\z \\
\z$L_{\rm Ly\alpha}/L_\star$\z &[-]& \z Lyman $\alpha$ luminosity  
                                                & 0.034 & \z 0.00021\z\\
$p_{\rm UV}$  & [-]       & UV powerlaw fit index     & 1.1     & 11  \\
\hline
&&&&\\[-2.2ex] 
$L_{X1}$     & \z[$10^{\,30}\rm erg/s$]\z
                             & \z hard X-ray luminosity$^{(3)}\z$
                                                 & 0.27 & \z 0.099\z\\
$L_{X2}$     & \z[$10^{\,30}\rm erg/s$]\z
                             & \z total X-ray luminosity$^{(4)}\z$
                                                 & 1.74   & 0.63\\
$N_{\rm H}$   & $10^{22}\,\rm cm^{-2}$ 
                             & X-ray absorption column 
                                                 & 0.06  & 0.15\\  
$T_X$        & [$10^{\,6}$\,K] & X-ray emission temps.$^{(5)}\z$   
                             & 2.3, 7.9, 21.5
                             & 2.0, 7.6 \\ 
$\langle T_X\rangle$ & [$10^{\,6}$\,K] & mean X-ray temp.$^{(6)}\z$  
                             & 3.7 & 4.4 \\
$T_{X,\rm fit}$ & [$10^{\,6}$\,K] & fitted X-ray temp.$^{(7)}\z$
                                                 & 18    & 8.2\\ 
\hline
\end{tabular}}\\[1mm]
\begin{minipage}{90mm}{\tiny 
$^{(0)}$: assumed;\hspace*{2mm}
$^{(1)}$: $L_{\rm UV1}$ integrated between 91.2\,nm and 111\,nm;\hspace*{2mm}
$^{(2)}$: $L_{\rm UV2}$ integrated between 91.2\,nm and 205\,nm;\hspace*{2mm}
$^{(3)}$: $L_{\rm X1}$ integrated between 1\,keV and 10\,keV;\hspace*{2mm}
$^{(4)}$: $L_{\rm X2}$ integrated between 0.3\,keV and 10\,keV;\hspace*{2mm}
$^{(5)}$: X-ray emission components fitted to the data;\hspace*{2mm}
$^{(5)}$: mean value, weighted by component emission masses;\hspace*{2mm}
$^{(7)}$: best fit obtained with bremsstrahlungs-fit (Eq.\,\ref{Xray-fit}).}
\end{minipage}
\label{tab:stellar}
\end{table}

\subsection{Stellar UV irradiation}
\label{app:StellarUV}

The UV irradiation by the central star is much more difficult to
determine, and hampered by the lack of high-quality UV data, 
especially in the hard FUV region ($<\!130$\,nm), and for Ly$\,\alpha$. 
We have systematically scanned and collected UV data from IUE, FUSE, 
HST/STIS, HST/COS, and HST/ACS. These data have been re-binned and 
collated, using the inverse square of the flux uncertainties as weighting
factors, following the idea used by \citet{Valenti2000,Valenti2003}. The 
results of this data collection will be described elsewhere 
(Dionatos\etal 2015, in prep.).

For many objects, the UV data is poor and incomplete, and we have to
use template stars or other tools to complete it. One option is to use
a simple powerlaw $I_\lambda\!\propto\!\lambda^{p_{\rm UV}}$, or
\begin{equation}
  I_\nu \propto \lambda^{p_{\rm UV}+2} \ ,
  \label{UV-fit}
\end{equation}
where the proportionality constant and the UV powerlaw index 
$p_{\rm UV}$ can be roughly fitted to the existing data, or in order 
to fill in gaps in the data.  For TW\,Hya and AB\,Aur, however, the UV data 
quality is excellent, and we can directly use the data to determine the UV 
irradiation, see Fig.\,\ref{fig:irradiation} and Table~\ref{tab:stellar}.

In case of the Herbig\,Ae/Be stars, the UV data seems mostly
photospheric in character, still having absorption lines down to
wavelengths of about 150\,nm and below, but at even shorter
wavelengths, the spectrum changes character, is dominated by emission
lines and is in excess to photospheric models. The soft part of the UV
data can then be used to improve the fits of the stellar parameters
and $A_V$.  For AB\,Aur, we do not need to pick $T_\star$ from the
literature. We find good agreement between the UV data and our
photospheric model with $T_\star\!=\!9550$\,K down to about 130\,nm,
in excellent agreement with \citep{DeWarf2003}.

\subsection{Stellar X-ray irradiation}
\label{app:StellarXrays}

X-ray data were obtained by using archival and new data from the two
X-ray observatories \textit{XMM-Newton} \citep{Jansen2001} and
\textit{Chandra} \citep{Weisskopf2000}. The spectra were extracted
using the software of either science centres (SAS and CIAO). To get
the source spectra, we select a circular extraction region around the
center of the emission, while the background area contained a large
source-free area on the same CCD. The extraction tools ({\it EVSELECT}
for XMM and {\it SPECEXTRACT} for Chandra) delivered the source and
background spectra as well as the redistribution matrix and the
ancilliary response files.  

The extracted spectra were then fitted with the X-ray emission model
{\it XSPEC} \citep{Arnaud1996} assuming a collisionally ionised plasma
({\it VAPEC}) and a model for an absorption column ({\it WABS}).  The
element abundance values in the {\it VAPEC} models were set to typical
values for pre-main sequence stars, as chosen by the XEST project
\citep{Gudel2007}.  Either a one component (1T), a two component (2T)
or a three component (3T) emission model is fitted to the data. Highly
absorbed sources or scarce data allow only for 1T fits. The fit
delivers an absorption column density towards the source $N_{\rm H}$,
and a plasma emission temperature $T_X$ and an emission measure EM for
each component.  Finally, the unabsorbed spectrum is calculated after
setting the absorption column density parameter to zero, and the flux
is derived by integrating over the energy range 0.3-10 keV.

In cases where no detailed X-ray data are available, a more simple
two-parameter approach is used considering a free-free
(bremsstrahlung) continuum with total luminosity $L_X$ and a 
fitted X-ray emission temperature $T_{X,\rm fit}$ as
\begin{equation}
  I_\nu \propto \frac{1}{\nu} \exp\left(-\frac{h\nu}{k T_{X,\rm fit}}\right) \ .
  \label{Xray-fit}
\end{equation}
Table~\ref{tab:stellar} shows that $T_{X,\rm fit}$ results to be close
to the highest X-ray component temperatures found, whereas the 
mean temperature $\langle T_X\rangle$ (linear mean of component
temperatures, weighted by component emission masses) would result in
a very bad fit and should {\em\,not\,} be used in Eq.\,(\ref{Xray-fit}).
The unabsorbed X-ray emission spectrum is finally converted to units
of surface intensities $I_\nu(r\!=\!R_\star)\ \rm[erg/cm^2/s/Hz/sr]$,
using the previously determined values for $d$ and $R_\star$, and
merged with the UV data and photospheric model spectrum.

\subsection{Background radiation}
\label{app:ISM}

Protoplanetary disks are irradiated not only by the central star, but
also from the environment. We assume an isotropic interstellar (IS)
background radiation field $I_\nu^{\rm ISM}$ with 3 components, the IS
UV-field (created by distant O-stars), the cosmic background (CMB,
approximated by a 2.7\,K Planckian), and an infrared background
radiation field $I_\nu^{\rm IR}$ (created by distant stars and
molecular clouds).

\begin{figure*}
\vspace*{-7mm}
\centering
\begin{tabular}{ll}
  \hspace*{-5mm}\includegraphics[width=87mm,trim= 0 -25 37 0,clip]
                {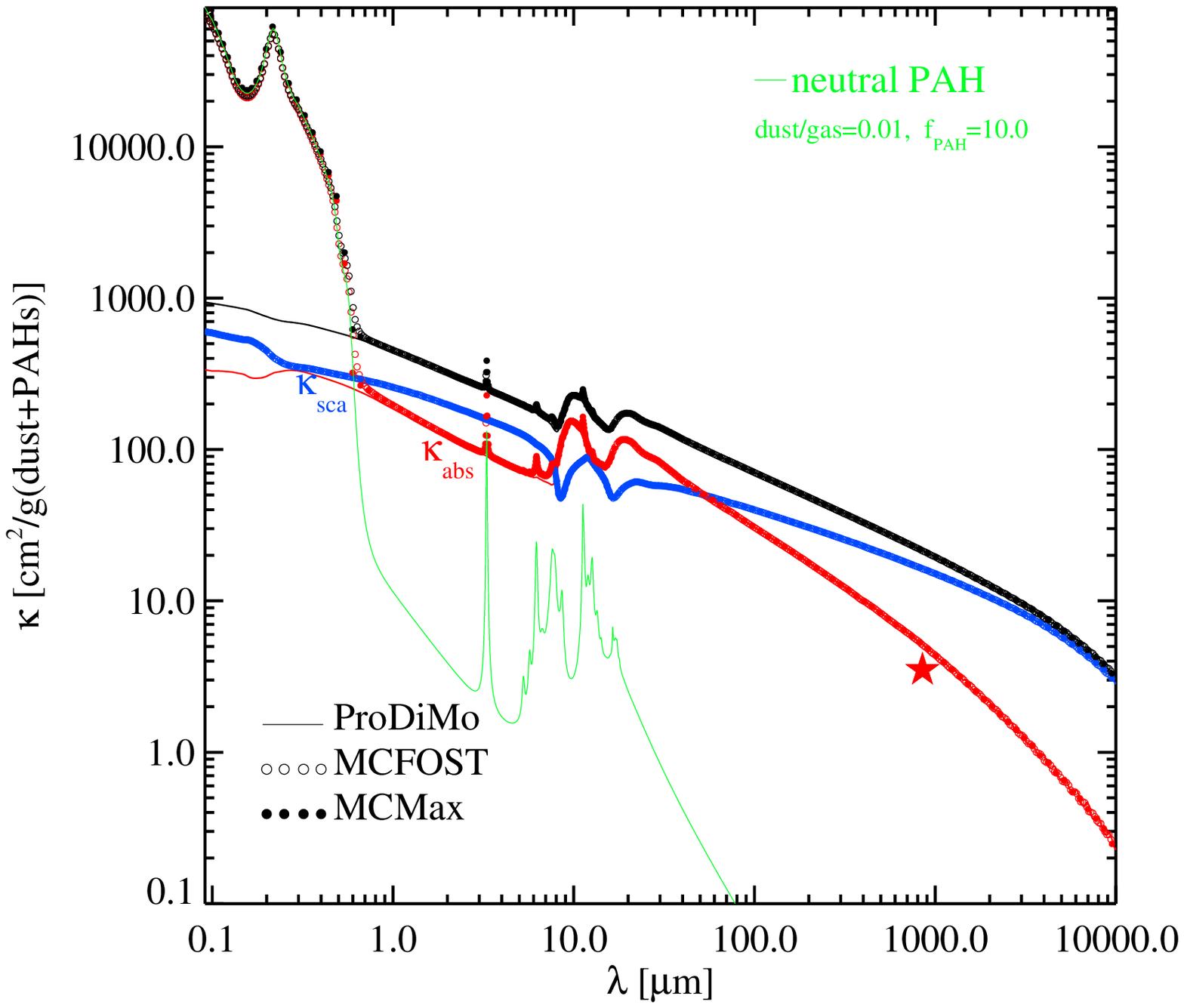}
  \hspace*{-5mm}\includegraphics[width=106mm,trim=0 20 10 0,clip]
                {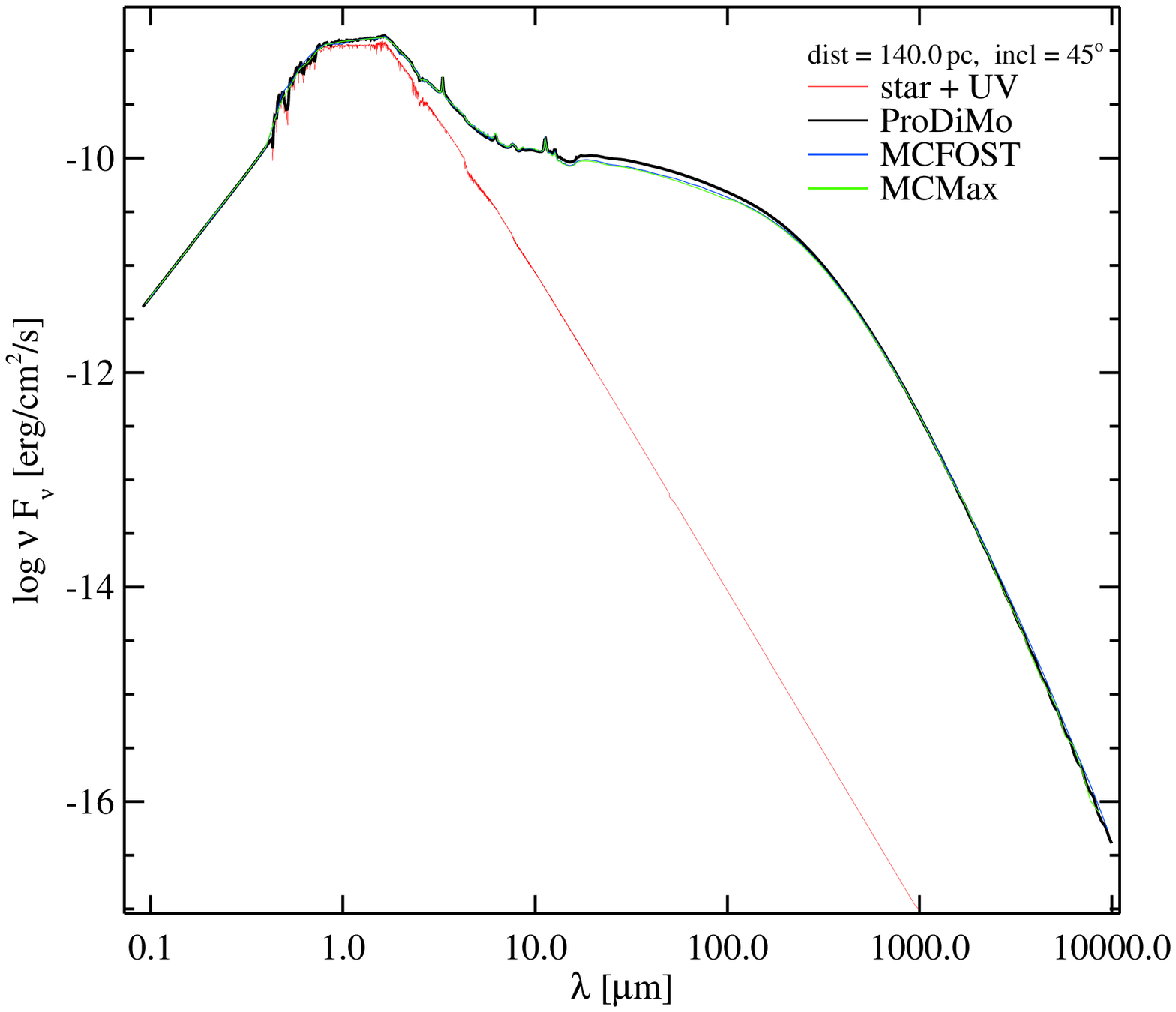}\\*[-62mm] 
  \hspace*{94mm}\includegraphics[width=55mm,trim=25 40 0 0,clip]
                {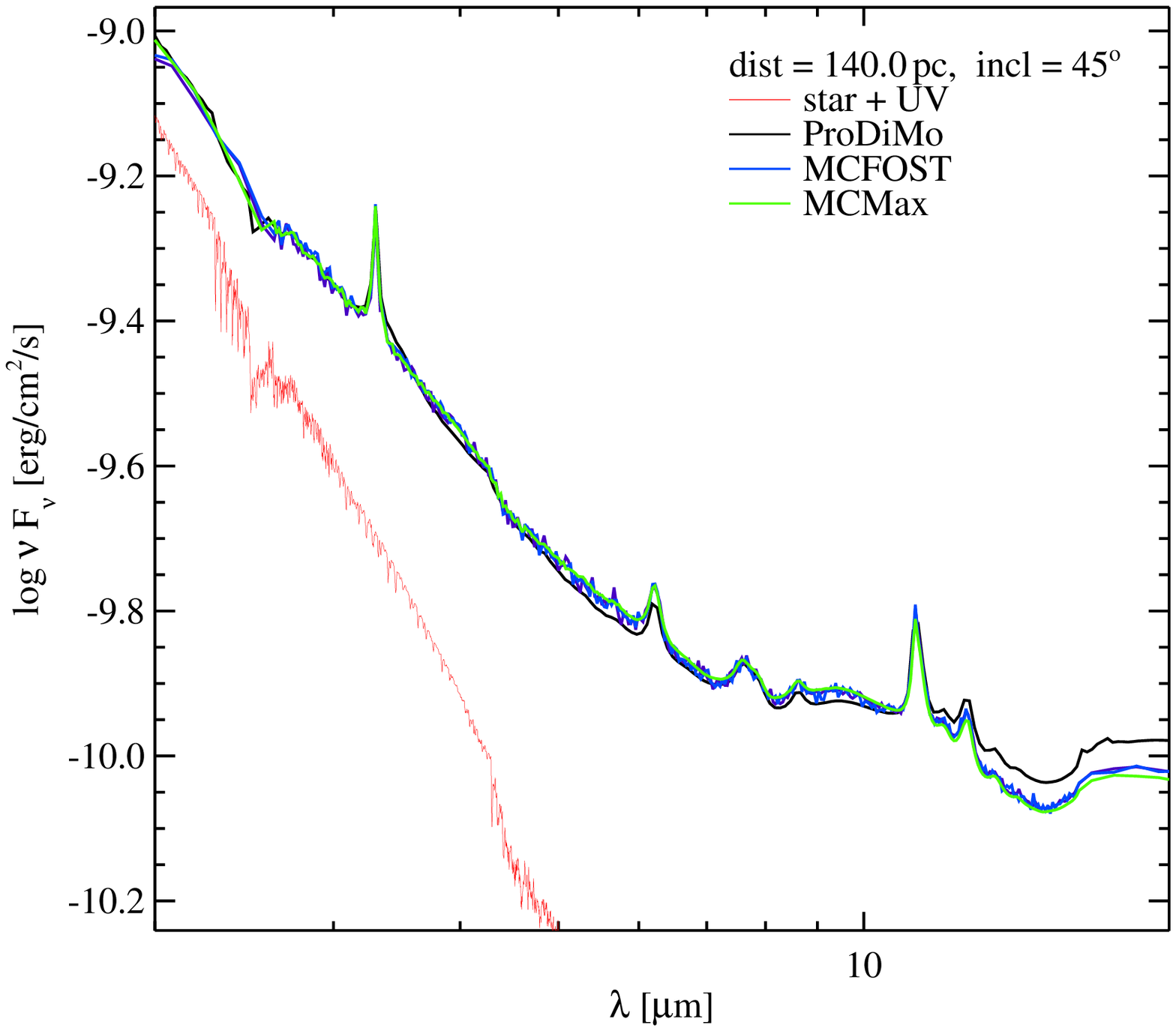}\\*[11mm]
\end{tabular}\\[-2mm]
\begin{tabular}{ccc}
  \large$\rho\rm\,[g/cm^3]$ &
  \large$T_{\rm dust}$\,[K] &
  \large$T_{\rm PAH}$\,[K]\\[-2.5mm]
  \hspace*{-4mm}\includegraphics[height=66mm,trim=39 0 0 0,clip]
                {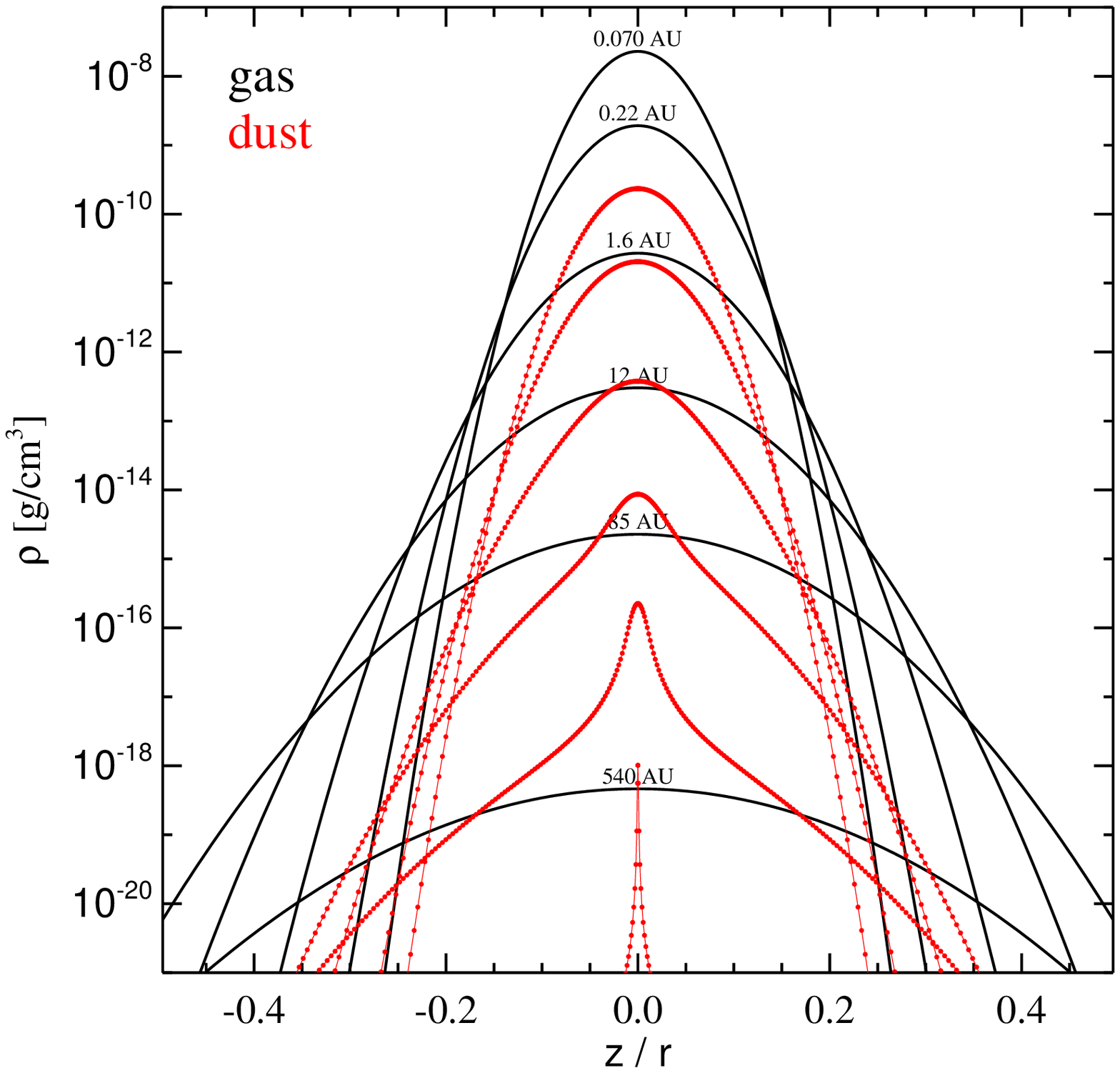} &
  \hspace*{-7mm}\includegraphics[height=66mm,trim=39 0 0 0,clip]
                {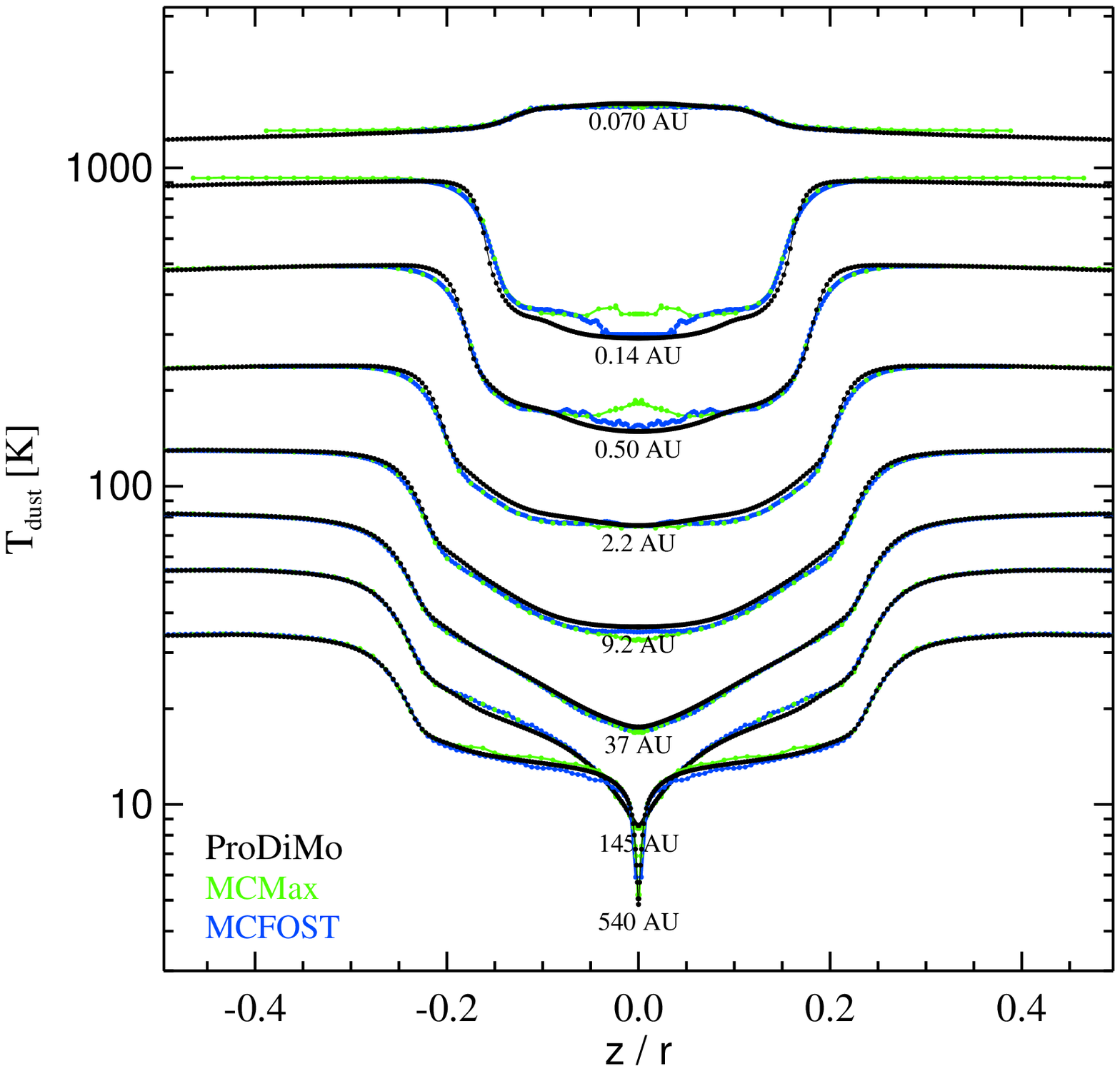} &
  \hspace*{-7mm}\includegraphics[height=66mm,trim=39 0 0 0,clip]
                {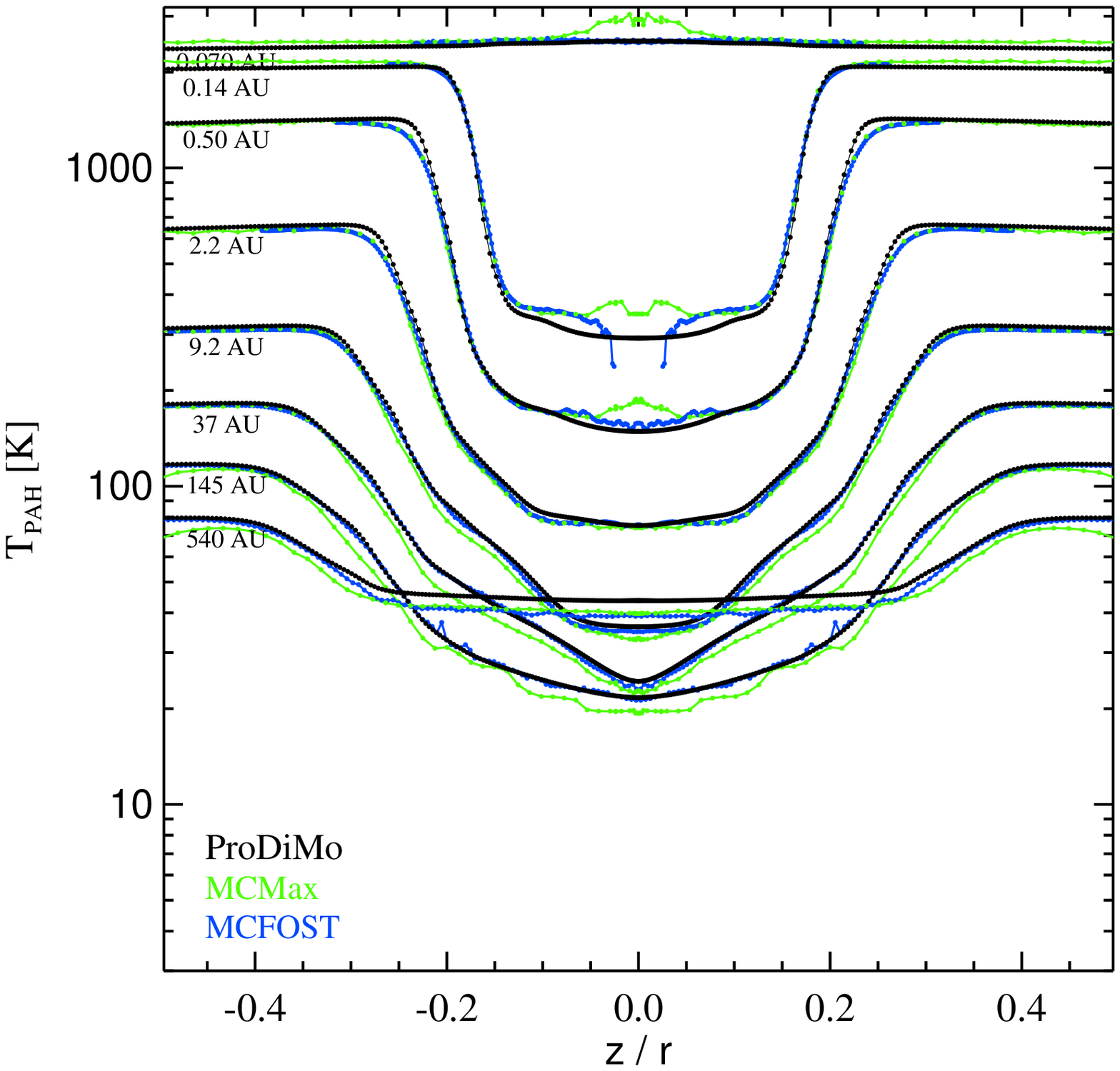}\\[-3mm] 
\end{tabular}
  \caption{Results for the T\,Tauri reference model with {\em
      \,artificially large PAH abundance\,} ($f_{\rm PAH}\!=\!10$),
    showing independent results from {\sl ProDiMo}, {\sl MCMax} and
    {\sl MCFOST} with PAHs in radiative equilibrium. The {\bf upper
      left} figure shows the well-mixed dust and neutral PAH
    opacities. {\sl ProDiMo} opacities are shown by lines, separately
    for dust (black\,$=$\,extinction, red\,$=$\,absorption,
    blue\,$=$\,scattering) and PAHs (green, only absorption), {\sl
      MCMax} and {\sl MCFOST} opacities are shown by dots (combined
    dust$+$PAH opacities). The {\bf upper right} figure shows the
    computed SEDs with a zoom-in on the prominent PAH emission
    features at 3.3\,$\mu$m, 6.2\,$\mu$m, 7.6\,$\mu$m, 8.6\,$\mu$m,
    11.3\,$\mu$m and 13.5\,$\mu$m. The {\bf lower left} plot shows the
    assumed gas densities $\rho_{\rm gas}$ (black) and derived
    (settled) dust densities $\rho_{\rm dust}$ (red) for a series of
    vertical cuts at radius $r$ as labelled. The PAH densities are
    given by $\rho_{\rm PAH}\!\approx\!0.00132\times\rho_{\rm gas}$
    for $f_{\rm PAH}\!=\!10$. The {\bf lower middle and lower right}
    plots compare the computed dust and PAH temperatures between {\sl
      MCMax} (green), {\sl MCFOST} (blue), and {\sl ProDiMo} (black).}
  \label{fig:PAHmodels}
  \vspace*{-2mm}
\end{figure*}

\begin{figure*}
  \vspace*{-1mm}
  \begin{tabular}{ll}
  \hspace*{0mm}\includegraphics[height=85mm]{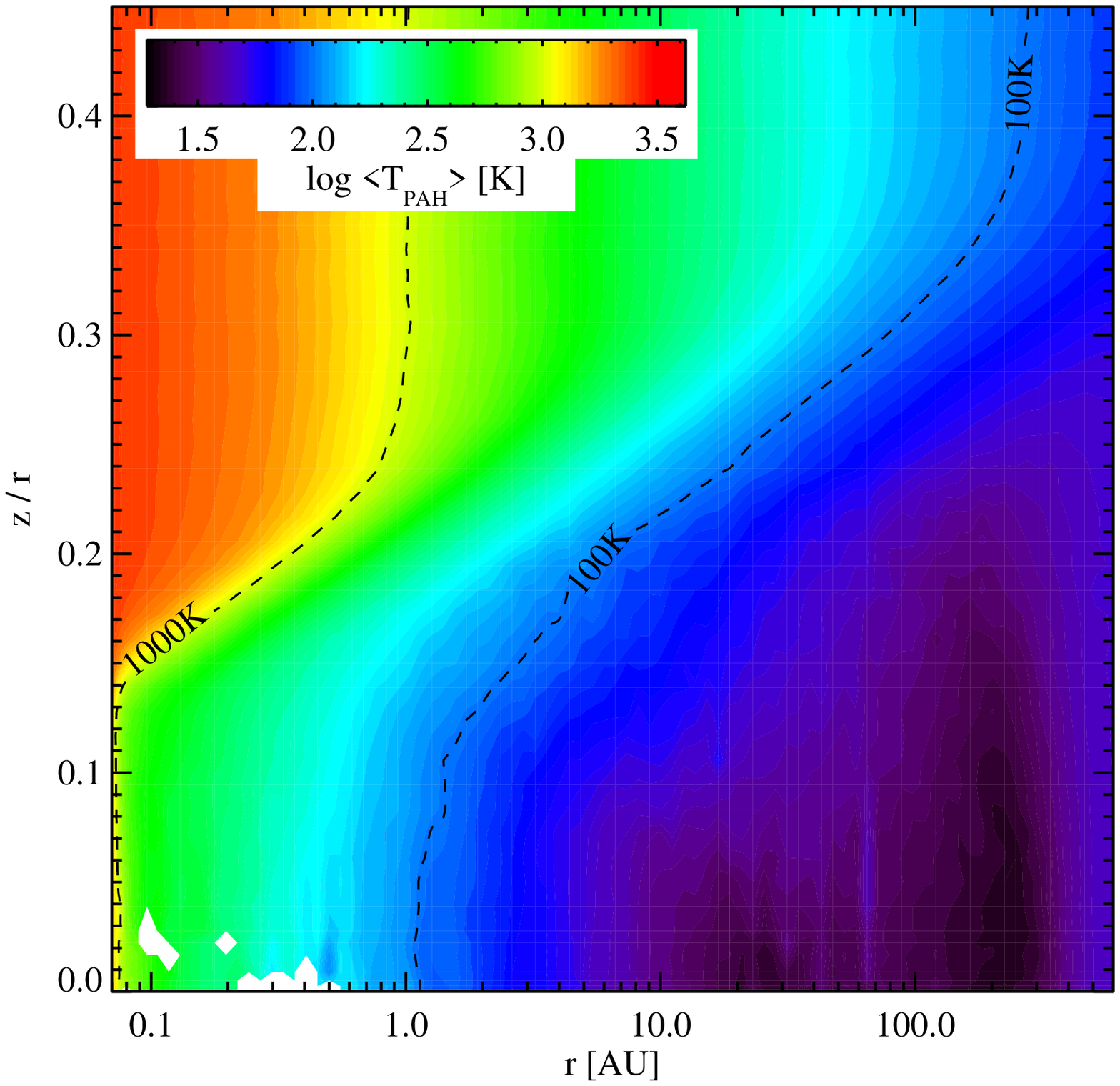} &
  \hspace*{-4mm}\includegraphics[height=85mm]{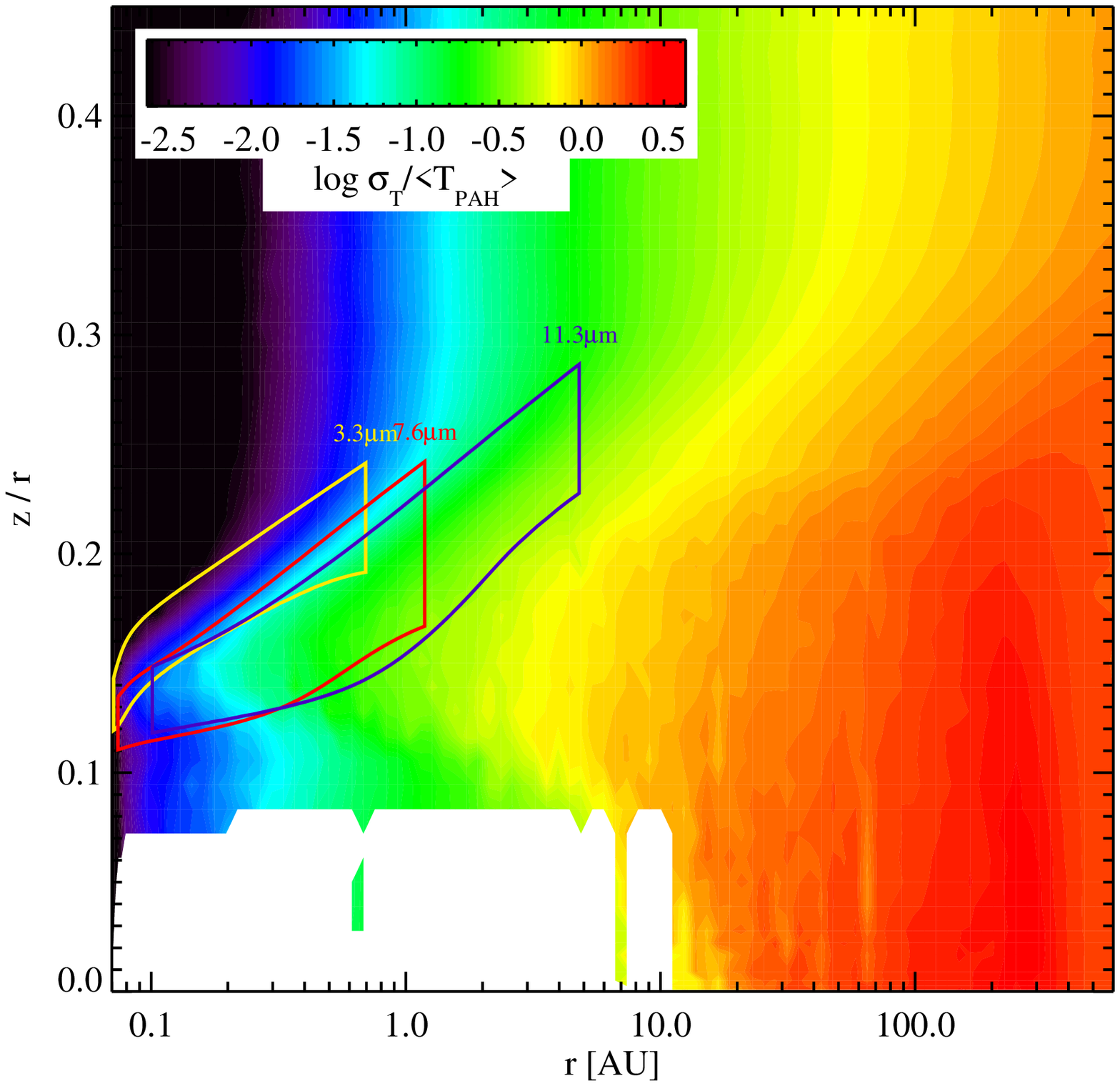} \\[-7mm]
  \hspace*{-1mm}\includegraphics[height=84mm]{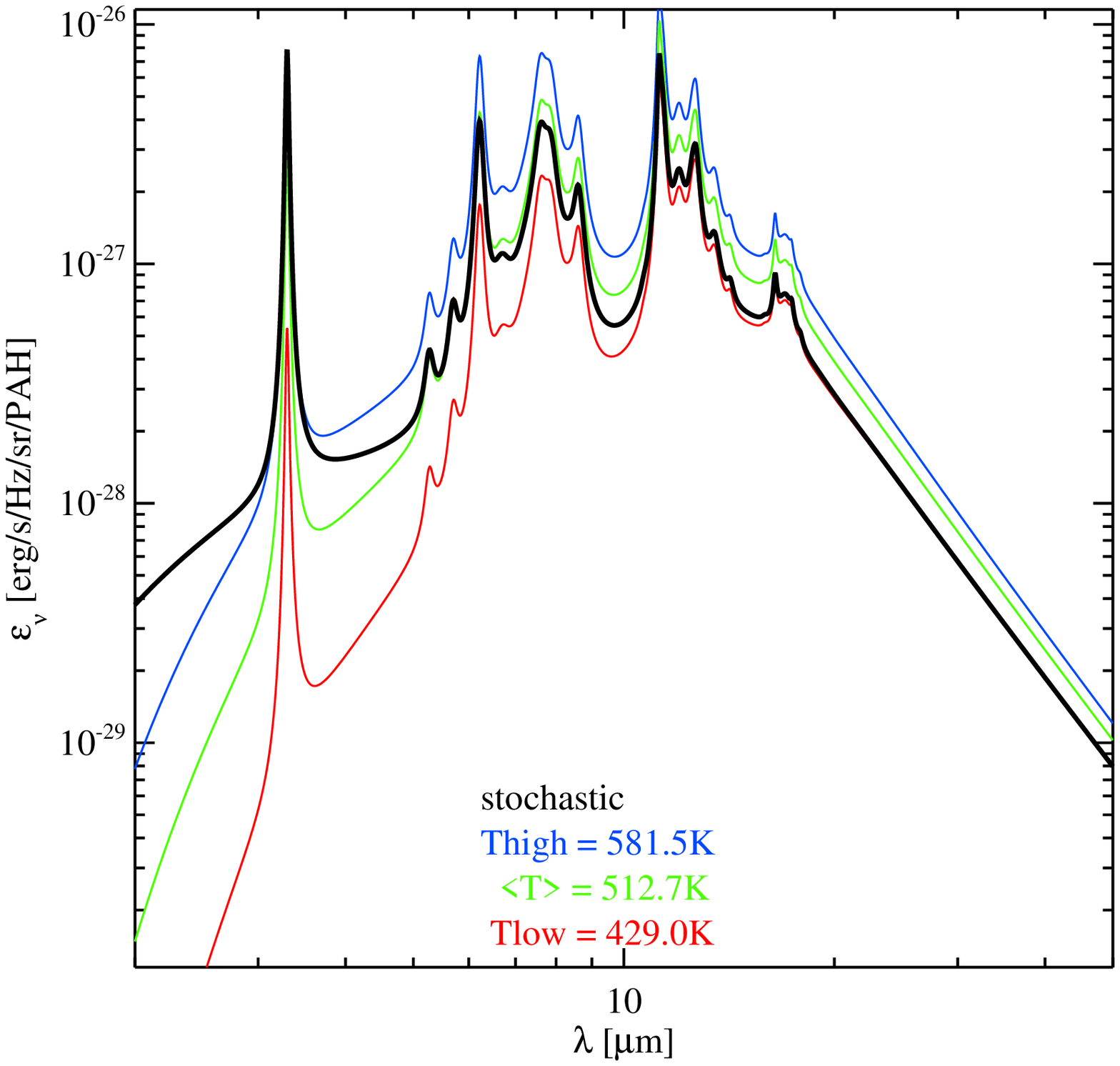} &
  \hspace*{-8mm}\includegraphics[height=86mm,trim=0 15 0 0,clip]
                                                   {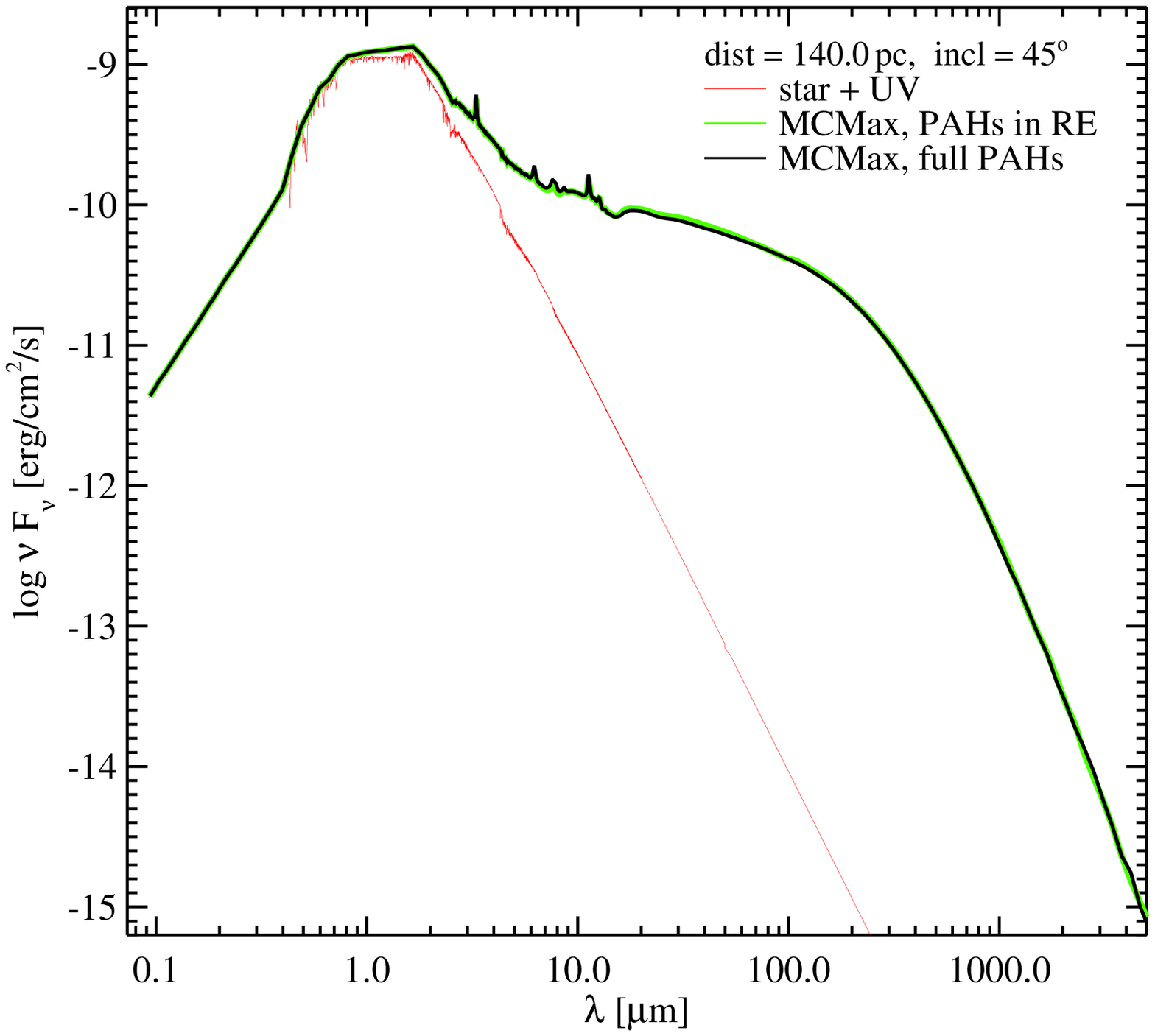}\\[-53mm]
  & \hspace*{7mm}\includegraphics[height=42mm,trim=30 40 0 0,clip]
                                                   {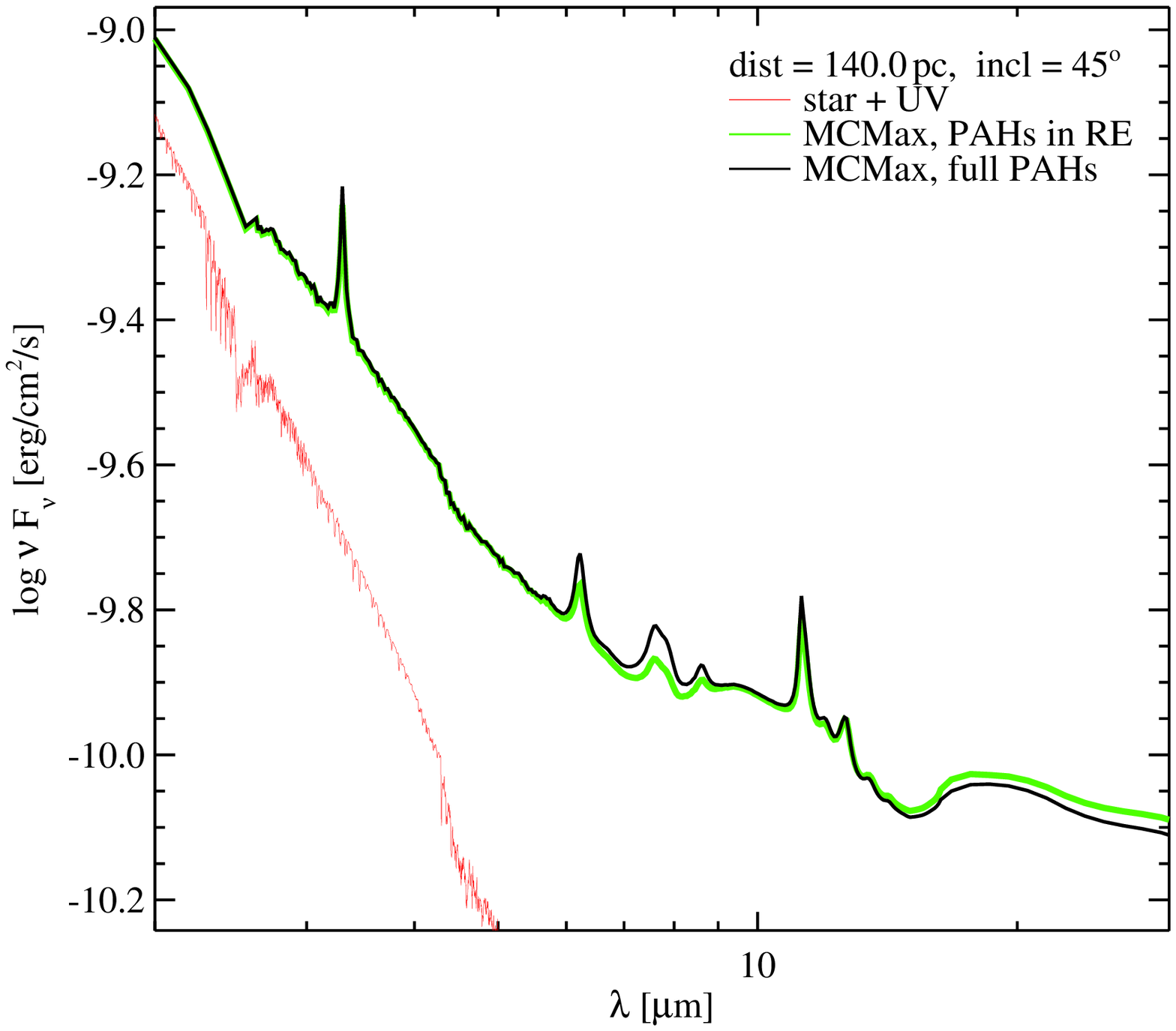}\\[7mm]
  \end{tabular}
  \caption{Comparison of PAH results between full treatment with
    stochastic quantum heating, and simplified method assuming the
    PAHs to be in radiative equilibrium. The reference model with
    artificially increased PAH abundance $f_{\rm PAH}\!=\!10$ is
    considered. The {\bf upper left} plot shows the mean PAH
    temperature $\langle T_{\rm PAH}\rangle$, and the {\bf upper right}
    plot shows the relative width of the PAH temperature distribution
    function $\sigma_T/\langle T_{\rm PAH}\rangle$ where
    $\sigma\!=\!\frac{1}{2}(T_{\rm high}-T_{\rm low})$. The coloured
    boxes on the right encircle the disk regions which emit 50\% of
    the PAH features at 3.3\,$\mu$, 7.6\,$\mu$m and 11.3\,$\mu$m. In
    the blank regions, the Monte Carlo statistics is too poor to
    compute the PAH temperature distribution function $p(T)$. The {\bf
      lower left} plot explains how we define $T_{\rm low}$ and
    $T_{\rm high}$ by bracketing the local PAH emission coefficient
    $\epsilon_\nu$ by two temperatures. The {\bf lower right} plot
    compares the SED results obtained with the two different methods
    using {\sl MCMax}, with a zoom-in between $2\,\mu$m and
    $30\,\mu$m. There is barely any difference.}
  \label{fig:PAHtemp}
  \vspace*{-1mm}
\end{figure*}

The interstellar UV field is approximated by a 20000\,K-black-body
with a tiny dilution factor $W_{\rm dil}=9.85357\times10^{-17}$ such
that $\chi^{\rm ISM}\!=\!1$ corresponds to the integrated standard IS
UV field of \citet{Draine1996}
\begin{equation}
  I_\nu^{\rm ISM} 
        = \chi^{\rm ISM} 1.71 W_{\rm dil}\,B_\nu(20000{\rm\,K})
      \,+\, \chi^{\rm ISM}_{\rm IR} I_\nu^{\rm IR} 
      \,+\, B_\nu(2.7{\rm\,K}) \ .
\end{equation}
For the infrared background field $I_\nu^{\rm IR}$, we take the
spectral shape from \citep[][their tables A3 and B1]{Mathis1983}, and
adjust the parameter $\chi^{\rm ISM}_{\rm IR}$ such that the
integrated background intensity equals
\begin{equation}
  \int I_\nu^{\rm ISM}\,d\nu = \frac{\sigma}{\pi} T_{\rm back}^4 \ .
\end{equation}
A blackbody would accomodate its temperature to $T_{\rm back}$ is this
background radiation field. $T_{\rm back}$ limits the dust
temperatures in the disk, because when including the star as
additional light source, the temperatures can only increase.  Without
the infrared contribution $(\chi^{\rm ISM}_{\rm IR}\!=\!0)$ the
background temperature results to be 2.97\,K, but we can increase
$\chi^{\rm ISM}_{\rm IR}$ to simulate a disk in close proximity to, e.g.,
a star formation region which provides additional IR background
radiation, with expected impact on the mm-slope (see
Sect.~\ref{sec:mm-slope}). For example, to achieve $T_{\rm
  back}\!=\!5\,$K and 10\,K, values for $\chi^{\rm ISM}_{\rm
  IR}\!=\!5.1$ and 92 are required, respectively. The original work of
\citet[][\ie $\chi^{\rm ISM}_{\rm IR}\!=\!1$]{Mathis1983} is derived
from stars and molecular clouds at a distance at 10\,kpc from the
Galactic centre.

When calculating spectral fluxes from a model with background radiation, 
we need to subtract the background as
\begin{equation}
  F_\nu = \int (I_\nu-I_\nu^{\rm ISM})\,d\Omega \ .
  \label{eq:backsubtract}
\end{equation}
The background subtraction (Eq.\,\ref{eq:backsubtract}) is important
in particular at long wavelengths where the CMB is bright. For example,
at $\lambda\!=\!1$cm, the CMB multiplied by solid angle
$\Omega\!=\!\pi R_{\rm out}^2 \cos(i)/d^2$ is $15\times$ stronger than
the disk signal from the reference model. Eq.\,(\ref{eq:backsubtract})
is also necessary to make the results independent of considered image
size (as long as the disk is well contained in the image). We note,
however, that Eq.\,(\ref{eq:backsubtract}) can lead to negative
fluxes, for example an edge-on disk in the UV or whenever the disk
appears darker than the background at the considered wavelength, for
example the ``silhouette disks'' in the Orion nebula.

\section{The PAH temperature distribution}
\label{app:PAHtemp}

Figure~\ref{fig:PAHmodels} shows some results for the T\,Tauri
reference model introduced in Sect.~\ref{sec:Results}, but, in order
to demonstrate the effects, with an artificially large PAH abundance
of $f_{\rm PAH}\!=\!10$.  The dust in the outer disk parts is strongly
concentrated towards the midplane, whereas the PAH molecules stay
co-spatial with the gas by assumption. In all optically thin regions,
and in the surface of the inner rim, the PAH temperatures are higher
by a large factor $\sim\!1.5-2$ as compared to the dust
temperatures, simply due to the very blue absorption characteristic of
the PAHs which facilitates radiative heating. In contrast, in the
optically thick midplane of the disk, we find $\TPAH\approx\Td$ as
expected. The PAH emission features result from the large temperature
contrast between PAHs and dust in the optically thin upper and inner disk
regions $<(1-10)\,$AU, depending on wavelength, see
Fig.~\ref{fig:PAHtemp}. In the outer optically thin disk regions, the
PAH temperature re-increases due to the interstellar UV irradiation.

The detailed physical modelling of the PAH molecules with the Monte
Carlo codes allows for a treatment of the quantum heating by single
photons absorption events by the PAHs, followed by radiative cooling, which
leads to a stochastic PAH temperature distribution $p(T)$ at every
point in the model.  In the following, we can thereby check the
validity of our simplified treatment of the PAHs in radiative
equilibrium, as outlined in Sect.~\ref{sec:PAHs}. The PAH emissivity
$\rm[erg/s/Hz/sr/$PAH-molecule] is given by
\begin{equation}
  \epsilon_\nu^{\rm PAH} = \kappa_\nu^{\rm PAH} \int B_\nu(T)\,p(T)
  \,dT \ ,
\end{equation}
where $\kappa_\nu^{\rm PAH}\,\rm[cm^2/$PAH-molecule] is the PAH
absorption cross section and $B_\nu(T)$ is
the Planck function. The temperature distribution function $p(T)$ is
normalised to $\int p(T)\,dT\!=\!1$.

In order to quantify the mean PAH temperature and the width of the
PAH temperature distribution, we consider  
\begin{eqnarray}
  \int \kappa_\nu^{\rm PAH} B_\nu\,\big(\langle T_{\rm PAH}\rangle\big)\,d\nu 
     &=& \int \epsilon_\nu^{\rm PAH} d\nu \\
  \kappa_{3.3\,\mu{\rm m}}^{\rm PAH} B_{3.3\,\mu{\rm m}}(T_{\rm high})
     &=& \epsilon_{3.3\,\mu{\rm m}}^{\rm PAH} \\
  \kappa_{30\,\mu{\rm m}}^{\rm PAH} B_{30\,\mu{\rm m}}(T_{\rm low})
     &=& \epsilon_{30\,\mu{\rm m}}^{\rm PAH} \\
  \sigma_T &=& \frac{1}{2}(T_{\rm high}-T_{\rm low}) \ ,
\end{eqnarray}
\ie we define a high temperature to match the 3.3\,$\mu$m PAH emission
and a low temperature to match the continuous 30\,$\mu$ PAH
emission. The mean temperature $\langle T_{\rm PAH}\rangle$ is
related to the total frequency integrated PAH emission. Usually, we
find $T_{\rm low}>\langle T_{\rm PAH}\rangle>T_{\rm high}$ in the
models, see an example in Fig.~\ref{fig:PAHtemp}, lower left part. In
the example shown, the relative width of the PAH temperature distribution
function is $\sigma_T/\langle T_{\rm PAH}\rangle\!\approx\!15\%$.

Figure~\ref{fig:PAHtemp} shows that $\sigma_T/\langle T_{\rm
  PAH}\rangle$ anti-correlates with $\langle T_{\rm
  PAH}\rangle$. Large average PAH temperatures imply a sharply peaked
temperature distribution function. The upper right plot shows which
spatial disk regions are responsible for the various PAH emission features.
The PAH 3.3\,$\mu$m feature originates in the innermost
1\,AU disk regions, whereas the 11.3\,$\mu$m emission region
stretches out to about 10\,AU. In all cases, the PAH emissions come
from borderline optically thin altitudes where the dust is not yet
optically thick, and where $\langle T_{\rm
  PAH}\rangle\gg\Td$. All these warm regions are characterised by 
quite a sharply peaked temperature distribution function
$\sigma_T/\langle T_{\rm PAH}\rangle \approx 0.2\%-30\%$,
which explains why the fast approximate method, assuming the PAHs
to be in radiative equilibrium, works so well for the SED modelling.

\begin{figure}
  \begin{center}
  \includegraphics[width=85mm,trim=0 39 0 0,clip]{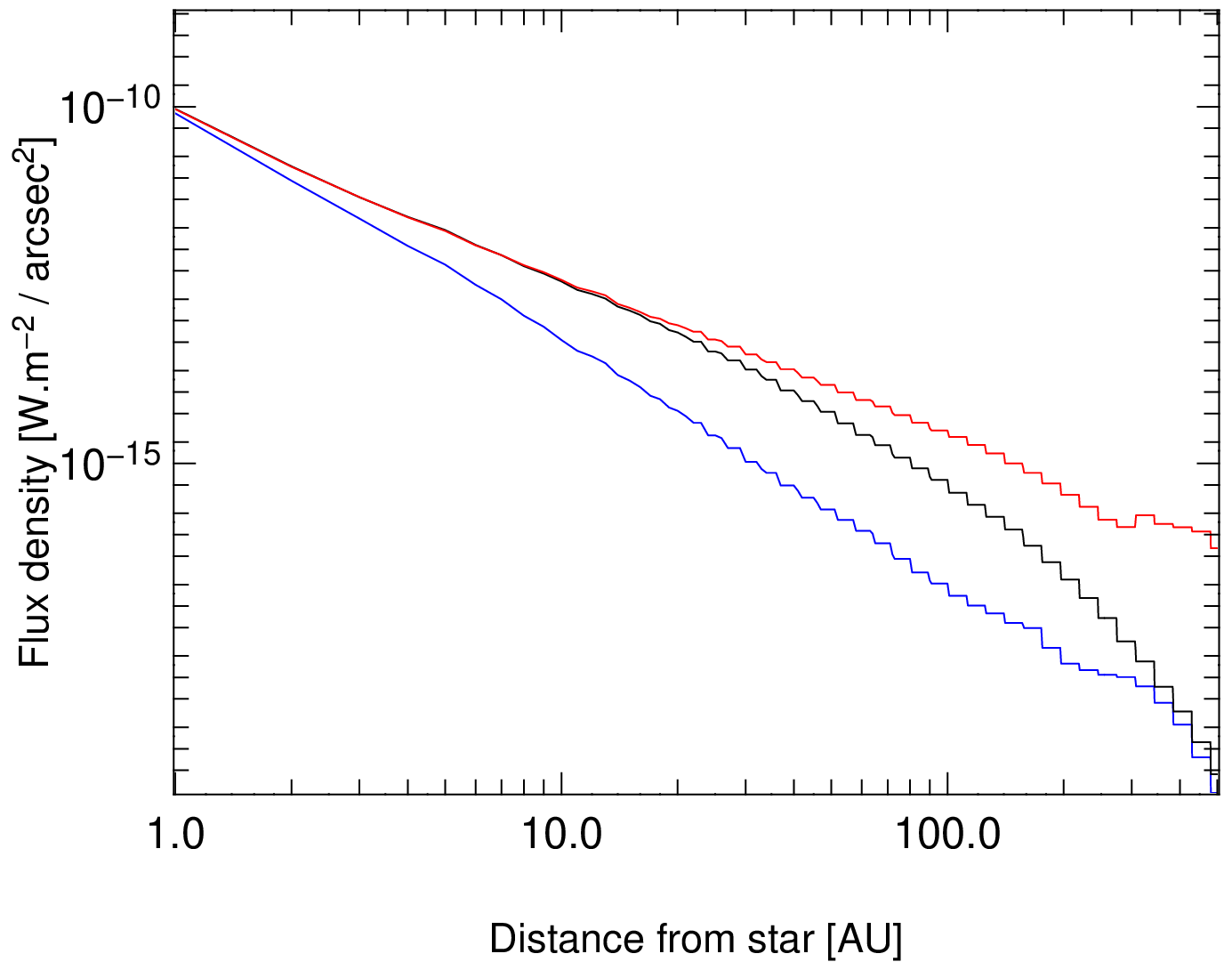}
  \end{center}
  \vspace*{-3mm}
  \hspace*{45mm}{\sf$r$\,[AU]}
  \caption{Comparison of radial intensity profiles in the 11.3\,$\mu$m
    PAH emission feature obtained from the reference T\,Tauri
    model. The blue lines shows the intensity profile if dust and PAHs
    are assumed to have a common radiative equilibrium
    temperature. The black line shows the simplified treatment with
    independent dust and PAH temperatures, both in radiative
    equilibrium. The red line shows the results obtained with the full
    PAH treatment with a stochastic PAH temperature distribution.}
  \label{fig:PAHprof}
\end{figure}

Figure~\ref{fig:PAHprof} shows, however, that there are substantial 
differences at larger radii, which are not important for the integral
emission of the PAH features. The blue model in Fig.~\ref{fig:PAHprof}
(assuming equal dust and PAH temperatures) fails completely to
predict the PAH emission features in the SED. The black model (PAHs in
radiative equilibrium) is good for the SED and sufficient to predict
the intensity profiles up to $\sim\!20$\,AU, but only the full
stochastic treatment of the PAHs (red model) can predict the intensity 
profile beyond $\sim\!20$\,AU.

\section{Time-dependent chemistry}
\label{app:tdep}

To investigate the effects of chemical evolution in the disk on the
observable gas emission lines, we have computed additional models
where the disk structure, the dust size distribution, dust settling
and opacities, the dust and gas temperatures, and the internal
radiation field are taken over from the reference model, but at each
grid point, the chemical rate network is advanced in time from zero to
5\,Myrs, starting from initial concentrations typical for the dark
cores of molecular clouds, see \citep{Helling2014} for details.

\begin{figure}[!b]
  \centering
  \includegraphics[width=90mm,trim=8 10 105 25,clip]{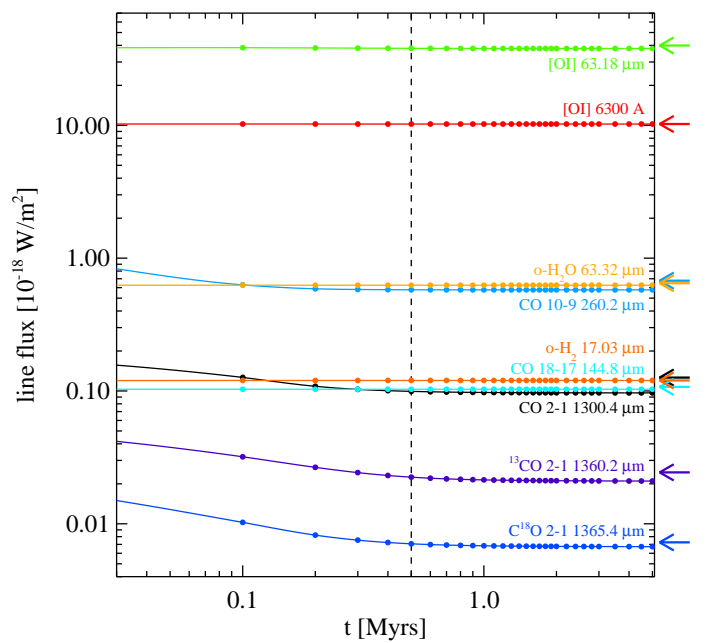}
  \caption{Computed line fluxes as function of disk age for the
    T\,Tauri reference model with time-dependent disk chemistry. Each
    vertical column of dots represents one disk model. The arrows on the
    r.h.s. indicate the results from the time-independent reference
    model. The vertical dashed line marks $t\!=\!0.5\,$Myrs, after
    which the line fluxes do not change significantly any more.}
  \label{fig:tdep}
\end{figure}

Figure~\ref{fig:tdep} shows the resulting line fluxes as function of
disk age.  After an initial relaxation phase which lasts a few
$10^5$\,yrs, the gas emission line fluxes become constant in the
model, and do not change significantly afterwards.  This behaviour is
a consequence of the relatively short chemical relaxation timescales
in most line forming regions situated well above the icy midplane,
compare Fig.~\ref{fig:LineOrigin}. The chemical relaxation timescale
$\tau_{\rm chem}(r,z)$ is introduced and discussed in \citep[][see
  Sect.~8.3 and Fig.~13 therein]{Woitke2009a}. The relaxation of the
optical and IR lines is indeed very quick; we measure mean values of
the chemical relaxation timescale as 1.2\,yrs, 90\,yrs, 120\,yrs,
4000\,yrs and 8000\,yrs in the line forming regions of [OI]\,6300\AA,
o-H$_2$\,17\,$\mu$m, CO\,$J$\,=\,18-17, CO\,$J$\,=\,10-9 and
[OI]\,63\,$\mu$m, respectively.

\begin{figure*}
  \centering
  \begin{tabular}{cc}
  $t=1$\,Myr & reference model\ \ ($t\!\to\!\infty$) \\
  \hspace*{-3mm}
  \includegraphics[width=85mm,trim=10 0 10 10,clip]{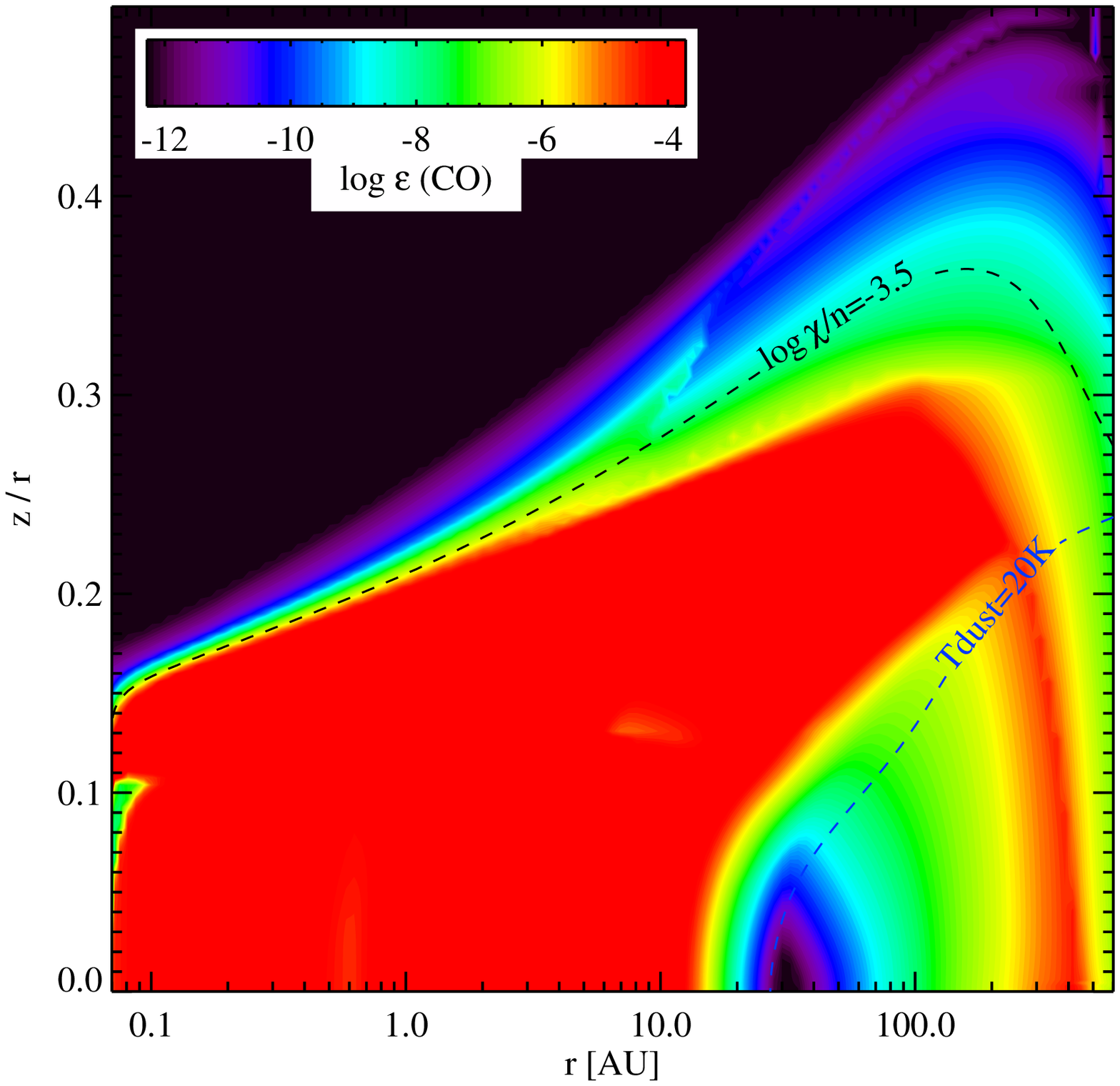} &
  \hspace*{-3mm}
  \includegraphics[width=85mm,trim=10 0 10 10,clip]{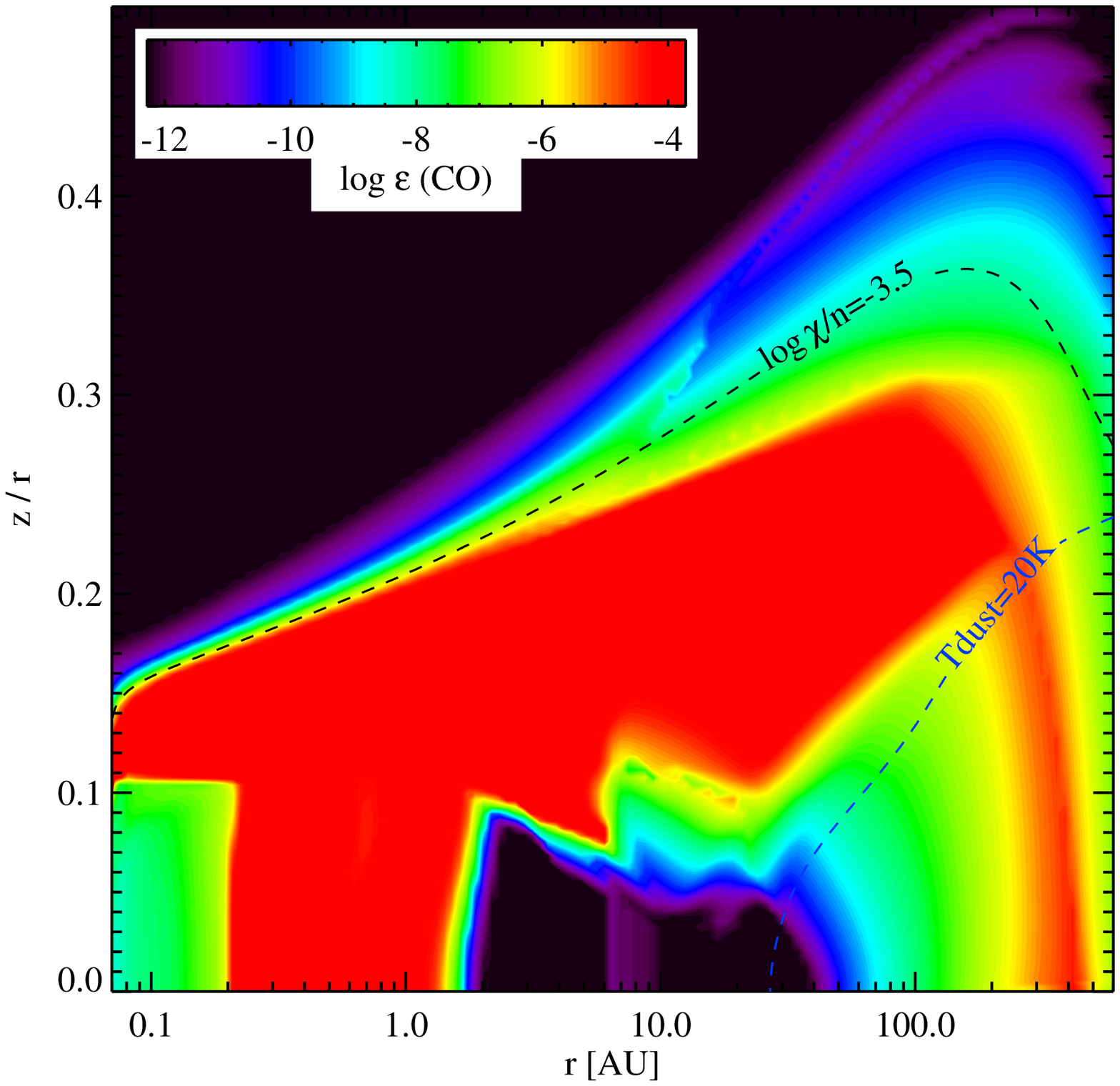}
  \end{tabular}
  \vspace*{-1mm}
  \caption{The CO concentration $\epsilon\rm(CO)\!=\!n_{\rm CO}/\nH$
    after 1\,Myrs (l.h.s.) compared to the CO concentration in 
    the time-independent reference model.}
  \label{fig:CO_tdep}
\end{figure*}

However, the CO mm-line fluxes do not fully converge to the results
obtained by the time-independent reference model, see arrows on the
r.h.s. of Fig.~\ref{fig:tdep}, even for integration times as long as
10\,Myrs. This is consistent with the very long chemical relaxation
timescales $\langle\tau_{\rm chem}\rangle$ we find for those lines,
namely 20\,Myrs, 50\,Myrs and 100\,Myrs in the line forming regions of
CO\,$J$\,=\,2-1, $^{13}$CO\,$J$\,=\,2-1, and C$^{18}$O\,$J$\,=\,2-1,
respectively.

The difference of the CO concentration between the $t\!=\!1\,$Myr disk
model and the time-independent reference model is shown in
Fig.~\ref{fig:CO_tdep}. All upper disk regions have practically
indistinguishable CO concentrations, however, in the dark icy midplane
(vertical visual extinction $A_{\rm V}\!>\!10$), CO cannot freeze out
directly where the dust temperature exceeds about 20\,K (at radii
$\la\!30\,$AU in this model). Instead, the CO is very slowly
dissociated in these regions by reactions with He$^+$ created by
cosmic rays. Most of the liberated oxygen atoms re-form CO, but a
small part can form other molecules with higher adsorptions energies
like water, and these will freeze out immediately. This mechanism provides
a slowly ticking chemical clock in the dark midplane regions of
protoplanetary disks, leading to gaseous carbon-to-oxygen ratios $\rm
C/O\!\approx\!1$ on Myr timescales between the water and the CO
ice-lines, and to $\rm C/O\!\gg\!1$ on even longer timescales, see
\citep{Oberg2011,Helling2014}.

The (sub-)mm CO isotopologue line fluxes are somewhat affected by
these differences, for example remainders of cold gaseous CO in
the dark icy midplane can partly re-absorb the CO line photons emitted by 
the warm surface of the disk on the opposite side which cross 
the midplane. In this way, the CO isotopologue line
fluxes of the time-independent model can be up to 30\% larger than
those of the time-dependent models. However, these effects are small
compared to the impact of disk mass, size and shape, dust size distribution
and settling, and the dust/gas ratio, and negligible for most optical
and near-IR to far-IR lines. 

We conclude that the emission lines of chemically robust gas tracers,
such as CO, H$_2$ and the oxygen atom, are little affected by
time-dependent chemical effects in our models. Disk evolution
will rather have an impact on those lines via the changing stellar
parameters, the changing shape of the disk, and physical gas and dust
evolution.  See e.g. \citet{Akimkin2013} for time-dependent disk
models which include those effects.

\section{Properties derived from the model}
\label{app:integrated}

The mean gas temperature in the disk $\langle \Tg\rangle$, the mean
dust temperature $\langle \Td\rangle$, the near-IR excess, the
$10\,\mu$m SED amplitude, the dust absorption mm and cm-slopes
$\beta_{\rm abs}$, and the millimetre and centimetre flux-slopes
$\alpha_{\rm SED}$, are calculated as
\begin{eqnarray}
  \langle \Tg\rangle &=& \frac{\int \Tg(r,z)\,\rho_{\rm gas}(r,z)\,dV}
                              {\int \rho_{\rm gas}(r,z)\,dV}
  \label{eq:Tgmean}\\
  \langle \Td\rangle &=& \frac{\int \Td(r,z)\,\rho_{\rm dust}(r,z)\,dV}
                              {\int \rho_{\rm dust}(r,z)\,dV}
  \label{eq:Tdmean}
\end{eqnarray}
\vspace*{-6mm}
\begin{eqnarray}
  \mbox{near-IR excess} &=& 4\pi\,d^2
                            \int_{2\,\mu{\rm m}}^{7\,\mu{\rm m}} 
                            \!\!\!(F_\lambda-F_\lambda^\star)\;d\lambda \\
  \mbox{10\,$\mu$m ampl.} &=& \frac{F_\nu(9.6\,\mu{\rm m})}
           {\sqrt{F_\nu(6.8\,\mu{\rm m})\cdot F_\nu(13.1\,\mu{\rm m})}}
\end{eqnarray}
\vspace*{-3mm}
\begin{eqnarray}
  \hspace*{8mm}
  \beta_{\rm abs}^{\rm mm} &=& -\frac{\log \kabs(0.85\,{\rm mm}) - \log \kabs(1.3\,{\rm mm})}
                           {\log 0.85\,{\rm mm} - \log 1.3\,{\rm mm}}
  \\
  \beta_{\rm abs}^{\rm cm} &=& -\frac{\log \kabs(5\,{\rm mm}) - \log \kabs(10\,{\rm mm})}
                           {\log 5\,{\rm mm} - \log 10\,{\rm mm}} 
\end{eqnarray}
\vspace*{-2mm}
\begin{eqnarray}
  \hspace*{8mm}
  \alpha_{\rm SED}^{\rm mm} &=& -\frac{\log F_\nu(0.85\,{\rm mm}) - \log F_\nu(1.3\,{\rm mm})}
                           {\log 0.85\,{\rm mm} - \log 1.3\,{\rm mm}}
  \label{eq:mm-slope}\\
  \alpha_{\rm SED}^{\rm cm} &=& -\frac{\log F_\nu(5\,{\rm mm}) - \log F_\nu(10\,{\rm mm})}
                           {\log 5\,{\rm mm} - \log 10\,{\rm mm}} 
  \label{eq:cm-slope}
\end{eqnarray}
where $\rho_{\rm gas}$ is the gas mass density, $\rho_{\rm
  dust}\!=\!\rho_{\rm gas}\cdot\delta$ the dust mass density,
$\delta$ the local dust-to-gas mass ratio, $dV\!=\!2\pi\,r\,dr\,dz$
the volume element, $d$ the distance,
$F_\lambda\!=\!\frac{\nu}{\lambda}F_\nu$ the spectral flux per
wavelength interval, $F_\lambda^\star$ the flux from the naked star
and $\kabs$ the dust absorption opacity $\rm[cm^2/g(dust)]$.

\section{Fluxes of optically thick emission lines}
\label{app:ThickLines}

It is noteworthy that all observable emission lines discussed in this
paper are optically thick in the reference model, with the only
exception being C$^{18}$O $J\!=\!2\!\to\!1$ which is borderline
optically thin.  For large optical depths we can use the
Eddington-Barbier approximation
\begin{equation}
  I_\nu \approx S_{\!\nu}\,(\tau_\nu\!=\!1)
        \approx\left\{\begin{array}{ll}
          S_{\!\nu}^{\rm cont}\,(\tau_\nu^{\rm cont}\!=\!1) &
            \mbox{,\ \ if $|\nu-\nu_0|\!\gg\!\Delta\nu$}\\[0.6ex]
          S_{\!\nu}^{\rm line}\,(\tau_\nu^{\rm line}\!=\!1) &
            \mbox{,\ \ if $|\nu-\nu_0|\!\leq \Delta\nu$}
        \end{array}\right. \ ,
  \label{eq:Inu}
\end{equation}
where $S_{\!\nu}$ is the general source function, $S_{\!\nu}^{\rm cont}$ the
continuum source function and $S_{\!\nu}^{\rm line}$ the line source
function. $\tau_\nu^{\rm cont}$ and $\tau_\nu^{\rm line}$ are the
continuum and line centre optical depth, $\nu_0$ and
$\Delta\nu\!=\rm\!FWHM/2$ are the line centre frequency and observed
frequency width, respectively. If the continuum is optically
thin $\tau_\nu^{\rm cont}<1$, its contribution can be neglected in
Eq.\,(\ref{eq:Inu}).

Integration over solid angle and frequency, and continuum subtraction,
results in the line flux
\begin{eqnarray}
  F_{\rm line} 
  &&\!\!\!\!=\,\int\!\!\!\int \big(I_\nu-I_\nu^{\rm cont}\big)\,d\nu\,d\Omega
  \nonumber\\
  \approx&& 2\Delta\Omega\,\Delta\nu
     \left(S_{\!\nu}^{\rm line}(\tau_{\rm line}\!=\!1)
          -S_{\!\nu}^{\rm cont}(\tau_{\rm cont}\!=\!1)\right) 
  \label{eq:Fline}\\
  \approx&& 2\Delta\Omega\,\Delta\nu
     \left(B_\nu[T_{\rm gas}(\tau_{\rm line}\!=\!1)]
          -B_\nu[T_{\rm dust}(\tau_{\rm cont}\!=\!1)]\right) 
  \label{eq:Flapprox} \,
\end{eqnarray}
where $\Delta\Omega$ is the solid angle occupied by the part of the
disk that is optically thick in the line, and $2\Delta\nu\!=\rm\!FWHM$
is the observed frequency full width half maximum. The first
approximation (Eq.\,\ref{eq:Fline}) is valid only if the excitation
conditions ($S_{\!\nu}^{\rm line}$, $S_{\!\nu}^{\rm cont}$) are about constant
in the line forming region, otherwise the integration over the solid
angle cannot be carried out this way. The second approximation
(Eq.\,\ref{eq:Flapprox}) is valid for LTE only.
Figure~\ref{fig:COvibUnderstand} shows the typical situation
encountered in disks where we look through a warm gas toward the cold,
optically thick midplane.

\section{Numerical convergence}
\label{app:NumConv}

\begin{figure}
\centering
\hspace*{-3mm}\includegraphics[width=96mm]{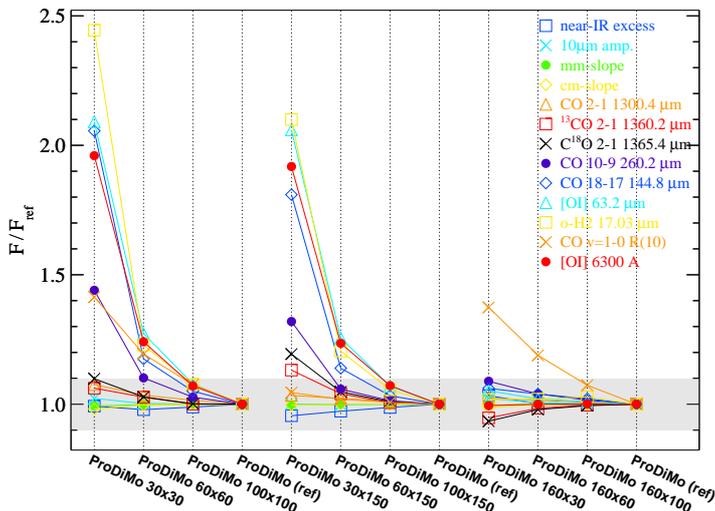}\\[-2mm]
  \caption{Various results of pure-ProDiMo models as function of
    spatial grid resolution. All results $F$ are plotted with respect
    to the results of the reference model $F_{\rm ref}$, which has
    $160\times150$ radial and vertical grid points, respectively. The
    quantities annotated with spectral lines are emission line
    fluxes.}
  \label{fig:NumConverge1}
\end{figure}

Figure~\ref{fig:NumConverge1} shows three series of {\sl ProDiMo} models
with increasing spatial resolution in the underlying numerical
grid. The resulting continuum predictions like near-IR excess,
10\,$\mu$m amplitude, millimetre and centimetre slopes etc.\ (see
Sect.~\ref{app:integrated}) are robust. Even quick $30\times30$
models are sufficient to predict the SED and derived quantities, with
an accuracy better than 5\% with respect to the results from the big
$160\times150$ reference model.

The grid resolution is more critical, however, when studying 
emission lines. A too coarse spatial grid usually
leads to an over-prediction of the emission line fluxes. Most
critical are lines which originate in a small portion of the disk
volume, like the weak o-H$_2$ and high-$J$ CO lines, but also
[OI]\,63.2\,$\mu$m and [OI]\,6300\,\AA. Here, the radial grid 
resolution is more important than the vertical grid resolution.
The only counter-example to this rule are the ro-vibrational CO lines
which are mostly emitted directly from the surface of the inner
rim in this model. Here, the vertical grid resolution is more important.
In summary, we need a grid resolution of about $100\times100$ to
achieve an accuracy better than 10\% for all predictions, see
grey shaded box in Figure~\ref{fig:NumConverge1}.

\begin{figure}[!t]
\centering
\hspace*{-3mm}\includegraphics[width=96mm]{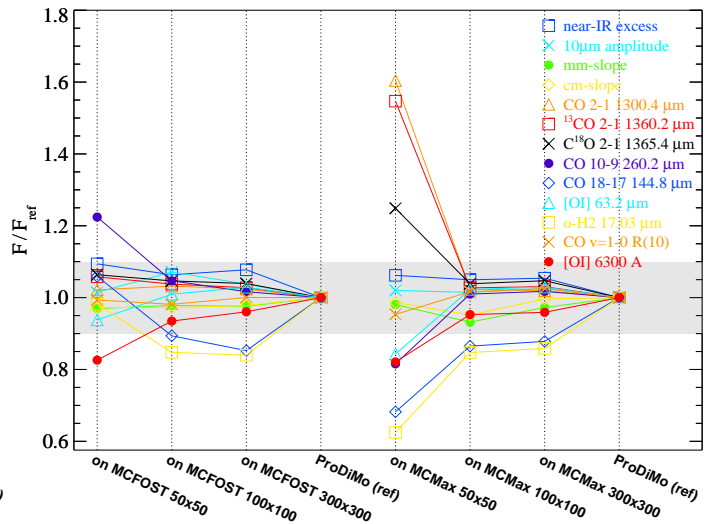}\\[-1mm]
  \caption{Various results of MC$\,\to\,$ProDiMo chain models as
    function of the spatial grid resolution in the Monte-Carlo (MC)
    programs {\sl MCFOST} and {\sl MCMax}. The continuum results are directly
    computed from the MC model output files. The line results are
    obtained by passing the MC model results (densities, opacities,
    $\Td$, radiation field, etc.) to a high-resolution
    ($160\times150$) {\sl ProDiMo} model. For further explanations, see
    caption of Fig.~\ref{fig:NumConverge1}.}
  \label{fig:NumConverge2}
\end{figure}

Figure~\ref{fig:NumConverge2} shows the numerical convergence of
MC$\,\to\,$ProDiMo chain models, when varying the spatial grid
resolution in the Monte-Carlo (MC) programs.  It is reassuring to see
that the MC$\,\to\,$ProDiMo chain models actually produce results that are
very similar as compared to the pure {\sl ProDiMo} models. For sufficient
spatial resolution in the MC models (again about $100\times100$ grid
points), the deviations in continuum results are smaller than 10\%,
and line results agree better than 15\%, where most critical are the
faint mid-far IR o-H$_2$ and high-$J$ CO lines. Other, \eg (sub-)mm
line results are more robust.  There is also no obvious asymmetry in
Fig.~\ref{fig:NumConverge2}, \eg some lines are weaker whereas others
are stronger when using the MC$\,\to\,$ProDiMo chain models.

For sufficient spatial resolution in the MC models, the {\sl MCFOST} and
{\sl MCMax} results show a similar pattern with respect to the
pure {\sl ProDiMo} reference results, \ie the deviations between
MCFOST$\,\to\,$ProDiMo and MCMax$\,\to\,$ProDiMo models are
actually smaller than the deviations between those models and pure
{\sl ProDiMo} models.

\section{Impacts of additional model parameters}
\label{app:figures}



\begin{figure*}
  \vspace*{-1mm}
  \hspace*{2mm}\resizebox{185mm}{!}{
  \begin{tabular}{lll}
  \hspace*{-7.0mm}\includegraphics[height=46mm,trim= 0 40 0 0,clip]{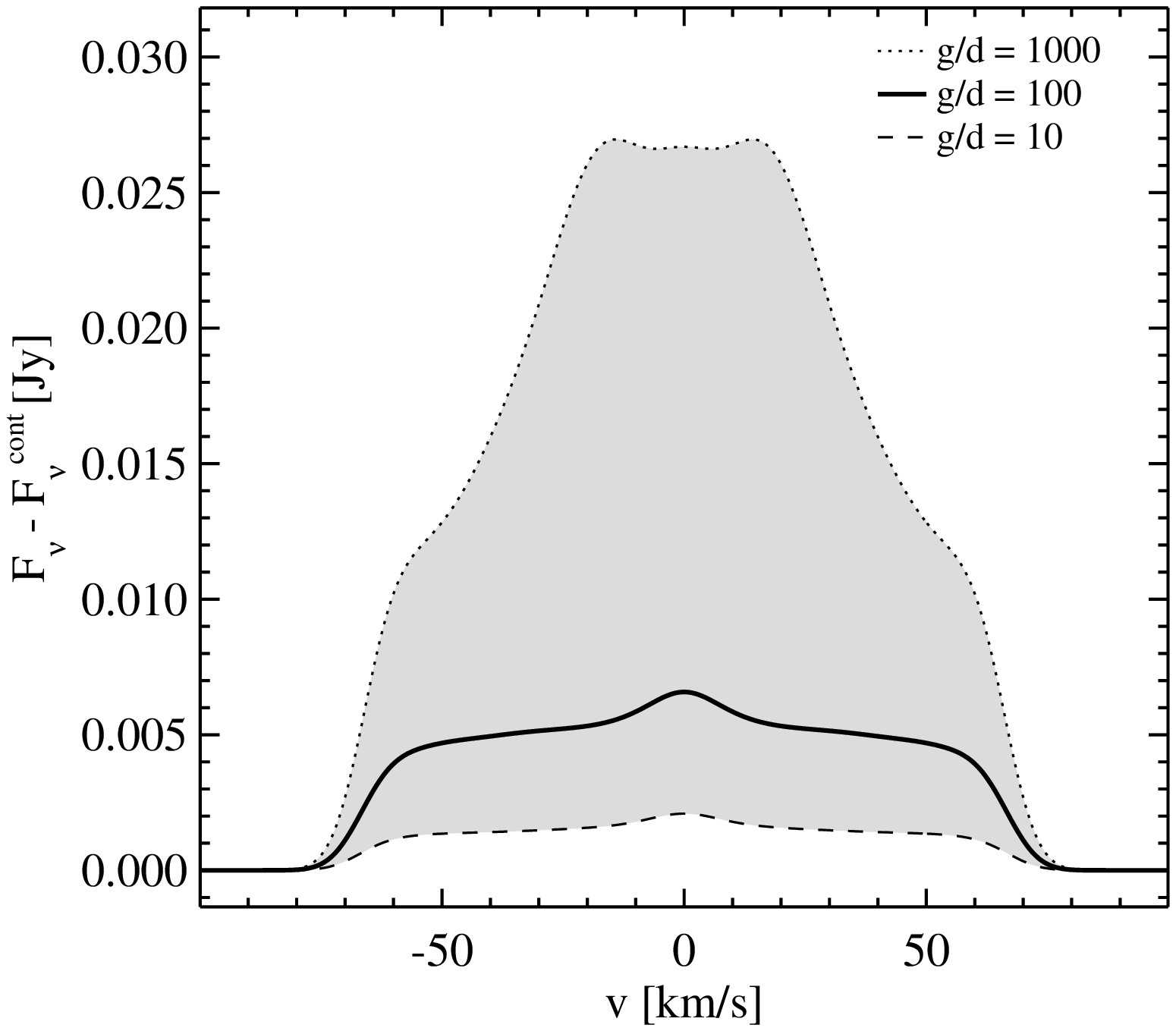}&
  \hspace*{-5.5mm}\includegraphics[height=46mm,trim=23 40 0 0,clip]{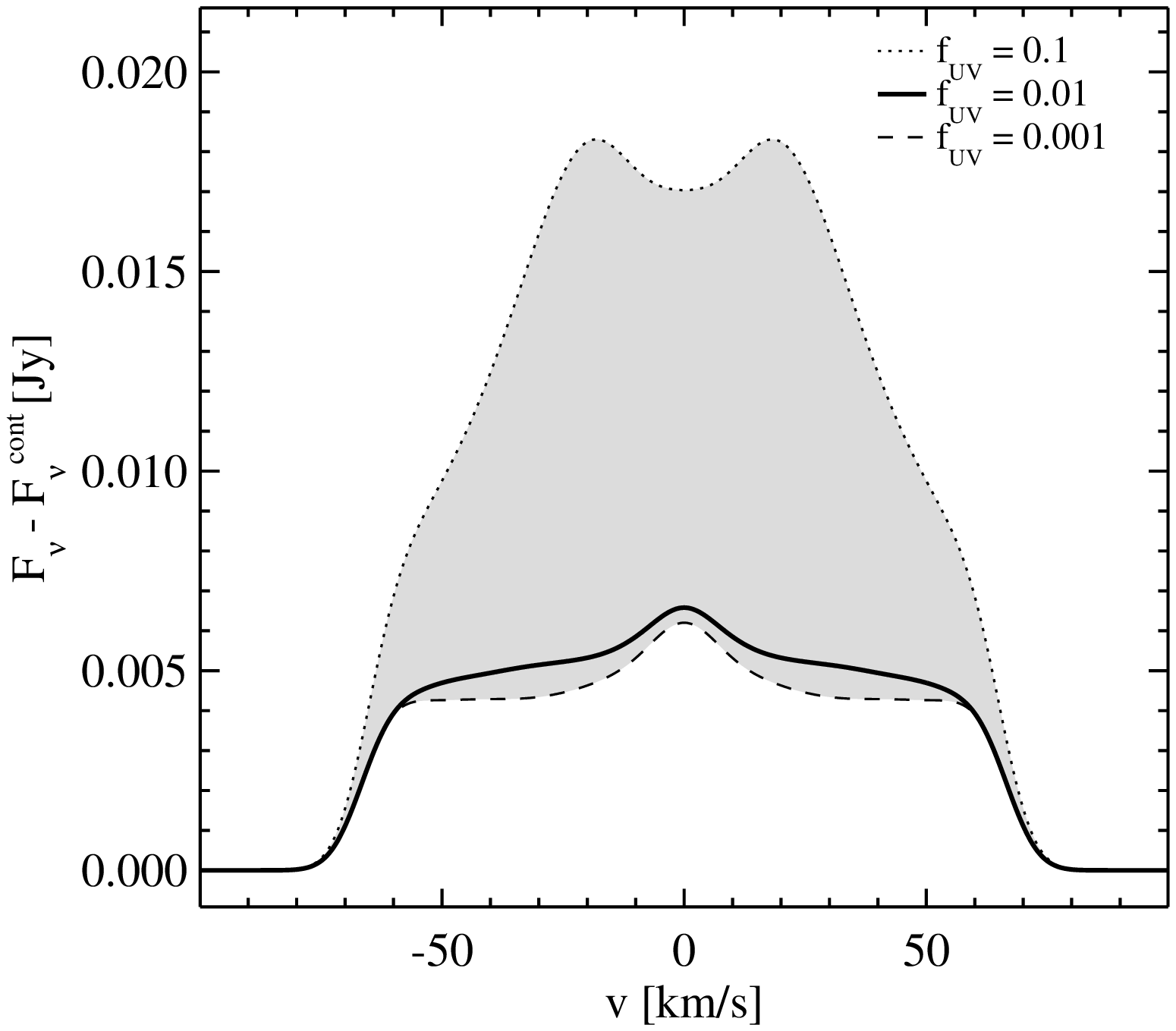}&
  \hspace*{-5.5mm}\includegraphics[height=46mm,trim=23 40 0 0,clip]{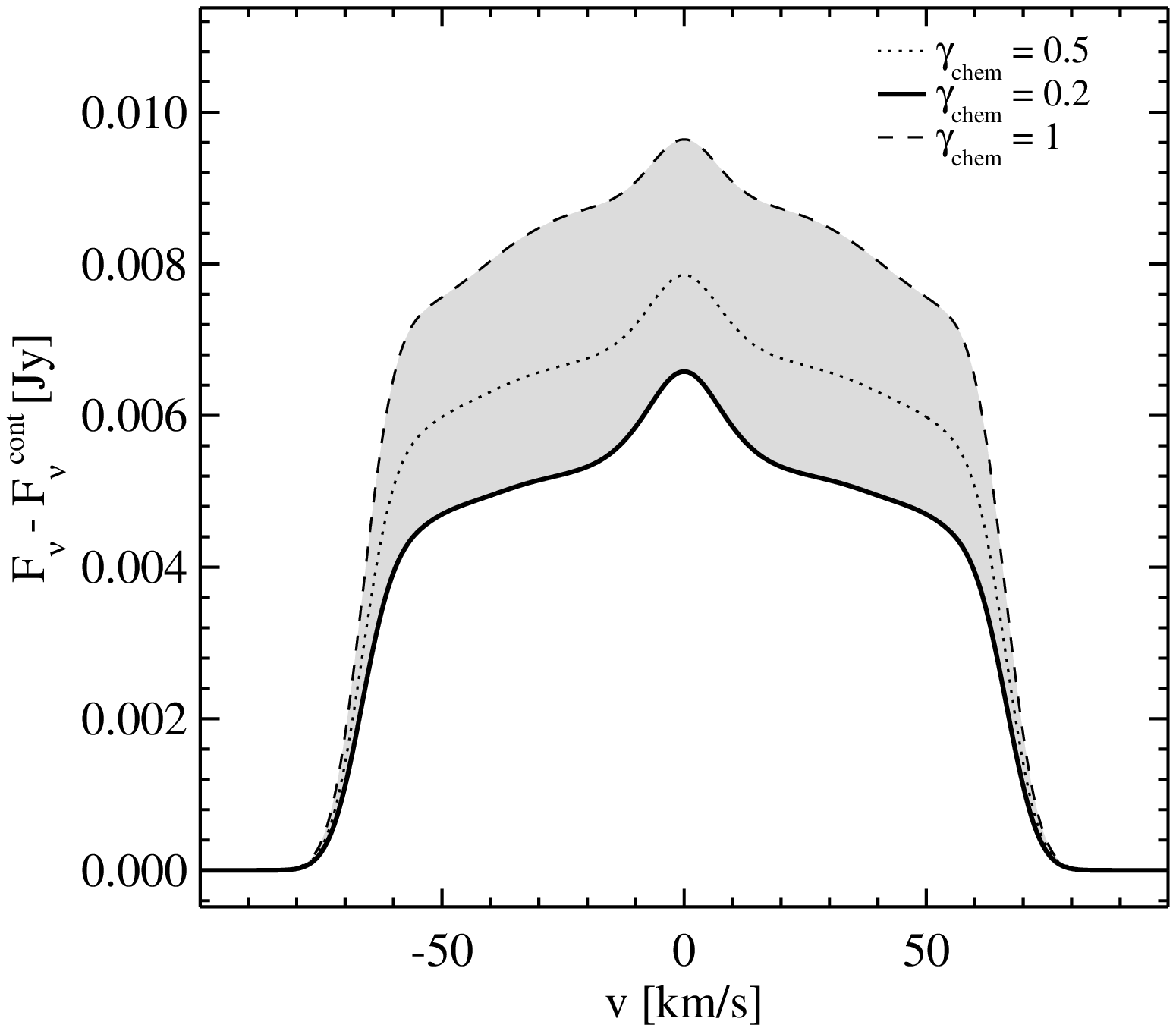} 
  \\[-45mm]
  \hspace*{5mm}\bf gas/dust ratio     & \hspace*{3.5mm}\bf stellar UV excess
                                      & \hspace*{3.5mm}\bf chem.\,heat.\,efficiency\\[38.5mm]  
  \hspace*{-7.0mm}\includegraphics[height=46mm,trim= 0 40 0 0,clip]{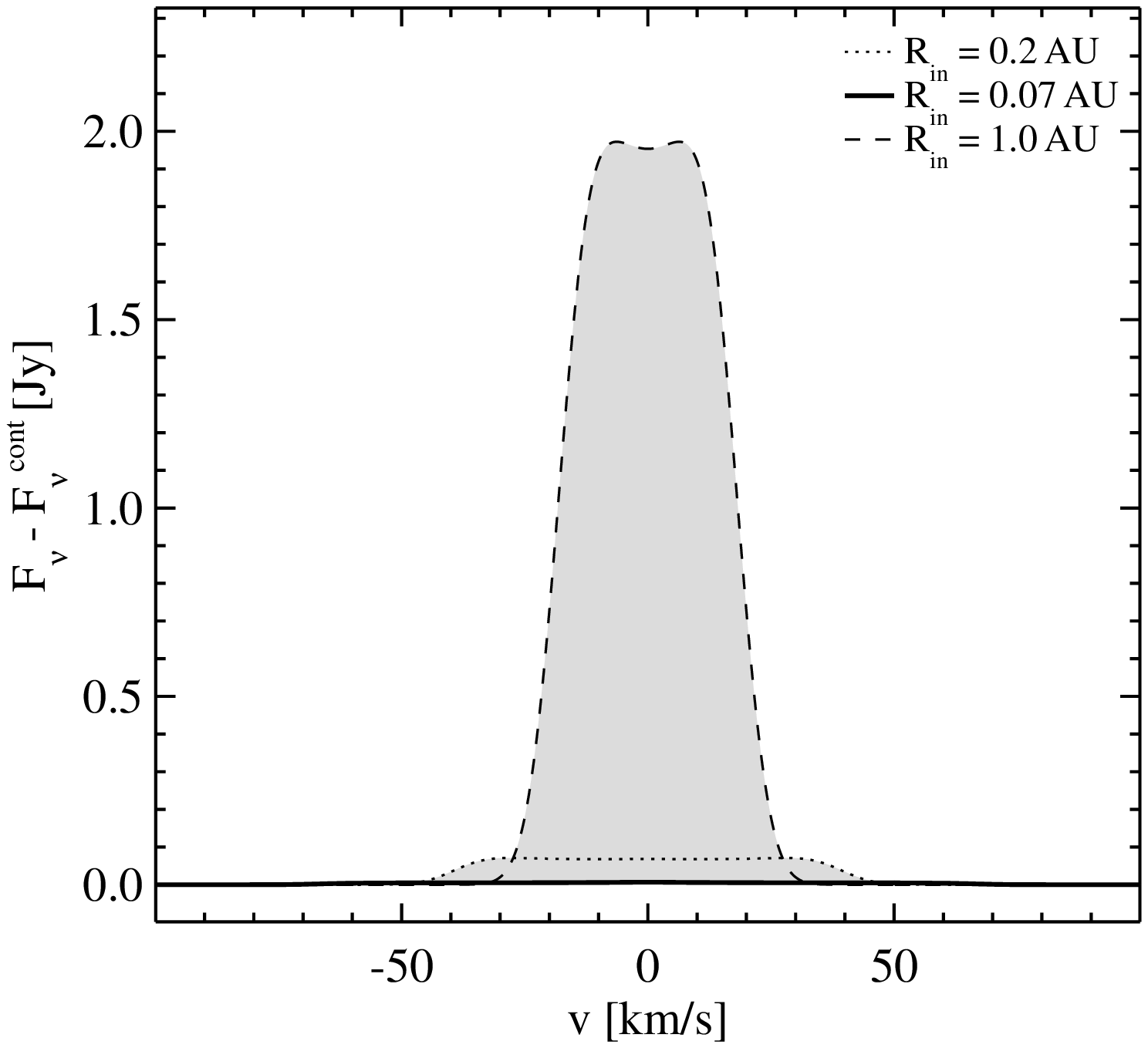}&
  \hspace*{-5.5mm}\includegraphics[height=46mm,trim=23 40 0 0,clip]{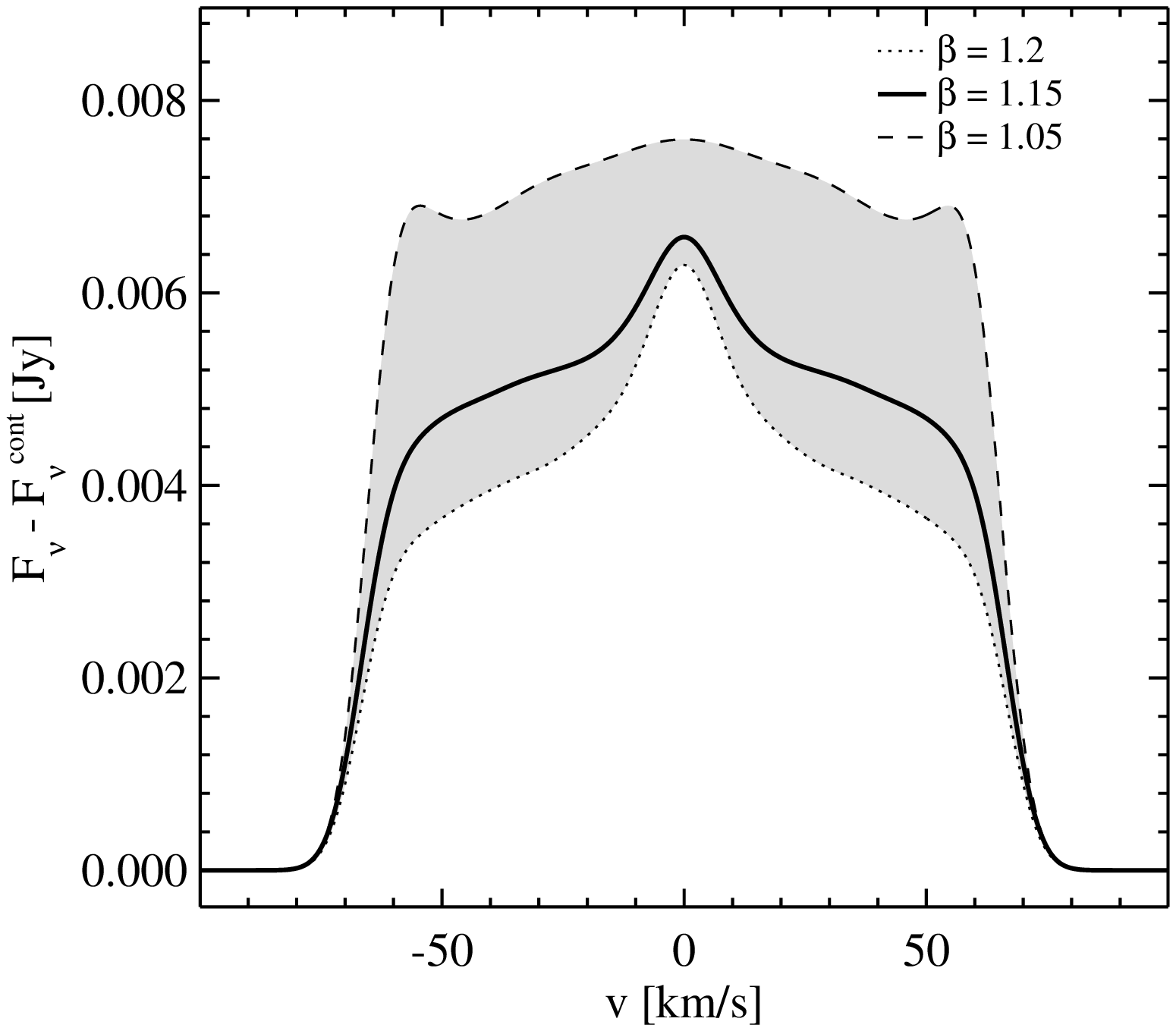}&
  \hspace*{-5.5mm}\includegraphics[height=46mm,trim=23 40 0 0,clip]{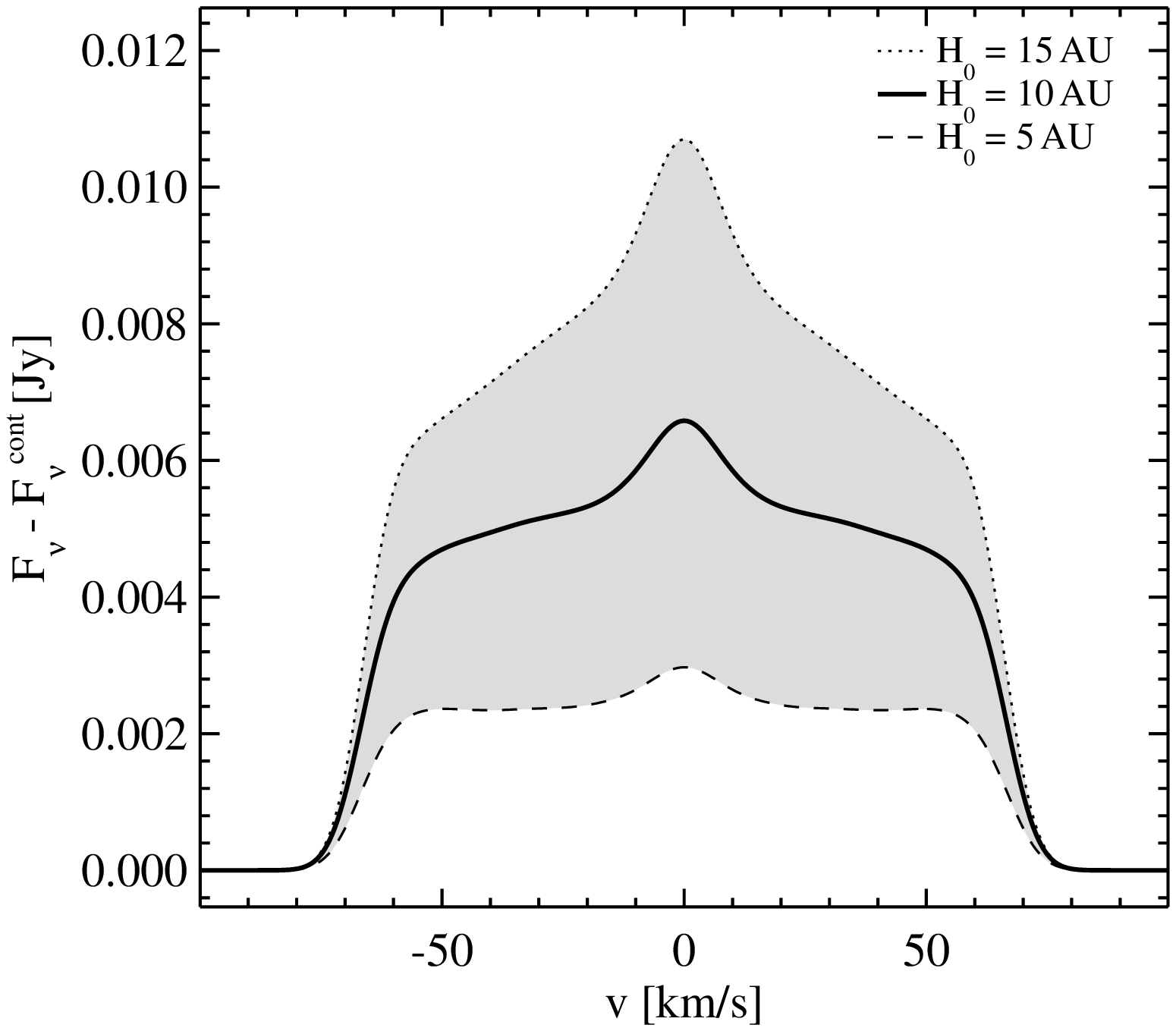} 
  \\[-45mm]
  \hspace*{5mm}\bf inner radius       & \hspace*{3.5mm}\bf flaring index
                                      & \hspace*{3.5mm}\bf scale height\\[38.6mm]  
  \hspace*{-7.0mm}\includegraphics[height=51.17mm,trim= 0 0 0 0,clip]{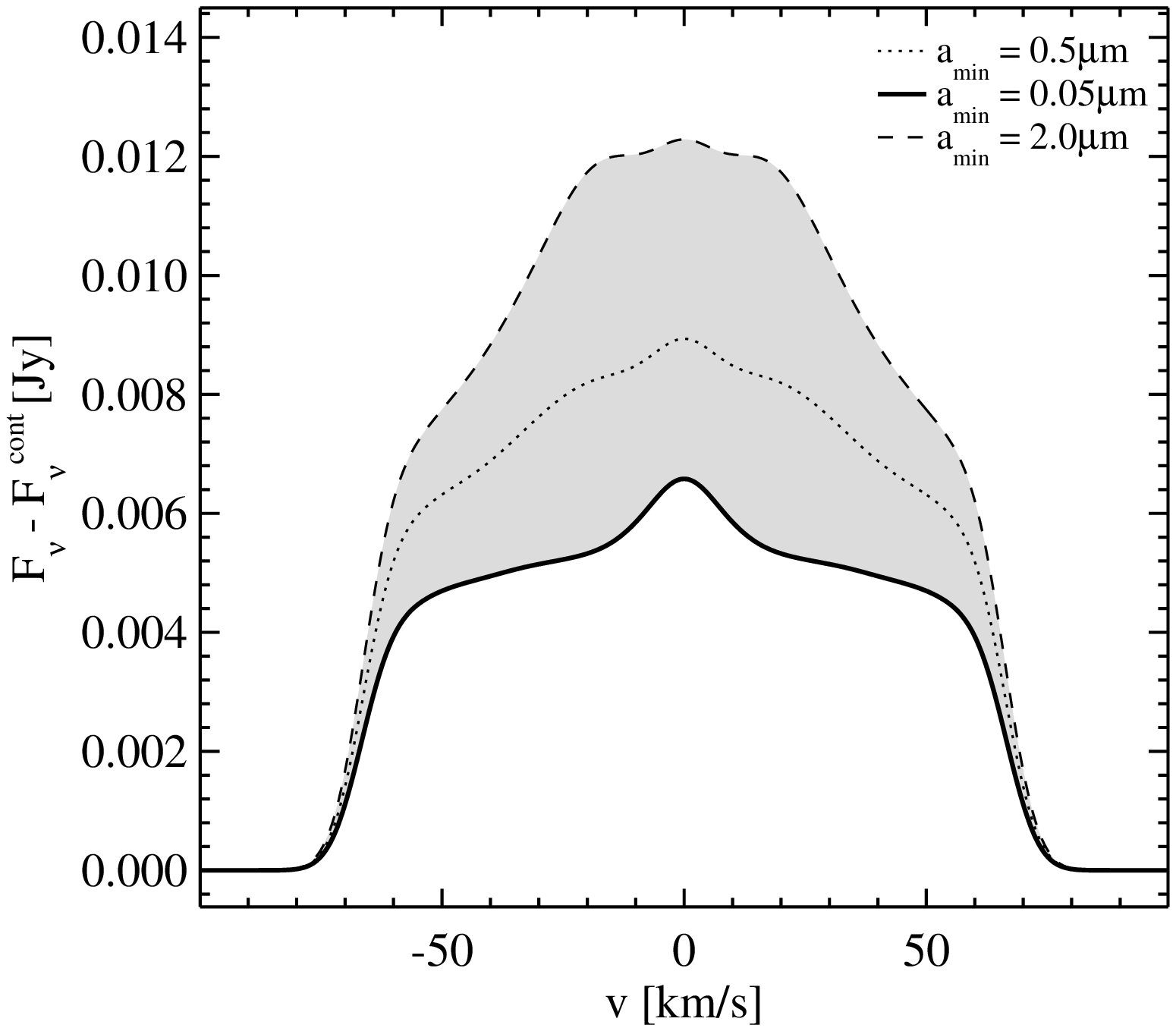}&
  \hspace*{-5.5mm}\includegraphics[height=51.17mm,trim=23 0 0 0,clip]{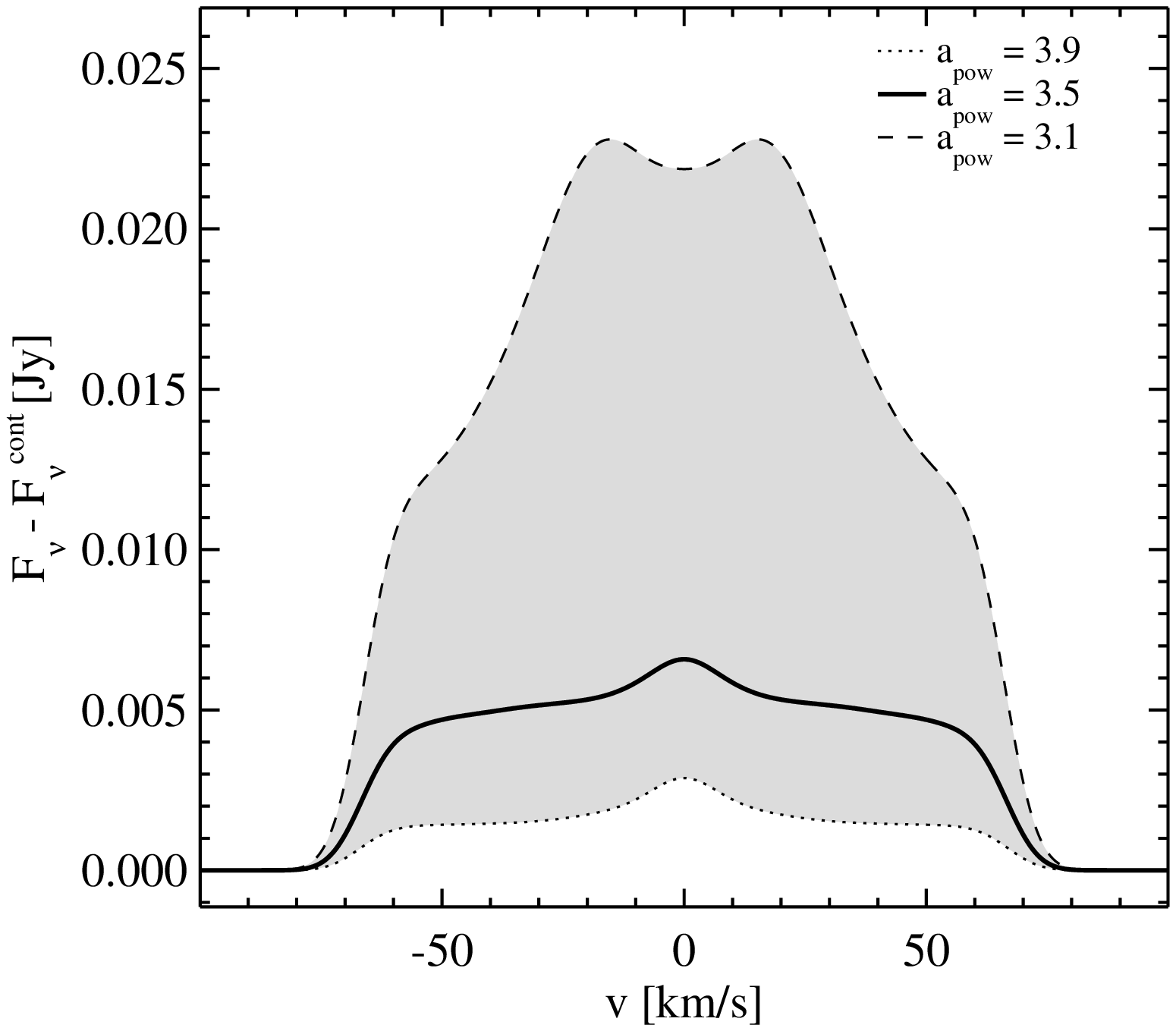}&
  \hspace*{-5.5mm}\includegraphics[height=51.17mm,trim=23 0 0 0,clip]{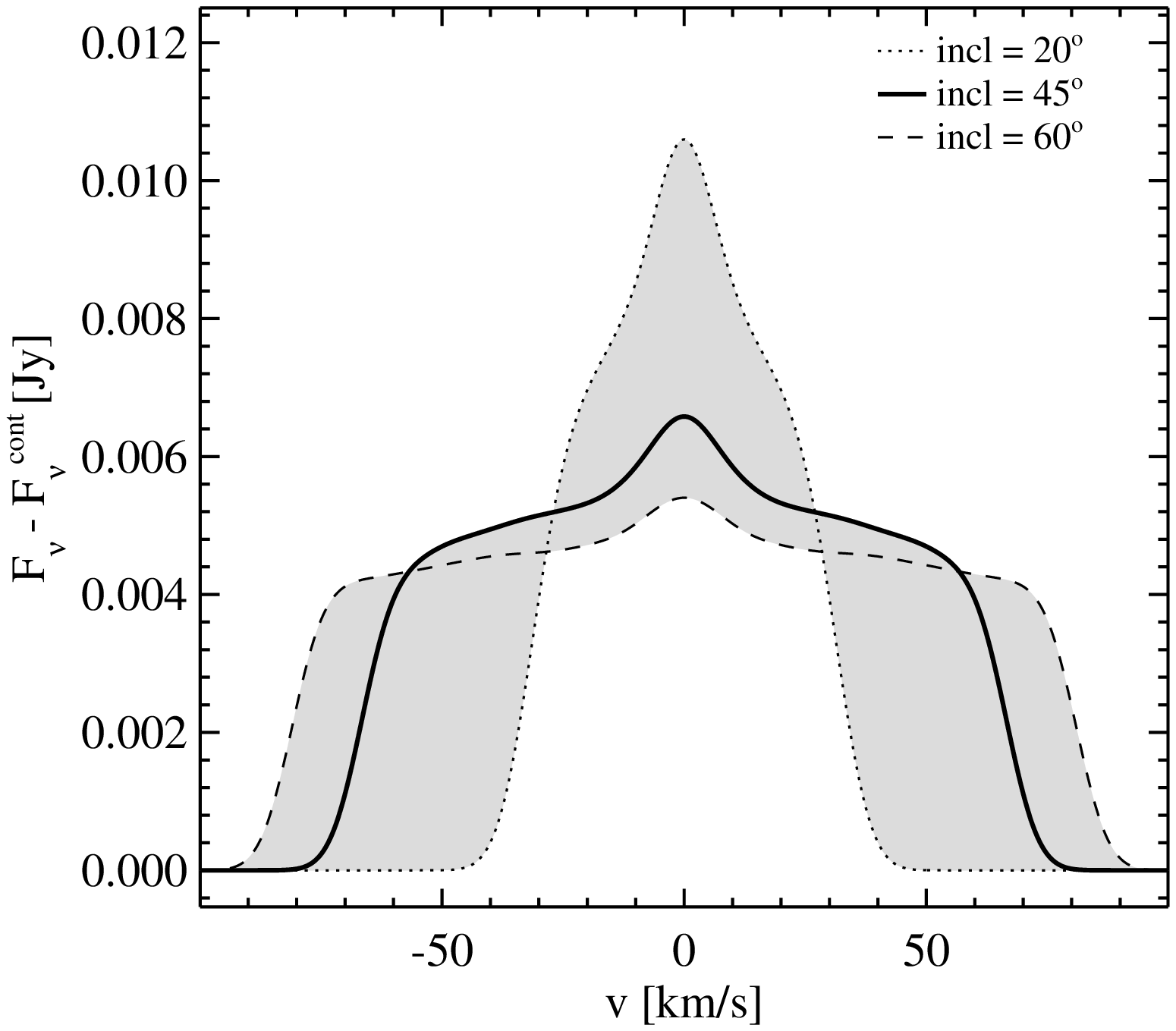} 
  \\[-50.15mm]
  \hspace*{5mm}\bf minimum dust size & 
  \hspace*{3mm}\resizebox{!}{1.9mm}{\bf dust size powerlaw index}
                                     & \hspace*{3.5mm}\bf disk inclination\\[38.5mm]  
  \end{tabular}}
  \vspace*{7mm}
  \caption{Effects of selected stellar, gas, dust and disk shape
    parameters on CO fundamental emission line profiles. Each part
    figure shows mean line profiles averaged over all computed
    CO $\upsilon\!=\!1\!\to\!0$ $R$-branch and $P$-branch emission lines,
    continuum subtracted and convolved with a 12\,km/s Gaussian
    (resolution $R\!\approx\!25000$).  The thick full lines show the
    reference model, identical in every part figure.  The shaded
    areas indicate the changes caused by single parameter variations,
    where the dashed and dotted lines correspond to the
    changed parameter values as annotated.  Non-depicted parameters
    have less influence on the CO fundamental emission, for example
    the X-ray luminosity $L_X$, compare
    Fig.~\ref{fig:COrovib_effect1}.}
  \label{fig:COrovib_effect2}
  \vspace*{-2mm}
\end{figure*}

\medskip\noindent
Figure~\ref{fig:COrovib_effect2} shows the impact of various model
parameters on the mean CO fundamental line emission strengths and
profiles. The results are discussed in Sect.~\ref{sec:COrovib}.

\begin{figure*}
  \vspace*{-2mm}
  \includegraphics[width=180mm]{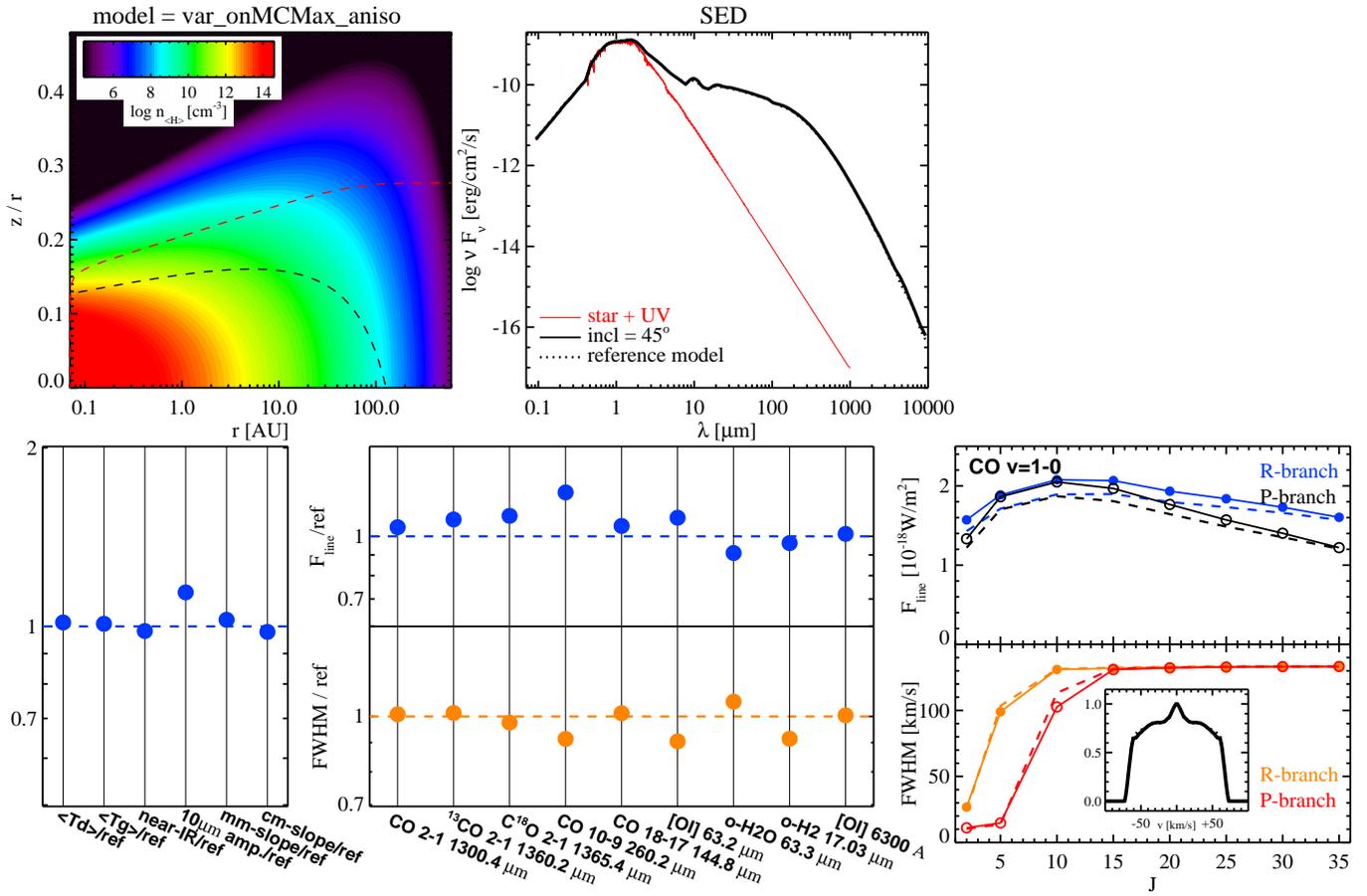}\\
  \vspace*{-6mm}
  \caption{Comparison of results between models using isotropic and
    anisotropic scattering. Both models are MCMax$\,\to\,$ProDiMo
    chain models. The results from the anisotropic model are shown
    with respect to the results obtained from the isotropic (reference)
    model. Depicted quantities are explained in
    Fig~\ref{fig:refmodel}. The visibility plot is omitted here.}
  \label{fig:aniso}
\end{figure*}

\medskip\noindent
Figure~\ref{fig:aniso} compares the results obtained from a
model using anisotropic scattering to the reference model.
The largest difference concerns the amplitude of the 10\,$\mu$m
silicate emission feature, the anisotropic model has a stronger
amplitude by about 14\%, making the feature clearly more visible in
the SED plot as compared to the reference model.

Concerning the gas emission lines, we see only little effects, and no
clear trend. The disk is mostly UV illuminated from the top, which
requires at least one scattering event in the borderline optically
thin disk surface layers. Since an angle-dependent treatment of dust
scattering favours forward scattering, one would expect the UV
illumination of the disk from above to be reduced, because $\approx
90^{\rm o}$ scattering events seem to be required.  However, this
does not seem to be entirely true. The question whether a scattered UV
photon reaches the disk or not is simply determined by whether the
photon is scattered upwards or downwards, which is a 50\%/50\% chance
even for anisotropic scattering.
With anisotropic scattering, we rather redistribute the entry points
where the scattered photons enter the disk, favouring the outer disk
regions, and their initial entry angle. Multiple scattering also
reduces the effects.
 
We measure line flux and FWHM differences of order 10\%, with no clear
trend, the only exception being CO $J\!=\!10\!\to\!9$ with is actually
enhanced by 28\%. These results are close to the ``noise level''
expected from various numerical effects in the models, compare
Figs.~\ref{fig:NumConverge1} and \ref{fig:NumConverge2}.

\end{document}